\documentclass[11pt,a4paper]{article}            
               
\usepackage[skins,theorems]{tcolorbox}
\tcbset{highlight math style={enhanced,
  colframe=red,colback=white,arc=0pt,boxrule=1pt}}
\usepackage[bookmarksopen, bookmarksnumbered, bookmarksopenlevel=2]{hyperref}
\usepackage{tikz}
\usepackage{tikz-3dplot}
\usetikzlibrary{calc}
\usetikzlibrary{decorations} %
\usepackage[UKenglish]{babel}
\usepackage[toc,page]{appendix}
\usepackage{amsmath}
\usepackage{amssymb}
\usepackage{graphicx}
\usepackage{hhline}
\usepackage[bf]{caption}
\usepackage{cite}
\usepackage[vcentermath]{youngtab}
\usepackage{geometry}
\usepackage{slashed}
\usepackage{color}
\usepackage{stackrel}
\usepackage{tikz}
\usepackage{tikz-cd} 
\usepackage{empheq}

\newcommand{\bqa}{\begin{eqnarray}}
\newcommand{\eqa}{\end{eqnarray}}

\usepackage{multirow}

\newcommand{\Ub}{{b}}

\def\unit{{1\kern-.65ex {\rm l}}}
\def\1{{1\kern-.65ex {\rm l}}}

\def\IZ{\mathbb{Z}}

\hypersetup{
    pdftitle={},
    pdfauthor={},
    pdfsubject={}
}
\numberwithin{equation}{section}
\numberwithin{table}{section}

\makeatletter

\newtheorem{definition}{Definition}

\newcommand{\be}{\begin{equation}}
\newcommand{\ee}{\end{equation}}
\newcommand{\ba}{\begin{aligned}}
\newcommand{\ea}{\end{aligned}}

\newcommand{\bea}{\begin{eqnarray}}
\newcommand{\eea}{\end{eqnarray}}

\newcommand{\fg}{\mathfrak{g}}

\newcommand*\widefbox[1]{\fbox{\hspace{2em}#1\hspace{2em}}}

\def\unit{{1\kern-.65ex {\rm l}}}
\def\1{{1\kern-.65ex {\rm l}}}

\newcount\hour \newcount\minute
\hour=\time \divide \hour by 60
\minute=\time
\def\now{%
\ifnum \hour<13
  \ifnum \hour=0 \advance \hour by 12 \number\hour:\else \number\hour:\fi%
     \ifnum \minute<10 0\fi%
     \number\minute%
\ A.M.%
\else \advance \hour by -12 \number\hour:%
  \ifnum \minute<10 0\fi%
  \number\minute%
  \ P.M.%
\fi%
}

\makeatother

\begin{document}

\baselineskip=14pt
\parskip 5pt plus 1pt

\vspace*{-1.5cm}
\begin{flushright}    
  {\small
CERN-TH-2018-126  
  }
\end{flushright}

\vspace{2cm}
\begin{center}        
  {\LARGE TASI Lectures on F-theory} \vspace{2mm}\\
  \end{center}

\vspace{0.75cm}
\begin{center}        
Timo Weigand$^{1,2}$
\end{center}

\vspace{0.15cm}
\begin{center}        
$^1$\emph{CERN, Theory Division, \\
             CH-1211 Geneva 23, Switzerland}
           \vspace{2mm}\\
$^2$\emph{Institut f\"ur Theoretische Physik, Ruprecht-Karls-Universit\"at, 
\\       Philosophenweg 19, 69120,
Heidelberg, Germany   \\[0.15cm] 
           }
          
\end{center}
\vspace{2cm}


\begin{abstract}
\noindent 

F-theory is perhaps the most general currently available approach to study non-perturbative string compactifications in their geometric, large radius regime.
It opens up a wide and ever-growing range of applications and connections to string model building, quantum gravity, (non-perturbative) quantum field theories in various dimensions and mathematics.
Its computational power derives from the geometrisation of physical reasoning, establishing a  deep correspondence between fundamental concepts in gauge theory and beautiful structures of elliptic fibrations.
These lecture notes, which are an extended version of my lectures given at TASI 2017, introduce some of the main concepts underlying the recent technical advances in F-theory compactifications and their various applications. 
The main focus is put on explaining the F-theory dictionary between the local and global data of an elliptic fibration and the physics of 7-branes in Type IIB compactifications to various dimensions via duality with M-theory.
The geometric concepts underlying this dictionary include the behaviour of elliptic fibrations in codimension one, two, three and four, the Mordell-Weil group of rational sections, and the Deligne cohomology group specifying gauge backgrounds.

\end{abstract}

\thispagestyle{empty}
\clearpage
\tableofcontents
\thispagestyle{empty}


\newpage
\setcounter{page}{1}

\section{Introduction}

F-theory \cite{Vafa:1996xn,Morrison:1996na,Morrison:1996pp} occupies a  remarkable sweetspot in the landscape of geometric, large radius compactifications of string theory: It is general enough to incorporate the regime non-perturbative in the string coupling $g_s$ while at the same time it is sufficiently well-controlled for reliable computations to be performed. 
As a result, F-theory offers a number of fascinating connections between physics in various dimensions and mathematics, most notably algebraic (and even arithmetic) geometry. 
Recent years have seen significant progress in our understanding of this correspondence between geometry and physics, along with numerous applications ranging from particle physics model building to quantum gravity to non-perturbative quantum field theory and back to mathematics. 
It is the purpose of these lectures to give a pedagogical introduction to some of the techniques to describe compactifications of F-theory and its manifold applications.

From its outset, F-theory can be understood as a supersymmetric compactification of Type IIB string theory with 7-branes. The backreaction of the 7-branes generates a holomorphically varying profile of the axio-dilaton $\tau = C_0 + i/g_s$ on the compactification space. Since the imaginary part of $\tau$ is the inverse string coupling, the resulting compactifications inevitably  include regions that are inherently strongly coupled. The variation of the string coupling can be made sense of thanks to the non-perturbative $SL(2,\mathbb Z)$ invariance of Type IIB string theory \cite{Schwarz:1995dk}. This formulation naturally  leads to the structure of an elliptic fibration over the compactification space and hence makes contact with most beautiful concepts in algebraic geometry. 
Duality with M-theory  \cite{Vafa:1996xn,Witten:1996bn} is a crucial tool to analyze  the resulting compactification at a quantitative level much in the spirit of the geometric engineering programme of Type II string theory.
The dual M-theory probes an a priori singular geometry, which is identified with the same elliptic fibration over the physical compactification space of Type IIB string theory.
Wrapped M2-branes along vanishing cycles in the geometry engineer massless matter states which give rise to non-abelian degrees of freedom. These complete the spectrum obtained from the supergravity modes in the long wavelength limit. 
In special cases, the F-theory compactification enjoys yet another duality with the heterotic string \cite{Vafa:1996xn,Morrison:1996na,Morrison:1996pp}. Indeed, it is often quoted that F-theory combines two attractive properties of both Type II compactifications with branes and compactifications of the heterotic string: From the first the hierarchy of localisation of gauge degrees of freedom along the branes compared to gravitational physics in the bulk of spacetime, and from the second the appearance of exceptional gauge symmetry. This has been exploited heavily amongst other things in the context of string model building \cite{Donagi:2008ca,Beasley:2008dc, Beasley:2008kw,Donagi:2008kj}, and was one of the motivations for the revived interest in F-theory in the past decade which has lead to significant, ongoing  activities with numerous far-reaching insights.

It is probably fair to say that its -  in many ways ideal - location at the intersection of various dual M-theory corners makes F-theory the most generic currently controllable framework for studying string vacua in their geometric regime.
At the same time, it must be kept in mind that F-theory is by definition most powerful in the long wavelength limit; here truly stringy effects, which appear at higher order in an expansion of $\ell_s/R$ (with $R$ a typical radius and $\ell_s$ the string length), are suppressed and other techniques are required to analyze this parameter regime. In this sense, F-theory addresses string vacua in their supergravity limit, and furthermore in their geometric regime. Interestingly enough, though, such compactifications can encapsulate properties of non-geometric string vacua by duality with the heterotic string \cite{Malmendier:2014uka,Font:2016odl,Garcia-Etxebarria:2016ibz,Font:2017cya}.

Among the many fascinating aspects of F-theory is the emergence of a clear dictionary, summarized in Table \ref{tab-dictionary}, between fundamental concepts in theoretical physics  and 
beautiful structures in algebraic geometry. Further developing this dictionary has been a source of continuous inspiration both for physics and mathematics:
Many challenging considerations about the physics  of string compactifications can be translated into purely geometric questions, which are, in favorable circumstances, solvable!
An example of this is the classification of six-dimensional superconformal field theories in F-theory via the classification of F-theory base spaces with contractible curves \cite{Morrison:2012np,Heckman:2013pva,Heckman:2015bfa,Heckman:2018jxk}.
A long-term goal is to put such potential classification schemes to direct use for physics in lower dimensions. This is particularly exciting when it comes to distinguishing between the landscape of string theory compactifications and the potential swampland of seemingly consistent field theories without a UV completion coupled to quantum gravity, as discussed at TASI in particular in \cite{Taylor:2011wt,Brennan:2017rbf}.

The dictionary of Table \ref{tab-dictionary} gives a direct meaning to sophisticated concepts in algebraic and arithmetic geometry and can therefore also act as a source of inspiration for physical mathematicians. As an example of this reverse use of the dictionary, let us note that studying anomalies can lead to interesting new insights on the algebraic geometry of F-theory and reveal new geometric identities which may be difficult to  prove in generality in pure mathematics \cite{Grassi:2000we,Grassi:2011hq,Bies:2017abs,Grassi:2018rva}.

\begin{table}
\begin{tabular}{| l || ll |}
\hline
Physics of effective theory in $\mathbb R^{1,9-2n}$ & Geometry of elliptic fibration $Y_{n+1}$& \\ \hline \hline
non-abelian gauge algebra    & codim.-one singular fibers & (sec. \ref{sec_singcodim1}) \\ \hline
localised charged matter representation     & codim.-two singular fibers  & \multirow{2}{*}  {(sec. \ref{sec_Codim-two})}  \\
localised uncharged matter     & $\mathbb Q$-factorial terminal singularities in codim. two   &  \\ \hline 
triple Yukawa interactions (4d/2d) & codim.-three singular fibers&  \multirow{2}{*}  {(sec. \ref{sec_Codimhigher})  } \\
quartic Yukawa interactions (2d) & codim.-four singular fibers & \\ \hline
abelian gauge algebra & free part of Mordell-Weil group&  \multirow{2}{*}  {(sec. \ref{sec_MWGROUP})}    \\
global structure of gauge group & torsional part of Mordell-Weil group& \\ \hline
 \multirow{2}{*}  {discrete abelian gauge group } & torsional cohomology/ & \multirow{2}{*}  {(sec. \ref{sec_genusone})} \\
   & Tate-Shafarevich group &\\ \hline
gauge background & Deligne cohomology group & \multirow{2}{*}  {(sec. \ref{sec_Gaugebackgrounds})}\\ 
matter multiplicities & sheaf cohomology groups& \\ \hline \hline
\end{tabular}
\caption{Dictionary between the physics of F-theory compactified on an elliptic fibration $Y_{n+1}$ of complex dimension $n+1$ and the geometry of the fibration $Y_{n+1}$. \label{tab-dictionary}}
\end{table}

The purpose of these lecture notes is to flesh out the dictionary between geometry and physics via F-theory in an introductory, pedagogical manner that can serve as an entry point to this active and ever-growing field of research. 
Given the maturity of the field on the one hand, which has seen more than 20 years of constant progress, and the limitation of space on the other, such an introduction is necessarily far from a complete account of the material worth covering and hence reflects a certain choice of topics rather than the actual state of the literature.
We begin by introducing F-theory both from the perspective of Type IIB string compactifications with general $[p,q]$ 7-branes and via the duality to M-theory in section \ref{sec_introducingF}.
In either of the two approaches, the occurrence of an elliptic fibration encoding the physics of the compactification is very natural, even though it appears for completely different reasons.
The mother of all F-theory compactifications is the Weierstrass model, which is further introduced in section \ref{secFonsmooth}, focussing, for starters, on the smooth case and its interpretation as a compactification with 7-branes carrying a  trivial gauge algebra. 
We then move on to discussing, in section \ref{sec_singcodim1}, the origin of non-abelian gauge algebras in F-theory due to singular fibers over codimension-one loci in the base of the elliptic fibration. An important role is played by the resolution of the singularities, which is always possible crepantly in codimension one. This process is interpreted in the dual M-theory as moving along the Coulomb branch of the gauge theory. Non-abelian gauge bosons and matter are most directly understood via wrapped M2-branes in M-theory.
This is also key to understanding the appearance of localised matter at the intersection of two 7-brane stacks: The weight lattice of representations is in one-to-one correspondence with the fibral curves of the resolution including the codimension-two singular fibers, see section \ref{sec_Codim-two}. However, beginning with codimension two, the singularities of the Weierstrass model are not always resolvable without breaking the Calabi-Yau condition of the elliptic fibration, leading to interesting new effects.
Localised charged matter interacts in a holomorphic way via Yukawa couplings, whose structure can be read off from the higher codimension fibers as reviewed in section \ref{sec_Codimhigher}.
The structure discussed up to this points has been mostly local, in the sense that it is associated with the different strata of the discriminant locus of the fibration in codimension one, two and higher. By contrast, understanding both the abelian sector of a compactification and the structure of its gauge group (as opposed to merely the algebra) requires global data. The data in question is furnished by the Mordell-Weil group of rational sections, which is the topic of section \ref{sec_MWGROUP}. This group is a freely generated abelian group; its non-torsional part encodes the abelian gauge symmetries, and together with its torsional part it determines in addition the global structure of the gauge group. 
Discrete gauge symmetries can likewise be traced back to geometric structure in F-theory: As we will explain in section \ref{sec_genusone}, an abelian $\mathbb Z_k$ gauge symmetry arises if the Weierstrass model has a suitable torsional cohomology group ${\rm Tor}(H^3(Y, \mathbb Z)) = \mathbb Z_k$; such a Weierstrass model is necessarily singular, and related to $k-1$ smooth genus-one fibrations without a section which together form a $\mathbb Z_k$ Tate-Shafarevich group. 
In section  \ref{sec_Gaugebackgrounds}, we discuss aspects of the important topic of gauge backgrounds in F-theory. These backgrounds are necessary to stabilize complex structure moduli and crucially determine the spectrum of massless matter in compactificaitons to four and two dimensions.  We conclude in section \ref{sec_applications} with a brief overview of some of the applications of F-theory in the more recent literature to particle physics model building, string landscape analysis, non-perturbative quantum field theory in various dimensions and to mathematics.

\section{Setting the stage for F-theory} \label{sec_introducingF}

There are three ways to approach F-theory \cite{Vafa:1996xn,Morrison:1996na,Morrison:1996pp}. 
By definition, F-theory is a non-perturbative formulation of Type IIB compactifications with general $[p,q]$ 7-branes backreacting on the geometry.
This theory can be described, in full generality, via duality with M-theory \cite{Vafa:1996xn,Witten:1996bn}.
Furthermore, a subset of F-theory compactifications are dual to a special class of heterotic compactifications \cite{Vafa:1996xn,Morrison:1996na,Morrison:1996pp}.\footnote{F-theory on an elliptic fibration $Y_{n+1}$ which is also a K3-fibration over a base $B_{n-1}$ is dual to the heterotic string on an elliptic fibration over $B_{n-1}$ with a suitable vector bundle. For reasons of space (and time) we will, unfortunately, not be discussing this duality further in the following lecture notes.}

This introductory chapter  approaches F-theory from the perspective of Type IIB compactifications and from the M-theory viewpoint. 
This foundational material has been explained in many reviews and books in the literature, including \cite{Lerche:1999de,Johnson:2003gi,Denef:2008wq,Weigand:2010wm,Taylor:2011wt,Blumenhagen:2013fgp}.

\subsection{Type IIB string theory, $SL(2,\mathbb Z)$ duality and  $[p,q]$ 7-branes}   \label{sec_pq }   

Consider type IIB string theory, approximated in the long wavelength regime by the ten-dimensional (10d) Type IIB supergravity action with bosonic part
\be
\label{IIB10d-democratic}
\ba
\frac{1}{2\pi }S_{\rm IIB, dem.} &=&  \int d^{10}x\ e^{-2 \phi} \sqrt{-g} \left(R + 4 \partial_\mu \phi \, \partial^\mu \phi \right) -\frac{1}{2} \int e^{-2 \phi}  H_3 \wedge \ast H_3   \cr
&& - \frac{1}{4}   \sum_{p=0}^4  \int   F_{2p+1}  \wedge \ast F_{2p+1} - \frac{1}{2} \int C_4 \wedge H_3 \wedge F_3   \,.
\ea
\ee
We set the string length $\ell_s = 2 \pi \sqrt{\alpha'} \equiv 1$.
The field strengths appearing in the above democratic formulation are defined as 
\be
\ba
& H_3 = dB_2, \quad F_1 = dC_0, \quad F_3 = dC_2 - C_0 \, dB_2, \cr 
& F_5 = dC_4 - \frac12 C_2 \wedge dB_2 + \frac12 B_2 \wedge dC_2 \,,
& F_9 = \ast F_1,  \qquad F_7 = - \ast F_3 \,.
\ea
\ee
This action is a pseudo-action to the extent that it must be supplemented by the duality relation
\be
F_5 = \ast F_5,
\ee
at the level of equations of motion. 

Of prime interest for us are D7-branes, which are magnetic sources for the IIB Ramond-Ramond (RR) axion $C_0$.
The effective action of a D7-brane is the sum of the usual Dirac-Born-Infeld action and the Chern-Simons action controlling in particular its coupling to the RR fields $C_{2p}$.
In the above conventions\footnote{The factor of $\frac{1}{2}$ in front of (\ref{CScouplingsFtheory}) is needed because we are working in the democratic formulation where the kinetic terms for the RR fields  are normalised as in (\ref{IIB10d-democratic}). The overall minus sign is chosen such that the 7-brane couples to the complex field $\tau= C_0 + i  e^{-\phi}$ rather than to $-C_0 + i  e^{-\phi}$. The CS-action of a D3-brane has a relative overall minus sign with respect to that of the D7-brane.},
\be \label{CScouplingsFtheory}
\ba
S_{\rm CS} = -\frac{2 \pi}{2} \int_{D7} {\rm Tr} \, e^{i\cal F}  \, \sum_{2p} C_{2p} \, \sqrt{  \frac{ \hat A({\rm T}D7)}{\hat A({\rm N}D7)}}  \,
\ea
\ee
with ${\cal F}   =  i(\mathbf{ F}+2\pi  B_2 )$ the gauge invariant  combination of the Yang-Mills field strength $F$ along the brane and the Kalb-Ramond 2-form $B_2$ (pulled back to the brane). The terms in brackets refer to the A-roof genus of the tangent and normal bundle.

The most important aspect of this action for us is that a D7-brane is an electric source for the IIB RR 8-form $C_8$ and hence a magnetic source for its magnetic dual, the axion $C_0$. Together with the dilaton $\phi$, $C_0$ appears in the complex axio-dilaton field
\be
\tau= C_0 + i e^{-\phi} \,, \qquad \quad g_s = e^{\phi} \,.
\ee
With the above conventions, the Bianchi identity for $F_9$ in the presence of a D7-brane along $\mathbb R^{1,7} \subset \mathbb R^{1,9} \simeq \mathbb R^{1,7} \times \mathbb C$ implies 
\bea
\int_{S^1}\ast F_9 = \int_{S^1} d C_0 = 1 \,.
\eea
Here $S^1$ is a circle around the 7-brane in the normal space $\mathbb C$. The location of the D7-brane in the normal $\mathbb C$-plane with complex coordinate $z$ be at $z = z_0$.
As a non-trivial input to determine the behaviour of $\tau(z)$ in the presence of such a magnetic source, we need use the fact that a D7-brane in flat space preserves sixteen supercharges, and that supersymmetry requires $C_0$ to appear holomorphically in the complex field $\tau$ such that $\bar\partial \tau = 0$ away from the source \cite{Greene:1989ya}. This determines
\bea \label{tauprofile}
\tau(z) = \frac{1}{2 \pi i} {\rm ln} (z - z_0) + (\rm{terms \, \, regular \, \, at \, \, } z_0) \,.
\eea
The logarithmic branch-cut induces a monodromy 
\be \label{taumono}
\tau \rightarrow \tau + 1
\ee
 as we encircle $z_0$.

To understand the meaning of the monodromic behaviour, we need to recall that the Type IIB supergravity action is invariant under an $SL(2,\mathbb R)$ duality group, whose $SL(2,\mathbb Z)$  subgroup is conjectured to be preserved in the full non-perturbative Type IIB string theory. This duality group is manifest if we write the Type IIB effective action in a slightly different form, abandoning the democratic formulation (\ref{IIB10d-democratic}) with both electric fields and their magnetic duals appearing in favour of the (Einstein frame) action
\be
\label{IIB10d-nondemo}
\ba
\frac{1}{2\pi}S_{\rm IIB} &=&  \int d^{10}x  \sqrt{-g} \left( R  - \frac{\partial_\mu \tau \partial^\mu \bar \tau}{2 ({\rm Im} \tau)^2}  - \frac{1}{2}   \frac{|G_3|^2}{{\rm Im}\tau }- \frac{1}{4} |F_5|^2  \right) 
 + \frac{1}{4i}  \int \frac{1}{{\rm Im}\tau} C_4 + G_3 \wedge \bar G_3   
\ea
\ee
with $G_3 = dC_2 - \tau d B_2$ and $|F_p|^2 = \frac{1}{p!} F_{\mu_1 \ldots \mu_p} F^{\mu_1 \ldots \mu_p} $.
This action is indeed invariant under an $SL(2,\mathbb Z)$ transformation \cite{Schwarz:1995dk}  \footnote{The action (\ref{IIB10d-nondemo}) is even invariant under a corresponding $SL(2,\mathbb R)$ transformation, and the breaking to $SL(2,\mathbb Z)$ is due to $D(-1)$-instanton effects, which involve terms of the form $e^{2 \pi i \tau}$.}
 \bea 
\label{SL2Zform}
\tau &\rightarrow& \frac{a \tau + b}{ c \tau + d}\,,
\qquad \begin{pmatrix}C_2\\ B_2 \end{pmatrix} \rightarrow  M \begin{pmatrix}C_2\\ B_2 \end{pmatrix}, 
\qquad M = \begin{pmatrix} a & b \\ c & d \end{pmatrix} \in SL(2, \mathbb Z) \\
C_4 &\rightarrow& C_4, \qquad \quad \quad   g_{\mu \nu} \rightarrow g_{\mu \nu} \,.
\eea

The monodromy (\ref{taumono}) in the presence of one D7-brane can therefore be made sense of by interpreting it as an
 ${ SL}(2, \mathbb Z)$ monodromy with
 \be
 M_{[1,0]} =   \begin{pmatrix}  1 & 1 \\ 0 & 1 \end{pmatrix} \,.
 \ee
 Consistency requires that this monodromy acts not only on $\tau$, but also on $B_2$ and $C_2$ as dictated by (\ref{SL2Zform}).
  
There are different types of 7-branes which induce a more general $SL(2,\mathbb Z)$ monodromy on the background.
Let us call the D7-brane with Chern-Simons couplings (\ref{CScouplingsFtheory}) a $[1,0]$ brane.
By definition, it is the object on which a fundamental string, called $(1,0)$ string, ends.
In this notation, a D1-brane is a $(0,1)$ string, and 
a $(p,q)$  string is a BPS bound state of $p$ fundamental strings and $q$ D1-strings, which exists as a supersymmetric bound state for $p$ and $q$ coprime \cite{Witten:1995im}. 
A $(p,q)$ string therefore couples to the combination $p B_2 + q C_2$.
If we assemble $C_2$ and $B_2$ into a row vector $\Phi^a$ as in (\ref{IIB10d-nondemo}), then it is natural to associate to a $(p,q)$
string a column vector $Q_a = (q,p)$ such that the coupling is described by $Q_a \Phi^a = \epsilon_{ab} Q^b \Phi^a$. Here  we have used the $SL(2, \mathbb Z)$ invariant 2-tensor $\epsilon_{ab}$ (with $\epsilon_{12} = -1 = - \epsilon_{21}$) to raise and lower $SL(2,\mathbb Z)$ indices.
This coupling is manifestly $SL(2,\mathbb Z)$ invariant if the objects $Q^a$ and $\Phi^a$ (with indices up) both transform as $SL(2,\mathbb Z)$ vectors by multiplication from the left.
To summarize, a $(p,q)$ string is  associated with the $SL(2,\mathbb Z)$ charge vector
\bea \label{chargedefi}
Q_a = (q,p) = \epsilon_{ab} Q^b, \qquad Q^b =  \begin{pmatrix} p\\-q \end{pmatrix}  \,.
\eea


A $(p,q)$  string ends, by definition, on a $[p,q]$ 7-brane.
The monodromy induced by a more general $[p,q]$ 7-brane can be derived by noting that a $(p,q)$   string can be reached from a $(1,0)$ 
 string by acting on the corresponding $SL(2,\mathbb Z)$ doublet vector $Q^a$ with an $SL(2,\mathbb Z)$ matrix  $g_{[p,q]}$,
 \be
 \begin{pmatrix}p\\  - q \end{pmatrix} = g_{[p,q]} \begin{pmatrix}1\\ 0 \end{pmatrix} \,\qquad 
 g_{[p,q]}  =  \begin{pmatrix}  p & r \\  - q & s \end{pmatrix} \,.
 \ee
  Here $r, s$ are not determined uniquely by the requirement that  $g_{[p,q]} \in SL(2,\mathbb Z)$  \cite{Douglas:1996du}, but this ambiguity drops out of all physical quantities. The $SL(2,\mathbb Z)$ monodromy induced by a general $[p,q]$ 7-brane is then given by
\bea \label{MpqfromM10}
M_{[p,q]} & = & g_{[p,q]} M_{[1,0]} g^{-1}_{[p,q]}  = \begin{pmatrix}  1 + p q  & p^2 \\ -q^2 &  1 - p q  \end{pmatrix}    \,.
\eea
Consistently, a $(p,q)$ string is invariant under the monodromy induced by a $[p,q]$ 7-brane. 
It is important to keep in mind, when one compares with results in the literature, that the specific form of the monodromy matrices depends on  the chosen  definition of the charge vectors, which in our case is as in (\ref{chargedefi}).

Note that every $[p,q]$ 7-brane can be transformed into a $[1,0]$ 7-brane by the  $SL(2,\mathbb Z)$ transformation inverse to (\ref{MpqfromM10}).
In this sense, {\it locally} every single 7-brane can be thought of as a D7-brane, but two 7-branes of different $[p,q]$ type can in general not be {\it simultaneously} transformed into a [1,0] brane.
If a $(p,q)$ string undergoes a non-trivial monodromy around a $[p',q']$ 7-brane, the $[p,q]$ and $[p',q']$ brane are said to be mutually non-local. This is the case if and only if their monodromy matrices do not commute.
Two such mutually non-local 7-branes in flat space can in general not be brought on top of each other in a supersymmetric way, but certain bound states exist and realise in particular the simply laced ADE Lie groups in flat space.
A basis of 7-branes sufficient to generate all ADE groups in eight dimensions (i.e. along the 7-branes) can be taken to be  \cite{Sen:1996vd} (modulo some arbitrariness due to an overall $SL(2,\mathbb Z)$ transformation and changes in the conventions)
\bea
A: [1,0] \,, \qquad B: [3,1] \,, \qquad C:[1,1] \,,
\eea
whose monodromies can be read off from (\ref{MpqfromM10}). 
With this notation, the ADE groups are obtained from 7-brane stacks of the following types  \cite{Gaberdiel:1997ud}:
\be
SU(N): A^N \, \qquad SO(2N): A^N B C \,, \qquad E_k:  A^{k-1} B C^2 \,, \quad k = 6,7,8 \,. 
\ee
The monodromies are obtained by multiplying the monodromies of the individual 7-branes in the given order. As will be elaborated on more below, this identifies a $BC$ bound state, whose monodromy is 
\be
M_{BC} = M_{[3,1]}  M_{[1,1]} =   \begin{pmatrix}   -1   & 4  \\ 0 &  -1  \end{pmatrix}  \,,
\ee
 with the perturbative O7-plane \cite{Sen:1996vd}.\footnote{More precisely, this object is the O7$_+$-plane; for an F-theory interpretation of other types of O7-planes see \cite{Bhardwaj:2018jgp} and references therein.} The exceptional brane stacks were first discussed in \cite{Johansen:1996am}. Much more information on the rules which allow us to combine mutually non-local 7-branes and the nature of $(p,q)$ strings furnishing the adjoint representation of the associated Lie algebras can be found in \cite{Gaberdiel:1997ud}.\footnote{Our conventions differ slightly from \cite{Gaberdiel:1997ud}, and agree with those used in section 18.6 of \cite{Blumenhagen:2013fgp}, to which we also refer for more details.}

Out of the many interesting properties of these $[p,q]$ brane systems let us merely stress the following:
The monodromy around an $A^4 BC$ brane stack, corresponding to 4 D7-branes on top of an O7-plane, is easily seen to be $M ={ \rm diag}(-1,-1)$.
This generates a $\mathbb Z_2$ subgroup of $SL(2,\mathbb Z)$  which reverses the orientation of a $(1,0)$ string and correspondingly sends $(C_2,B_2)^T \rightarrow (-C_2,-B_2)^T $. This is nothing but the $\mathbb Z_2$ involution of a perturbative Type IIB orientifold. The charge and tension of the 4 D7 (i.e. $A$-type) branes is locally cancelled by the negative charge and tension of the O7-plane system corresponding to the $BC$ stack. Therefore $C_0$ does not shift, and this action leaves the axio-dilaton $\tau$ invariant. 
On the other hand, the monodromy matrices induced by brane stacks of the form $E_6$, $E_7$, $E_8$ give rise to $SL(2,\mathbb Z)$ monodromies of the following form:
\bea \label{So8Ek}  
&SO(8):   \quad M =   \begin{pmatrix}   -1   & 0  \\ 0 &  -1  \end{pmatrix} \qquad & \langle M \rangle = \mathbb Z_2, \qquad \tau_0 =  {\rm arbitrary} \\
&E_6:   \quad M =   \begin{pmatrix}   -1   & -1  \\ 1 &  0  \end{pmatrix} \qquad & \langle M \rangle = \mathbb Z_3, \qquad \tau_0 =  e^{ \pi i/3} \\
&E_7:   \quad M =   \begin{pmatrix}   0   & -1  \\ 1 &  0  \end{pmatrix} \qquad & \langle M \rangle =\mathbb Z_4, \qquad \tau_0 = e^{2 \pi i/4} = i \\
&E_8:  \quad M =  \begin{pmatrix}   1   & -1  \\ 1 &  0  \end{pmatrix} \qquad & \langle M \rangle =\mathbb Z_6, \qquad \tau_0=e^{ \pi i/3} \label{E8mono} \,.
\eea
Here $\langle M \rangle$ denotes the finite order $SL(2,\mathbb Z)$ subgroup generated by $M$ and $\tau_0$ is the value of the axio-dilaton fixed under the associated $SL(2,\mathbb Z)$ transformation (\ref{SL2Zform}). 
The theory around an exceptional brane stack can hence {\it locally} be described as  a non-perturbative generalisation of an orientifold \cite{Dasgupta:1996ij}. The gauge coupling is in the truly non-perturbative regime where $|\tau| = 1$; from a Type IIB perspective, all $D(-1)$ instanton effects contribute without any suppression.

\subsection{Elliptic curves, $SL(2,\mathbb Z)$ bundle and  elliptic fibrations}  \label{sec_Frompqtoell}

The most important conclusion from the previous section is that due to the backreaction the axio-dilaton varies in the directions normal to a 7-brane as in the local expression (\ref{tauprofile}). This forces us to study compactifications with non-trivial field profiles, which is the hallmark of F-theory \cite{Vafa:1996xn}.\footnote{Models with constant $\tau$ are very special, but they do exist. For instance, there exist globally consistent compactifications to eight dimensions  involving  only brane stacks of the type (\ref{So8Ek}). These are globally of  (non-perturbative) orientifold form and  $\tau$ is constant \cite{Dasgupta:1996ij}.} As we will now review, the variation of the axio-dilaton gives rise, in a most natural way, to an elliptic fibration over the physical spacetime.

Let us structure the discussion in two parts. First we will give a geometric interpretation to the supergravity field $\tau$ as the complex structure of an elliptic curve. Second, we will turn in more detail to the variation of $\tau$ over spacetime to pass from elliptic curves to elliptic fibrations.

\subsubsection*{Elliptic curves and $SL(2,\mathbb Z)$}

Formally, the transformation (\ref{SL2Zform}) of the Type IIB field $\tau$ under an $SL(2,\mathbb Z)$ duality transformation is identical to the 
behaviour of the complex structure of an elliptic curve  $\mathbb E_\tau$ under a modular transformation. At this stage, this is merely a mathematical analogy, but we will see in section \ref{sec_Mtheorytoell} that this identification is deeply rooted in duality with M-theory \cite{Schwarz:1995dk}. 

An elliptic curve $\mathbb E_\tau$ with complex structure $\tau$ is a torus with a marked point  called the origin ${\cal O}$. It can be represented as the lattice $\mathbb C/\Lambda$, 
\bea
\mathbb E_\tau = \mathbb C/\Lambda = \{ w \in \mathbb C:  w \simeq w + (n + m \tau)\}  \,, \qquad n, m \in \mathbb Z, \qquad  \tau = \tau_1 + i \tau_2 \in \mathbb  H \,.
\eea
The origin ${\cal O}$ of $\mathbb E_\tau$ is identified with the point $w=0$, and the complex structure parameter $\tau$ takes values in the complex upper half-plane $\mathbb H$.
A transformation  $\tau \rightarrow \frac{a \tau + b}{ c \tau +d}$ with $ad-bc=1$ and $a,b,c,d \in \mathbb Z$ leaves the lattice $\Lambda$ and hence the shape of the torus invariant. This transformation of course defines an $SL(2,\mathbb Z)$ matrix $M$ as in (\ref{SL2Zform}). More information on this modular group action on the torus can be found e.g. in section 6.2 of \cite{Blumenhagen:2013fgp}. 
The message to take away from this is that we can think of the Type IIB supergravity field $\tau$ as representing the shape modulus of an elliptic curve $\mathbb E_\tau$, and identify an $SL(2,\mathbb Z)$ duality transformation (as far as its effect on the field $\tau$ is concerned) by a modular transformation of 
$\mathbb E_\tau$. 

There are many other ways to represent an elliptic curve, for instance as a hypersurface in a suitable complex space of complex dimension two or as a complete intersection within a higher-dimensional ambient space.
A representation which will play a particularly important role for reasons explained later is the so-called 
Weierstrass form. In Weierstrass form, the elliptic cure $\mathbb E_\tau = \mathbb C/\Lambda$ is described as the vanishing  locus of the polynomial
\bea \label{Weier-firsttime}
P_W := y^2 - (x^3 + f x  z^4 + g z^6) 
\eea
within $\mathbb P_{231}$. Here
$[x : y : z]$ are homogeneous coordinates of $ \mathbb P_{231}$, defined as the space  $\mathbb C^3 \setminus \{(0,0,0)\}$ modulo  the equivalence relation 
\bea \label{P231rescaling}
(x,y,z) \simeq (\lambda^2 x, \lambda^3 y, \lambda z), \qquad \lambda \in \mathbb C^* \,.
\eea
Many details on elliptic curves can be found in standard works in the mathematics literature, for instance in the review \cite{2009arXiv0907.0298S} and references therein. In the sequel we collect some facts of particular relevance for us without proof.

The map from $\mathbb C/\Lambda$ to the Weierstrass model is as follows:
\begin{itemize}
\item
There exists a unique meromorphic function $\wp(w;\tau)$ doubly periodic on $\mathbb C/\Lambda$ with double poles at the lattice points,
\be
\wp(w;\tau) := \frac{1}{w^2} + \sum_{(m,n) \in \mathbb Z \neq (0,0)}   \left( \frac{1}{(w + m + n \tau)^2)} - \frac{1}{(m + n \tau)^2}  \right) \,.
\ee
The fact that it is doubly period means that it has the property
\be
\wp(w;\tau) =  \wp(w + \tau;\tau) \,.
\ee
It satisfies the differential equation
\bea \label{pwid}
(\wp(w;\tau)')^2 = 4 \wp(w;\tau)^3 - g_2(\tau) \wp(w;\tau) - g_3(\tau)
\eea
where $g_2$ and $g_3$ are the Eisenstein series
\bea
g_2(\tau) &=& 60 \sum_{(m,n) \in \mathbb Z \neq (0,0)} (m + n \tau)^{-4} \\
g_3(\tau) &=& 140  \sum_{(m,n) \in \mathbb Z \neq (0,0)} (m + n \tau)^{-6} \,.
\eea
\item
We can now consider the map
\bea \label{CtoEtau}    \mathbb C &\rightarrow& \mathbb P_{231} \vspace{1mm}\cr
w &\mapsto& \begin{cases}  [4^{2/3} \wp(w; \tau) : 2 \wp(w;\tau)' : 1]    &w \notin \Lambda \\
                                                     [1 : 1 : 0]                                                    &   w \in \Lambda     \end{cases}
\eea

and define
\bea
f(\tau) : =  - 4^{1/3} g_2(\tau), \qquad g(\tau) : =  - 4 g_3(\tau) \,.
\eea
Then the identity (\ref{pwid}) translates into the Weierstrass equation
\bea
y^2 - (x^3 + f x z^4 + g z^6) = 0 \,.
\eea
\item
Conversely, given a Weierstrass model, we can deduce $\tau$ from $f$ and $g$ via the Jacobi $j$-function
\bea \label{jtau1}
j(\tau) = 4 \,  \frac{24^3 f(\tau)^3}{\Delta} \,, \qquad \quad \Delta = 4 f^3(\tau) + 27 g^2(\tau) \,.
\eea
This function is a bijection from the fundamental domain of ${SL}(2, \mathbb Z)$ to the upper half plane and enjoys the expansion
\bea \label{jtauexpansion}
j(\tau) = e^{-2 \pi i \tau} + 744 + 196884 e^{2 \pi i \tau} + \ldots \,.
\eea

Note that $\tau$ can also be expressed as the ratio
\bea \label{BAcycles}
\tau = \frac{\oint_B \Omega_1}{\oint_A \Omega_1} \,,
\eea
where 
\bea
\Omega_1 = \frac{d x}{y} = \frac{d x}{\sqrt{x^3 + f x z^4 + g z^6}}
\eea
represents the holomorphic 1-form (which is unique up to rescaling) on the elliptic curve and $A$ and $B$ denote a symplectic basis of 1-cycles on the elliptic curve.
These two one-cycles transform as an $SL(2, \mathbb Z)$ doublet, i.e. the transformation  $\tau \rightarrow \frac{a \tau + b}{ c \tau + d}$ corresponds to the transformation
\bea \label{BAmono}
\begin{pmatrix}B\\ A \end{pmatrix} =  \begin{pmatrix}  a & b \\ c & d \end{pmatrix}  \begin{pmatrix}B\\ A \end{pmatrix} \,.
\eea

\item
Importantly, under an ${\rm SL}(2, \mathbb Z)$ transformation $\tau \rightarrow \frac{a \tau + b}{ c \tau + d}$, the functions $f$ and $g$ transform as 
\be
\begin{split}
f &\rightarrow (c \tau + d)^4 f  \label{ftrafo}\cr
g &\rightarrow (c \tau + d)^6 g  \,.
\end{split}
\ee
\end{itemize}

To summarize this first part of the discussion, specifying the value of the axio-dilaton $\tau$ of a Type IIB background is equivalent to specifying the complex structure of an elliptic curve $\mathbb E_\tau$, which in turn is equivalent to specifying the complex parameters $f$ and $g$ of a Weierstrass model. 
If we consider a  background with varying axio-dilaton, this correspondence must be applied pointwise and hence defines a family of elliptic curves with varying complex structure. This is what will concern us next.

\subsubsection*{Varying $\tau$, $SL(2,\mathbb Z)$ bundle and elliptic fibrations}

Consider a Type IIB compactification on a spacetime of the form
\be
 {\cal M}^{1,9} = \mathbb R^{1, 9 - 2n} \times B_n \,,
\ee
where $B_n$ is a compact manifold of complex dimension $n$.
Our background includes spacetime filling 7-branes wrapping a suitable cycle $\Sigma_{n-1} \subset B_n$ of complex codimension one. The 7-brane  worldvolume therefore takes the form
\be
7-{\rm brane}: \qquad  \mathbb R^{1, 9 - 2n} \times \Sigma_{n-1} \,.
\ee
We are interested in compactifications which preserve the maximal possible amount of supersymmetry in the respective dimensions.
On general grounds this requires that  $B_n$ be a complex K\"ahler manifold and that 
the 7-brane cycle $\Sigma_{n-1}$ be a holomorphic cycle. Furthermore, as noted already, $\tau$ must vary in a holomorphic way, i.e. $\bar\partial \tau = 0$ away from the 7-brane sources.
The Einstein equations relate the curvature of $B_n$ to the variation of the dilaton $\phi$, which appears in the imaginary part of $\tau$, as
\bea \label{Einstein1}
R_{a \bar b} = \nabla_a \nabla_{\bar b}  \phi  \neq 0 \,.
\eea
The last inequality holds in the presence of 7-branes. In particular, the compactification space $B_n $ cannot be Calabi-Yau. This is a consequence of the gravitational backreaction of the 7-branes.

The holomorphic variation of $\tau$ defines a holomorphic line bundle ${\cal L} $ over $B_n$.
Let us sketch this construction, following the lucid presentation in  \cite{Bianchi:2011qh} (see also \cite{Douglas:2014ywa} for related aspects).
\begin{itemize}
\item Given the holomorphically varying field $\tau$ on $B_n$ we can define a 1-form 
\bea
A = \frac{i}{2} \frac{d(\tau + \bar \tau)}{\tau - \bar \tau } = \frac{i}{2} (\bar \partial \phi - \partial \phi) \,.
\eea
Since $\tau$ transforms under an $SL(2,\mathbb Z)$ transformation as in (\ref{SL2Zform}), also the 1-form $A$ transforms accordingly.
By explicit computation one verifies that $A$ transforms in the correct way such that one can identify it with the 
 connection of a complex line bundle $L$ over $B_n$ with transition function ${\rm exp}(i \, {\rm arg}(c \tau + d))$.
This means the following:
As we encircle a 7-brane in its normal space, the Type IIB supergravity fields transform in a manner dictated by the type of the 7-brane. As a result we cannot define the Type IIB  fields as global functions on $B_n$, but only as local functions. 
I.e. we can cover $B_n$ with open neighborhoods $U_\alpha$ such that the Type IIB supergravity fields in each $U_\alpha$  
 are in a certain $SL(2,\mathbb Z)$ frame (they are locally defined functions on $U_\alpha$). 
 The fields in different open patches differ by an $SL(2,\mathbb Z)$ transformation.
 To take this into account, in the overlap $U_\alpha \cap U_\beta$ we transform the fields from one frame to another by an $SL(2, \mathbb Z)$ transformation  (\ref{SL2Zform}) with matrix $M_{\alpha \beta}$, which is determined by the 7-brane background. 
 Let us parametrise this matrix as in (\ref{SL2Zform}) as
 \be \label{Malphabeta}
 M_{\alpha \beta} = \begin{pmatrix} a_{\alpha \beta}  &b_{\alpha \beta}    \\  c_{\alpha \beta} & d_{\alpha \beta}  \end{pmatrix} \,.
 \ee
 
On the other hand, a complex line bundle on $B_n$ is defined by specifying its complex transition functions for the transformation $\hat t_{\alpha \beta}$ on each overlap $U_{\alpha} \cap U_\beta$ such that a section $\hat h$ of the line bundle transforms as 
\be \label{hath}
\hat h|_{U_\alpha}  = \hat t_{\alpha \beta} \, \hat h|_{U_\beta} \,.
\ee
If we parametrise the $SL(2, \mathbb Z)$ transformation on the overlap $U_\alpha \cap U_\beta$ by the above matrix $M_{\alpha \beta}$,
then we can define a complex line bundle $L$ by setting  
\be \label{hatt}  
\hat t_{\alpha \beta} = {\rm exp}(i \, {\rm arg}(c_{\alpha \beta} \tau + d_{\alpha \beta})) \,.
\ee
One can check that 
 a connection of this complex line bundle $L$ transforms in the same way as the 1-form $A$ transforms under an $SL(2,\mathbb Z)$ duality transformation on $\tau$ induced by $M_{\alpha \beta}$. 
 
 \item 
 To every complex line bundle with a connection whose curvature is of (1,1) type, one can associate a {\it holomorphic} line bundle (see e.g. \cite{Green:1987mn}, p. 454ff) with the properties that it allows for holomorphic sections transforming with {\it holomorphic} transition functions.
 Since $d A$ is a (1,1) form, our background therefore defines a holomorphic line bundle ${\cal L}$ with transition function $c_{\alpha \beta} \tau + d_{\alpha \beta}$: If $\hat h$ is a section of $L$, i.e. it transforms as in (\ref{hath}) with transition function (\ref{hatt}), then    
\be
h := {\rm Im}(\tau)^{-1/2} \hat h
\ee
transforms from patch to patch as
\be \label{equh}
 h|_{U_\alpha}  = t_{\alpha \beta} \,  h|_{U_\beta}\,, \qquad h_{\alpha \beta} = c_{\alpha \beta} \tau + d_{\alpha \beta} \,.
\ee
This transition function is indeed holomorphic. 
\end{itemize}

Furthermore, the Einstein equation (\ref{Einstein1}) can be shown to be equivalent to the relation   \cite{Bianchi:2011qh}
\bea \label{c1Bc1L}
c_1(B_n) = c_1({\cal L})  \,,
\eea 
where $c_1({\cal L})= \frac{1}{2\pi} F \in H^{1,1}(B_n)$ is the first Chern class of ${\cal L}$.

The crucial insight is now that the line bundle ${\cal L}$ over $B_n$ together with a choice of a section of ${\cal L}^4$ and of ${\cal L}^6$ uniquely defines an elliptic fibration over $B_n$ with varying elliptic parameter $\tau$. Indeed, (\ref{equh}) rings a bell - it is (up to the powers) the transformation behaviour (\ref{ftrafo}) of the Weierstrass parameters $f$ and $g$ of a Weierstrass model (\ref{Weier-firsttime}) under an $SL(2,\mathbb Z)$ transformation. 
To construct the elliptic fibration associated with ${\cal L}$, we promote the coordinates $[x : y : z]$ of the Weierstrass model as well as the complex parameters  $f$ and $g$ to sections of a suitable line bundle over $B_n$ such that $\mathbb E_\tau$ varies over $B_n$ to form an elliptic fibration over $B_n$. Comparison with the transformation behaviour (\ref{equh}) for a section of the line bundle ${\cal L}$ identifies $f$ and $g$ as holomorphic sections of $ {\cal L}^4$ and $ {\cal L}^6$, i.e. 
\bea \label{fgdata}
f \in \Gamma(B_n, {\cal L}^4), \qquad \quad g \in \Gamma(B_n, {\cal L}^6) \,.
\eea
Let us furthermore make the ansatz $x \in \Gamma(B_n, L_x)$, $y \in \Gamma(B_n, L_y)$, $z \in \Gamma(B_n, L_z)$ with $L_{x,y,z}$ holomorphic  line bundles over $B_n$. Then consistency, i.e. homogeneity, of the Weierstrass equation $P_W$ in (\ref{Weier-firsttime})  requires
\be \begin{split} \label{calLLz}
L_x &= {\cal L}^2 \otimes L_z^2 \cr
L_y &= {\cal L}^3 \otimes L_z^3 \cr
L_z &= {\cal O} \otimes    L_z  \,.
\end{split}
\ee

In this sense a choice of $f$ and $g$ as in (\ref{fgdata}) and (\ref{calLLz}) defines an
 elliptic fibration
\bea
\pi :\quad \mathbb{E}_\tau \ \rightarrow & \  \ Y_{n+1} \cr 
& \ \ \downarrow \cr 
& \ \  B_n 
\eea
We have shown that the elliptic fibration $Y_{n+1} $ is in one-to-one correspondence with a holomorphic $SL(2,\mathbb Z)$ bundle ${\cal L}$ over $B_n$ together with a choice of sections of ${\cal L}^4$ and ${\cal L}^6$.
By standard methods in algebraic geometry one shows that for general duality bundle ${\cal L}$, its first Chern class is related to the curvature on $B_n$ via
\bea \label{c1Yn+1}
c_1(Y_{n+1}) = c_1(B_n) - c_1({\cal L}) \,.
\eea
Since supersymmetry and the Einstein equations require (\ref{c1Bc1L}) this implies that $Y_{n+1}$ in F-theory is Calabi-Yau,
\bea
c_1(Y_{n+1}) = 0 \,.
\eea
To sketch the proof of (\ref{c1Yn+1}) note that $Y_{n+1}$ is a hypersurface $P_W=0$ in a $\mathbb P_{231}$-bundle over $B_n$ given by 
\be \label{P231bundle}
\mathbb P_{231}({\cal E}) = \mathbb P_{231}( {\cal L}^2 \oplus  {\cal L}^3 \oplus {\cal O}) \,.
\ee
The three summands are associated with the homogenous fiber ambient coordinates $[x : y :z]$, which transform as the sections  (\ref{calLLz}), and the notation means that we have to take the projectivisation of this bundle.
The total Chern class of this bundle is 
\be
c(\mathbb P_{231}({\cal E})) = \left( (1 + c_1({\cal L}^2) +  c_1(L_z^2)) (1 + c_1({\cal L}^3) +  c_1(L_z^3)) (1 + c_1(L_z)) \right) \, c(B) \,,
\ee
and by the adjunction formula, the total Chern class of the hypersurface $P_W=0$ therein follows from this as
\bea
c(Y_{n+1})  = \frac{c(\mathbb P_{231}({\cal E}))}{1 + 6(c_1(L_z) + c_1({\cal L}))} \,.
\eea
In particular,
\bea
c_1(Y_{n+1}) &=& c_1(\mathbb P_{231}({\cal E})) - c_1([P_W]) = \\
&=&  c_1(B_n) + 6 c_1(L_z) + 5 c_1({\cal L}) - 6(c_1(L_z) + c_1({\cal L})) \\
&=&  c_1(B_n) - c_1({\cal L}) \,.
\eea

The F-theory paradigm consists in using this one-to-one correspondence between supergravity backgrounds with 7-branes and Calabi-Yau elliptic fibrations in order to study the first using insights on the latter. 
Remarkably, the F-theory geometry automatically sums up the effect of $D(-1)$ instantons \cite{Billo:2010mg}. We will come back to this in section \ref{NPQFT}, where we will point out that the holomprphically varying profile of $\tau$ over the base $B_n$
as determined by the elliptic fibration gives the full quantum corrected answer e.g. for the gauge coupling on a D3-brane probing this geometry. 
This can be checked in particular in 8d F-theory compactifications on K3, where the profile of $\tau$ on the base $\mathbb P^1$ can, in favorable circumstances, be explicitly extracted from the fibration.

The idea of reading off  the physics of 7-branes from the geometry of elliptic fibrations becomes particularly powerful if we use duality with M-theory as will be introduced in the next section.


 \subsection{From M-theory to elliptic fibrations} \label{sec_Mtheorytoell}

The duality between Type IIB string theory compactified on a circle and M-theory on a torus provides a useful viewpoint on the origin and meaning of the elliptic fibration \cite{Schwarz:1995dk}.
To appreciate this, consider M-theory in its long-wavelength limit of 11d supergravity with bosonic field content the metric $g_{MN}$ and the 3-form gauge potential $C_3$.
The theory is invariant under a 3-form gauge transformation $C_3 \rightarrow C_3 + d \Lambda_2$ which leaves the field strength $G_4 = d C_3$ unchanged.
The bosonic part of the action is  - up to the most relevant order in the 11d Planck length $\ell_{11}$ - 
\bea \label{Mtheoryaction}
S = \frac{2\pi}{ \ell_{11}^9} \left(  \int_{\mathbb R^{1,10}} \sqrt{-g} R - \frac{1}{2} \int_{\mathbb R^{1,10}} d C_3 \wedge \ast d C_3 - \frac{1}{6} \int_{\mathbb R^{1,10}} C_3 \wedge G_4 \wedge G_4  + \ell_{11}^6 \int_{\mathbb R^{1,10}} C_3 \wedge I_8  \right) \,,
\eea
with the topological  higher curvature term \cite{Duff:1995wd}
\bea
I_8 = \frac{1}{(2\pi)^4} \left( - \frac{1}{768} ({\rm tr} R^2)^2  + \frac{1}{192} {\rm tr} R^4 \right) \,.
\eea
The gauge potential $C_3$ couples electrically to M2-branes via
\bea \label{SM2-gen}
S_{\rm M2} = \frac{2 \pi}{\ell_{11}^3} \int_{\rm M2} \sqrt{-g} +  \frac{2 \pi}{\ell_{11}^3} \int_{\rm M2} C_3 \,.
\eea
The magnetically dual potential $C_6$ couples electrically to M5-branes.

The duality with Type IIB theory involves compactifying M-theory on $\mathbb R^{1,8} \times T^2$ with 
\be
T^2 = S^1_A \times S^1_B.
\ee
In very broad brushes, the picture is as follows:
First, we interpret the circle $S^1_A$ with coordinate $x$ as  the M-theory circle with radius $R_A$. As $R_A \rightarrow 0$ we approach Type IIA string theory on $\mathbb R^{1,8} \times S^1_B$. In particular the metric components $g_{x \ast}$ become the RR field $C_1$ of Type IIA supergravity, while $g_{xx}$ is related to the Type IIA dilaton.
T-duality along $S^1_B$ with coordinate $y$ takes us to Type IIB string theory on $\mathbb R^{1,8} \times \tilde S^1_B$. The dual circle $\tilde S^1_B$ with dual coordinate $\tilde y$ has radius 
\be
\tilde R_B = \frac{\ell_s^2}{R_B}
\ee
in terms of the string length $\ell_s$.
In the limit $\tilde R_B \rightarrow \infty$ we recover Type IIB string theory on $\mathbb R^{1,9}$. The  components  $(C_1)_y$ of the Type IIA 1-form dualize to the Type IIB axion $C_0$.

In all we arrive at the duality
\bea
{\rm M-theory \,\,  on} \quad \mathbb R^{1,8} \times (S^1_A \times S^1_B) |_{R_A, R_B \rightarrow 0} \quad  \simeq \quad  {\rm Type \, \,  IIB \, \, theory \, \,  on} \quad \mathbb R^{1,9} \,.
\eea
In particular the limit requires that $V : = {\rm vol}(T^2) \rightarrow 0$ with $T^2 = S^1_A \times S^1_B$. 
A very careful tracing of the effective action and the metric in M-theory and Type IIB through the limit can be found on p. 23-25 of  \cite{Denef:2008wq} and reveals, in more detail, a duality between the two theories as follows:

\begin{minipage}{8cm}
\underline{M-theory on ${\mathbb R}^{1,8} \times T^2$  } \\
\end{minipage}
\begin{minipage}{2cm}
\phantom{aaa}
\end{minipage}
\begin{minipage}{10cm}
\underline{ Type IIB on ${\mathbb R}^{1,8} \times \tilde S^1_B$ }\\
\end{minipage}
\begin{minipage}{8cm}
\begin{itemize}
\item
$T^2$ complex structure $\tau = \tau_1 + i \tau_2$  \\
\end{itemize}
\end{minipage}
\begin{minipage}{2cm}
\phantom{aaa}
\end{minipage}
\begin{minipage}{10cm}
\begin{itemize}
\item
axio-dilaton $\tau = C_0 + i e^{-\phi}$  \\
\end{itemize}
\end{minipage}  
\begin{minipage}{8cm}
\begin{itemize}
\item
$T^2$ volume      $ V$\vspace{2mm}  \\
\phantom{aaaaaa}\\
\end{itemize}
\end{minipage} 
\begin{minipage}{2cm}
\phantom{aaa}
\end{minipage}
\begin{minipage}{10cm}
\begin{itemize}
\item
Einstein frame metric \vspace{2mm}  \\
$ds^2_{\rm IIB} = ds^2_{\mathbb R^{1,8}} + \frac{\ell_s^4}{V} d \tilde y^2$, \qquad  $\tilde y \simeq \tilde y+ 1$ \\
\end{itemize}
\end{minipage} 
In particular, the limit $V \rightarrow 0$ therefore restores full ten-dimensional Poincar\'e invariance of the dual Type IIB theory. 

This duality explains why it is no accident that the Type IIB duality group acts on the axio-dilaton in the same way as the modular parameter $\tau$ of an elliptic curve is transformed by a modular $SL(2,\mathbb Z)$ transformation  \cite{Schwarz:1995dk}. In the dual M-theory, the elliptic curve $\mathbb E_\tau$ introduced in section \ref{sec_Frompqtoell} merely acts as a book-keeping device for the axio-dilaton is part of the physical spacetime.

The duality can be promoted to a fiberwise duality by considering M-theory on $\mathbb R^{1,8-2n} \times Y_{n+1}$, where $Y_{n+1}$ is a torus fibration over $B_n$ \cite{Vafa:1996xn}.
Supersymmetry requires $Y_{n+1}$ to be Calabi-Yau, which agrees with our findings around (\ref{c1Yn+1}), albeit in a more direct way. Applying the duality fiberwise implies that M-theory on $\mathbb R^{1,8-2n} \times Y_{n+1}$ is dual to Type IIB string theory compactified on the background locally of the form $B_n \times \tilde S^1_B$ with Einstein frame metric
\bea
ds^2 = ds^2_{\mathbb R^{1, 8-2n}} + ds^2_{B_n} + \frac{\ell_s^4}{V} d\tilde y^2 \,.
\eea 
As before $\tilde y \simeq \tilde y + 1$ is the periodic coordinate on $\tilde S^1_B$, and it is in the limit $V\rightarrow 0$ that  we recover Type IIB string theory compactified on $B_n$. The coordinate $\tilde y$ then becomes part of the uncompactified spacetime on the Type IIB side. Note furthermore that in F-theory, the volume $V$ is not a dynamical modulus because we are considering the limit $V \rightarrow 0$. Only the complex structure $\tau$ is a dynamical modulus in the F-theory limit.

Equivalently, the idea of F/M-theory duality can be expressed like this:
The effective action in $\mathbb R^{1,8-2n}$ of M-theory compactified on $Y_{n+1}$ with fiber volume $V$ is dual to the circle reduction 
of the effective action of Type IIB theory on $B_n$ on a circle $\tilde S^1_B$ with radius $\tilde R_B  \sim \frac{1}{V}$.
The F-theory limit $V \rightarrow 0$ corresponds to the decompactification limit of the Type IIB theory. This is depicted in the following diagram:
\be\label{MFDuality}
\begin{array}{ccc}
\hbox{M-theory on $Y_{n+1}$}\qquad \qquad 
& \xrightarrow[]{\ {\rm Vol} (\mathbb{E}_{\tau}) \rightarrow 0\ } 
&\qquad \hbox{Type IIB on $B_{n}$}
\cr \cr 
\downarrow\qquad \qquad 
& 
&\qquad \downarrow\cr \cr 
\hbox{{\rm Eff.  action  in} \,  $\mathbb R^{1,8-2n}$ }\qquad \qquad 
& \xrightarrow[]{\ R_A \sim {1\over \tilde R_B} \,\rightarrow \, 0 \ } 
&\qquad \hbox{  Eff. action in $\mathbb R^{1,9-2n}$}
\end{array}
\ee


We summarize, in table \ref{tab-MFgen}, the different possibilities of compactifying F-theory and its dual M-theory to various dimensions, and also indicate the amount of supersymmetry preserved.

\begin{table}
\begin{tabular}{|c|c|c|l|l| }
\hline
fibration & base                             &  {7-brane} on $B_n$    &  F-theory    & M-theory   \\ \hline \hline
$Y_2 = K3$   & $B_1 = \mathbb P^1$&  point                           & 8d ${\cal N}=1$ (16)    & 7d ${\cal N}=2 $ (16)    \\ \hline
$Y_3 $   & $B_2$&  complex curve                           & 6d ${\cal N}=(1,0)$ (8)    & 5d ${\cal N}=2 $ (8)    \\ \hline
$Y_4 $   & $B_3$&  complex surface                           & 4d ${\cal N}=1$ (4)    & 3d ${\cal N}=2 $ (8)    \\ \hline
$Y_5 $   & $B_4$&  complex 3-fold                          & 2d ${\cal N}=(2,0)$ (2)    & 1d ${\cal N}=2$ supermechanics  (2)    \\ \hline
\end{tabular}
\caption{F- and M-theory in various dimensions. The number in brackets in column 4 and 5 gives the number of real supercharges. \label{tab-MFgen}}
\end{table}

\section{ F-theory on a smooth elliptic fibration} \label{secFonsmooth}

After this general introduction to  F-theory we now begin taking a closer look at the geometry of an elliptic fibration and its physics interpretation. 
In this section we focus on smooth elliptic fibrations as described by a non-singular Weierstrass model. We will first establish some of the geometric properties of such smooth elliptic fibrations in section \ref{sec_smoothWeier} and then explain the importance of the degenerate fibers for the study of 7-branes in section \ref{sec_singfibs}. 
The discriminant locus over which these fibers occur will be identified, in section \ref{sec_PhysicsIIB1}, with the location of the 7-branes. This is in agreement with  the perturbative  limit briefly discussed in the same section.
We develop further the duality with M-theory in section \ref{sec_firstlook}.

\subsection{The smooth Weierstrass model}\label{sec_smoothWeier}

An elliptic fibration is a torus fibration 
\bea \label{ellfibrationpi}
\pi :\quad \mathbb{E}_\tau \ \rightarrow & \  \ Y_{n+1} \cr 
& \ \ \downarrow \cr 
& \ \  B_n 
\eea
with a rational section $s_0$, i.e. a meromorphic map 
from the base to the fiber,
\be
\begin{split}\label{sigmo0def}
s_0:  \quad B_n  &\rightarrow \mathbb{E}_\tau  \cr
 b &\mapsto s_0(b) \,.
\end{split}
\ee
Meromorphic here means that $s_0$ is a rational function in the function field of the base $B_n$.
We furthermore require that the map $\pi$ be equi-dimensional or flat, i.e. the pre image $\pi^{-1}(b)$ is of complex dimension one for each point $b \in B_n$.

If no rational section exists, (\ref{ellfibrationpi}) defines merely a genus-one or torus fibration, as opposed to an elliptic fibration.
 For now we assume the existence of a section, and discuss more general torus  fibrations in section \ref{sec_genusone-b}. 
 
We have already noted that there are many ways to model the torus fiber as a hypersurface or complete intersection (or more general constructions) in a suitable fiber ambient space. Of special importance, however, is the description of an elliptic fibration as a Weierstrass model. This is due to the general fact that   
every elliptic fibration
is {\it birationally equivalent} (i.e. isomorphic up to higher codimension loci) to a Weierstrass model, which  has been introduced in section \ref{sec_Frompqtoell} as the hypersurface 
\bea
P_W := y^2 - (x^3 + f \, x  \, z^4 + g \, z^6) = 0 \subset \mathbb P_{231}({\cal E}) \equiv X_{n+2} \
\eea
with $f \in \Gamma(B_n, {\cal L}^4)$, $g \in \Gamma(B_n, {\cal L}^6)$ and the fiber ambient space coordinates $x$, $y$, $z$ transforming as the sections (\ref{calLLz}). 
Note that $x$, $y$ and $z$ are not allowed to vanish simultaneously because they are locally coordinates on the fiber ambient space space  $ \mathbb P_{231} = \mathbb C^3 \setminus \{(0,0,0)\}$ modulo the projective identification (\ref{P231rescaling}).  
The total space of the bundle $P_{231}({\cal E})$ introduced in (\ref{P231bundle}) is a complex $(n+2)$-fold which we will oftentimes denote as the ambient space $X_{n+2}$ of $Y_{n+1}$. 
Recall that $Y_{n+1}$ is Calabi-Yau if and only if $c_1({\cal L}) = c_1(B_n)$, i.e. the line bundle ${\cal L}$ coincides with the anti-canonical bundle of $B_n$. Oftentimes we will use the same symbol for the anti-canonical bundle of the base and its associated (first Chern) class, e.g. when we  write $c_1({\cal L}) = \bar K_{B_n}$. While we can without loss of generality focus on Weierstrass models to analyze elliptic fibrations, we must keep in mind that other, birationally equivalent models may differ in interesting ways as far as the structure of the fiber in higher codimension is concerned. In particular the Weierstrass model may not always be the most practical model for the elliptic fibration.
Early works studying non-Weierstrass representations include \cite{Klemm:1996hh,Berglund:1998va}, and more generally such models will play a major role in section \ref{sec_MWGROUP}.

In the sequel we will always take $Y_{n+1}$ to be Calabi-Yau. This requires $B_n$ to be a K\"ahler manifold with
\bea
h^{i,0}(B_n) = 0 \quad \,  \forall \, i = 1,\ldots, n \,.
\eea
Otherwise the corresponding cohomology groups would pull back to the full fibration, in disagreement with the Calabi-Yau property. Indeed the latter requires that $h^{i,0}(Y_{n+1}) = 0$ for $i=1,\ldots,n$. Unless stated otherwise we take $B_n$ to be smooth as this is the physical compactification space of the non-perturbative Type IIB string theory we are studying in the language of F-theory.

A Weierstrass model possesses a {\it holomorphic} section, i.e. for all values of $f$ and $g$ there exists a rational point in the fiber over any point $b \in B_n$ whose coordinates $[x : y : z]$ are described by a holomorphic function of the base coordinates. In the Weierstrass model this point is cut out from the fiber by setting $z=0$ in the Weierstrass equation and is hence given by the point $[x : y : z] = [1 : 1 : 0]$ on $P_W = 0$.
Indeed if $z=0$ we can use the $\mathbb P_{231}$ scaling relation to set $x=1$, and the Weierstrass equation reduces to $y^2=1$ with solution $y = \pm1$. We still have the freedom to rescale the coordinates with $\lambda =-1$ without affecting the choice $x=1$ and can hence set $y=1$. This is the single point $z=0$ in the Weierstrass model, and the holomorphic section is given by the holomorphic map
\bea \label{sigma0Weier}
s_0:  \quad   b \mapsto [1 : 1 : 0] \,.
\eea
The divisor 
\bea \label{S0divisor}
S_0: \{ z=0\}
\eea
therefore intersects the fiber over each $b \in B_n$ in this one point $s_0(b)$.
Since the defining equation of the section is a holomorphic, rather than a meromorphic, function, the section is called {\it holomorphic} as opposed to {\it rational}.\footnote{The difference between a holomorphic and a rational section only appears in higher codimension, where the meromorphic function defining the rational section hits a pole. Given an elliptic fibration with merely a rational section, it is therefore still birational to a Weierstrass model even though the section of the latter is holomorphic.}
The meaning of the point  defined by the zero-section as the zero element in the additive group law on the elliptic fiber will be discussed in more detail in section \ref{rationalShioda}.

As summarized in Appendix \ref{app-notation}, we denote by (\ref{S0divisor}) both the concrete holomorphic cycle on $Y_{n+1}$ corresponding to the vanishing locus of the coordinate $z$ and its divisor class. Unless stated otherwise, we will be working on simply-connected spaces which are smooth projective varieties; on such a space $X$, the divisor class group coincides with the N\'eron-Severi group ${\rm NS}(X)$, as reviewed in appendix \ref{app_divisors}. In this sense we take $S_0$ to be an element in  ${\rm NS}(X)$. Its Poincar\'e dual cohomology class in $H^{1,1}_{\mathbb Z}(Y_{n+1}) := H^{1,1}(Y_{n+1}) \cap H^{2}(Y_{n+1},\mathbb Z) $ will usually be denoted by $[S_0]$. On a smooth, simply-connected complex algebraic variety $X$ there is essentially no difference between  $H^{1,1}_{\mathbb Z}(X)$ and ${\rm NS}(X)$ (see  again Appendix  \ref{app_divisors}). In particular,
\bea
{\rm rk}({\rm NS}(Y_{n+1})) =    h^{1,1}(Y_{n+1}) \,.
\eea

Let us now restrict to smooth elliptic fibrations of complex dimension $3$ or higher.
On the smooth Weierstrass model $Y_{n+1}$, $n \geq 2$, there are  two types of divisors: The section $S_0$ and 
the vertical divisors, which are the pullback of divisors on the base $B_n$. We will use the following notation for the 
\bea
{\rm generators \, \,  of} \, \, {\rm NS}(B_n): \qquad D_{\alpha}^{\rm b} \,, \quad \alpha= 1, \ldots,  h^{1,1}(B_n) \,.
\eea
The pre-image $\pi^{-1}( D_{\alpha}^{\rm b})$ of such a divisor with respect to the projection defines a divisor on $Y_{n+1}$; it is itself elliptically fibered, with base $D_{\alpha}^{\rm b}$.
Then the divisor group (or equivalently $H^{1,1}_{\mathbb Z}(Y_{n+1})$) is  generated by
\bea \label{NSYn+1}
{\rm NS}(Y_{n+1}) = \langle S_0, \pi^{-1}(D_{\alpha}^{\rm b}) \rangle
\eea
and its rank equals
\bea \label{h2Yn}
h^2(Y_{n+1}) = h^{1,1}(Y_{n+1}) = 1 + h^{1,1}(B_n) \,.
\eea
 This is a special case of the Shioda-Tate-Wazir theorem, which we will encounter in more generality in later sections.\footnote{The most general form of this equation is (\ref{STW}). This version also holds for $n=1$, i.e. for elliptic K3s. The point is that on a K3 surface, extra sections do not induce singularities as these occur in codimension two on the base and hence on a smooth Weierstrass model of complex dimension $2$,  (\ref{h2Yn}) need not hold.}

The cohomological intersection ring on $Y_{n+1}$ (see Appendix \ref{app-notation} for our notation) has the structure
\bea
{} [S_0] \cdot \pi^\ast[D^{\rm b}_{\alpha_1}] \cdot \ldots \cdot \pi^\ast[D^{\rm b}_{\alpha_n}] &=&  [D^{\rm b}_{\alpha_1}] \cdot_{B_n} \ldots \cdot_{B_n} [D^{\rm b}_{\alpha_n}] \label{S0int1} \\
{}  [S_0] \cdot  [S_0] \cdot \pi^\ast[D^{\rm b}_{\alpha_1}] \cdot \ldots \cdot \pi^\ast[D^{\rm b}_{\alpha_{n-1}}]   &=&  -c_1(B_n) \cdot_{B_n} [D^{\rm b}_{\alpha_1}]  \cdot_{B_n}\ldots \cdot_{B_n} [D^{\rm b}_{\alpha_{n-1}}] \label{SSwint}  \\
{}  \pi^\ast[D^{\rm b}_{\alpha_1}] \cdot \ldots \cdot \pi^\ast[D^{\rm b}_{\alpha_{n+1}}] &=& 0 \label{S0int3} \,.
\eea
The first equation holds because the section $S_0$ defines an embedding of the base $B_n$ as a holomorphic $n$-cycle of $Y_{n+1}$. As the notation suggests, the intersection product $"\cdot_{B_n}"$ is to be taken on $B_n$. For a top-form this equals integration over the total space. Equivalently, the relation follows from the fact that $S_0$, being a section, intersects the generic fiber in one point. (\ref{S0int3}) holds for dimensional reasons because of verticality of the pullback divisors. 
Concerning (\ref{SSwint}), note that since $S_0$ is a holomorphic, as opposed to merely a rational, section, it satisfies in fact
 \bea
[S_0] \cdot [S_0] = -  [S_0]  \cdot c_1(B_n)
\eea
in the cohomology ring of $Y_{n+1}$. 
This can be computed as follows:
In fiber ambient space bundle $X_{n+2} = \mathbb P_{231}({\cal E})$ into which $Y_{n+1}$ is embedded, we have the relation
\bea
0 = [x] \cdot_{X_{n+2}} [y] \cdot_{X_{n+2}} [z] = c_1(L_z) \cdot_{X_{n+2}}   c_1(L_z^3 \otimes {\cal L}^3) \cdot_{X_{n+2}}  c_1(L_z^2 \otimes {\cal L}^2)  \,. 
\eea
This is a consequence of the fact that the coordinates $x$, $y$, $z$ (which transform as the sections (\ref{calLLz})) are not allowed to vanish simultaneously on the fiber ambient space.
Since the Weierstrass equation is a section of $L_z^6 \otimes {\cal L}^6$, we can interpret one of the last two factors as enforcing the restriction of the ambient space intersection numbers to $Y_{n+1}$. On the elliptic fibration, we hence obtain
\bea
c_1(L_z) \cdot_{Y_{n+1}} (c_1(L_z) + c_1({\cal L})) = 0 \,.
\eea
But $c_1(L_z)$ is the class of the zero-section $S_0$ cut out by $z=0$, and due to the Calabi-Yau condition $c_1({\cal L}) = c_1(B_n)$, which proves the claim.

Having described the intersection structure of the divisors, let us briefly remark on the types of curves on $Y_{n+1}$:
Apart from  the generic elliptic  fiber $\mathbb E_\tau$, the independent curve classes of the smooth Weierstrass model $Y_{n+1}$ are the independent curve classes on the base $B_n$. Both fiber and base curves can be written as intersections of the divisors $S_0$ and $\pi^{-1}( D_\alpha^{\rm b})$: For any collection of $n$ base divisors with intersection number $k$ on $B_n$, i.e. $[D_{\alpha_1}^{\rm b}] \cdot_{B_n} \ldots \cdot_{B_n} [D_{\alpha_n}^{\rm b}] = k$, we have on $Y_{n+1}$
\be
\pi^\ast [D_{\alpha_1}^{\rm b}] \cdot \ldots \cdot  \pi^\ast [D_{\alpha_n}^{\rm b}] = k \,  [\mathbb E_\tau] \,.
\ee
Furthermore, if $n-1$ divisors intersect on the base in a curve $C$, $[D_{\alpha_1}^{\rm b}] \cdot_{B_n} \ldots \cdot_{B_n} [D_{\alpha_{n-1}}^{\rm b}] = [C]$, then on $Y_{n+1}$ the corresponding base curve is
\bea
[S_0] \cdot \pi^\ast [D_{\alpha_1}^{\rm b}] \cdot \ldots \cdot  \pi^\ast [D_{\alpha_{n-1}}^{\rm b}]  = [C_{\rm base}]  \in H_2(\hat Y_{n+1}) \,.
\eea
We call it a base curve because it has the property that its intersection with a pullback divisor is non-zero. 
Indeed, the intersection numbers (\ref{S0int1})  - (\ref{S0int3}), together with the fact that a section intersects the fiber in one point,  imply that 
\be
 \begin{aligned}
  & [S_0] \cdot [\mathbb E_\tau] &= 1, \qquad  &\pi^{\ast}[D_\beta^{\rm b}]  \cdot [\mathbb E_\tau] = 0 \cr
 & [S_0] \cdot [C_{\rm base}] &= - c_1(B_n) \cdot_{B_n} [C], \qquad & \pi^{\ast}[D_\beta^{\rm b}]  \cdot [C_{\rm base}] = [D_\beta^{\rm b}]  \cdot_{B_n} [C]  \,.\\
\end{aligned}
\ee

\subsection{Singular fibers on the smooth Weierstrass model} \label{sec_singfibs}

Of special importance for us will be the singular fibers of the elliptic fibration because these contain information about the location and the $[p,q]$ type of the 7-branes on the base $B_n$.
The fiber ${\mathbb E}_\tau$ of  the Weierstrass model becomes singular whenever the discriminant 
\bea
\Delta = 4 f^3 + 27 g^2
\eea
vanishes. 
To see this, note that a hypersurface $P=0$ is singular whenever its gradient vanishes, i.e. whenever    $dP=0$ along with $P=0$.
The singularity cannot sit at $z=0$ since in this case the Weierstrass equation reduces to $y^2 = x^3$. The only singularity of this equation occurs at $x=y=0$, but the point $x=y=z=0$ is not in $\mathbb P_{231}$ and hence not on the elliptic curve.\footnote{In particular, the holomorphic zero-section of the Weierstrass model is therefore always smooth.} We can therefore restrict ourselves to the patch where $z=1$, and rewrite the 
Weierstrass model as the hypersurface
\bea
P_W:   y^2 = F(x) \,,   \qquad F(x) =x^3 + f x + g \equiv  \prod_{i=1}^3 (x - x_i) \,.
\eea
Then the gradient with respect to $x$ is
\bea
\frac{d F(x)}{dx} = \sum_{j=1}^3 \prod_{i \neq j} (x - x_i) \,.
\eea
It follows that $P=0=dP$ whenever two roots $x_i$ and $x_k$ coincide. By definition, this happens whenever the discriminant of $F(x)$ vanishes. The discriminant of the special cubic $F(x)= x^3 + f x + g$
 is  $\Delta  = 4 f^3 + 27 g^2$, which proves the claim.  

We have therefore shown that the singularities of the elliptic fiber ${\mathbb E}_\tau$ of the Weierstrass model occur at a point of the form  $[x : y : z] = [\ast : 0 : 1]$ in the fiber over $\Delta = 0$, and hence away from the section (\ref{sigma0Weier}).
As we will discuss in the next section, the singularity can be thought of as due to the vanishing, or shrinking to zero, of a certain 1-cycle in the fiber.
Note that in order for this point to be a singularity not only of the fiber ${\mathbb E}_\tau$, but of the elliptic fibration $Y_{n+1}$ as an $(n+1)$-fold, also the gradient with respect to the base coordinates must vanish. A smooth Weierstrass model is one where the singularities in the fiber over the discriminant are {\it not} singularities of $Y_{n+1}$. 

For future reference, we will refer to the divisor on the base along which $\Delta$ vanishes as the discriminant locus or discriminant divisor\footnote{If not only $\mathbb E_\tau$, but also $Y_{n+1}$ is singular, $\Sigma$ must be defined as in (\ref{Simgadef}).}
\bea
\Sigma := \{\Delta = 0 \} \subset B_{n} \,.
\eea

In the remainder of this section, we assume that the Weierstrass model is generic, i.e. the defining sections $f$ and $g$ are maximally generic functions. In this case, the discriminant locus $\Sigma$ is an irreducible divisor on $B_n$ in homology class $[\Sigma]  = 12 \bar K_{B_n}$.
At a generic point on this divisor, $f$ and $g$ do not vanish simultaneously. The fiber $\mathbb E_\tau$ over such generic discriminant points  has a singularity at $[x : y : z] = [\ast : 0 : 1]$ and $\mathbb E_\tau$ forms a {\it nodal curve}, i.e. a curve with a generic self-intersection at this point.
The type of singularity is called Kodaira-type $I_1$ and obeys the following criterion: If we define by ${\rm ord}(f,g,\Delta)$ the order of the zero of the respective functions at a given point, then we have for
\bea \label{I1type}
{\rm Kodaira-type} \, \, I_1:  \qquad  {\rm ord}(f,g,\Delta) = (0,0,1) \,.
\eea
Note that even though the elliptic curve $\mathbb E_\tau$ is singular at $[x : y : z] = [\ast : 0 : 1]$, the elliptic fibration $Y_{n+1}$ is smooth: For generic $f$ and $g$, $dP/df\neq 0$ and $dP/dg \neq 0$ at $\Delta = 0$.

At special points on $\Delta =0$, both $f=0$ and $g=0$. This occurs in complex codimension one on $\Delta$ and hence altogether in complex codimension two on $B_n$. At these points, the elliptic fiber forms a so-called {\it cuspidal curve} $y^2 = x^3$, with a non-generic point of self-intersection at $[x: y :z] =  [0: 0 : 1]$. 
Since $f$ and $g$ are, by assumption, maximally generic, the vanishing orders are ${\rm ord}(f,g,\Delta) = (1,1,2)$. More generally, cusps occur for fibers of 
\bea \label{IItype}
{\rm Kodaira-type} \, \, II:  \qquad  {\rm ord}(f,g,\Delta)  = (\geq 1,1,2) \,.
\eea
The notation means that $f$ vanishes to order $1$ or higher, and $g$ and $\Delta$ to orders $1$ and $2$, respectively.
Again, at such points $Y_{n+1}$ is smooth as an $(n+1)$-fold. 
It so happens that the points of Kodaira type $II$ singularities in the fiber, the discriminant $\Delta = 4 f^3 + 27 g^2$ itself develops a cusp (since $f=g=0$).

Singularities in the fiber are singularities of $Y_{n+1}$ only if the vanishing orders exceed (\ref{I1type}) or (\ref{IItype}). This occurs only for non-generic $f$ and $g$ and will be discussed in section \ref{sec_singcodim1}. 

\subsection{Singular fibers due to 7-branes and Type IIB limit} \label{sec_PhysicsIIB1}

What is the physics interpretation of the fibral singularities?
It turns out that $\Delta =0$ is nothing but the divisor wrapped by the 7-branes in the Type IIB picture. 
This is of course not unexpected because of the identification of the fiber complex structure with the axio-dilaton $\tau$: The singularity in the fiber should translate into a singular value of this field at the location of its source.

Indeed, the zeroes of $\Delta$ imply such special values for $\tau$ according to formula (\ref{jtau1}).  From the expansion (\ref{jtauexpansion}) we infer that if $j({\tau}) \rightarrow \infty$, the axio-dilaton $\tau \rightarrow i \infty$ (recall that $j$ is defined over the upper half-plane since $\tau \geq 0$).
This is precisely what happens at an $I_1$-locus. At this locus, consequently $g_s \rightarrow 0$. Such behaviour matches the value of $g_s$ near a D7-brane, where $\tau$ has a profile of the form (\ref{tauprofile}).

Equivalently, one can use the following property of the fibral singularities: 
The singular fiber is characterized by the vanishing of a  linear combination $\Pi = p A + q B$ of the two elements of the basis of one-cycles introduced in (\ref{BAcycles}).
If one transports the two 1-cycles $A$, $B$, which exist away from the singular locus, around the singularity in the basis, they undergo  an $SL(2,\mathbb Z)$ monodromy  (\ref{BAmono}) with monodromy matrix $ M_{[p,q]}$ as in (\ref{MpqfromM10}).
This is a special instance of the 
the Picard-Lefshetz theorem, which tells us about the monodromic behaviour of middle-dimensional cycles in a complex variety around a special point in the complex structure moduli space where a combination of them vanishes. The variety here is the elliptic curve, whose complex moduli space is identified with the base.
Then a general 1-cycle $\gamma$ in the fiber undergoes a monodromy around the vanishing locus of $\Sigma$ given by
\bea
\gamma \rightarrow \gamma - (\gamma \cdot \Pi) \,  \Pi \,.
\eea
With $A \cdot B = 1 = - B \cdot A$ (and all others vanishing) this reproduces the monodromy (\ref{MpqfromM10}) \cite{Blumenhagen:2013fgp}.

We conclude that a singularity at which the cycle $\Pi = p A + q B$ vanishes in the fibre occurs precisely over the location of a $[p,q]$ 7-brane on the base.
For an $I_1$-singularity, the associated monodromy matrix is $M_{[1,0]}$, in agreement with our interpretation of the $I_1$ locus as due to a 7-brane of type $[1,0]$.
More subtle is the behaviour at a Kodaira type $II$ singularity, where $j(\tau) = \frac{0}{0}$ . A careful analysis \cite{Aluffi:2009tm} reveals that this codimension-two locus is the remnant of the O7-plane intersection with the D7-brane in the weakly coupled Type IIB orientifold uplift.

Note that for generic $f$ and $g$, corresponding to a smooth Weierstrass model, the discriminant locus $\{\Delta =0\}$ describes a single irreducible divisor. Indeed, F-theory on a smooth Weierstrass model  has the same brane content as a Type IIB orientifold compactification on a Calabi-Yau space $X_n$ with a single D7-brane together  with an O7-plane. The associated Type IIB orientifold picture is inferred by performing the Sen limit \cite{Sen:1997gv}. The Type IIB orientifold is defined on the Calabi-Yau double cover $X_n$ of $B_n$. 
The starting point of this construction is to parametrise
\bea
f &=& - 3 h^2 + \epsilon \,  \eta \\
g &=& - 2 h^3 + \epsilon h \, \eta - \frac{\epsilon^2}{12} \chi 
\eea
with $h \in \Gamma(B_n,\bar K^2_{B_n})$, $\eta \in \Gamma(B_n,\bar K^4_{B_n})$ and $\chi \in \Gamma(B_n,\bar K^6_{B_n})$ generic sections of indicated degree.
The perturbative limit corresponds to taking $\epsilon \rightarrow 0$. In this limit, the discriminant factorises as 
\bea
\Delta = - 9 \epsilon^2 h^2 \, (\eta^2 - h \chi) + {\cal O}(\epsilon^3) \,.
\eea
The D7-brane is located at $\eta^2 - h \chi$, while $h=0$ describes the O7-plane.
The Type IIB Calabi-Yau $n$-fold is obtained by adding a local coordinate to $\xi$ to the local coordinates of $B_n$ and by considering the hypersurface 
\bea
X_n: \quad \xi^2 = h \,.
\eea
This space is indeed Calabi-Yau and admits an orientifold involution $\sigma: \xi \rightarrow - \xi$.
The divisor $\{h=0\}$  is the fixed-point locus of the involution, which is wrapped by the O7-plane. Uplifting the 7-brane locus to $X_n$ results in a single D7-brane along the divisor 
\bea
{\rm D7}: \quad \{ \eta^2 - \eta^2 \chi = 0 \}  \subset X_n \,,
\eea
which is invariant under the orientifold involution.
The gauge group of this system is trivial because the $U(1)$ gauge symmetry on the D7-brane is projected out by the orientifold action. 
The geometry of the D7-brane is that of a so-called Whitney umbrella \cite{Collinucci:2008pf}. 
The Sen limit can be interpreted as a stable degeneration limit of a family of elliptic fibrations  \cite{Clingher:2012rg}. It has been studied in great detail in the more recent F-theory literature, including \cite{Collinucci:2008zs,Collinucci:2009uh,Blumenhagen:2009up,Aluffi:2009tm,Esole:2011cn,Esole:2012tf,Krause:2012yh,MayorgaPena:2017eda}.

We have thus concluded that a smooth Weierstrass model describes the physics of a single 7-brane in F-theory, with trivial gauge group along the 7-brane.
We will momentarily arrive at 
the same conclusion via duality with M-theory.

\subsection{M-theory picture (II)} \label{sec_firstlook}

Consider M-theory compactified on a smooth Weierstrass model $Y_{n+1}$. How would we describe the gauge symmetry along the 7-branes in the effective theory in $\mathbb R^{1,8-2n}$?
Abelian massless vector fields in the M-theory effective action arise by expanding the M-theory 3-form gauge potential $C_3$ along a basis of harmonic 2-forms on $Y_{n+1}$.
For a smooth and generic Calabi-Yau Weierstrass model $Y_{n+1}$ with $n \geq 2$, the group of divisors is generated by the zero-section $S_0$ and the pullback divisors $\pi^{-1}(D^{\rm b}_\alpha)$ as in (\ref{NSYn+1}). After taking the Poincar\'e dual, this generates a basis of the space of harmonic 2-forms.

For our purposes, it will turn out more convenient to define a basis of $H^{1,1}(Y_{n+1})$ in terms of the shifted divisor  \cite{Park:2011ji,Bonetti:2011mw}
\bea
 [\tilde S_0] := [S_0] - \frac{1}{2} [K_{B_{n}}]
\eea
along with the vertical divisors $\pi^{-1}(D^{\rm b}_\alpha)$. 
Expansion of $C_3$ along this basis of $H^{1,1}(Y_{n+1})$ as
\bea \label{C3exp1}
C_3 = \tilde A^0 \wedge [\tilde S_0] + \sum_\alpha A^\alpha \wedge \pi^\ast[D^{\rm b}_\alpha]
\eea
identifies the gauge group in the M-theory effective action in $\mathbb R^{1,8-2n}$ as $U(1)^{h^{1,1}(B_n) + 1}$.

According to the general paradigm summarized in section \ref{sec_Mtheorytoell}, this theory is related to the effective theory of F-theory on $Y_{n+1}$ in $\mathbb R^{1,9-2n}$ by compactifying the latter on a circle $\tilde S^1_B$.
Before (wrongly!) concluding that the F-theory gauge group should be $U(1)^{h^{1,1}(B_n) + 1}$, however, note that 
a vector field on $\mathbb R^{1,8-2n}$ can have several origins from the perspective of the F-theory effective action.
A careful explanation of this relation can be found in \cite{Grimm:2010ks}. 

Indeed, the vectors $A^\alpha$ must be interpreted as the dimensional reduction of the 2-form fields $b^{(2)}_\alpha$ which are obtained in Type IIB/F-theory  on $B_n$ by expanding
\bea
C_4 = \sum_\alpha b_{(2)}^\alpha \wedge [D_\alpha^{\rm b}] \,.
\eea
More precisely, $A^\alpha$ is obtained by reducing $b_{(2)}^\alpha = A^\alpha \wedge [e_1]$ with $[e_1]$ the 1-form along $\tilde S^1_B$.\footnote{The tensor $b^{(2)}_\alpha$ with no legs along  $\tilde S^1_B$ becomes a tensor field in M-theory, which can be obtained by reduction of the gauge potential $C_6$ magnetically  dual to $C_3$ in M-theory.}
Hence the subset of $h^{1,1}(B_n)$ vector fields $A^\alpha$ in M-theory correspond to tensors in the dual F-theory effective action and are not related to a 1-form gauge symmetry in F-theory.
In compactifications to six dimensions, $h^{1,1}(B_n) - 1$ of these tensors in F-theory sit in anti-self dual tensor multiplets, and the remaining tensor is self-dual and part of the gravitational multiplet. In compactifications to four dimensions, the tensors are dual to axionic scalars which complexify  the $h^{1,1}(B_n)$ K\"ahler moduli. In F-theory on K3, the tensor from the base $B_1 = \mathbb P^1$ sits again in the 8d gravitational multiplet.

The interpretation of the vector field $\tilde A^0$, on the other hand, is rather different: 
It describes the Kaluza-Klein $U(1)$ gauge field which appears universally in the circle reduction along $\tilde S^1_B$ from F-theory to M-theory. This M-theory gauge potential becomes part of the metric in the dual F-theory.

To understand this statement, note that the objects charged electrically under the abelian gauge fields in M-theory are M2-branes wrapping holomorphic or anti-holomormphic curves on $Y_{n+1}$.
A wrapped M2-brane corresponds to a particle in the M-theory effective action of mass
\bea \label{mcformula}
| m(C) | \simeq {\rm vol(C)} =  | \int_C J |
\eea
with $J$ the K\"ahler form of $Y_{n+1}$. The second equality follows from the fact that $C$ is holomorphic or anti-holomorphic.
 If we collectively denote the abelian gauge fields in the expansion (\ref{C3exp1}) as 
\bea
C_3 = \sum_j A^j \wedge w_j
\eea
then in view of the coupling (\ref{SM2-gen}) the $U(1)_j$ charge of an M2-brane along a curve $C$ is
\bea
q_j = \int_C w_j = [C] \cdot w_j \,.
\eea
Consider now an M2-brane wrapping the generic fiber $\mathbb E_\tau$ with wrapping number $n$. Its charges are
\bea \label{KKcharges}
q_0 &=& [\tilde S_0] \cdot n [\mathbb E_\tau] = n \\
q_\alpha &=& \pi^\ast[D^{\rm b}_\alpha]  \cdot n  [\mathbb E_\tau]  = 0 \,.
\eea
These M2-branes give rise to massive states in the M-theory effective action with mass $|m| \simeq V$ with $V$ the volume of the generic fiber. Such states become massless in the limit $V \rightarrow 0$, which coincides with the decompactification limit of the circle $\tilde S^1_B$. They must therefore be interpreted as Kaluza-Klein states. More precisely, each of the supergravity fields of the F-theory vacuum decomposes, upon circle reduction on $\tilde S^1_B$, into a Kaluza-Klein zero mode $\psi_0$ and a tower of Kaluza-Klein states $\psi_n$ with $n \in \mathbb Z$. The zero-modes are to be matched with zero-modes of the supergravity reduction of the dual M-theory, while the Kaluza-Klein tower is associated with M2-branes wrapping $n [\mathbb E_\tau]$ with $n \in \mathbb Z$;  a negative wrapping number corresponds to negative orientation in the wrapping of the holomorphic curve  $[\mathbb E_\tau]$. The charges (\ref{KKcharges}) are in agreement with the identification of $ [\tilde S_0] $ as the generator of the Kaluza-Klein $U(1)$. The claim that M2-branes wrapping a curve in the class $n [\mathbb E_\tau] $ indeed reproduce the full spectrum of Kaluza-Klein towers can be justified further by computing the Gromov-Witten invariants for the curve class $n [\mathbb E_\tau] $. The Gromov-Witten invariants compute the Euler number of the moduli space of holomorphic curves in the given class. As found in \cite{Klemm:1996hh}, for the curve class $[\mathbb E_\tau]$ this Euler number indeed agrees with a suitable index of supergravity states in F-theory.

The requirement that  $[D]\cdot n [\mathbb E_\tau] = n$  for a divisor associated with the Kaluza-Klein $U(1)$ alone does not fix the divisor class $[D]$ uniquely because every vertical divisor has vanishing intersection with $[\mathbb E_\tau]$.
The reason for the exact definition as $[\tilde S_0]$  is explained in \cite{Park:2011ji,Bonetti:2011mw} (see also \cite{Grimm:2011sk}) by a detailed match of the M-theory and the F-theory effective action. A special role in this match is played by the Chern-Simons couplings: Dimensionally reducing the topological coupling $\int_{\mathbb R^{1,10}} C_3 \wedge G_4 \wedge G_4$ in (\ref{Mtheoryaction}) gives rise to Chern-Simons couplings in the M-theory effective action. For instance, in M-theory compactifications on $Y_3$, these are of the form 
\be
S_{\rm CS}   = \kappa^{(M)}_{\Lambda \Gamma \Sigma }\int_{\mathbb R^{1,4} } A^\Lambda \wedge F^\Gamma \wedge F^\Sigma\, \qquad \Lambda  = 0, \alpha
\ee
with $A^\Lambda$ and $F^\Lambda$ the abelian vectors and their field strengths as obtained from the M-theory reduction. On the other hand, if one reduces the dual F-theory effective action on a circle, similar such Chern-Simons terms, with coefficients  $\kappa^{( F)}_{\Lambda \Gamma \Sigma }$, are induced as a quantum effect which is exact at 1-loop in perturbation theory \cite{Witten:1996qb, Intriligator:1997pq,Aharony:1997bx}.
The requirement that the classical M-theory Chern-Simons terms and the 1-loop induced Chern-Simons  terms of the circle reduction agree fixes the correct normalization of the abelian gauge fields \cite{Park:2011ji,Bonetti:2011mw}. Such Chern-Simons terms have been considered in great detail in the more recent F-theory literature \cite{Grimm:2011fx,Bonetti:2013ela,Bonetti:2013cza}.

Coming back to the physical interpretation of F-theory on the smooth Weierstrass model, we conclude that none of the $1 + h^{1,1}(B_n)$ abelian gauge fields in the M-theory effective action on $Y_{n+1}$ uplifts to a gauge field in F-theory, in agreement with our findings in section \ref{sec_PhysicsIIB1}. 
To describe an F-theory vacuum with non-trivial gauge algebra we therefore need to move on to more complicated elliptic fibrations with extra elements in $H^{1,1}(Y_{n+1})$. 
Given a divisor ${\cal D}$ with associated cohomology class $[\cal D]$, we can already give the criterion for the vector field $A^{\cal D}$ appearing in the reduction $C_3 = A^{\cal D} + \ldots \wedge [{\cal D}]$ to correspond to a gauge field in F-theory: The class $[\cal D]$ must satisfy the transversality conditions
\bea
 [{\mathfrak D}]     \cdot [S_0] \cdot  \pi^\ast (w^{\rm b}_{2n-2}) = 0& \qquad         &\forall \quad w^{\rm b}_{2n-2} \in H^{2n-2}( B_{n}) \label{transversal1}\\
 {}  [{\mathfrak D}]  \cdot  \pi^\ast (w^{\rm b}_{2n}) = 0                       &         \qquad &\forall \quad w^{\rm b}_{2n} \in H^{2n}( B_{n})  \label{transversal2}  \,.
\eea
Condition (\ref{transversal1}) is the statement that the intersection number between  $[{\mathfrak D}] $ and any curve class on the base vanishes. M2-branes wrapping such curve classes uplift, as we have just learned, to D3-branes wrapping the same curve in F-theory (more precisely, they correspond to the string obtained by wrapping a D3-brane on a base curve in F-theory and wound along the $S^1$ in going from F to M-theory).
The abelian gauge potential in M-theory with respect to which these states are charged uplift to tensor fields in F-theory.
Condition (\ref{transversal2}) ensures that the intersection number with the class of the full generic fiber $[\mathbb E_\tau]$ vanishes; since M2-branes wrapping the latter are Kaluza-Klein-modes in the reduction from F- to M-theory and 
if (\ref{transversal2}) is not imposed, the M-theory $U(1)$ has admixture from the Kaluza-Klein $U(1)$. 
Divisor classes satisfying both transversality conditions can be either due to extra sections  - see section \ref{sec_MWGROUP} -  or due to a more severe enhancement of the singularity in codimension one.

Before coming to this let us note how to implement the F-theory limit (\ref{MFDuality}) of vanishing fiber volume in more detail geometrically.
For definiteness we focus on F-theory compactifications on Calabi-Yau 4-folds. The key idea is that as the fiber volume shrinks to zero, the volume of the base must be scaled up in such a way that the volume of the vertical divisors stays constant  \cite{Witten:1996bn}.
Otherwise, if the volume of a pullback divisor $\pi^{-1}(D^{\rm b})$ would go to zero, the contribution of an M5-brane instanton wrapping this divisor to the effective action would be unsuppressed. This is clearly an unphysical result: The M5-brane instantons along $\pi^{-1}(D^{\rm b})$ should rather be matched to the non-perturbative effects which in Type IIB language are due to Euclidean D3-branes wrapping a divisor on the base \cite{Witten:1996bn}. 
Let us expand the K\"ahler form of $Y_{n+1}$ as
\bea
J = J^0 [S_0] + J^\alpha \,  \pi^{\ast}[D_{\alpha}^{\rm b}] \,.
\eea
Since $J^0$ measures the volume of the fiber, we should rescale it to zero. 
The volume of $\pi^{-1}(D^{\rm b})$ is proportional to  $J^3  \cdot \pi^{\ast}[D^{\rm b}]$. This stays finite in the F-theory limit if we rescale
\bea \label{Frescaling4d}
J^0 \rightarrow \epsilon \, J^0  \,, \qquad J^\alpha \rightarrow \epsilon^{-1/2} \, J^\alpha \,,
\eea
where the 4d F-theory limit is obtained by taking $\epsilon \rightarrow 0$.
This rescaling is explained very carefully from a supergravity point of view in \cite{Grimm:2010ks} for F-theory on Calabi-Yau 4-folds and in \cite{Bonetti:2011mw} for F-theory on Calabi-Yau 3-folds.

\section{Codimension-one singularities and non-abelian gauge algebras} \label{sec_singcodim1}

In this section we describe F-theory compactifications with a non-trivial non-abelian gauge algebra.
We begin in section \ref{sec_KodairaNeron} by reviewing the classification of codimension-one singularities on elliptically fibered surfaces due to Kodaira and N\'eron. The classification assigns to the singular fibers a simply laced, i.e. A-D-E Lie algebra in a natural way.
The singularities can be understood both at the level of the Weierstrass model, and in terms of its resolution.
On higher-dimensional elliptic fibrations, monodromies can affect the global structure of codimension-one singular fibers, leading also to non-simply laced Lie algebras, as discussed in section \ref{sec_Codimoneandgroup}.
The physics interpretation of these Lie algebras as the gauge algebra on a stack of 7-branes is derived in section \ref{sec_Nonabeliangaugealgebras}. The process of resolving the singular Weierstrass model corresponds to moving in the Coulomb branch of the dual M-theory compactification, as described in section \ref{sec_Coulomb}. We provide an example and discuss the significance of global Tate models in section \ref{subsec_TateI2-1}. The counting of massless matter along 7-branes is the topic of section \ref{subsec_Zeromodeson7brane}.

\subsection{The classification of Kodaira and N\'eron} \label{sec_KodairaNeron}

Consider a Weierstrass model and allow for its
sections $f$ and $g$ to be non-generic polynomials of their given degree. Depending of the non-generic form of $f$ and $g$, the vanishing order of the discriminant polynomial $\Delta$ increases and the singularity type in the fiber over $\Delta$ enhances. As a result, the Weierstrass model becomes singular itself.
In this section we study the structure of singularities appearing over {\it generic}  points on the discriminant divisor, called codimension-one singularities.
For elliptic surfaces, the codimension-one fibers have been classified in seminal work by Kodaira \cite{Kodaira2} and N\'eron \cite{Neron}. This classification carries over, {\it mutatis mutandis}, to 
codimension-one singularities on higher dimensional Weierstrass models.

To appreciate the meaning of the classification, note that if 
 $Y_{n+1}$ is singular, one can consider instead its resolution $\hat Y_{n+1}$. Mathematically, a resolution $\hat X$ of a singular variety $X$ is a morphism
\bea
\rho: \hat X \rightarrow X
\eea 
such that $\hat X$ is smooth and is isomorphic to $X$ away from the singular loci of $X$.
A resolution $\hat Y_{n+1}$ of a singular Weierstrass model is again an elliptic fibration since its generic fibers are elliptic curves. By slight abuse of notation, we stick to the same symbol for the projection map
\bea
\pi :\quad \mathbb{E}_\tau \ \rightarrow & \  \ \hat Y_{n+1} \cr 
& \ \ \downarrow \cr 
& \ \  B_n \,.
\eea 
However, the fibers of $\hat Y_{n+1}$ over the discriminant locus are degenerate, and the precise form of these degenerate fibers is one way to characterise the original singularity of the Weierstrass model.

Kodaira and N\'eron classified the possible degenerate fibers occurring in a smooth {\it minimal} elliptic surface $\hat Y_2$ (not necessarily Calabi-Yau). The smooth surface $\hat Y_2$ can be thought of as the resolution of a singular Weierstrass model $Y_2$. 
Minimality means that $\hat Y_{2}$ contains no $(-1)$-curve, i.e. no curve of self-intersection $C \cdot C = -1$. On a complex surface such $(-1)$ curves are the only curves which can be blown down to a smooth point without changing the canonical bundle of the surface. Hence minimality is the requirement that $\hat Y_2$ cannot be obtained from another surface by blowing up a smooth point. 
Note that a surface which is Calabi-Yau, i.e. topologically of type K3, can never contain a $(-1)$ curve.
Therefore the classification of minimal smooth surfaces provides for us, in particular, a classification of the possible singularities of a Weierstrass model of complex dimension two with $c_1(Y_2)=0$ which admit a resolution $\hat Y_2$ which is still Calabi-Yau. Resolutions which do not change the canonical bundle are called crepant. 

The result of this classification is as follows: Apart from a few outliers, the degenerate fibers of $\hat Y_2$ take the form of the extended Dynkin diagrams associated with the Lie algebras of type A-D-E.
This means that the fibers are unions of rational curves 
\be
\mathbb P^1_i, \qquad  i=0,1, \ldots, {\rm rk}(\mathfrak g) \,,
\ee
with $\mathfrak g$ one of the A-D-E Lie algebras. The extended node $\mathbb P^1_0$ is singled out as the fiber component intersected once by the zero-section $S_0$. The rational curves $\mathbb P^1_i$ intersect like the nodes of the affine Dynkin of $\mathfrak g$. The curves appear with a multiplicity $a_i$ which coincides with the dual Kac label (or co-mark) of the corresponding node. In particular, the classes of the generic fiber and  of the degenerate fiber components are related as
\bea \label{EtauP1relation}
[\mathbb E_{\tau}] = \sum_{i=0}^{{\rm rk}(\mathfrak g)} a_i \, [\mathbb P^1_i]
\eea
with $a_0=1$.  
The few outliers mentioned above are the Kodaira type $III$, corresponding to two rational curves touching each other in an intersection point of order two, and type $IV$, consisting of three rational curves intersecting in one point. 

Contracting all fiber components but $\mathbb P^1_0$, which is intersected by the zero-section $S_0$, to a point corresponds to the blowdown from $\hat Y_2$ to the singular elliptic surface $Y_2$. The singularity sits at a single point in the fiber away from the zero-section, and the singularity is a hypersurface singularity whose local equation is of A-D-E type. This is true also for the outliers of type $III$ and type $IV$, which correspond to hypersurface singularities of type $A_1$ and $A_2$, respectively. 

The vanishing orders for $f$, $g$ and $\Delta$ leading to the various types of Kodaira fibers are listed in table \ref{KTtable}.
For the singular fibers on elliptic surfaces, we are to ignore the monodromy cover in column 6, and the Lie algebra $\mathfrak{g}$ associated with the fiber is the maximal one in column 7 (corresponding to the maximally split case).

\begin{table}
\begin{tabular}{|c|c|c|c|c|c|l|c|}
\hline
type & ${\rm ord}(f)$ &  ${\rm ord}(g)$ &  ${\rm ord}(\Delta)$ & sing.      & monodromy cover & $\mathfrak{g}$  & split  \\ \hline \hline 
  $I_0$ &        $ \geq 0 $            &                $\geq 0$      &                      $0$        & $-$ &   $  -$                        &    $- $  & \\ \hline
  $I_1$ &        $0$             &                $0$      &                      $1$        &$- $ &     $-  $                      &   $ -$ &  \\ \hline 
  $II$ &          $ \geq1$             &                $1$      &                      $2$         &$-$&      $- $                       &    $-$&  \\ \hline 
   $III$ &        $1$             &                $\geq 2$      &                      $3$         &$A_1$&           $ -$                   &   $\mathfrak{su}(2)$  &    \\ \hline 
  \multirow{2}{*}{  $IV$} &      \multirow{2}{*}{  $ \geq 2$  }           &            \multirow{2}{*}{    $2$  }    &              \multirow{2}{*}{        $4$   }      &\multirow{2}{*}{$A_2$}&   \multirow{2}{*}{  $ \psi^2 - \frac{g}{w^2}|_{w=0}$    }                    & 1-comp: $\mathfrak{sp}(1)$ &  $IV^{ns}$ \\ 

        &&&&&& 2-comp:  $\mathfrak{su}(3)$ &$IV^s$ \\ \hline
\multirow{2}{*}{$I_m$} &      \multirow{2}{*}{  0      }       &             \multirow{2}{*}{   0   }   &                   \multirow{2}{*}  { $m$ }     &      \multirow{2}{*}  { $A_m$ }     &      \multirow{2}{*}  {  $\psi^2 + \frac{9 g}{2 f}|_{w=0} $     }         &  1-comp:  $\mathfrak{sp}([\frac{m}{2}])$&  $I_m^{ns}$ \\ 
           &&&&& &   2-comp:   $ \mathfrak{su}(m)$& $I_m^{s}$ \\ \hline

\multirow{3}{*}{$I_0^\ast$} &      \multirow{3}{*}{  $\geq 2 $     }       &             \multirow{3}{*}{  $ \geq 3 $  }   &                   \multirow{3}{*}  { $6$ }     &      \multirow{3}{*}  { $D_4$ }     &      \multirow{3}{*}  {  $\psi^3 + \psi  \frac{f}{w^2}|_{w=0}  +  \frac{g}{w^3}|_{w=0}$     }         &  1-comp:  $\mathfrak{g}_2$ &$I_0^{\ast ns}$\\ 
           &&&&& &   2-comp:   $ \mathfrak{so}(7)$ & $I_0^{\ast ss}$ \\ 
           &&&&& &   3-comp:   $ \mathfrak{so}(8)$ & $I_0^{\ast s}$ \\  \hline

 $I^\ast_{2n-5}$, &      \multirow{2}{*}{  $2$      }       &             \multirow{2}{*}{   $3$   }   &                   \multirow{2}{*}  { $2n+1$ }     &      \multirow{2}{*}  { $D_{2n-1}$ }     &      \multirow{2}{*}  {  $\psi^2 +\frac{1}{4} (\frac{\Delta}{w^{2n+1}})(\frac{2 w f}{9 g})^3|_{w=0} $     }         &  1-comp:  $\mathfrak{so}(4n-3)$ &$I_{2n-5}^{\ast ns}$ \\ 
   $n\geq 3$        &&&&& &   2-comp:   $ \mathfrak{so}(4n-2)$ &$I_{2n-5}^{\ast s}$  \\ \hline

$I^\ast_{2n-4}$, &      \multirow{2}{*}{  $2$      }       &             \multirow{2}{*}{   $3$   }   &                   \multirow{2}{*}  { $2n+2$ }     &      \multirow{2}{*}  { $D_{2n}$ }     &      \multirow{2}{*}  {  $\psi^2 + (\frac{\Delta}{w^{2n+2}}) (\frac{2 w f}{9 g})^2|_{w=0} $     }         &  1-comp:  $\mathfrak{so}(4n-1)$ & $I_{2n-4}^{\ast ns}$\\ 
    $n\geq 3$       &&&&& &   2-comp:   $ \mathfrak{so}(4n)$ &$I_{2n-4}^{\ast s}$ \\ \hline         
           
       \multirow{2}{*}{  $IV^\ast$} &      \multirow{2}{*}{  $ \geq 3$  }           &            \multirow{2}{*}{    $4$  }    &              \multirow{2}{*}{        $8$   }      &\multirow{2}{*}{$E_6$}&   \multirow{2}{*}{  $ \psi^2 - \frac{g}{w^4}|_{w=0}$    }                    & 1-comp: $\mathfrak{f}_4$ &$IV^{\ast ns}$   \\ 
 &&&&& &   2-comp:   $ \mathfrak{e}_6$ &$IV^{\ast s}$ \\ \hline

   $III^\ast$ &       $ 3$             &            $\geq 5$      &             $9$        &$E_7$&   $-$                 & $\mathfrak{e}_7$   & \\ \hline

   $II^\ast$ &       $ \geq 4$             &            $5$      &             $10$        &$E_8$&   $-$                 & $\mathfrak{e}_8$  &  \\ \hline

   non-min. &       $ \geq 4$             &            $\geq 6$      &             $\geq 12$        & non-can. &   $-$                 & $-$  &  \\ \hline

 \end{tabular}
\caption{Kodaira-Tate table for singular fibers of the Weierstrass model. The monodromy cover is taken from \cite{Grassi:2011hq}, Table 4. The gauge algebra depends on the number  of its irreducible components as indicated. In the last column, the superscript $s$ ('split'), $ns$ ('non-spilt'), $ss$ ('semi-split') refers to the refined Tate fiber type.  \label{KTtable}}
\end{table}

From the table we read off that $f$, $g$ and $\Delta$ must vanish to the given order along a codimension-one locus on the base. 
If we describe this divisor locally by the vanishing of a local coordinate  $w=0$, then except for all
 Kodaira type fibers except the  $I_m$ series, $f$ and $g$ must factorise (in the given patch) as 
\bea
f = w^k \, \tilde f, \qquad g = w^l \, \tilde g \,\quad \quad \quad  k = {\rm ord}(f), \,l = {\rm ord}(g)   \,.
\eea
Here $\tilde f$ and $\tilde g$ are sufficiently generic such that the discriminant has the prescribed vanishing order.
As a result of the specialiation of $f$ and $g$, the discriminant polynomial factorises as 
\bea
\Delta = w^m  \, \Delta_0\, \qquad m = {\rm ord}(\Delta) \,.
\eea
In particular, the value $m$ in the table is the one obtained for completely generic $\tilde f$ and $\tilde g$. 
The $I_m$ series is more complicated: Even though $f$ and $g$ have no zeroes at $w=0$, they are of a non-generic form such that cancellations in $\Delta$ lead to a zero of order $m$. 
The general procedure is explained for instance in \cite{Katz:2011qp}: The starting point is a general ansatz (valid locally near the divisor)
\bea \label{ffiggi}
f = \sum_i f_i \, w^i, \qquad g= \sum_i g_i \,  w^i
\eea
with generic $f_i$ and $g_i$. The associated discriminant takes the form
\bea
\Delta = (4 f_0^3 + 27 g_0^2) + (12 f_1 f_0^2 + 54 g_0 g_1) w + {\cal O}(w^2) \,.
\eea
To obtain an $I_1$ fiber we need to choose $f_0$ and $g_0$ such that the term in the first brackets vanishes. 
The possible types of tunings leading to this behaviour may depend on whether or not the divisor is smooth.
For instance, if the divisor $w=0$ is non-singular, \cite{Katz:2011qp} shows that one can locally find a function $u_0$ such that 
\bea
f_0 = - \frac{1}{3} u_0^2 + {\cal O}(w), \qquad g_0 = \frac{2}{27} u_0^3 + {\cal O}(w) \,,
\eea
hence tuning an $I_1$ singularity. This procedure can be repeated order by order in $w$ to arrive at the higher $I_m$ types.

If ${\rm ord}(f) \geq 4$ and at the same time 
${\rm ord}(g) \geq 6$ (implying that ${\rm ord}(\Delta) \geq 12$), no {\it minimal} smooth elliptic surface $\hat Y_2$ exists. In particular, there exists no crepant resolution of a Weierstrass model with this property. The singularity type in the fiber is non-canonical (see section \ref{sec_terminalco2} for a definition) and sits at infinite distance in moduli space. In this case one can use the freedom to rescale the homogeneous coordinates of the Weierstrass model such as to arrive at a new Weierstrass model for which $f$ and $g$ satisfy the minimality bound. 
Concretely, suppose that 
\bea
f =  w^{4n} \, \tilde f, \qquad g = w^{6n}  \, \tilde g
\eea
with $n \in \mathbb N$ such that $\tilde f$ and $\tilde g$ are holomorphic sections whose vanishing orders $({\rm ord}(\tilde f)|_{w=0}, {\rm ord}(\tilde g)|_{w=0})$ do not equal or exceed $4$  and $6$.
The scaling assignment (\ref{calLLz}) allows us to rescale
\bea \label{xyfgshift}
x \rightarrow x \,  w^{-2n} =:\tilde x, \qquad y \rightarrow  y \,  w^{-3n} =: \tilde y, \qquad f \rightarrow  f \,  w^{-4n} = \tilde f, \qquad g \rightarrow g \,  w^{-6n}  = \tilde g \,.
\eea
As a result, the Weierstrass equation becomes
\bea
\tilde y^2 = \tilde x^3 + \tilde f \, \tilde x \, z^4 + \tilde g \, z^6 \,,
\eea
and satisfies the minimality constraint. However, the operation (\ref{xyfgshift}) corresponds to a shift of the defining bundle ${\cal L}$ appearing in  (\ref{calLLz}) to $\tilde {\cal L}  = {\cal L} \otimes W^{-n}$, where $W$ denotes the line bundle with $c_1(W) = [\{w=0\}]$.
If $c_1({\cal L})  = \bar K_{n}$ in the original Weierstrass model according to the Calabi-Yau condition, then this condition is no longer satisfied by the new, minimal Weierstrass model.
In this sense, vanishing orders of $f$ and $g$ of $4$ and $6$ and beyond are really incompatible with the Calabi-Yau condition. 

Finally, let us note that the monodromies around the location of the singular fibers coincide with the monodromies computed in section \ref{sec_pq }, especially eqns. (\ref{So8Ek}) - (\ref{E8mono}); the monodromies of fibers of Type $II$, $III$, $IV$ are the inverse of those of $II^\ast$, $III^\ast$, $IV^\ast$.

\subsection{General structure of codimension-one fibers and relation to group theory} \label{sec_Codimoneandgroup}

The results of Kodaira and N\'eron carry over to elliptic fibrations over higher dimensional base spaces as follows:
Over {\it generic} points of the discriminant divisor  $\Sigma$ (i.e. in codimension one), a smooth elliptic fibration $\hat Y_{n+1}$ compatible with the requirement of flatness (or equi-dimensionality) of the projection $\pi$ allows for {\it local} fiber types of the same form as in the case $n=1$. 
Globally along $\Sigma$, however, a new effect occurs for $n >1$ in that the fibers may undergo monodromies. Taking into account these monodromies corresponds to folding the A-D-E dynkin diagram associated with the local fibers such as to produce the Dynkin diagrams associated with non-simply laced Lie algebras of type $B_n$, $C_n$, $G_2$ and $F_4$. 

Whether or not the monodromy occurs can be determined already at the level of the singular Weierstrass model $Y_{n+1}$, i.e. without considering a resolution $\hat Y_{n+1}$. This is the content of Tate's algorithm \cite{Tate}, as explained in the physics literature in \cite{Bershadsky:1996nh,Grassi:2011hq,Katz:2011qp}. 
The last two columns in table \ref{KTtable}  summarize the explanation of Tate's algorithm in \cite{Grassi:2011hq}, to which we refer for a derivation. From the table we infer that the existence of monodromy depends on whether or not a certain monodromy cover factorises globally.
Consider e.g. the case of a Type $IV$ singularity: The monodromy cover to consider is associated with the equation
\bea
\psi^2  - \frac{g}{w^2}|_{w=0} = 0  
\eea
with $\psi$ a formal variable.
Locally around the divisor $w=0$ carrying the  type $IV$ singularity we can make the ansatz (see column 3)
\bea
g = g_2 w^2 + g_3 w^3 + \ldots \,.
\eea
For generic $g_2$, the monodromy cover takes the form
\bea
\psi^2  - \frac{g}{w^2}|_{w=0} = \psi^2 - g_2 \,.
\eea
Above equation does not factorise (it is '1-component' in the notation of the table). This indicates, according to the general algorithm, a monodromy in the fiber over $w=0$ corresponding to the breaking $\mathfrak{su}(3) \rightarrow \mathfrak{sp}(1)$.
In less generic situations, more precisely if 
$g_2 = \phi^2$, the monodromy cover takes the form
\bea
\psi^2  - \frac{g}{w^2}|_{w=0} = \psi^2 - \phi^2 =( \psi+ \phi)( \psi- \phi) 
\eea
and hence factorises into two components (called '2-comp' in the table). This means that there is indeed no monodromy in the fiber. 
This criterion can be applied to all other cases. For $I_0^\ast$ one has to distinguish three situations, depending on whether the monodromy cover factorises into 3 components (no monodromy, called split fiber), into two components (monodromy  $\mathfrak{so}(8) \rightarrow \mathfrak{so}(7)$ with so-called semi-split fibers) or does not factorise at all (monodromy  $\mathfrak{so}(8) \rightarrow \mathfrak{g}_2$, called non-split fiber).
Elliptic fibrations with non-simply laced algebras and their resolutions have been worked out in detail in particular in \cite{Esole:2017rgz,Esole:2017qeh,Esole:2018mqb}, and \cite{Grassi:2018wfy} provides an in-depth study of the realisation of such algebras in terms of $(p, q)$ strings.  

The information about the monodromies is automatically included in a slightly different representation of the Weierstrass model as a global Tate model, which is, however, not always possible to obtain {\it gobally} \cite{Bershadsky:1996nh,Katz:2011qp}. More details will be given in section \ref{subsec_TateI2-1}, and the interested reader can directly jump to this section. 

Depending on the base space, it may happen that even the most generic choice of polynomials of degree $4 \bar K_{B_n}$ and $6 \bar K_{B_n}$ for $f$ and $g$ unavoidably leads to a singularity.
The underlying reason is that in such geometries $f$ and $g$ necessarily vanish to a certain order along one or even several divisors on the base. For a base of complex dimension two, the resulting types of singularities have been completely classified in \cite{Morrison:2012np} and are called 'non-Higgsable' clusters. The same phenomenon on Calabi-Yau fourfolds \cite{Morrison:2014lca,Halverson:2015jua} is still far less systematically understood, but it seems to be a generic property of base spaces for elliptic fibrations \cite{Taylor:2015ppa,Halverson:2017ffz,Taylor:2017yqr,Wang:2018rkk}, as discussed at this TASI school in \cite{Halverson:2018xge}.

Let us now analyze the appearance of monodromy in the resolution $\hat Y_{n+1}$ of the Weierstrass model.
An early systematic study includes \cite{Aspinwall:2000kf}, and many more details and derivations can be found in \cite{Park:2011ji}.
We need to carefully distinguish between the component curves of the local fiber and the resolution divisors.
In the most general situation the discriminant $\Delta$  factorises as 
\bea \label{DeltaFactorisation}  
 \Delta = \Delta_0  \, \prod_{I=1}^N (\Delta_I)^{p_I} \,,
  \eea
  where $\Delta_0$ and $\Delta_I$ describe irreducible polynomials and 
  $p_I$ is the multiplicity with which the polynomial $\Delta_I$ appears.
 The vanishing locus of each of the polynomials $\Delta_0$ and $\Delta_I$ corresponds to an irreducible divisor on $B_n$ which we denote by $\Sigma_0$ and $\Sigma_I$. Their union gives the discriminant divisor 
 \bea \label{Simgadef}
 \Sigma = \Sigma_0 \cup \Sigma_1 \cup \ldots \cup \Sigma_N  \,.
 \eea
 An example will be given in (\ref{fgDeltaI_2}) and (\ref{fgDeltaI_3}) in section \ref{subsec_TateI2-1}, which can be read in parallel.
 
 Over generic points of $\Sigma_0$ the fiber is of Kodaira type $I_1$, and no monodromies are to be considered.
 Suppose that over generic points of $\Sigma_I$
  the topology of a fiber is {\it locally} associated with the affine Dynkin diagram of an A-D-E Lie algebra $\tilde{\mathfrak{g}}_I$ and denote the components of the local fiber as $\tilde{\mathbb P}^1_{a_I}$, $a_I=0,1, \ldots, {\rm rk}(\tilde{\mathfrak{g}}_I)$. Monodromies along $\Sigma_I$ may map some of these local components to one another.
We make the following definitions:
\bea
&\tilde{\mathbb P}^1_{a_I}: &\text{the components of the {\it local} fiber} \\
&\mathbb P^1_{i_I}: &  \text{the {\it independent} rational curves in the fiber}  \\
&C_{i_I}: & \text{the {\it invariant orbits} of fiber components}
\eea
If a curve $\tilde{\mathbb P}^1_{k_I}$ is invariant by itself under monodromies, the corresponding orbit contains only this invariant rational curve. In particular, the fibre component $\tilde{\mathbb P}^1_{0_I}$ intersected by the zero-section forms an orbit by itself.
More generally, an orbit is a union of two or even three\footnote{This occurs for monodromies folding ${\mathfrak{so}}(8)$ to $\mathfrak{g}_2$.} rational curves which are related by monodromies. 
Fibering the orbits $C_{i_I}$ over $\Sigma_I$ produces a divisor $E_{i_I}$ of $\hat Y_{n+1}$ sometimes called {\it resolution divisor} or {\it Cartan divisor}. Here the label $i_I$ takes values $0, 1, \ldots, {\rm rk}({\mathfrak g}_I)$ with $\mathfrak{g}_I$ an in general non-simply laced simple Lie algebra.

To summarise:
\begin{itemize}
\item
In the simply-laced case (i.e. in absence of monodromies) the resolution divisors $E_{i_I}$ are rationally fibered over $\Sigma_I$, and their fibers are the rational curves $\mathbb P^1_{i_I}$, $i_I = 0, 1, \ldots, {\rm rk}({\mathfrak g}_I)$.
\item
In the non-simply laced case, the fibers of the resolution divisors $E_{i_I}$ are  invariant orbits $C_{i_I}$, $i_I = 0, 1, \ldots, {\rm rk}({\mathfrak g}_I),$ of rational curves. Over any given point on $\Sigma_I$, the invariant orbit $C_{i_I}$ splits into several rational curves $\tilde{\mathbb P}^1_{a_I}$, which are transformed into one another by global monodromies.
The locally defined rational curves $\tilde{\mathbb P}^1_{a_I}$, $a_I = 0, 1, \ldots, {\rm rk}(\tilde{{\mathfrak g}}_I),$ intersect like the Dynkin diagram of a simply laced Lie algebra $\tilde{\mathfrak{g}}_I$, which is a covering algebra of the non-simply laced algebra ${\mathfrak g}_I$.
The resolution divisors are, in this case, not themselves rationally fibered over $\Sigma_I$, but only over a branched cover $\Sigma_I'$ of $\Sigma_I$. Furthermore, the independent rational curves (not necessarily invariant) are denoted by $\mathbb P^1_{i_I}$, $i_I = 0, 1, \ldots, {\rm rk}({\mathfrak g}_I)$.
\end{itemize}

The key observation is that the intersection structure of the resolution divisors with one another and with the independent fibre components $\mathbb P^1_{j_J}$ encode the Lie algebra $\mathfrak{g}_I$  in the sense that
\bea 
[E_{i_I}] \cdot [E_{j_J}] \cdot \pi^*(\omega_{2n-2}) &=&  - \delta_{IJ} \,  {\mathfrak C}_{i_I j_I} \,  [ \Sigma_I] \cdot_{B_n} \omega_{2n-2}  \qquad   \forall \,  \omega_{2n-2}  \in H^{2n-2}(B_n) \label{fundrelEalpha1}    \\
{} [E_{i_I}] \cdot  [\mathbb P^1_{j_J}] &=& - \delta_{IJ} \, C_{i_I j_I} \label{fundrelEalpha2} \\
{} [S_0] \cdot  [\mathbb P^1_{j_J}] &=& 0 \label{fundrelSalpha2}   \,.
\eea
Here $C_{i_I j_I}$ is the Cartan matrix of $\mathfrak{g}_I$ and ${\mathfrak C}_{i_I j_I}$ is related to this object as in (\ref{calCartan}) below.  
Furthermore, the generalisation of (\ref{EtauP1relation}) is the relation
\bea \label{piSigmaI}
\pi^{-1} (\Sigma_I)=  E_{0_I} + \sum_{i_I = 1}^{{\rm rk}(\fg_I)}  a_{i_I} E_{i_I}
\eea
between the resolution divisors and the vertical divisor $\pi^{-1}(\Sigma_I)$.
These general facts can be proven by carefully analyzing the fibral intersection theory for the case of each different Lie algebra.

The intersection numbers (\ref{fundrelEalpha1}) and   (\ref{fundrelEalpha2}) suggest that we should identify the resolution divisors $E_{i_I}$ with the coroots of the algebra $\mathfrak{g}_I$ and the fibral curves $\mathbb P^1_{i_I}$ with (the negative of ) the simple roots. 
Since this  identification lies at the heart of all that follows, let us take a step back and briefly review these group
 theoretic concepts in a small

\subsubsection*{Group Theoretic Interlude (I):}

Consider a simple Lie algebra $\mathfrak{g}$. Its Cartan subalgebra $\mathfrak{h}$ is the maximal commuting subalgebra and is generated by the 
\bea
{\rm Cartan \,\,  generators}: \quad  { T}_{\cal I}, \quad {\cal I}=1, \ldots, {\rm rk}({\mathfrak{g}}),
\eea
normalised such that 
\bea \label{canonnorm}
{\rm tr}_{\rm fund}(T_{\cal I} T_{\cal J}) = \delta_{\cal IJ} \,.
\eea
We can now find a basis $\{ {T}_{\cal I}, e_\alpha \}$ of 
 $\mathfrak{g}$ such that 
 \bea
 [{T}_{\cal I}, {T}_{\cal J}] &=&0 \\
{} [{ T}_{\cal I}, e_\alpha] &=& \alpha_{\cal I} \,  e_\alpha \,.
 \eea
 The objects $\alpha_{\cal I}$ denote the {\it roots} or {\it root vectors}, which we can think of as vectors in $\mathbb R^{ {\rm rk}(\mathfrak{g})}$ endowed with an inner product $\langle \cdot ,  \cdot  \rangle$.
The basis of $\mathfrak{g}$ can be specified further as 
$\{{ T}_{\cal I}, e_{\alpha^+}, e_{\alpha^-} \}$
such that every positive root ${\alpha^+}$ is expressible as a non-negative linear combination of the so-called 
\bea
{\rm simple \, \, roots}: \,  ( {\alpha_j})_{\cal I} \quad   j = 1, \ldots, {\rm rk}(\fg),
\eea
and similarly every negative root ${\alpha^-}$ as a non-positive linear combination of ${\alpha_j}$.
Let us furthermore define the objects
\bea
{\cal T}_i =  \frac{2  \sum_{\cal I} (\alpha_i)_{\cal I}  T_{\cal I}}{\langle \alpha_i, \alpha_i  \rangle}    \equiv  \frac{2 \alpha_i \cdot T}{\langle \alpha_i, \alpha_i  \rangle} \,,
\eea
where $\alpha_i$ are the simple roots.
In view of this definition  the generators $e_{\alpha_j}$  associated with the simple roots satisfy the important relation
\bea \label{Tialphaj}
[{\cal T}_i, e_{\alpha_j}] = C_{ij} \, e_{\alpha_j}  \quad {\rm with} \quad C_{ij} = \frac{2 \langle\alpha_i,\alpha_j\rangle}{\langle \alpha_i, \alpha_i  \rangle} \,.
\eea
The {\it Cartan matrix} $C_{ij}$ has diagonal entries $+2$ for all simple Lie algebras $\fg$, while the off-diagonal entries differ from algebra to algebra.
It is symmetric only for the simply-laced Lie algebras of A-D-E type as for these $\langle \alpha_i, \alpha_i  \rangle$ takes the same value for every simple root.
The relation (\ref{Tialphaj}) gives rise to a non-degenerate pairing between the Cartan generators and the Lie algebra generators associated with the simple roots. In fact, the linear combinations  ${\cal T}_i$ of the Cartan generators form the so-called {\it co-root lattice}.
They are normalised, as a result of (\ref{canonnorm}), such that
\bea \label{calCartan}
 {\rm tr}_{\rm fund} {\cal T}_i \, {\cal T}_j = \lambda \,  \mathfrak{C}_{ij} \qquad {\rm with} \quad  \mathfrak{C}_{i j}=\frac{2}{\lambda}  \frac{1}{\langle  \alpha_{j}, \alpha_{j}  \rangle} \, C_{i j}  \,.
\eea
Here 
\bea
\lambda = \frac{2}{\langle  \alpha_{\rm max}, \alpha_{\rm max} \rangle} \, ,
\eea
with $\langle  \alpha_{\rm max}, \alpha_{\rm max} \rangle$ the length of the longest root of $\fg$, denotes the {\it Dynkin index of the fundamental representation} as collected in Table \ref{Table_lam}. Note that for the simply-laced Lie algebras of A-D-E type, $C_{ij} = \mathfrak{C}_{ij}$, but more generally the two differ.

 \begin{table}[t!]
\begin{center}
\begin{tabular}{|c||c|c|c|c|c|c|c|c|c|}
\hline
$ {\mathfrak g}$ & $A_n$ & $D_n$ & $B_n$ & $C_n$ & $E_6$ & $E_7$ & $E_8$ & $F_4$ & $G_2$ \\ \hline
$ \lambda$         &   $1$   &    $2$  &  $2 $     &  $1$   &   $6$     & $12$   & $60$  & $6$   & $2$   \\ \hline
 \end{tabular}
\caption{Dynkin index of the fundamental representation, $\lambda$,  for the simple Lie algebras. \label{Table_lam}} 
 \end{center} 
 \end{table}

Coming back to the study of singularities in elliptic fibrations, the intersection relations (\ref{fundrelEalpha1}) and (\ref{fundrelEalpha2})  suggest the following remarkable identification between the resolution divisors $E_{i_I}$ and the independent fiber components $\mathbb P^1_{j_J}$,

\begin{subequations} \label{dictionaryF-theory1}
\begin{empheq}[box=\widefbox]{align}
 E_{i_I}                    & \qquad \simeq   &   \hspace{-4cm } \text{ co-roots} \, \, {\cal T}_i   \label{identif1}\\
 \mathbb P^1_{j_I}  & \qquad \simeq   & \hspace{-4cm }    - (\text{simple roots})     =  - \alpha_{j_I}      
           \end{empheq}
\end{subequations}

\subsection{Non-abelian gauge algebras in M- and F-theory } \label{sec_Nonabeliangaugealgebras}

The identification between geometric and group theoretic entities may a priori come as a surprise from the perspective of pure mathematics. In the geometric classification of the singular fibers due to Kodaira and N\'eron, the simple Lie algebras appear for purely combinatorical reasons because the allowed intersection structure of the fiber codimension in codimension-one happens to agree with that of nodes of an affine Dynkin diagram. In this sense one associates a Lie algebra to a singular fiber, but the deeper reason behind this is elusive. 

String theory provides a beautiful rationale for this connection between geometry and group theory.
The Lie algebras $\mathfrak{g}_I$ appearing in the Kodaira-Tate table are identified with the gauge algebra along the stack of 7-branes wrapped along the discriminant component $\Sigma_I$.
There are two different ways to come to this conclusion.
The first uses the language of $[p,q]$ 7-branes and identifies the Picard-Lefshetz monodromy around the singularity of the elliptic fibration with the $SL(2,\mathbb Z)$ monodromy induced by a stack of 7-branes. Having identified the brane stack in this way, the gauge algebra is read off from the spectrum of $(p,q)$ strings starting and ending on the 7-brane.
The second method uses duality with M-theory, and this is the language we shall be focussing on in the sequel.

Indeed from our discussion in section \ref{sec_firstlook} we concluded that the appearance of extra divisors on an elliptic fibration $\hat Y_{n+1}$ beyond the zero-section and the base divisors leads to extra abelian gauge fields in the M-theory effective action. 
Consider the resolution $\hat Y_{n+1}$ of a Weierstrass model $Y_{n+1}$ with $f$ and $g$ such that $\Delta = \Delta_0 \times \prod_I (\Delta_I)^{p_I}$ and otherwise generic.\footnote{The requirement of maximal genericity of $f$ and $g$ compatible with the vanishing order of $\Delta$ precludes the existence of extra rational sections, which will be discussed in section \ref{sec_MWGROUP}.} Then 
\bea \label{STW-nosection}
h^{1,1}(\hat Y_{n+1}) = 1+ h^{1,1}(B_n) + \sum_I {\rm rk}(\fg_I) \,.
\eea
The extra contribution $\sum_I {\rm rk}(\fg_I)$ compared to (\ref{h2Yn}), valid for a generic Weierstrass model, is precisely due to the independent Cartan divisors $E_{i_I}$, $i_I = 1,\ldots,{\rm rk}(\fg_I)$.
Expanding the M-theory 3-form
\bea
C_3 = \tilde A^0 \wedge [\tilde S_0] + \sum_\alpha A^\alpha \wedge \pi^\ast[D^{\rm b}_\alpha] + \sum_{i_I} A^{i_I} \wedge [E_{i_I}]
\eea
gives rise to extra massless abelian vector fields $A^{i_I}$ in the M-theory effective action in $\mathbb R^{1,8-2n}$. Unlike their cousins  $\tilde A^0$ and $A^\alpha$, these vectors do uplift, in the dual F-theory, to massless abelian gauge fields. 
The reason is that the resolution divisors $E_{i_I}$ satisfy both transversality conditions (\ref{transversal1}) and (\ref{transversal2}). The first follows from the fact that the zero-section does not intersect the resolution curves in the fiber, and the second holds because the fiber of each resolution divisor is contained inside the generic fiber and thus has vanishing intersection product with it.
The interpretation of  the vector fields $A^{i_I}$ is that they represent the gauge fields associated with the Cartan subalgebras  $\mathfrak{h}_I$.

To corroborate this further, we should be able to reproduce in addition the vector fields associated with the non-Cartan generators of the full Lie algebras $\fg_I$.
For simplicity, let us first restrict the discussion to the simply-laced Lie algebras of A-D-E type.
Again, the general idea is clear already from section \ref{sec_firstlook}: M2-branes wrapping a holomorphic or anti-holomorphic curve $C$ in the fiber give rise to particles in the M-theory effective action of mass (\ref{mcformula}) and charge 
\bea
q_{i_I} = \int_C E_{i_I} = [E_{i_I}] \cdot  [C] \,.
\eea
In view of (\ref{fundrelEalpha2}) and (\ref{Tialphaj}), an M2-brane wrapping one of the independent fiber components $\mathbb P^1_{i_I}$  therefore yields a particle whose charges we identify with the negative of a simple root, i.e. with $ - \alpha_{i_I}$, of $\fg_I$. Wrapping an M2-brane along the same curve with opposite orientation, or equivalently wrapping an anti-M2-brane with positive orientation, gives rise to a particle associated with  $\alpha_{i_I}$. Each of the non-simple $\mp$-ve roots is formed as a non-negative linear combination of $\mp \sum_{i_I} n_{i_I} \alpha_{i_I}$. A particle corresponding to this state is obtained by wrapping an M2-brane along a fibral curve in class $\pm \sum_{i_I} n_{i_I} \mathbb [\mathbb P^1_{i_I}]$. 

This way one reproduces states in the full adjoint representation of the A-D-E algebra $\fg_I$, and it is indeed not hard to show that these are the only types of 1-particle states which are associated with wrapped M2-branes in the fiber.  All states except for the abelian vectors $A^{i_I}$ are massive on the fully resolved space $\hat Y_{n+1}$ and become massless in the singular limit. We will discuss this point more carefully in section \ref{sec_Coulomb}.

If the fiber over a discriminant component is subject to monodromies, we must distinguish between the algebra $\tilde \fg_I$ corresponding to the {\it local} fiber type and the actual Lie algebra 
$\fg_I$ relevant for the gauge theory on $\Sigma_I$.  The gauge algebra $ \fg_I$ is obtained by suitable identifications of the nodes of the Dynkin diagram of $\tilde \fg_I$. Mathematically, 
$\fg_I$ is a subalgebra of $\tilde \fg_I$ which is fixed under a finite outer automorphism of some finite order. The representations present in this situation are obtained by decomposing the adjoint representation of $\tilde \fg_I$ into irreducible representations of the gauge algebra $\fg_I$,
\bea
{\rm \bf adj}(\tilde \fg_I) = {\rm \bf adj}(\fg_I) \oplus  \tilde \rho_0, \qquad \tilde\rho_0 =  \bigoplus_k {\bf R}^{\oplus \tilde n_k}_k  \,,
\eea
where some of the irreducible representations ${\bf R}_k$ may occur with a multiplicity $ \tilde n_k$. 
As we will see momentarily, the $ {\fg_I}$ gauge theory contains states in ${\rm \bf adj}(\fg_I)$ and in addition in the representation $\rho_0 =  {\bf R}^{\oplus  n_k}_k$. Note that  the extra representations ${\bf R}_k$ may occur with a smaller multiplicity $n_k < \tilde n_k$ if the order of the outer automorphism is bigger than $2$. This is the case, in fact, only for algebra $\mathfrak{g}_2$, as summarized in Table 
\ref{Table_rho_0}. For more information we refer to \cite{Grassi:2000we}, p.24/25.

 \begin{table}[t!]
\begin{center}
\begin{tabular}{|c|c|c|c|}
\hline
$ {\mathfrak g}$         & $\tilde \fg $                              & $ \tilde \rho_0$                                                &   $\rho_0$  \\ \hline
$\mathfrak{sp}(k)$     & $\mathfrak{su}(2k)$                &  ${\bf \Lambda^2} $                                         &${\bf \Lambda^2} $  \\ \hline
$\mathfrak{sp}(k)$     & $\mathfrak{su}(2k+1)$            &  ${\bf \Lambda^2} + {\rm fund}^{\oplus 2}$      &${\bf \Lambda^2} + {\rm fund}^{\oplus 2}$  \\ \hline
$\mathfrak{so}(2k-1)$&$\mathfrak{so}(2k)$                 &  ${\rm vect}$                                                    & ${\rm vect}$   \\ \hline
$\mathfrak{g}_2$       & $\mathfrak{so}(8)$                  &   ${\bf 7}^{\oplus 2}$                                        & ${\bf 7}$ \\ \hline
$\mathfrak{f}_4$        & $\mathfrak{e}_6$                    & ${\bf 26} $                                                        & ${\bf 26} $ \\ \hline
 \end{tabular}
\caption{Representations of non-simply laced algebras \cite{Grassi:2000we}.  \label{Table_rho_0}} 
 \end{center} 
 \end{table}

So far we have only understood the charges of the particles in the M-theory effective theory from wrapped M2-branes, but not yet their spacetime quantum numbers.
These must be determined by quantizing the moduli of the wrapped M2-brane states. For M-theory compactifications on an elliptic Calabi-Yau 3-fold $\hat Y_3$ this has been discussed in \cite{Witten:1996qb}, p. 13/14, confirming previous results obtained by a topological twist in \cite{Katz:1996ht}. The moduli space of the wrapped M2 branes is $\mathbb R^{1,8-2n} \times \Sigma_I$ since the M2-branes can freely move along the component $\Sigma_I$ of the discriminant divisor. According to the arguments of \cite{Witten:1996qb} this gives rise to 
\begin{enumerate}
\item
a full vector-multiplet in the M-theory effective action in $\mathbb R^{1,8-2n}$ in the adjoint representation of $\fg_I$. At the bosonic level, this includes a gauge field in $(9-2n)$ dimensions and a real scalar. These modes lift in the dual F-theory effective action to a corresponding vector multiplet;
\item if ${\rm dim}(\Sigma_I) \geq 1$ (as is the case for F-theory on $\hat Y_{n+1}$ for $n \geq 2$), extra scalar fields and their fermionic superpartners filling suitable multiplets in the representation ${\rm \bf adj}(\fg_I) \oplus  \rho_0$ of $\fg_I$. The representation $\rho_0$ is present only for non-simply-laced algebras and given by the last column of table \ref{Table_rho_0}.
\end{enumerate}

Wrapped M2-brane states hence constitute matter charged under the Cartan subalgebra $\mathfrak{h}_I$. To form complete  representations of the Lie algebra $\mathfrak{g}_I$, extra, uncharged matter states are required. These are not due to wrapped M2-branes, but come directly from the supergravity sector. We have already seen this for the case of the vector fields themselves, in that the Cartan gauge fields originate in the M-theory 3-form $C_3$. 
 As for the matter multiplets, the uncharged fields arise from the complex structure moduli sector of $\hat Y_{n+1}$, i.e. from suitable modes of the holomorphic $(n,0)$-form.

For M-theory compactified on $\hat Y_2$, the resulting matter content is that of a 7d ${\cal N}=2$ (i.e. 16 supercharges) vector multiplet. By duality with F-theory, we find a corresponding vector mutiplet in 8d.
In this case, according to the classification of Kodaira and N\'eron, no non-simply laced gauge algebras can occur. 
Interestingly, any  non-abelian supersymmetric gauge theory with 16 supercharges  in eight dimensions with gauge algebra other than of A-D-E type is inconsistent due to global anomalies\footnote{\label{FootnoteG2}As of this writing this holds possibly up to one exception as the jury is still out there  for the case of $\fg_2$ \cite{Garcia-Etxebarria:2017crf}.} and hence in eight dimensions F-theory (almost) exhausts the list of consistent gauge theories \cite{Garcia-Etxebarria:2017crf}.
In lower dimensions,  also matter multiplets of the above type can arise.
We will elaborate more on the precise counting of these extra multiplets in addition to the vector multiplet in section \ref{subsec_Zeromodeson7brane}.

Of the many aspects worth mentioning of the effective action of the resulting gauge theory, let us at least point out that the inverse gauge coupling of the non-abelian gauge theory in F-theory is set by the volume of the associated discriminant component 
\be \label{gaugecouplinggeneral}
\frac{1}{g^2_I} \simeq {\rm vol}(\Sigma_I) \,.
\ee
This is intuitively clear from the perspective of a 7-brane stack wrapping $\Sigma_I$. 
The same relation can also be derived in M-theory. We perform this derivation in section \ref{app_Gaugecoupl} in the context of an abelian gauge theory, and the same steps can be easily repeated here.

\subsection{The M-theory Coulomb branch} \label{sec_Coulomb}

On the smooth fibration $\hat Y_{n+1}$,  ${\rm vol}(\mathbb P^1_{i_I}) \neq 0$ and hence all states with non-trivial charges under the Cartan generators are massive. Therefore the part of the gauge symmetry related to the fibers in the M-theory effective action  on $\hat Y_{n+1}$  is broken to its Cartan subgroup $\mathfrak{h}_I$.    
In the limit where all ${\rm vol}(\mathbb P^1_{i_I}) \rightarrow 0$, this abelian gauge algebra enhances to the full non-abelian Lie algebra $\fg_I$.   
We  identify the moduli ${\rm vol}(\mathbb P^1_{i_I})$ on $\hat Y_{n+1}$ with the Coloumb branch parameters associated with the gauge theory {\it in the M-theory effective action}. Indeed, in a supersymmetric theory in $\mathbb R^{1,8-2n}$, the vector multiplet always contains a real scalar field. 
The modulus 
\bea
\xi_{i_I} = {\rm vol}(\mathbb P^1_{i_I}) = \int_{\mathbb P^1_{i_I}} J
\eea
is identified with the real scalar field in the vectormultiplet associated with the Cartan generator ${\cal T}_{i_I}$. A non-trivial vacuum expectation value of $\xi_{i_I}$ breaks the gauge symmetry to the commutant of this generator, and hence if all ${\rm vol}(\mathbb P^1_{i_I}) \neq 0$, $\fg_I$ is broken to its maximal commuting subalgebra $\mathfrak{h}_I$. 
Our geometry-group theory dictionary as obtained so far can hence be summarized as
\begin{subequations} \label{AnomaliesF-theory}
\begin{empheq}[box=\widefbox]{align}
 E_{i_I}                    & \qquad \simeq   &   \hspace{-2cm } \text{ Cartan generators} \, \, {\cal T}_{i_I}   \label{identif1}\\
 \mathbb P^1_{j_I}  & \qquad \simeq   & \hspace{-2cm }    - (\text{simple roots})     =  - \alpha_{j_I}      \\
{\rm vol}(\mathbb P^1_{i_I})   & \qquad \simeq   & \hspace{-2cm }  \text{Coulomb \, branch parameters \, } \, \,  \xi_{i_I}  
           \end{empheq}
\end{subequations}
The origin of the Coulomb branch is attained in the singular limit of blowing down all fiber components of the degenerate fibers in codimension one except $\mathbb P^1_{0_I}$, which is intersected by the zero-section. This realises the blowdown map back to the singular fibration $Y_{n+1}$,
\bea
\oplus_I  \, \mathfrak{u}(1)^{\oplus {\rm rk}(\fg_I)}   \xrightarrow[]{\hat Y_{n+1} \rightarrow   Y_{n+1}}  \oplus_I {\mathfrak g}_I \,.
\eea
Conversely, the mathematical procedure of resolving the singularities of $Y_{n+1}$ by passing to $\hat Y_{n+1}$ corresponds to moving along the Coulomb branch in M-theory.

From the perspective of the dual F-theory in $\mathbb R^{1,9-2n}$, the scalar fields $\xi_{i_I}$ play the role of the Wilson line degrees of freedom along the circle $\tilde S^1_B$. In the circle reduction to M-theory the gauge field in F-theory decomposes as 
\bea \label{Axigen}
\mathbb A_{i_I} = (A_{i_I}, \xi_{i_I}) \qquad {\rm with} \quad \xi_{i_I} = \int_{\tilde S^1_B} (\mathbb A_{i_I})_{\tilde y} \,.
\eea
Since in the $(9-2n)$-dimensional effective action of F-theory the vector multiplet does not contain any scalars, this means that the Coulomb branch is only accessible in the dual M-theory on $\mathbb R^{1,8-2n}$ in a Lorentz invariant way. 
Indeed, it is clear that the F-theory limit of vanishing fiber volume, ${\rm vol}({\mathbb E_\tau}) \rightarrow 0$, implies the limit ${\rm vol}(\mathbb P^1_{i_I}) \rightarrow 0$ and hence  an enhancement of the gauge symmetry,
\bea
\oplus_I  \, \mathfrak{u}(1)^{\oplus {\rm rk}(\fg_I)} \quad \xrightarrow[ {\rm Vol} (\mathbb{E}_{\tau}) \rightarrow 0 ]{\rm F-theory \, limit }   \quad   \oplus_I {\mathfrak g}_I \,.
\eea

We have already noted in section \ref{sec_firstlook} that upon circle reduction on $\tilde S^1_B$, a field $\psi$ in F-theory decomposes into a Kaluza-Klein zero-mode $\psi_0$ together with an entire tower of Kaluza-Klein states $\psi_n$,
\bea
\psi(x, \tilde y)  = \sum_{n=-\infty}^\infty \psi_n(x) e^{i \frac{n}{R} \tilde y} \,.
\eea
Consider a field associated with one of the roots of ${\mathfrak g}_I$. Its KK zero mode in M-theory is described by the particle wrapping one of fibral curves $C$ described above. Since by construction, these curves do not intersect the zero-section $S_0$ and not  any of the vertical divisors, the KK charge $[\tilde S_0] \cdot [C] =0$ as required.
The tower of KK states is obtained by including M2-branes wrapping a curve in the class 
\bea
[C_n] = [C] + n [\mathbb E]_\tau \quad {\rm  with}  \quad q_{KK} = [\tilde S_0] \cdot [C_n] =n.
\eea

The tower of KK states is an important ingredient when it comes to matching the F-theory and the M-theory effective action.
Since by construction the M-theory effective action  is on its Coulomb branch as long as we are compactifying a smooth resolved space $\hat Y_{n+1}$, the relevant modes in the low-energy effective action include only the massless, uncharged
modes. The effect of the massive states has been integrated out and summed up in the classical effective action.
To match this effective action with the F-theory dynamics one dimension higher one must integrate out the KK states in the circle reduction along $\tilde S^1_B$.
We have already alluded to this in the context of the Chern-Simons terms in section \ref{sec_firstlook}, and this effect plays an even more important role in presence of matter, such as the matter along the 7-branes. 
The detailed match between the classical M-theory Chern-Simons terms and the F-theory loop induced Chern-Simons terms contains valuable information about the F-theory spectrum. This has been analyzed from various perspectives in F/M-theory duality in 6d/5d \cite{Witten:1996qb, Intriligator:1997pq,Park:2011ji,Bonetti:2011mw,Grimm:2011fx,Bonetti:2013ela,Bonetti:2013cza}, in 4d/3d/ \cite{Aharony:1997bx,Cvetic:2012xn,Corvilain:2017luj} and in 2d/1d \cite{Schafer-Nameki:2016cfr,Lawrie:2016rqe}.

\subsection{Tate models and resolutions} \label{subsec_TateI2-1}

It is high time to exemplify the geometric structure and its physics interpretation analyzed so far in a concrete example. 
In the vicinity of a codimension-one singularity a Weierstrass model can locally be brought into the so-called Tate form by means of a general algorithm \cite{Tate} placing the singularity in the fiber at the points $[x : y : z] = [0 : 0 : 1]$. This algorithm has been introduced to the physics community in \cite{Bershadsky:1996nh} and is analyzed in further depth in particular in \cite{Grassi:2011hq,Katz:2011qp}. 
The algorithm automatically distinguishes between the {\it split Kodaira} fibers, where no monodromies occur, and the {\it non-split} (and semi-split) fibers with (partial) monodromy (see the last column in table \ref{KTtable}).

Under certain conditions, an elliptic fibration with a singularity along a single divisor $W$ on $B_n$ can in fact be globally described in Tate form, i.e. as the vanishing locus of the hypersurface polynomial
\bea \label{PT}
P_T:= y^2 + a_1 x y z + a_3 y z^3 - x^3 - a_2 x^2 z^2 - a_4 x z^4 - a_6 z^6  \quad {\rm with} \quad a_i \in \Gamma(B_n, K_{B_n}^{-n}) \,.
\eea
The polynomial $P_T$ is the most general polynomial of degree $6$ in $\mathbb P_{231}$ with homogenous coordinates $[x : y : z] $.
The space $\mathbb P_{231}$ is the fiber ambient space. The fibration of this space over the base $B_n$ defines the ambient space $X_{n+2}$, into which the elliptic fibration $Y_{n+1}$ is embedded as the hyersurface 
\bea
Y_{n+1}:    \{ P_T= 0 \} \subset X_{n+1} \,.
\eea
The fact that the polynomial (\ref{PT}) contains all possible monomials compatible with the projective scaling relation means that it defines a toric hypersurface model.

Given such a Tate model, a Weierstrass model  can be obtained from $P_T$ by completing the square in $y$ and the cube in $x$. After relabeling the coordinates, this gives rise to a Weierstrass equation $P_W$ with 
\bea \label{fg-Weier1}
f &=& - \frac{1}{48} (b_2^2 - 24 b_4), \qquad g= \frac{1}{864} (b_2^3 - 36 b_2 b_4 + 216 b_6) 
\eea
in terms of
\bea \label{fg-Weier2}
b_2 = 4 a_2 + a_1^2, \qquad b_4 = 2 a_4 + a_1 a_3, \qquad b_6 = 4 a_6 + a_3^2 \,.
\eea
A generic Tate model with generic $a_i  \in \Gamma(B_n, K_{B_n}^{-i})$ and a generic Weierstrass model with generic $f$ and $g$ are in fact equivalent.
For specific $f$ and $g$, leading to a certain enhancement pattern, on the other hand, it may in general not be possible to write the Weierstrass model \emph{globally} in Tate form. This is in particular the case for Weierstrass models with enhancements $I_n$ for $n=6,7,8,9$ or $I_3^\ast$, and for certain choices of base spaces $B_n$. For more details on the potential global obstructions we refer to \cite{Katz:2011qp}.  In these cases, the Tate form (\ref{PT}) (or a variant thereof \cite{Katz:2011qp}) can be obtained only locally. Note that  \cite{Katz:2011qp} is working under the hypothesis that each discriminant component carrying non-abelian gauge enhancement is itself smooth. For singular divisors, additional restrictions can occur. This becomes particularly relevant for the structure of codimension-two singularities, as will be discussed later.

Tate's algorithm systematically describes the specializations of the Tate polynomials $a_i$ leading to the various Kodaira types.
The vanishing orders are summarized in Table 2 of \cite{Bershadsky:1996nh},  an updated version of which can be found in \cite{Katz:2011qp}.\footnote{In particular, the gauge algebra associated with the entries labeled 'unvconvent.' in Table 2 of \cite{Bershadsky:1996nh} is the one listed in table (\ref{KTtable}).}
 As the simplest example with non-trivial gauge algebra, consider a Tate model with an $I_2$ Kodaira fiber over a divisor $W: \{w=0\}$ on a base $B_n$.
From Table 2 in  \cite{Bershadsky:1996nh} we read off that this singularity type is achieved by the specializations 
\bea \label{aispecialisationSU2}
a_1 \, \,  {\rm generic}, \quad a_2 = a_{2,1} w, \quad a_3 = a_{3,1} w, \quad a_4 = a_{4,1} w, \quad  a_6 = a_{6,2} w^2 \,.
\eea
The polynomials $a_{k,l}$ are to be taken to be generic polynomials of degree $k [\bar K_{B_n}] - l [W]$ on the base.
All that follows is independent of a concrete choice of base as long as the existence of sufficiently generic such polynomials is guaranteed.

From (\ref{fg-Weier1}) and (\ref{fg-Weier2}) we find
\be \label{fgDeltaI_2}
\begin{split}
f &= \frac{1}{48} (-a_1^4 + {\cal O}(w)), \qquad g= \frac{1}{864}(a_1^6 + {\cal O}(w)), \cr
\Delta &= \Delta_0 \, \Delta_1^2 \qquad \Delta_0 = \frac{1}{16} (a_1^4(-a_1 a_{3,1} a_{4,1} - a_{4,1}^2 + a_1 a_{6,2}) + {\cal O}(w)), \quad  \Delta_1 = w \,.
\end{split}
\ee
In the notation of (\ref{DeltaFactorisation}) and (\ref{Simgadef}), the discriminant divisor $\Sigma$ splits into
\be \label{fgDeltaI_3}
\begin{split}
\Sigma = \Sigma_0 \cup  \Sigma_1, \qquad 
\Sigma_0 = \{ \Delta = 0 \}, \qquad \Sigma_1 =  \{\Delta_1=0\} \equiv  \{w =0\} \,.
\end{split}
\ee
Clearly ${\rm ord}(f,g,\Delta)|_{\Sigma_1} = (0,0,2)$ as befits an $I_2$-fiber. In fact, the fiber over $\Sigma_1$ is of split type $I_2^{\rm s}$, i.e. there are no monodromies over $\Sigma_1$. The non-split case $I_2^{\rm ns}$ with monodromies would correspond to generic $a_2$. This is confirmed by testing if the monodromy cover in table \ref{KTtable} factorises. Since $\mathfrak{sp}(1) = \mathfrak{su}(2)$ the gauge algebras agree in both cases, but we will find traces of the split versus non-split nature of the fibers in the geometry below. 

The fiber is singular when $P_T = d P_T=0$, which happens at the point $[x : y : z] = [0 : 0 : 1]$ for $w=0$.
To resolve this singularity into a globally defined fibration $\hat Y_{n+1}$, we follow \cite{Krause:2012yh} and perform a blow-up,  replacing
\bea
(x,y,w) \rightarrow (x e_1, y e_1, e_0 e_1) =: (\tilde x, \tilde y, \tilde w).
\eea
Plugged into $P_T$, this replacement leads to
\bea
P_T \rightarrow e_1^2  \, \hat P_T
\eea
with the {\it proper transform} 
\bea
\hat P_T = (y^2 + a_1 x y z + a_{3,1} y z^3 e_0) - (x^3 e_1 + a_{2,1} x^2 z^2 e_0 e_1 + a_{4,1} x z^4 e_0 + a_{6,2} e_0^2 z^6) 
\eea
representing the hypersurface equation of the resolved space.
The ambient space coordinates are subject to the scaling relations
\bea
(x,y,z,e_0,e_1) &\simeq& (\lambda^2 x, \lambda^3 y, \lambda z, e_0, e_1) \label{scaling1}    \\
&\simeq&  (\mu x, \mu y, z, \mu e_0, \mu^{-1} e_1) \, \qquad \lambda,\sigma \in \mathbb C^\ast \label{scaling2}   \,.
\eea
The first is just the old scaling relation of $\mathbb P_{231}$, and the second relation derives from the fact that the new coordinates $ (\tilde x, \tilde y, \tilde w)$ by which we replace $(x,y,w)$ must be invariant under a rescaling of $e_1$. 
Clearly, due to the introduction of the extra scaling relation the total dimension of the ambient space has not changed even though we have introduced a new coordinate, the blow-up coordinate $e_1$. 
The resolved elliptic fibration $\hat Y_{n+1}$ is now given by the hypersurface 
\bea
\hat Y_{n+1}:  \{ \hat P_T = 0 \} \subset \hat X_{n+2} \,,
\eea
where $\hat X_{n+2}$ is the blown-up ambient space.
The ambient space with the scaling relations (\ref{scaling1}) and (\ref{scaling2}) can be understood in an elementary manner by interpreting the coordinates as the fields in a two-dimensional Gauged Linear Sigma Model (GLSM) \cite{Witten:1993yc} with gauge group $U(1) \times U(1)$. The scalings of the coordinates correspond to the abelian charges of the associated fields. A to-the-point review of important properties of such models and their geometric interpretation can be found e.g. in section 5 of \cite{Denef:2008wq}. 
The hypersurface equation translates into a superpotential for the matter fields.
The D-term conditions of this 2d gauge theory allow for two different types of solutions, each describing a different topological phase of the associated ambient space geometry. In the language of toric geometry, these different phases correspond to the two possible triangulations of the toric space.
These phases are distinguished by the Stanley-Reisner (SR) ideal generated by certain monomials in the toric coordinates: Each monomial describes a combination of coordinates which are not allowed to vanish simultaneously. Their zero-locus is hence absent from the geometry. 
In the present situation, the two different topological phases of the toric ambient space are encoded in the SR ideal of the ambient space generated by \cite{Krause:2012yh}
\bea
x y z, \, x  e_0 y, \, y ze_1, \begin{cases}   x\, e_0 & {\rm phase} \, 1 \\
                                                           z \, e_1 &         {\rm phase} \, 2                  \end{cases}
\eea
Note that $x=e_0 = \hat P_T= 0$ and $z = e_1 = \hat P_T=0$ both imply $y=0$, but $xye_0$ and $z y e_1$ are in the SR ideal in each of the two phases. Hence the Stanley-Reisner ideals of both phases, once restricted to the hypersurface $\hat P_T = 0$, reduce to the same 
\bea
{\rm SR-ideal} = \{xyz, \, ze_1, \, x e_0\} \,.
\eea
This is a special property of the current simple model. An interpretation of the different resolutions of the same Weierstrass model will be given at the end of section \ref{subsec_Mori}. 
This is crucial because $\hat P_T$ and $d \hat P_T$ continue to vanish at $x=y=e_0=0$, but this locus is absent from the geometry because $x$ and $e_0$ must not vanish simultaneously. We have therefore  succeeded in resolving the singularity over generic points of $w=0$. In fact, $\hat P_T=0$ is smooth.  
The single blow-up has increased the rank of the Picard group by one,
\bea
h^{1,1}(\hat Y_{n+1}) = 2 + h^{1,1}(B_n) \,.
\eea

Let us now investigate in more detail the degenerate fibers.
The resolution divisor
\bea
E_1:   \{e_1 =0 \}   \quad {\rm on} \,\,  \{\hat P_T = 0\}
\eea
can be described on the ambient space $\hat X_{n+2}$ as follows: Since $z e_1$ is in the SR-ideal (as it must be because the original singularity was at $[x:y:z] = [0: 0:1]$ and hence away from $\{z=0\}$), we can set $z\equiv1$ and evaluate explicitly
\bea
E_1:  \{\hat P_T = 0\}  \cap \{e_1 =0 \}   =  \{-y^2 - a_1 x y - a_{3,1} e_0 y + a_{4,1}x e_0 + a_{6,2} e_0^2 = 0\} \cap \{e_1=0\} \,.
\eea
Since $ \hat P_T |_{e_1=0} $ is a quadratic polynomial, this describes a rational curve $\mathbb P^1_1$ fibered over the locus $\{w=0\}$ on $B_n$.
Similarly,
\bea
E_0:  \{\hat P_T = 0\}  \cap \{e_0 =0 \}   =  \{e_1 - y^2 - a_1 y z = 0\} \cap \{e_0=0\} \,,
\eea
where we have set $x \equiv 1$ since $x e_0$ is in the SR ideal. This describes the rational curve $\mathbb P^1_0$ fibered over $\{w=0\}$.
Unlike $\mathbb P^1_1$, this fiber component is intersected once by the zero-section $S_0 = \{z=0\}$.
The two fiber components $\mathbb P^1_0$ and $\mathbb P^1_1$ intersect at two distinct points in the fiber because
\bea \label{E0E1inter1}
E_0 \cap E_1 =    \{\hat P_T = 0\}  \cap \{e_0 =0 \}  \cap \{e_1 =0 \} = \{y (y+a_1) = 0 \}   \cap \{e_0 =0 \}  \cap \{e_1 =0 \} \,.
\eea
This reproduces the Dynkin diagram of $\mathfrak{su}(2)$, as expected.
It is interesting to contrast this to the non-split case with fiber $I_2^{\rm ns}$, where $a_2$ is generic: In this case the two intersection points are exchanged by a monodromy because 
\bea
E_0 \cap E_1|_{\rm non-split} &=&    \{\hat P_T|_{\rm non-split} = 0\}  \cap \{e_0 =0 \}  \cap \{e_1 =0 \}  \\
&=& \{y (y+a_1) - a_{2} = 0 \}   \cap \{e_0 =0 \}  \cap \{e_1 =0 \} \,.
\eea

For future convenience, let us introduce the following notation to describe spaces of the type encountered above:
The vanishing locus of a number of polynomials $p_1$, $p_2$, \ldots, $p_k$ on $\hat X_{n+2}$ will be denoted as $V(p_1, p_2, \ldots, p_k)$. In this sense,

\be
\begin{split}
E_0 &= V(e_1 - y^2 - a_1 y z, e_0), \qquad E_1 = V(-y^2 - a_1 x y - a_{3,1} e_0 y + a_{4,1}x e_0 + a_{6,2} e_0^2, e_1),  \cr
E_0 \cap E_1 &= V(y (y+a_1) - a_{2}, e_0, e_1) \,.
\end{split}
\ee

The polynomials $p_i$, which take values in the coordinate ring of $\hat X_{n+2}$, form an ideal, whose associated vanishing locus is the indicated space.

This geometry is clearly the simplest possible example both of a non-trivial gauge algebra and of the resolution of a singular Weierstrass model. 
The fact that the Weierstrass model is formulated as a global Tate model makes it amenable to toric methods \cite{Candelas:1996su,Candelas:1997eh,Bouchard:2003bu}: We have already stressed that the Tate polynomial is the most generic hypersurface of degree six in $\mathbb P_{231}$; in toric language, the fiber ambient space is described by a two-dimensional reflexive polygon, and the generic 
 monomials appearing in the smooth hypersurface equation are encoded in the dual polygon. 
There are, in fact, sixteen different realizations of a genus-one curve as a hypersurface in a toric ambient space, and $\mathbb P_{231}$ corresponds to polygon 10 in the enumeration of \cite{Bouchard:2003bu}. The remaining polygons describe genus-one fibers with either no rational point at all or with several such points, as will be discussed in more detail in section \ref{sec_generalisations}.
 The specialization (\ref{aispecialisationSU2}) is enforced by setting some of the monomials of the fibration to zero. The resolution of the resulting singularities can be understood in the language of a toric top, as introduced originally by Candelas and Font \cite{Candelas:1996su}. This process can be repeated for all possible gauge algebras.
The data for the associated toric tops encoding the resolution of the global Tate model are listed in \cite{Bouchard:2003bu}. 
 
Models which do not have the property that the gauge algebra is achieved simply by setting suitable monomials to zero are called {\it non-toric} or {\it non-canonical}: Here the singularity is the effect of tunings of the polynomials of the hypersurface equation relying on non-trivial cancellations between them.
An example is the tuning of an $I_n$ singularity in a Weierstrass model as sketched around (\ref{ffiggi}), and for $n=6,7,8,9$ such models cannot be brought into generic Tate form globally \cite{Katz:2011qp}.
The distinction between generic and non-generic models becomes even more subtle in the presence of extra sections (see section \ref{sec_generalisations}).
There exists by now a large F-theory literature devoted specifically to the systematic resolution of singular elliptic fibrations, both of toric and non-toric type, including \cite{Blumenhagen:2009yv,Esole:2011sm,Marsano:2011hv,Krause:2011xj,Krause:2012yh, Lawrie:2012gg,Hayashi:2013lra,Braun:2014kla, Braun:2015hkv,Huang:2018gpl}. We will survey them more in section \ref{sec_combab-nonab}.

\subsection{Zero-mode counting along the 7-brane} \label{subsec_Zeromodeson7brane}

We now address in more detail  the question of how to count the massless spectrum of charged modes propagating along a stack of 7-branes in F-theory compactified on an elliptic fibration $\hat Y_{n+1}$.
While determining the gauge quantum numbers depends only on the structure of the fiber of the elliptic fibration, the actual zero-mode counting is sensitive to the details of the base including its dimension.
There are two possible derivations of the massless 7-brane 'bulk' spectrum. The first proceeds by quantization of the moduli space of wrapped M2-branes in the dual M-theory, as performed for M/F-theory on an elliptic 3-fold $\hat Y_{3}$ in  \cite{Witten:1996qb}. 
The spectrum of the extra scalar fields and their superpartners in the dual F-theory  can alternatively be determined by viewing the effective F-theory as the result of compactifying  8d ${\cal N}=1$ Super-Yang-Mills theory along the divisor $\Sigma_I$. This requires a partial topological twist studied for F-theory compactifications to four dimensions in \cite{Donagi:2008ca,Beasley:2008dc} and for F-theory compactifications to two dimensions in \cite{Schafer-Nameki:2016cfr,Apruzzi:2016iac}. 

We now collect the main results of these two approaches as they have appeared in the literature so far, treating the different cases $n=2,3,4$ separately. 



\subsubsection*{F-theory on  $\mathbb R^{1,5} \times \hat Y_3$}

In M-theory compactification on an elliptic threefold $\hat Y_3$, the methods of \cite{Witten:1996qb} predict, in addition to a 5d ${\cal N}=2$ vector multiplet, a number of $g(\Sigma_I)$ hypermultiplets in the adjoint representation of the Lie algebra $\fg_I$, where $g(\Sigma_I)$ is the genus of the curve $\Sigma_I$.
If $\fg_I$ is non-simply laced, extra hypermultiplets in representation $\rho_0$ occur as listed in Table \ref{Table_rho_0}. Their number is given by 
\bea
n(\rho_0) =   g(\Sigma_I') - g(\Sigma_I) \,.
\eea
Here $\Sigma_I'$ is a (branched) multi-cover of $\Sigma_I$ such that the elliptic fibration over $\Sigma_I$ can be viewed as the quotient of an elliptic fibration over $\Sigma_I'$ with generic fiber type $\tilde \fg$ (and without monodromies) \cite{Aspinwall:2000kf}.
By Hurwitz's theorem,
\bea
g(\Sigma_I') - g(\Sigma_I) = (d-1) (g(\Sigma_I) - 1) + \frac{1}{2} {\rm deg}(r) \,,
\eea
where $d$ is the degree of the multi-covering and $r$ the ramification divisor of the covering. Note that $d=2$ for all non-simply laced algebras except $\mathfrak{g}_2$, for which $d=3$.
For more information, especially on the ramification divisor of the multi-covering,  we refer to \cite{Grassi:2000we}.

\subsubsection*{F-theory on $\mathbb R^{1,3} \times \hat Y_4$}


Compactifications of F-theory to four dimensions preserve  ${\cal N}=1$ supersymmetry. Let us briefly recap the derivation of the massless spectrum along a 7-brane stack using the topological twist of \cite{Donagi:2008ca,Beasley:2008dc}.
For simplicity we restrict this discussion to the case of a simply-laced Lie algebra $\mathfrak{g}_I$.
The starting point is the spectrum of an 8d ${\cal N}=1$ gauge theory in flat space $\mathbb R^{1,7}$. Its vector multiplet  contains the 8d vector potential $A_m$, one complex scalar $\Phi$ as well as an 8d Weyl spinor in the ${\bf 16}$ of $SO(1,7)$. All of these modes transform  in the adjoint of $\mathfrak{g}_I$. The complex scalar $\Phi$ parametrizes the motion of the 7-brane in the two normal directions. 
Compactifying this theory on a complex K\"ahler surface $\Sigma_I$ requires a topological twist along $\Sigma_I$ in order to preserve four real supercharges in $\mathbb R^{1,3}$.
The details of this twist can be found in section 3.2 of \cite{Beasley:2008dc}.
The twist ensures that  the bosonic and fermionic modes obtained by decomposing the fields in the vector multiplet organize themselves into full ${\cal N}=1$ supermultiplets.
At the bosonic massless level, the fluctuations of the 8d gauge potential $A_m$ along the non-compact directions $\mathbb R^{1,3}$ give rise to the 4d components $A_\mu$ of the 4d gauge potential, while the internal fluctuations of $A_m$ along $\Sigma_I$ contribute 4d scalar fields called 'Wilson line moduli'.
The complex scalar $\Phi$ contributes extra scalar fields associated with the brane deformations in the directions normal to $\Sigma_I$ in $B_3$.

The Wilson line and the deformation moduli each form the bosonic part of massless ${\cal N}=1$ chiral and anti-chiral multiplets in the adjoint of ${\mathfrak{g}_I}$ propagating along $\Sigma_I$. They are counted by the following cohomology groups:
\begin{equation}
\begin{split} \label{4dbulknoflux}
        {\rm chiral}: \quad &  H^1(\Sigma_I, {\cal O}_I)     \oplus   H^0(\Sigma_I, {\cal O}_I \otimes K_{\Sigma_I}) \cr
 {\rm anti-chiral}:\quad &  H^2(\Sigma_I,{\cal O}_I)     \oplus   H^1(\Sigma_I, {\cal O}_I \otimes K_{\Sigma_I})  \,.
\end{split}
\end{equation}
Here  ${\cal O}_I$ refers to the trivial bundle along $\Sigma_I$; we have included it here because it will be replaced by a more general gauge bundle in the presence of a gauge background, as will be discussed in section \ref{sec_Gaugebackgrounds}.
For now, we stick to the situation of a trivial gauge bundle. Note that
\bea
{\rm dim}(H^1(\Sigma_I, {\cal O}_I)) &=& {\rm dim}(H^1(\Sigma_I, {\cal O}_I \otimes K_{\Sigma_I}))  = h^{0,1}(\Sigma_I) \\
{\rm dim}(H^2(\Sigma_I, {\cal O}_I)) &=& {\rm dim}(H^0(\Sigma_I, {\cal O}_I \otimes K_{\Sigma_I}))  = h^{0,2}(\Sigma_I) \,.
\eea 
The second equality in each line uses the Serre duality formula for cohomology groups of a vector bundle $V$ on a complex space $X$ of dimension $n$,
\bea \label{Serre}
H^i(X, V)  =  [ H^{n-i}(X, V^\vee \otimes K_X) ]^* \,,
\eea
where $V^\vee$ denotes the dual vector bundle. In particular, the dual of a line bundle $L$ is the line bundle such that $L \otimes L^\vee = {\cal O}$ and hence $c_1(L^\vee) = - c_1(L)$. Clearly the dual of the trivial bundle is again trivial.

The CPT conjugate of a 4d chiral multiplet in representation ${\bf R}$ is an {\it anti-chiral} multiplet in the conjugate representation ${\bf \bar R}$. Since the adjoint representation is self-conjugate, the fields counted by the first and second line of (\ref{4dbulknoflux}) are not independent, but CPT conjugate to one another. The modes counted by $h^{0,1}(\Sigma_I)$ represent the Wilson line moduli. The remaining modes counted by $h^{0,2}(\Sigma_I)$ correspond to 
the brane deformation moduli.
 In summary, the bulk modes along a 7-brane stack in a 4d F-theory compactification contribute $h^{0,1}(\Sigma_I) + h^{0,2}(\Sigma_I)$ massless ${\cal N}=1$ chiral multiplets in the adjoint representation of ${\mathfrak{g}_I}$, in addition to one vector multiplet. For vanishing gauge background, the spectrum is non-chiral: To each chiral fermion in representation ${\bf R} = {\rm \bf adj}(\mathfrak{g}_I)$ there exists an anti-chiral fermion in the {\it same} representation.

\subsubsection*{F-theory on  $\mathbb R^{1,1} \times \hat Y_5$}

Compactifying F-theory on an elliptic 5-fold $\hat Y_5$ gives rise to a 2d theory with ${\cal N}=(0,2)$ supersymmetry. 
The 7-brane now wraps a complex K\"ahler 3-fold $\Sigma_I$ on the base $B_4$. The zero-modes and effective action involving the bulk modes along the 7-brane can again be determined by a topological twist \cite{Schafer-Nameki:2016cfr,Apruzzi:2016iac}.
The resulting supersymmetry is chiral, much like in 6d, and our conventions are that the supercharges are given by two {\it chiral} Majorana-Weyl fermions. The chiralities of the fermions below are counted with respect to this choice.
As before the 8d vector potential contributes both the gauge potential along the extended directions and Wilson line moduli, while the complex scalar $\Phi$ gives rise to brane deformation moduli.
The Wilson line moduli continue to be counted by $H^{1}(\Sigma_I, {\cal O}_I) = H^{(0,1)}(\Sigma_I)$ and the brane deformation moduli take values in $H^{0}(\Sigma_I, K_{\Sigma}) =  H^{(0,3)}(\Sigma_I)$.
The 8d gaugino $\Psi$ furnishes the respective superpartners. This leads to one 2d ${\cal N}=(0,2)$ vector multiplet (with an anti-chiral Majorana-Weyl fermion as the gaugino) as well as $h^{0,1}(\Sigma_I)$ 2d $(0,2)$ chiral multiplets counting the Wilson line degrees of freedom and $h^{0,3}(\Sigma_I)$ chiral multiplets counting the brane deformation moduli. The fermions in these multiplets are chiral Weyl spinors. 
A peculiarity of 2d ${\cal N}=(0,2)$ supersymmetry is the existence of Fermi multiplets consisting only of  an anti-chiral Weyl spinor with no scalar superpartner. Indeed, decomposition of $\Psi$ yields, in addition to the above fermionic modes, the degrees of freedom of $h^{0,2}(\Sigma_I)$ such Fermi multiplets in the topologically twisted theory. All of these modes transform in the adjoint representation and are accompanied by their CPT conjugate fields. Note that in 2d, the CPT conjugate of a chiral fermion in representation ${\bf R}$ gives a {\it chiral} fermion in representation ${\bf \bar R}$. The adjoint representation is self-conjugate. The independent bulk modes are then counted as follows:
\begin{equation}
\begin{split} \label{2dbulknoflux}
  {\rm vector \,\,  multiplets}: \quad & H^0(\Sigma_I, {\cal O}_I)       \cr
  {\rm chiral \,\,  multiplets }: \quad & H^1(\Sigma_I, {\cal O}_I)       \cr
 {\rm Fermi \, \,  multiplets}:\quad &  H^2(\Sigma_I,{\cal O}_I)    \cr  
  {\rm chiral \,\,  multiplets}:\quad &  H^3(\Sigma_I,{\cal O}_I)  \,.
\end{split}
\end{equation}
Unlike in the 4d case discussed above, this spectrum exhibits a net chirality because the CPT conjugate multiplets (transforming in ${\bf \bar R} = {\bf \rm adj}(\mathfrak{g}_I) = {\bf R}(\mathfrak{g}_I)$)  have identical chirality. A measure for the chirality of the spectrum is the index (assuming a smooth divisor $\Sigma_I$)
\bea
- \chi(\Sigma_I,{\cal O}) &=& - \sum_{i=0}^3  (-1)^i {\rm dim}(H^i(\Sigma_I, {\cal O}_I)) = - \int_{\Sigma_I} {\rm ch}({\cal O}_I) {\rm Td}(\Sigma_I)  \\
&=& - \frac{1}{24} \int_{\Sigma_I} c_1(\Sigma_I) \, c_2(\Sigma_I) \,.
\eea
We have included an overall minus sign to comply with our convention that the vector and the Fermi multiplets contain negative chirality Weyl spinors.
The second equality in the first line uses the Hirzebruch-Riemann-Roch index theorem, which is applicable in this form as long as the 7-brane divisor is smooth. 
The chirality of the bulk spectrum - even in absence of non-trivial gauge backgrounds - leads to chiral anomalies, whose consistent cancellation is discussed in detail in \cite{Schafer-Nameki:2016cfr,Apruzzi:2016iac,Lawrie:2016rqe,Weigand:2017gwb}.

\section{Codimension-two singularities and localised charged matter} \label{sec_Codim-two}

Having understood the connection between codimension-one singularities and non-abelian gauge algebras in F-theory, we now turn to the behaviour of elliptic fibrations in codimension two on the base.
The special loci of interest describe the intersection of two 7-brane stacks. Here new types of $(p,q)$ strings stretched between the intersecting 7-branes localize and give rise to massless matter in F-theory.
In the dual M-theory, these states are due to M2-branes wrapping new curve components appearing in the codimension-two fiber.  

In section \ref{subsec_gen} we approach the codimension-two singularities from the Weierstrass perspective and anticipate the general pattern of representations via the Katz-Vafa method. The structure of the codimension-two fibral components is analysed in more detail in section \ref{subsec_Mori} both in geometric and group theoretic terms.
These general patterns are illustrated in an example in section \ref{SU2Tatematter}.
The counting of the charged massless localised matter is the topic of the subsequent section \ref{subsec_Countingloc-1}.
Two interesting obstructions to the existence of a smooth, flat fibration are discussed in section \ref{subsec_CFT}  and \ref{sec_terminalco2}.

\subsection{Codimension-two singularities in the Weierstrass model and Katz-Vafa method} \label{subsec_gen}

Over special loci in complex codimension-one on the discriminant divisor $\Sigma$ of a Weierstrass model $Y_{n+1}$, the singularity in the fiber enhances. 
These loci are of complex codimension two on the base $B_n$ and the associated fibers are oftentimes called codimension-two singular fibers.
The change in singularity type is 
 indicated by an increase in the vanishing orders $({\rm ord}(f), {\rm ord}(g),{\rm ord}(\Delta))$ in the Weierstrass model. If we 
  adopt the notation (\ref{DeltaFactorisation}) for the dicriminant polynomial and  (\ref{Simgadef}) for the associated discriminant divisor,
the codimension-two loci in question are given by the intersection loci
\bea
C_{IJ} = \Sigma_{I} \cap \Sigma_J \,.
\eea
As a special case, we explicitly allow for the possibility that a divisor $\Sigma_I$ self-intersects \cite{Morrison:2011mb,Klevers:2017aku}. In this case $\Sigma_I$ is necessarily singular as a divisor on $B_n$.
Oftentimes $C_{IJ}$ decomposes into several loci, each characterized by a different type of fiber enhancement. In this case we shall write
\bea \label{CIJunion}
C_{IJ} = \cup_r \, C^{(r)}_{IJ} \,.
\eea
If the base is of complex dimension $n=2$, then each $C^{(r)}_{IJ}$ is a set of points on $B_2$, while  for  $n=3$ and $n=4$, each $C^{(r)}_{IJ}$ is an irreducible curve on $B_3$ or surface on $B_4$, respectively. 

From the perspective of the Weierstrass model the first step in analyzing the enhancement loci is to determine the vanishing orders and to associate a 'naive fiber type'  to each $C^{(r)}_{IJ}$ from the first four columns of table \ref{KTtable},
\bea
{\rm ord}(f, g, \Delta)|_{C^{(r)}_{IJ}}    \Rightarrow {\rm naive \, \, Kodaira \, \, type} \,.
\eea
This can serve as a first indication of the matter which is expected to be localized at ${C^{(r)}_{IJ}}$ even though special care has to be applied, as will be discussed in section \ref{subsec_Mori}. 
From the table one can read off the Lie algebra 
\be
\mathfrak{h}_{IJ, r} \supset \mathfrak{\tilde g}_I \oplus  \mathfrak{\tilde g}_J
\ee
associated with this vanishing behaviour. 
Here $\mathfrak{\tilde g}_I$ represents the simply-laced covering algebra of the gauge algebra $\mathfrak{g}_I$ along $\Sigma_I$ in case the latter is not simply laced. 
The Lie algebra $\mathfrak{h}_{IJ, r}$ does not correspond to a gauge algebra in the effective action, but it contains information about the expected representations. As a crude rule of thumb, the decomposition of the adjoint of this algebra into irreducible representations of $\mathfrak{g_I}$ and 
$\mathfrak{g_J}$ gives, apart from the respective adjoints, a number of new representations of $\mathfrak{g}_I$ and $\mathfrak{g}_J$ (plus extra singlets). The charged states are in a first approximation the types of representations expected.
The underlying reasoning is this: 
The gauge theories along $\Sigma_I$ and $\Sigma_J$ can be viewed as a deformation of a mother gauge theory with algebra $\mathfrak{h}_{IJ, r}$ by the VEV $\langle \varphi \rangle$ of a Higgs field in the adjoint of $\mathfrak{h}_{IJ,r}$. This VEV varies over the 7-brane loci and vanishes at the intersection ${C^{(r)}_{IJ}}$. Note that this is a local picture, and in particular for the different enhancement loci the associated Lie algebras $\mathfrak{h}_{IJ, r}$ differ.
Since at the location  of $C^{(r)}_{IJ}$ the Higgs VEV vanishes, it is here that remnants of the full adjoint $\mathfrak{h}_{IJ, r}$ are localised. The remnants are precisely the matter states in the extra charged representations in the decomposition
$\mathfrak{h}_{IJ,r} \rightarrow \mathfrak{ g}_{I} \oplus \mathfrak{ g}_{J}$. 

The above procedure is the Katz-Vafa picture \cite{Katz:1996xe}, which was spelled out and formalized  in detail for F-theory compactification on elliptic 3-folds in \cite{Grassi:2000we,Grassi:2011hq} (see for instance Assignment 8.21 in \cite{Grassi:2018rva} for a more precise formulation than we have given above).
In particular, this method allows one to determine the representations and, in compactifications to six dimensions, their multiplicities even without studying a full resolution. 
The possible enhancement types of the Weierstrass model in codimension two and the associated matter representations have been classified in \cite{Grassi:2000we,Grassi:2011hq} for all Weierstrass models which satisfy a certain genericity assumption, namely where
the discriminant divisor is of the form $\Sigma = \Sigma_0 \cup \Sigma_1$ with only one non-abelian gauge algebra along a {\it smooth} divisor $\Sigma_1$, but the model is otherwise maximally generic.\footnote{The behaviour at the non crepantly resolvable codimension two loci of these models has been specified in \cite{Grassi:2018rva}, see section \ref{sec_terminalco2}.}

New types of representations can occur in this manner if we drop requirement that the divisor $\Sigma_I$ is smooth: 
For instance, in the case of a model with gauge algebra $\mathfrak{su}(n)$, if the gauge divisor is smooth, the only representations which occur in codimension two are the fundamental and 2-index  antisymmetric ones (which are possible also in perturbative Type IIB orientifolds) as well as, for $\mathfrak{su}(6)$, $\mathfrak{su}(7)$, $\mathfrak{su}(8)$, three-index anti-symmetric representations if the enhancement is tuned to higher rank \cite{Morrison:2011mb,Anderson:2015cqy}. This is a truly non-perturbative effect which involves enhancements to an exceptional gauge algebra $\mathfrak{h}_{IJ, r}$; the resulting representation is composed of $(p,q)$ strings not available in the weak coupling limit. 
A description of codimension-two matter directly in terms of such $(p,q)$ string junctions is given in the recent works \cite{Grassi:2013kha,Grassi:2014sda,Grassi:2014ffa,Grassi:2016bhs,Grassi:2018wfy}, which also contain references to earlier work on $(p,q)$ strings.
By contrast, at self-intersections of $\Sigma_I$ other representations can occur including symmetric tensor representations \cite{Sadov:1996zm,Kumar:2010am} (two-index symmetric representations were studied in \cite{Cvetic:2015ioa,Anderson:2015cqy} and three-index symmetric ones in \cite{Klevers:2016jsz}) or more exotic representations such as box representations \cite{Morrison:2011mb,Klevers:2017aku}. Here the fact that the divisor is non smooth implies that the coordinate ring in a neighborhood is a non-UFD (unique factorization domain) \cite{Klevers:2017aku}. This makes possible very non-generic enhancements of the vanishing order due to intricate cancellations in $f$ and $g$.

\subsection{The relative Mori cone and the weight lattice} \label{subsec_Mori}

In the resolution $\hat Y_{n+1}$, the topology of the fiber changes compared to the fibers over generic points of the discriminant. A complete classification of the possible fiber types in complex codimension two is not yet available in full generality in the mathematics literature. In particular the classification of Kodaira-N\'eron is a priori valid only for elliptic surfaces and, with the modifications due to global effects as described, over {\it generic} points in complex codimension one on more general elliptic fibrations. 
However, insights from F/M-theory and the physics interpretation of the codimension-two loci described in the previous section  allow for a classification of the {\it expected} fiber types \cite{Hayashi:2014kca,Braun:2014kla,Esole:2014bka,Esole:2014hya,Braun:2015hkv}. 
For minimal elliptic threefolds mathematical theorems are proven in \cite{Cattaneo:2013vda}.

Let us assume that the singularity types of a Weierstrass model $Y_{n+1}$ in codimension two are such that a flat (i.e. equidimensional), smooth Calabi-Yau resolution $\hat Y_{n+1}$ exists.
In particular, this implies that the vanishing orders of $(f, g)$  do not simultaneously exceed $(4,6)$ in codimension two. We will come back to what happens if the assumption of a smooth, flat Calabi-Yau resolution in codimension two fails in sections  \ref{subsec_CFT}  and \ref{sec_terminalco2}. 
With a few exceptions\footnote{For instance an enhancement or the vanishing orders $(0,0,1) \rightarrow (1,1,2)$, corresponding to a change of the naive Kodaira fibers, does not give rise to such an increase.} the cone of effective curves in the fiber not intersecting the zero-section $S_0$ becomes larger at the special fibers. 
If one approaches the special fibers from one of the discriminant components $\Sigma_I$, the enhancement is 
 due to a splitting of one or more of the fibral curves $\mathbb P^1_{i_I}$ into two or several curves over the special loci. As a result, effective curves exist in codimension-two which cannot be holomorphically transported away from the special loci. An M2-brane wrapping any combination of curves involving one or several of these special fiber components 
gives rise to a state localised in codimension two, called localised matter for this reason. 
In all examples studied so far, the intersection pattern of the codimension-two fibers of $\hat Y_{n+1}$ reproduces the extended Dynkin diagram of a Lie algebra.
The Lie algebra is the one associated with the Kodaira fiber which one would naively attribute to the vanishing order of $(f, g,\Delta)$. However, various monodromy effects can delete the fiber components associated with one or several of the nodes of the Dynkin diagram. In this way, the fibers in codimension two can be of 'non-Kodaira' type, i.e. differ from the list of possible fibers in codimension-one. 
Indeed such behaviour has been exemplified in the mathematics literature by Miranda \cite{Miranda83}, and first appeared in the F-theory context in \cite{Morrison:2011mb,Esole:2011sm}. Cattaneo proves in \cite{Cattaneo:2013vda} that for every smooth, flat Calabi-Yau resolution of a Weierstrass model of complex dimension three, the non-Kodaira fibers must always be of this form.

What is important for the physics interpretation is that, similar to our treatment of codimension-one fibers, one can form the intersection product of the resolution divisors $E_{i_I}$ 
with the new curves in the fibers. These compute the charges of the corresponding M2-brane states with respect to the Cartan $U(1)_{i_I}$ and hence identify the representation of the wrapped M2-brane states. To understand this in more detail, we again indulge in a small

\subsubsection*{Group Theoretic Interlude (II):}

Given a Lie algebra $\fg$, we associate to an irreducible representation ${\bf R}$ of $\fg$ 
a weight vector $\beta^a({\bf R})$, $a = 1, \ldots, {\rm dim}({\bf R})$. Each entry is itself a vector of dimension ${\rm rk}({\fg})$. It contains  the charges of a state in the given representation with respect to the generators ${\cal T}_i$ of the Cartan subalgebra, $i=1, \ldots, {\rm rk}({\fg})$, i.e.
 \bea
{\cal T}_i |  \beta^a ({\bf R}) \rangle = \beta^a_i({\bf R}) |  \beta^a ({\bf R}) \rangle \,.
\eea
A representation is characterized by its highest weight $\beta^1({\bf R})$, and the full weight vector $\beta^a({\bf R})$ can be reconstructed from the highest weight by adding suitable linear combinations of simple roots,
\bea
\beta^a({\bf R}) = \beta^1({\bf R}) + \sum_i  n^a_i  \alpha_i   \quad {\rm for \,  \, some\, } \quad n^a_i \in \mathbb Z.
\eea
For example, the fundamental representation ${\bf fund}$ of $\fg = \mathfrak{su}(N)$ is characterized as
\be \label{fundweights}
\begin{split}
\beta^1({\bf fund}) &= (1,0,0,\ldots,0) \cr
\beta^2({\bf fund}) &= \beta^1({\bf fund}) - \alpha_1 = (-1,1,0,\ldots,0) \cr
\beta^3({\bf fund}) &= \beta^2({\bf fund}) - \alpha_2 = (0,-1,1,0,\ldots,0) \cr
\ldots & \cr
\beta^n({\bf fund}) &= \beta^{n-1}({\bf fund}) - \alpha_n = (0,0,\ldots,0,-1) \,.
\end{split}
\ee

Let us now anaylse the structure of the fibers over one of the loci $C_{IJ}^{(r)}$ in (\ref{CIJunion}) in more detail. 
The material of this section is a condensate of the findings of \cite{Morrison:1996xf}\cite{deBoer:1997kr}\cite{Aharony:1997bx} \cite{Intriligator:1997pq} \cite{Diaconescu:1998ua}\cite{Grimm:2011fx}\cite{Hayashi:2013lra}\cite{Hayashi:2014kca} \cite{Braun:2014kla} \cite{Esole:2014bka}\cite{Esole:2014hya} \cite{Braun:2015hkv}. We will follow mostly the presentation in \cite{Grassi:2018rva}.

The fiber over $C_{IJ}^{(r)}$ contains a union of rational curves. We have already alluded above to the relative Mori cone ${\rm NE}(C_{IJ}^{(r)})$ as the cone of effective curve classes  in this fiber with vanishing intersection number with the class of the zero-section $[S_0]$. 
This means that the curve classes have as representatives holomorphic curves not intersected by $S_0$.
The generators of ${\rm NE}(C_{IJ}^{(r)})$ are the classes associated with the curves $\mathbb P^1_{i_I}$ and $\mathbb P^1_{j_J}$ in the fiber over generic points of $\Sigma_I$ and $\Sigma_J$, together with all the curves $C^{(k)}_{\rm sp}$ which arise by a splitting of these curves. 

Then the first key fact is that the intersection numbers of the split curves with the resolution divisors $E_{i_I}$ reproduce a weight vector of some representation $\mathbf{R}$ of $\mathfrak{g}_I$, i.e.
\bea \label{ECbeta}
[E_{i_I}] \cdot [C^{(k)}_{\rm sp}] =  \beta_{i_I}^a(\mathbf{R})     
\eea
for  some  $a \in \{1, \ldots, {\rm dim}(\mathbf{R}) \}$. 
The complete weight vector is found by adding suitable integer linear combinations of positive simple roots $\alpha_{i_I}$,
\bea
\beta^b(\mathbf{R}) =  \beta^a(\mathbf{R})  + \sum_{i_I} n^b_{i_I} \alpha_{i_I} \,.
\eea
In view of the identification (\ref{dictionaryF-theory1}),   consider therefore the linear combination $C_{\rm sp} + \sum_{i_I} n^b_{i_I} (-{\mathbb P}^1_{i_I})$ of curves. If $C$ is a holomorphic curve, we mean by $-C$
the curve with opposite orientation such that $[-C]$ is anti-effective. 
The intersection number of $C_{\rm sp} + \sum_{i_I} n^b_{i_I} (-{\mathbb P}^1_{i_I})$ with the resolution divisor $E_{i_I}$ gives the weight  $\beta^b_{i_I}(\mathbf{R})$ by construction.

The second non-trivial fact is that the class of this curve is either effective or anti-effective, depending on the integers  $n^b_{i_I}$. We symbolize this by the notation
\bea
C_{\epsilon^b} (\beta^b({\bf R}); C^{(k)}_{\rm sp}) = C^{(k)}_{\rm sp}  + \sum_{j_I} n^b_{j_I} (-{\mathbb P}^1_{j_I})
\eea
where $\epsilon^b = 1$ ($\epsilon^b = -1$) indicates that $[C_{\epsilon}(\beta^b({\bf R}))]$ is effective (anti-effective).

We can thus form a set $M(C^{(k)}_{\rm sp})$ of $\mathrm{dim}(\mathbf{R})$  curves with effective or anti-effective classes,

\bea \label{curvesetM}
M(C^{(k)}_{\rm sp}) := \{  C_{\epsilon^b}(\beta^b({\bf R}); C^{(k)}_{\rm sp}), \quad  \beta = 1, \ldots, {\rm dim}(\mathbf{R}) \}\,,
\eea
with the property that
\bea
[E_{i_I}] \cdot [C_{\epsilon^b} (\beta^b({\bf R}); C^{(k)}_{\rm sp})] = \beta^b(\mathbf{R}) \,.
\eea
An M2-brane wrapping a holomorphic (if $\epsilon^b = 1$) or anti-holomorphic (if $\epsilon^b = -1$)   curve  in this set gives rise to a BPS state with the corresponding weight; altogether this  realises the full representation ${\bf R}$. An anti-M2 brane wrapping the same curve, or equivalently an M2-brane wrapping the orientation reversed curve, gives rise to a BPS state with the negative weight, corresponding to the complex conjugate representation ${\bf \bar R}$. The set of curves associated with ${\bf \bar R}$ is hence
\bea
 - M(C^{(k)}_{\rm sp}) := \{   - C_{\epsilon^b}(\beta^b({\bf R}); C^{(k)}_{\rm sp}), \quad  \beta = 1, \ldots, {\rm dim}(\mathbf{R}) \}\,.
\eea
If $M(C^{(k)}_{\rm sp}) = -M(C^{(k)}_{\rm sp})$, the representation ${\bf R}$ is identical to its conjugate.
We can now continue this process of building complete representations, starting from each of the split curves in the fiber.
This way we can realize several distinct copies of the same representation $\mathbf{R}$, or possibly even several distinct representations.
On the other hand, 
if for $C^{(l)}_{\rm sp} \neq C^{(k)}_{\rm sp}$, the corresponding sets are equal in the sense that $M(C^{(k)}_{\rm sp}) = M(C^{(l)}_{\rm sp})$ or $M(C^{(k)}_{\rm sp}) =  - M(C^{(l)}_{\rm sp})$, then we have simply constructed the same set of curves in both cases. Keeping only the representations associated with the distinct sets we systematically obtain the full set of representations. 
Note that the states obtained in this way over $C^{(r)}_{IJ}$ form a representation of $\mathfrak{g}_I \oplus \mathfrak{g}_J$.

In this sense, the relative Mori cone ${\rm NE}(C_{IJ}^{(r)})$ of the elliptic fibration generates the weight lattice of the gauge theory. This  includes the states in codimension one. 

Different resolutions of the same singular elliptic fibration are birationally equivalent and give rise to the same assignment of representations.\footnote{While intuitively clear from a physics perspective, it is not obvious from a purely mathematical point of view that the representations assigned in this way to the singular fibers are birational invariants. This is proven in \cite{Grassi:2018rva} in the context of elliptic threefolds subject to the genericity assumptions of \cite{Grassi:2000we}.} However, they differ by the collection of signs ${\epsilon^b}$
indicating which weights are realized by an effective versus anti-effective curve. 
The different choices of signs, which are geometrically related by flops in the fiber, correspond to different sub wedges in the Coulomb branch of the M-theory effective action \cite{Morrison:1996xf,deBoer:1997kr,Aharony:1997bx,Intriligator:1997pq,Diaconescu:1998ua}. This one-to-one correspondence between the classical Coulomb branch phases and the web of birational resolutions has been studied in detail in the recent literature \cite{Grimm:2011fx,Hayashi:2013lra,Hayashi:2014kca,Braun:2014kla,Esole:2014bka,Esole:2014hya,Braun:2015hkv}.

\subsection{Example: SU(2) Tate model} \label{SU2Tatematter}

Let us illustrate these ideas in an example. For simplicity we continue analysing the $SU(2)$ Tate model introduced in section \ref{subsec_TateI2-1}. 
From (\ref{fgDeltaI_2}), (\ref{fgDeltaI_3}) we recall that the discriminant factorises as
\bea
\Delta = \Delta_0  \, \Delta_1^2, \qquad \Delta_1 =  w, \quad \Delta_0 = \frac{1}{16} (a_1^4  \, P + {\cal O}(w)), \qquad P = (-a_1 a_{3,1} a_{4,1} - a_{4,1}^2 + a_1 a_{6,2})  \,.
\eea
The intersection $C_{01} = \Sigma_0 \cap \Sigma_1$ factorises into two different loci with Weierstrass vanishing orders
\be
\begin{split}
 \label{vanishingordersI2model}
&C_{01}^{(1)} = \{w = 0 \} \cap  \{ a_1 = 0 \}:  ( {\rm ord}(f), {\rm ord}(g), {\rm ord}(\Delta) ) =  (1,2,3)   \cr
&C_{01}^{(2)} = \{w = 0 \} \cap  \{ P = 0 \} \,:       ( {\rm ord}(f), {\rm ord}(g), {\rm ord}(\Delta) ) =  (0,0,3) \,.
\end{split}
\ee

The vanishing orders along $C_{01}^{(1)}$ indicate a change of fiber type from  $I_2$ to  $III$. We have stressed that in codimension-two and higher, the actual fiber type may differ from Kodaira's classification by deleting nodes in the corresponding Dynkin diagram. Here this is not the case. To determine the topology of the fiber over $C_{01}^{(1)}$ we follow the two generic fiber components $\mathbb P^1_{0}$ and $\mathbb P^1_{1}$ as we approach $a_1=0$.
Over generic points of $\Sigma_1$, the rational fibers of the two resolution divisors $E_0$  and $E_1$ can be understood as the complete intersection
\be
\begin{split}
\mathbb P^1_0 &= V(e_1 - y^2 - a_1 y z,e_0, d_1, \ldots, d_{n-1} ) \cr
\mathbb P^1_1 &= V(y^2 - a_1 x y - a_{3,1} e_0 y + a_{4,1}x + a_{6,2} e_0^2, e_1, d_1, \ldots, d_{n-1} ) \,,
\end{split}
\ee
where $D_i: \{d_1= 0 \}$, $i=1, \ldots, n-1$ is a collection of divisors on the base $B_n$ which intersect the discriminant component $\Sigma_1$ in one generic point\footnote{For simplicity we assume here the existence of such divisors. Their sole purpose is to single out a point on $\Sigma_1$.} and by abuse of notation we do not distinguish here between the divisor and its defining equation.
The behaviour of these two curves over $C_{01}^{(1)}$ can be studied by replacing one of the polynomials $d_i$  by $\{a_1=0\}$. As we can see, neither $\mathbb P^1_0$ nor $\mathbb P^1_1$ factorises over $\{a_1=0\}$, but (\ref{E0E1inter1}) shows that the two intersection points in the fiber characteristic for an $I_2$ fiber coalesce in a double intersection as we approach $a_1=0$. This reproduces the structure of a Kodaira Type $III$ fiber, in agreement with the naive expectations from the vanishing orders (\ref{vanishingordersI2model}). 
Since none of the fibral curves splits, we find no new representations at this locus in addition to the adjoint of $\mathfrak{su}(2)$ present over generic points of $\Sigma_1$.

On the other hand, over $C_{01}^{(2)}$, the fibral curve $\mathbb P^1_1$ splits into
\bea \label{P1splitI3}
\mathbb P^1_1 \rightarrow C_{\rm sp}^{(1)} \cup C_{\rm sp}^{(2)} \,.
\eea
We can see this already by going to a patch where $a_1 \neq 0$ and solving $P=0$ for $a_{6,2}$. Plugging this value into the defining hypersurface equation gives
\bea
\hat P_T |_{e_1= 0 \cap P = 0} = -\frac{1}{a_1^2} (a_1 y - a_{4,1} e_0 z^3) (a_1 y + a_1^2 x + a_1 a_{3,1} e_0 + a_{4,1} e_0) \,.
\eea
Hence, in the given patch, in which we can set $a_1 = 1$, we can define the holomorphic curves in the fiber
\be
\begin{split}
 \label{Cspequations}
 C_{\rm sp}^{(1)} &= V(y - a_{4,1} e_0 z^3, e_1, P, d_2, \ldots, d_{n-1}) \cr
  C_{\rm sp}^{(2)} &= V(y + a_1^2 x + a_1 a_{3,1} e_0 + a_{4,1} e_0 , e_1, P, d_2, \ldots, d_{n-1})  \,.
\end{split}
\ee
Both curves are in fact homologous since the defining vanishing polynomials are of equal degree,
\bea \label{Csp1homol2} 
[C_{\rm sp}^{(1)} ] =[ C_{\rm sp}^{(2)} ]  \,.
\eea
A more elegant way to describe the splitting (\ref{P1splitI3}) is to observe that the ideal generated by $\hat P_T$, $e_1$, $P$ and $d_2, \ldots, d_{n-1}$ decomposes into two primary ideals, whose associated vanishing locus describes the two curves over $P=0$. 

Note that the two split curves $C_{\rm sp}^{(i)}$ cannot be described as a complete intersection on $\hat Y_{n+1}$ itself, but only as a complete intersection in the ambient space $\hat X_{n+2}$ of $\hat Y_{n+1}$. This is because none of the defining equations in (\ref{Cspequations}) is the hypersurface equation $\hat P_T$ (restricted to a given locus), but $\hat P_T = 0$
is of course implied by (\ref{Cspequations}). 

The rational curve $\mathbb P^1_0$, on the other hand, does not split further. The fiber over $C_{01}^{(2)}$ therefore consists of three 
rational curves $\mathbb P^1_0$,  $C_{\rm sp}^{(1)}$,  $C_{\rm sp}^{(2)}$. By counting common points it is clear that they intersect pairwise and thus form a fiber of  Kodaira type $I_3$, again in agreement with (\ref{vanishingordersI2model}).

The next step in analyzing the fiber over $C_{01}^{(2)}$ consists in computing the intersection numbers of the split curves with the resolution divisors. These are given as
\bea
[E_1] \cdot [C_{\rm sp}^{(1)} ] = - 1, \qquad [E_1] \cdot [C_{\rm sp}^{(2)} ] = - 1 \,.
\eea
This follows from the characteristic intersections $[E_1] \cdot [\mathbb P^1_1] = -2$ along generic points of $\Sigma_1$ together with (\ref{Csp1homol2}).  
In view of (\ref{fundweights}) this intersection number is readily recognized as the weight $\beta^2$ of the fundamental representation ${\bf R} = {\bf 2}$ of $\mathfrak{su}(2)$, whose complete set of weights is
\bea
\begin{pmatrix} \beta^1({\bf 2}) \\ \beta^2({\bf 2}) \end{pmatrix}   = \begin{pmatrix} 1 \\ -1 \end{pmatrix} \,.
\eea
Let us now follow 
 the systematic procedure to identify the full set of fibral curves associated with the weight vector, starting with the effective curve $C_{\rm sp}^{(1)}$.
The other weight $\beta^1({\bf 2}) = \beta^2({\bf 2}) + \alpha_1$ (with $\alpha_1 = 2$ the simple positive root) is constructed from $C_{\rm sp}^{(1)}$ by adding the curve $ - \mathbb P^1_1 = - (C_{\rm sp}^{(1)}  + C_{\rm sp}^{(2)} )$. This curve class is anti-effective, and we therefore identify
\be
\begin{split} \label{SU2weights1}
C_{-1} (\beta^1({\bf 2}); C^{(1)}_{\rm sp})  &= C^{(1)}_{\rm sp} - \mathbb P^1_1   = - C^{(2)}_{\rm sp}  \cr
C_{+1} (\beta^2({\bf 2}); C^{(1)}_{\rm sp})  &=  C^{(1)}_{\rm sp}
\end{split}
\ee
and
\bea
M(C^{(1)}_{\rm sp}) = \{ C^{(1)}_{\rm sp} , - C^{(2)}_{\rm sp} \} \,.
\eea
Repeating this process starting with $C^{(2)}_{\rm sp}$ gives 
\bea
M(C^{(2)}_{\rm sp}) = \{ C^{(2)}_{\rm sp} , - C^{(1)}_{\rm sp} \} =  - M(C^{(1)}_{\rm sp}) \,.
\eea
Both sets are therefore not independent, and we really have found only one copy of a fundamental representation.

\subsection{Counting of localised zero modes} \label{subsec_Countingloc-1}

In the limit of vanishing fibral curve volume, the M2-branes wrapping the curves in the relative Mori cone give rise to BPS particles in $\mathbb R^{1,8-2n}$. By the logic spelled out in section \ref{sec_Coulomb}, these constitute the KK zero modes of corresponding charged matter fields in the dual F-theory in $\mathbb R^{1,9-2n}$. To determine the number of BPS states in M-theory (and hence the number of massless matter fields in F-theory) we must again quantize the moduli space of the wrapped M2-branes, as performed in \cite{Witten:1996qb} for the case $n=2$. 
For F-theory compactifications to four and two dimensions, it is more practical to count the zero-modes in the topologically twisted field theory approach.
These methods imply the following counting:

\subsubsection{Localised zero-mode counting for F-theory on $\mathbb R^{1,5} \times \hat Y_3$} 
The discriminant $\Delta$ factorises as in (\ref{DeltaFactorisation}), and the discriminant divisor $\Sigma$ is a union of complex curves $\Sigma_I$ intersecting at isolated points.
According to (\ref{CIJunion}), these are grouped into different sets of points $C^{(r)}_{IJ}$, distinguished by the structure of the fiber over them. 
Each $C^{(r)}_{IJ}$ is a set of $n_r$ points on the base, each of which gives rise to the same fiber type,
\bea
C^{(r)}_{IJ} = \cup_{i=1}^{n_r} Q_i^{(r)} \,.
\eea
Consider one such point $Q_i^{(r)}$ and form the independent sets of curves in the fiber $M(C^{(k)}_{\rm sp})$ as described around (\ref{curvesetM}).
The quantization argument of \cite{Witten:1996qb} implies that the M2-branes wrapping the independent sets of curves $M(C^{(k)}_{\rm sp})$   
give rise to a massless hypermultiplet in the associated representation ${\bf R}$.
If $M(C^{(k)}_{\rm sp}) = - M(C^{(k)}_{\rm sp})$, one merely obtains a half-hypermultiplet, which is possible only if the representation is self-conjugate, or more precisely quaternionic. 
Hence the total number of (half-) hypermultiplets in representation ${\bf R}$ is $m \, n_r$ where $m$ is the number of independent sets  $M(C^{(k)}_{\rm sp})$ giving rise to the same representation ${\bf R}$ in the fiber.  

\paragraph{Example: SU(2) Tate model}

In the $SU(2)$ Tate model of section \ref{SU2Tatematter}, $C_{01}^{(1)}$ consists of $n_1 = [a_1] \cdot   [\Sigma_1] = [\bar K] \cdot [\Sigma_1]$ points $Q_i^{(1)}$. However, none of these points carries extra massless matter.
Only the locus $C_{01}^{(2)}$ hosts hypermultiplets in a non-trivial representation. 
This locus consists of $n_2 = [P] \cdot [\Sigma_1] =   (8 \bar K - 2 [\Sigma_1]) \cdot [\Sigma_1]$ points $Q_i^{(2)}$, each of which carries one massless hypermultiplet in representation ${\bf R} = {\bf 2}$.

\subsubsection{Localised zero-mode counting for F-theory on $\mathbb R^{1,3} \times \hat Y_4$}  \label{subsec-zeromode4d-1}

The different loci $C^{(r)}_{IJ}$ are now  irreducible Riemann surfaces on the base $B_3$, corresponding to irreducible components of $C_{IJ} = \Sigma_I \cap \Sigma_J$ .
In general each such irreducible component may be described by the vanishing locus associated with a primary ideal, but for simplicity of presentation 
let us assume in the sequel  that the curve ${C^{(r)}_{IJ}}$ arises as a complete intersection locus  ${C^{(r)}_{IJ}} = D_I \cap D_J$ with $D_I$ and $D_J$ two holomorphic divisors on $B_3$. This assumption is correct, for instance, for the curve ${C^{(2)}_{01}} = \Sigma_1 \cap P$
 in the context of the $SU(2)$ Tate model.
As we have described above, to each curve $C^{(r)}_{IJ}$ we can associate  a certain representation ${\bf R}$ of the total gauge algebra. 
We will henceforth simplify notation a bit and denote the matter curves on $B_3$ as $C_{\bf R}$.

As will be discussed in more detail in section \ref{sec_Gaugebackgrounds}, in F-theory compactifications on a Calabi-Yau 4-fold, the counting of zero-modes depends on the gauge background. If the gauge background can be chosen to be trivial, then the counting is as follows:  
To each independent set $M(C^{(k)}_{\rm sp})$ of M2-branes one associates a number of 4d ${\cal N}=1$ chiral and anti-chiral supermultiplets in the respective representation, counted by 
\begin{equation}
\begin{split} \label{H0H1a}
{\rm chiral:}& \qquad H^0(C_{\bf R}, {\cal O}_{C_{\bf R}} \otimes \sqrt{K_{C_{\bf R}}}) \cr
{\rm anti-chiral:}& \qquad H^1(C_{\bf R}, {\cal O}_{C_{\bf R}} \otimes \sqrt{K_{C_{\bf R}}})  \,.
\end{split}
\end{equation}
Each of these mutliplets is accompanied by its CPT conjugate, where we recall that  the CPT conjugate of a 4d chiral multiplet in representation ${\bf R}$ is an anti-chiral multiplet in ${\bf \bar R}$. The trivial bundle ${\cal O}_{C_{\bf R}}$ can of course be omitted, and will be replaced by a suitable gauge sheaf or bundle in the presence of non-trivial gauge backgrounds in section \ref{sec_Gaugebackgrounds}.

(\ref{H0H1a}) has  been derived in the framework of the topologically twisted field theory on the 7-branes in F-theory \cite{Beasley:2008dc,Donagi:2008ca}, and in fact agrees with
its counterpart in perturbative intersecting B-type branes \cite{Katz:2002gh,Blumenhagen:2008zz}.
The starting point of the derivation is the 8d ${\cal N}=1$ Super Yang-Mills theory along the 7-branes coupled to a 6d defect consisting of a 6d ${\cal N}=(0,1)$ hypermultiplet in representation ${\bf R}$.
The topological twist of the 8d bulk theory compactified on the K\"ahler surface $\Sigma_I$ induces a compatible topological twist of the 6d defect theory compactified on the curve $C_{\bf R}$.
 Reduction of the scalars and fermions of the 6d hypermultiplet along $C_{\bf R}$ gives rise to the above spectrum in 4d upon applying the topological twist.

The formula (\ref{H0H1a}) reflects the intuitive idea that the wavefunctions describing the zero-modes transform as spinors on the matter curve and are hence sections of the spin bundle.
On a general Riemann surface of genus $g$, the notion of the spin bundle is highly ambiguous as there exist $2^{2g}$ inequivalent spin structures. These correspond to the possible boundary conditions (periodic or anti-periodic) along each of the $2g$ one-cycles on the complex curve. In the present context, the correct choice for the 
spin bundle on $C_{\bf R}$  is to pick the one 
compatible with the   embedding of  $C_{\bf R}$ as a holomorphic curve into the base $B_3$ \cite{Intriligator:2012ue,Bies:2014sra}.
To determine the spin bundle induced by this embedding let us first compute the canonical bundle on ${ C_{\bf R}}$. This is done with the help of the adjunction formula, which states the following: 
Consider the hypersurface  associated with a divisor $D$  within a complex space $X$. Then the canonical bundle on $D$ is computed in terms of the pullback of the canonical bundle on $X$ and the normal bundle of $D$ within $X$ as 
\bea
K_D = K_X |_D \otimes N_{D/X} = K_X |_D \otimes {\cal O}(D)|_X \,.
\eea
The second equality uses that by definition, the normal bundle to the divisor $D$ is a line bundle on $X$ with first Chern class $[D]$. If $X$ is simply-connected, a line bundle is uniquely determined by its first Chern class.

We can now apply the adjunction formula to  $C_{\bf R} = D_I \cap D_J$, viewed as a hypersurface within $D_J$, to write  
\bea
K_{ C_{\bf R}} = K_{D_I} |_{C_{\bf R}} \otimes N_{ C_{\bf R} / D_I} 
\eea
with 
\bea
N_{C_{\bf R} / D_I} = {\cal O}_{D_I}(D_J|_{D_I}) |_{C_{\bf R}} \,.
\eea
As for $K_{D_I}$, we use again the adjunction formula for $D_I$, viewed as a hypersurface on $B_3$,
\bea
K_{D_I}  = K_{B_3}|_{D_I} \otimes N_{D_I/B_3} = K_{B_3}|_{D_I} \otimes  {\cal O}_{B_3} (D_I)    |_{D_I} \,.      
\eea
Altogether this allows us to express the canonical bundle on $C_{\bf R}$ as the pullback of a line bundle from $B_3$, 
\bea
K_{ C_{\bf R}}  = {\cal M}|_{C_{\bf R}}, \qquad {\cal M} = K_{B_3} \otimes {\cal O}_{B_3}(D_I) \otimes {\cal O}_{B_3}(D_J) \,.
\eea 
Since $B_3$ is simply connected, the line bundle in question is uniquely determined by its first Chern class.
Furthermore, the assumption that we can switch on a trivial gauge background\footnote{More generally, this is guaranteed by the assumption that the gauge background can be chosen such that a trivial gauge background on $C_{\bf R}$ is consistent.} implies that the first Chern of the line bundle ${\cal M}$ on $B_3$ is even. Hence the line bundle $\sqrt{\cal M}$ on $B_3$ is the \emph{unique} line bundle with first Chern class $\frac{1}{2} c_1({\cal M}) = \frac{1}{2} (c_1(B_3) + [D_I] + [D_J]) \in H^2(B_3, \mathbb Z)$. The spin bundle on $C_{\bf R}$ induced by the holomorphic embedding is then 
\bea \label{spnbundlechoice}
\sqrt{K_{ C_{\bf R}} } =  \sqrt{{\cal M}} |_{C_{\bf R}} \,.
\eea

Note that the dimensions of the two cohomology groups in (\ref{H0H1a}) are the same. This is a consequence of the Serre duality formula (\ref{Serre}).
The spectrum in absence of gauge flux is therefore vector-like, meaning that the chiral index vanishes,
\bea
\chi({\bf R}) = h^0(C_{\bf R}, \sqrt{K_{C_{\bf R}}}) -  h^1(C_{\bf R}, \sqrt{K_{C_{\bf R}}})  = 0 \,.
\eea

If we are interested in evaluating the actual number of massless vectorlike pairs, we must apply techniques to compute the dimensions of the cohomology groups (\ref{H0H1a}). This amounts, in absence of gauge backgrounds, to counting the number of sections of the spin bundle, for the specific choice (\ref{spnbundlechoice}). Since this spin bundle is by construction the pullback of a line bundle from the base $B_3$ to $C_{\bf R}$ we can use for instance the methods of \cite{Blumenhagen:2010pv} whenever $B_3$ is embedded into a toric space as a hypersurface or complete intersection. We will come back to this at the end of section \ref{sec_localchargedmatterb}.

\subsubsection{Localised zero-mode counting for F-theory on $\mathbb R^{1,1} \times \hat Y_5$}  \label{subsec-zeromode2d-1}

In F-theory compactifications on Calabi-Yau 5-folds to two dimensions, the matter loci $C_{\bf R}$ represent complex K\"ahler surfaces.
The topological twisting procedure applied for the bulk modes in section \ref{subsec_Zeromodeson7brane} allows one to deduce the counting of localised matter \cite{Schafer-Nameki:2016cfr,Apruzzi:2016iac}. The latter is viewed as a 6d defect, which is dimensionally reduced on the surface $C_{\bf R}$. We again assume for now that the gauge background is trivial.
The independent 2d $(0,2)$ multiplets in representation ${\bf R}$ are counted as follows:
\be
\begin{split} \label{Y5localisedmatter}
{\rm chiral \, \, multiplets}:&   \qquad H^0(C_{\bf R}, {\cal O}_{C_{\bf R}} \otimes \sqrt{K_{C_{\bf R}}})  \cr
{\rm Fermi \, \, multiplets}:&   \qquad H^1(C_{\bf R}, {\cal O}_{C_{\bf R}} \otimes \sqrt{K_{C_{\bf R}}})  \cr
{\rm chiral \, \, multiplets}:&   \qquad H^2(C_{\bf R}, {\cal O}_{C_{\bf R}} \otimes \sqrt{K_{C_{\bf R}}}) \,, \cr
\end{split}
\ee
where we use the same conventions for the chirality of the 2d $(0,2)$  multiplets as in section \ref{subsec_Zeromodeson7brane}. 
These are accompanied by their CPT conjugates in representation ${\bf \bar R}$. Note, however, that CPT does not change the chirality of the 2d $(0,2)$ multiplets. 

The net chiral index associated with these modes is non-zero, even for trivial gauge bundle ${\cal O}_{C_{\bf R}}$, and is computed by the Hirzebruch-Riemann-Roch index
\bea
\chi(C_{\bf R}, {\bf R}) &=& \sum_{i=0}^2 (-1)^i h^i(C_{\bf R}, {\cal O}_{C_{\bf R}} \otimes \sqrt{K_{C_{\bf R}}})  \\
&=&\int_{C_{\bf R}}      c_1^2(C_{\bf R}) \left( \frac{1}{12} - \frac{1}{8} {\rm rk}({\cal O}_{C_{\bf R}})   \right) + \frac{1}{12} c_2({C_{\bf R}})  + \left(\frac{1}{2} c_1^2({\cal O}_{C_{\bf R}}) - c_2({\cal O}_{\bf R})\right) \\
&=& \int_{C_{\bf R}}     ( - \frac{1}{24}c_1^2(C_{\bf R}) + \frac{1}{12} c_2({C_{\bf R}})) \,.
\eea 
Application of this index requires the surface $C_{\bf R}$ to be smooth; otherwise the singular space must be normalised and above formula is modified by various correction terms \cite{Schafer-Nameki:2016cfr}.

\subsection{Conformal matter in codimension two} \label{subsec_CFT}    

As long as the vanishing orders of the Weierstrass sections $f$ and $g$  do not equal or exceed $4$ and $6$ in codimension one, a flat Calabi-Yau resolution of the singularities
over generic points of the discriminant is guaranteed to exist. 
But does this imply that also the singularities in higher codimension admit a Calabi-Yau resolution?

The answer is that two types of interesting complications can - and do - occur in codimension two.
The first complication is that, even though the Weierstrass sections satisfy the $(4,6)$-bound in codimension one, their vanishing orders can easily overshoot these values at the intersection of two components of the discriminant. 
For instance, at the intersection of two discriminant components $\Sigma_1$ and $\Sigma_2$ each carrying an ${\mathfrak e}_6$ gauge algebra, the enhancement pattern is of the form
\begin{equation}
\begin{split}
&( {\rm ord}(f), {\rm ord}(g),{\rm ord}(\Delta))|_{\Sigma_i} = (3,4,8), \quad i=1,2 \\
 \Longrightarrow \quad &  ({\rm ord}(f), {\rm ord}(g),{\rm ord}(\Delta))|_{\Sigma_1 \cap \Sigma_2} = (6,8,16) \,.
\end{split}
\end{equation}
In such a situation, a flat, i.e. equi-dimensional, resolution of the Weierstrass model is not possible without further surgeries on the base.
More precisely, consider an elliptic fibration over a two-dimensional base $B_2$ (i.e. F-theory in six dimensions), and assume
that the vanishing orders in codimension two satisfy the bound
\bea
( {\rm ord}(f), {\rm ord}(g),{\rm ord}(\Delta))|_{\rm codim-2}   \neq (8,12,24)  \quad {\rm or \, \, higher} \,.
\eea
Under this assumption one can perform of a finite number of blow-ups in the base such that the elliptic fibration over the blown-up base satisfies the minimality bound.
I.e. $( {\rm ord}(f), {\rm ord}(g),{\rm ord}(\Delta)$ are smaller than $(4,6,12)$ over every point after the blowup \cite{Grassi1991,GrassiEqui}.  
Note that the blow-up in the base introduces new divisors $\Sigma^{e}_i$ and in particular in general changes the canonical bundle of $B_2$, but the canonical bundle of the total elliptic fibration does not change. Depending on specifics of the vanishing orders, the new blow-up divisors $\Sigma^{e}_i$ may carry non-trivial gauge algebra. 

The original theory, prior to the blowup, can be interpreted as the limit in which the exceptional base divisors are blown-down. In view of the relation $1/g_i^2 \simeq {\rm Vol}(\Sigma^e_i)$ for the 7-brane gauge theory wrapping the divisor $\Sigma^{e}_i$ (see eqn (\ref{gaugecouplinggeneral})), this limit is the strong coupling limit of the gauge theory. In addition, string-like objects from D3-branes wrapping the exceptional divisor become tensionless. 
There is strong evidence that this tensionless limit defines a strongly coupled non-trivial superconformal field theory (SCFT), dubbed 6d $N =(1,0)$ SCFT. 
The superconformal fixed-point lies at the origin of the tensor branch of the gauge theory because the volume of the divisors $\Sigma^{e}_i$ is controlled by the VEV of the scalar field in the $N=(1,0)$ tensor multiplets coupling to the self-dual strings. We come back to this in section \ref{NPQFT}.
The realisation of such SCFTs in F-theory has been studied extensively in the recent literature, as reviewed in \cite{Heckman:2018jxk}.
The extra degrees of freedom hidden at codimension-two loci with vanishing orders beyond $(4,6)$ have been called conformal matter \cite{DelZotto:2014hpa}, in generalisation of the ordinary localised matter at the intersection of two discriminant components not exceeding this bound.

\subsection{Codimension-two singularities without a crepant resolution} \label{sec_terminalco2}     

A second, and rather frequent, phenomenon is that even though the vanishing orders in codimension two are minimal, the codimension-two singularities do nonetheless not admit a Calabi-Yau resolution.
Mathematically, the singularities of this type which occur in elliptic fibrations in codimension-two are $\mathbb Q$-factorial terminal, and they indicate the presence of localised matter uncharged under any continuous gauge group \cite{Arras:2016evy,Grassi:2018rva}.

To understand this phenomenon, we first need to introduce some terminology. We are following the presentation in \cite{Grassi:2018rva}, to which we refer for the detailed referencing of the individual mathematical results quoted.  General background on singularities is provided e.g. in \cite{Ishi}.
Given an algebraic variety $X$ with singularities, a resolution of $X$ is a birational morphism $\rho: \hat X \rightarrow X$ such that $\hat X$ is smooth and $\hat X$ and $\rho^{-1}X$ differ only along the exceptional set on $\hat X$. If the exceptional set contains codimension-one loci, the resolution is called a big resolution, with exceptional divisors $E_i$; otherwise it is called a small resolution. 
On a general singular variety $X$ it might not be possible to define the canonical bundle $K_X$ as a line bundle (but merely as a coherent sheaf). Equivalently the canonical divisor may not be a Cartier divisor (but merely a Weil divisor). We recall the difference between both notations of divisors in Appendix \ref{app_divisors}. If the canonical bundle {\it is} a line bundle, $X$ is called Gorenstein (equivalently, $X$ is said to have only Gorenstein singularities); more generally, if $r \, K_X$ is a line bundle for some $r \in \mathbb Z$, then $X$ is called $\mathbb Q$-Gorenstein and the minimal such $r$ is the index of the singularity. 

As we will see momentarily, for our purposes it is sufficient to focus on Gorenstein singularities, i.e. to the case $r=1$.
The canonical bundle of $\hat X$ and $X$ are then related as\footnote{For $\mathbb Q$-Gorenstein singularities, $r K_{\hat X} =  \rho^* (r K_X) + \sum_i (r  a_i) E_i$ with {\it rational} discrepancies $a_i \in \mathbb Q$.}
\bea
K_{\hat X} = \rho^* K_X + \sum_i a_i E_i \,.
\eea
The parameters $a_i \in \mathbb Z$ are called the discrepancies and they only depend on the type of singularity, not on the specifics of the birational resolution chosen. 
If all $a_i = 0$, the resolution is called {\it crepant} because there is no discrepancy between $K_{\hat X}$ and $\rho^* K_X$. More generally,  the singularity on $X$ is called
{\it terminal}   if all $a_i >0$, {\it canonical} if  all $a_i \geq 0$ and {\it log canonical} if  all $a_i \geq -1$.\footnote{For $\mathbb Q$-Gorenstein singularities with index $r \neq 1$, for which $a_i \in \mathbb Q$, another notion are the {\it klt} singularities (all $a_i >-1$).}
The importance of terminal singularities is that given a canonical singularity, one can always perform a {\it partial} crepant resolution to a space $\hat X$ such that $K_{\hat X} = \rho^\ast K_X$ and $\hat X$ has only terminal singularities.

Let us from now on focus on the case where $X$ is a complex 3-fold. $X$ is Calabi-Yau if its canonical bundle is the trivial line bundle, $K_X = {\cal O}_X$, and if $H^{1}(X,{\cal O}) = H^{2}(X,{\cal O})=0$.
Note that if $X$ is a Calabi-Yau, then it is Gorenstein ($r=1$) in the above sense.
We are particularly interested in determining whether a resolution into a smooth Calabi-Yau 3-fold $\hat X$ exists, i.e. whether a crepant resolution of the singularity exists. 

According to the above definitions, this is the case if either for any big resolution $a_i =0$ $\forall \, i$, or if there exists a small resolution; in the latter case, all exceptional divisors are trivial by definition.
To determine whether such a small resolution is possible, we need one last mathematical notion, due to the following fact:
A small resolution of a canonical singularity (or, if $X$ is only $\mathbb Q$-Gorenstein with $r \neq 1$, of a  klt singularity) exists if and only if $X$ is not $\mathbb Q$-factorial. An algebraic variety is $\mathbb Q$-factorial if every Weil divisor is also $\mathbb Q$-Cartier, i.e. if there exists some $k \in \mathbb Z$ such that $k \, D$ can be locally expressed as the vanishing locus of a single function on $X$. While on smooth spaces, this is always the case, this property need not hold in the presence of singularities. Determining if a singular variety is $\mathbb Q$-factorial requires global information.
As a familiar example, consider a singularity of conifold type. The singularity is {\it locally} a hypersurface singularity $x_1 x_2 - x_3 x_4 =0$. The so-defined hypersurface is not locally $\mathbb Q$-factorial because e.g. the divisor $x_1 = x_3 =0$ lies on the hypersurface even though it cannot be expressed as the intersection  of the defining hypersurface equation with the vanishing locus of another single holomorphic function. Correspondingly, a local small resolution exists. 
However, in the presence of higher order terms, such as in $x_1 x_2 - x_3 x_4 + x_1^3 + x_2^3 =0$, the singularity is globally $\mathbb Q$-factorial even though it is not locally $\mathbb Q$-factorial, and the local small resolution does not extend globally. 
In the above nomenclature, the conifold singularity is terminal in the sense that a big resolution leads to exceptional divisors with positive discrepancies $a_i >0$. But if the conifold singularity is not $\mathbb Q$-factorial globally, then a small resolution exists and hence the singularity is crepantly resolvable.

After this preparation, we come to the main theorem which makes the importance of these concepts clear: 
Consider a Weierstrass model $Y_3$ over a complex 2-dimensional base $B_2$ such that $f$ and $g$ satisfy the minimality bound. Then there exists, possibly after a succession of birational transformations (blowups) of the base $B_2$, an equi-dimensional partial crepant resolution $\hat Y_3$ of the Weierstrass model {\it up to the appearance of $\mathbb Q$-factorial terminal singularities} over isolated points on $B_2$ \cite{Grassi1991,GrassiEqui}. In particular, the end-point of the blowups in the base required to get rid of the non-minimal vanishing orders in codimension mentioned in the previous section may in general contain such types of singularities.

The appearance of $\mathbb Q$-factorial terminal singularities is quite common: Consider for instance  the set of otherwise generic Weierstrass models with the property that the discriminant factorises into two components $\Sigma = \Sigma_0 \cup \Sigma_1$ such that the elliptic fiber over $\Sigma_0$ is of $I_1$-type and the fibration over $\Sigma_1$ is minimal (no vanishing orders beyond $(4,6)$).
Among the $23$ different families of such models \cite{Grassi:2000we,Grassi:2011hq} (including infinite series such as the $I_n$ or $I_n^\ast$ families over $\Sigma_1$), three such families include $\mathbb Q$-factorial terminal singularities in codimension two, namely  \cite{Grassi:2018rva}
\begin{center}\begin{tabular}{llll}
& $\Sigma_1$: $I_1$   \quad &$I_1 \times I_1 \rightarrow I_2$    & ({\rm conifold}) \\
&$\Sigma_1: II $\quad &$I_1 \times II \rightarrow III$  &$(x_1^3 + x_2^2 + x_3^2 + x_4^2 = 0)$ \\
&$\Sigma_1: I_{2k+1}^{\rm n.s.} $& $I_1 \times I_{2k+1}^{\rm n.s.} \rightarrow I_{2k+2}$   &({\rm conifold})
\end{tabular}
\end{center}
Here we are listing the Kodaira fiber over $\Sigma_1$, the enhancement at the relevant collision with $\Sigma_0$ as well as the local form of the singularity. In the last case, one has to perform a partial resolution until the only remaining singularity is a $\mathbb Q$-factorial terminal singularity of conifold type in codimension-two. 
In fact, singularities which are locally of the first type $I_1 \times I_1 \rightarrow I_2$ occur quite generally in the Jacobian of genus-one fibrations describing abelian discrete symmetries \cite{Braun:2014oya}.   These will be discussed in detail in section \ref{sec_genusone}.
The physics of the first two of these and other examples of $\mathbb Q$-factorial terminal singularities has been studied in \cite{Arras:2016evy}. Other occurrences of $\mathbb Q$-factorial terminal singularities on elliptic threefolds in F-theory have appeared for instance in \cite{Font:2016odl,Morrison:2016lix}.

Even though a Weierstrass model $Y_3$ with $\mathbb Q$-factorial terminal singularities does not have a crepant resolution, essentially all of its properties of relevance to us can be computed by working on the singular space (or its partial resolution):
This is because Calabi-Yau 3-folds with $\mathbb Q$-factorial terminal singularities still enjoy rational Poincar\'e invariance \cite{Grassi:2018rva} and hence admit a non-degenerate intersection pairing in particular between the Cartan divisors $E_i$ of the partial resolution and the fibral curves. Therefore the assignment of Lie algebras and representations to the degenerate fibers in codimension one and two is not affected by the $\mathbb Q$-factorial terminal singularities. This is proven in detail in \cite{Grassi:2018rva}. 

What is affected by the appearance of the singularity,
on the other hand, is the precise counting and interpretation of {\it uncharged} massless matter. In F-theory on a smooth Calabi-Yau threefold $\hat Y_3$, the number of uncharged matter hypermultiplets is given by
$1 + h^{1,2}(\hat Y_3) = \frac{1}{2} b_3(\hat Y_3)$, where $h^{1,2}(\hat Y_3)$ counts the complex structure moduli and the first term accounts for the universal hypermultiplet. 
In the presence of singularities, even $\mathbb Q$-factorial terminal ones, the number of complex structure deformations is no longer counted by $h^{1,2}(\hat Y_2)$. In fact, the existence of a pure Hodge structure and Hodge duality, which underlies this reasoning in the smooth case, is in general not guaranteed on singular spaces. 
Rather, to compute the number of complex deformations of a singular Calabi-Yau 3-fold $X$ with $\mathbb Q$-factorial terminal singularities one first performs a smooth deformation of $X$ to a smooth Calabi-Yau space $X_t$; such a smoothing always exists for $\mathbb Q$-factorial terminal singularities. The counting of the complex deformations is then performed on the smooth space,
\bea
{\rm CxDef}(X) = {\rm CxDef}(X_t) = \frac{1}{2} b_3(X_t) = \frac{1}{2} (b_3(X) + \sum_P m(P)) \label{Cxdefsplit1} \,. 
\eea
Here we are using the relation between the Betti numbers of $X$ and $X_t$, which involves summing up the Milnor numbers $m(P)$ over each singular point $P$.
The Milnor number can be thought of as the number of independent 3-spheres inserted upon smoothing the singularity. For a given hypersurface singularity described by the vanishing of a function $f(x_1, \ldots, x_n)$ with local coordinates $x_i$, it can be computed as
\bea
m(P) = {\rm dim}(\mathbb C{[x_i]}/\langle \partial_1 f, \ldots, \partial_n f \rangle) \,.
\eea
Note that in general only $b_3(X) + \sum_P m(P)$ is even, but not $b_3(X)$ itself, in agreement with the expectation that a pure Hodge structure and Hodge duality is in general not available. 
The Milnor number is related to another invariant of the singularity, the Tyurina number defined as 
\bea
\tau(P) = {\rm dim}(\mathbb C{[x_i]}/\langle f,\partial_1 f, \ldots, \partial_n f \rangle) \,.
\eea
The Tyurina number coincides with the dimension of the space of versal deformations, i.e. localised complex structure deformations of $X$ which only affect the singularity at $P$. 
Importantly, $\tau(P) = m(P)$ whenever the singularity is a homogenous  weighted hypersurface singularity. All $\mathbb Q$-factorial terminal singularities that have been studied on Weierstrass models so far are of this type. Let us  write (\ref{Cxdefsplit1}) as
\bea
{\rm CxDef}(X) &=&\frac{1}{2} (b_3(X) - \sum_P m(P))  + \sum_P m(P) \label{Cxdefsplit2} \,.
\eea
 Whenever $\tau(P) = m(P)$, the first term,
 $\frac{1}{2} (b_3(X) - \sum_P m(P))$, counts the unlocalized uncharged hypermultiplets, while $\sum_P m(P)$ counts the localised uncharged hypermultiplets corresponding to the localised complex structure deformations. 
This interpretation of localised uncharged matter at the codimension-two $\mathbb Q$-factorial terminal singularities is also supported by examples with a Type IIB limit, in particular by the singularities of the form  $I_1 \times I_1 \rightarrow I_2$ \cite{Braun:2014oya,Braun:2014nva}. The counting is in agreement with the cancellation of gravitational anomalies in the 6d effective supergravity \cite{Arras:2016evy,Grassi:2018rva}.

To summarize, possibly after performing blowups of the base to remove non-minimal vanishing orders of the Weierstrass model in the codimension-two, an elliptic Calabi-Yau threefold always allows for a partial crepant resolution up to the appearance of $\mathbb Q$-factorial terminal singularities over isolated points on $B_2$. Each such singularity contributes $m(P)$ localised hypermultiplets which are uncharged under any global gauge group. In addition, the codimension-two point can carry extra charged hypermultiplets which are determined with the same methods as in the smooth case.

It is expected that a similar interpretation of  codimension-two $\mathbb Q$-factorial terminal singularities generalizes to elliptic 4-folds. In particular, a natural conjecture for the counting of the localised uncharged matter is that if the singularity occurs over a curve $C \subset B_3$, the number of corresponding vectorlike pairs of ${\cal N}=1$ chiral multiplets is given by $m(P) \, h^0(C, \sqrt{K_C})$ \cite{Arras:2016evy}, but the corresponding mathematical theorems have not yet been proven as of this writing.

\section{Higher-codimension singularities} \label{sec_Codimhigher}

If the base of the elliptic fibration is at least complex 3-dimensional, the singularity type in the fiber can enhance further in codimension three (or higher).
Typically, this occurs when two or three matter loci intersect. The enhancement loci reflect the existence of holomorphic matter couplings in the associated effective action.
We first describe this phenomenon in 4d compactifications on an elliptic 4-fold $\hat Y_4$, where the codimension-three singularities signal cubic Yukawa couplings between localised matter fields.
We then briefly comment on the structure and F-theory interpretation of codimension-three and -four singularities on elliptic 5-folds.

\subsection{Codimension-three singularities and Yukawa couplings on elliptic 4-folds} \label{Codim3Yuk}

 \subsubsection{Fiber degenerations in codimension three}

Consider the intersection of three matter curves $C_{{\bf R}_1}$, $C_{{\bf R}_2}$ and $C_{{\bf R}_3}$ on the base $B_3$ of an elliptic 4-fold $\hat Y_4$. 
Some of the representations (and their associated curves) can in fact be identical. 
Generically, the involved curves intersect in several isolated points. For instance, if two matter curves  $C_{{\bf R}_1}$ and $C_{{\bf R}_2}$ lie within the same discriminant component $\Sigma_I$ (such that ${\bf R}_1$ and ${\bf R}_2$ are both representations of the algebra $\mathfrak{g}_I$), they generically intersect at points within the complex surface $\Sigma_I$. 
These intersection points may fall into several groups distinguished by the local enhancement type in the fiber.
We focus for now on the neighborhood of one such intersection point $p$.

At the collision point $p$ the vanishing orders of the Weierstrass model sections and of the discriminant increase, signalling an enhancement of the fiber degeneration similar to the enhancement from codimension one to codimension two.
In this sense one can again formally associate a Kodaira singularity type and a corresponding Lie algebra to the enhancement point, by naive application of the Kodaira table. As in codimension two, there is no gauge theory associated with this Lie algebra, but there is a sense in which the Lie algebra does describe aspects of the effective action as will be described below.

On the resolved space $\hat Y_4$, as we approach $p$ along the intersecting matter loci $C_{{\bf R}_i}$ one or several of the fibral curves split. This is analogous to the splitting of fibral curves responsible for the enhanced degeneration of the fiber in codimension two, with an important difference: The result of the curve splitting as considered as we approach $p$ from one of the matter curves, say $C_{{\bf R}_1}$, leads to fibral curves at $p$ which already exist over generic points of one of the other matter curves meeting at $p$. In this sense, no new types of curves are produced in the cone of effective curves in the fiber.

The fiber degenerations obtainable in this manner have not been classified rigorously in the mathematics literature to date, but many examples have been studied in the physics literature. Indeed, for any given elliptic fibration together with a resolution, the analysis of the codimension-three enhancements is possible in practice. 
The enhanced fiber structure that has been observed is that the fibers correspond to Kodaira-type fibers, possibly with one or several of the nodes deleted. This phenomenon was first observed in \cite{Esole:2011sm} in an elliptic fibration with gauge algebra $\mathfrak{g} = \mathfrak{su}(5)$, which serves as our\footnote{Note that the $I_2$ fibration which we have studied in section \ref{subsec_TateI2-1} and \ref{SU2Tatematter}, no codimension-three fiber enhancements occur: There is only one matter curve $C_{\bf 2}$ and no enhancement at potential self-intersections. This is of course in agreement with the physics interpretation presented below.} 

\subsubsection*{Example: $I_5$ Tate model}

At the level of the Weierstrass model, the enhancements patterns and the resolution are particularly simple to study if we focus on a fibration in global Tate form (\ref{PT}), with vanishing orders leading to an $I_5$ enhancement in the fiber over the divisor $W: \{w=0\}$ \cite{Bershadsky:1996nh}
\be \label{TateSU5ai}
a_1 \, \, {\rm generic}, \qquad a_2 = a_{21} \, w, \qquad a_3 = a_{32} \, w^2, \qquad a_4 = a_{43} \, w^3, \qquad a_6 = a_{65} \, w^5 \,.
\ee
From the Weierstrass sections $f$ and $g$ (see (\ref{fg-Weier1}), (\ref{fg-Weier1})) 
 \bea
     f  = \frac{1}{48} (-a_1^4 - 8 a_1^2 a_{2,1} w + {\cal O}(w^2)), \qquad g =  \frac{1}{864} (a_1^6 + 12 a_1^4 a_{2,1} w + {\cal O}(w^2))
\eea
and the resulting discriminant
\bea
\Delta &=& \Delta_0 \, \Delta_1^5 \\
\Delta_1 &=& w, \quad \Delta_0 = \frac{1}{16 }(a_1^4 \, P + {\cal O}(w)), \quad P = a_{3,2} (a_{2,1} a_{3,2} - a_1 a_{4,3}) + a_1^2 a_{6,5} \label{defPSU5}
\eea
we immediately read off the following enhancements in codimension two, at the collision of $\Sigma_1 = \{w=0\}$ with $\Sigma_0 =\{\Delta_0=0\}$:
\bea
&C_{\bf 10} = \{w = 0\} \cap \{a_1=0\}:   &{\rm ord}(f,g,\Delta) = (2,3,6) \Rightarrow I_5 \times I_1 \rightarrow I_1^\ast \, \,  (D_5)  \\
&C_{\bf 5} = \{w = 0\} \cap \{P=0\}:   &{\rm ord}(f,g,\Delta) = (0,0,6) \Rightarrow I_5 \times I_1 \rightarrow I_6 \, \, (A_6)\,.
\eea
A detailed analysis of the resolution confirms that the curve $C_{\bf 10}$ and $C_{\bf 5}$ carry matter in representation ${\bf 10}$ and ${\bf 5}$, respectively, of $\mathfrak{su}(5)$.
The two curves intersect at two types of points on $\Sigma_1$,
\bea\label{p1p2}
 &p_1: \{w=0\} \cap \{a_1 = 0\} \cap \{a_{2,1}=0\}:   &{\rm ord}(f,g,\Delta) = (3,4,8) \Rightarrow {\rm type} \, IV^\ast \, \, (E_6)\\
 &p_2: \{w=0\} \cap \{a_1 =0 \} \cap \{a_{3,2}=0\}:   &{\rm ord}(f,g,\Delta) = (2,3,8) \Rightarrow {\rm type} \, I_2^\ast \, \, (D_6) \,.
 \eea
Here we are indicating the \emph{naive} Kodaira type and its associated Lie algebra by blindly applying the same rules for the vanishing orders of $f$, $g$ and $\Delta$ as in codimension one.
The global $I_5$-Tate model above allows for six inequivalent resolutions, which are all related by birational transformations. Since these are well documented in the literature \cite{Esole:2011sm,Marsano:2011hv,Krause:2011xj,Krause:2012yh}, we do not list them in detail here.
As it turns out, the so-called point of $E_6$-enhancement in the above model does not correspond to a full Kodaira-fiber, but rather to a  ${\rm type} \, IV^\ast$ fiber with one of its nodes  deleted from the graph. 
The precise location of the deleted node depends on the chosen resolution: In four of the six different resolutions, the  deleted node is one of the nodes of multiplicity two in the affine Dynkin diagram of $E_6$, and in the remaining two it is the central node of multiplicity three. We will give an interpretation to this deviation from the standard Kodaira fibers in section \ref{sec_YukcouplFT}. 
On the other hand, the enhancement at $p_2$ gives rise, in each of the birational resolutions, to a fiber of Kodaira type $I_2^\ast$, corresponding to the extended Dynkin diagram of $D_6$.

\subsubsection{Yukawa couplings from merging M2-branes}

We have already pointed out that the fiber degeneration at the special point $p$ is due to a splitting of fiber curves as one approaches $p$ from one of the involved matter curves.
The pattern is more precisely as follows: 
Suppose one of the splitting curves over $C_{{\bf R}_1}$ is ${\cal C}_{{\bf R}_1}^{a_1}$ in the sense that at $p$
\bea \label{splitting-Yukawa1}
{\cal C}_{{\bf R}_1}^{a_1} |_p = {\cal C}_{{\bf R}_2}^{a_2}|_p + {\cal C}_{{\bf R}_3}^{a_3}|_p \,,
\eea 
where ${\cal C}_{{\bf R}_2}^{a_2}$ and ${\cal C}_{{\bf R}_3}^{a_3}$ are fibral curves which exist also in the generic fiber over  $C_{{\bf R}_2}$ and $C_{{\bf R}_3}$ away from the intersection $p$.
The above equation can be considered either as a statement about a splitting of ${\cal C}_{{\bf R}_1}^{a_1}$ at $p$, or equivalently as the joining of the other two fibral curves. 
An M2-brane wrapping each of the three curves gives rise to a matter state carrying weight vector $\beta^{a_i}({\bf R}_i)$, localised in the fiber over the respective matter curves.
In fact, the weights are uniquely determined by the homology classes of the wrapped curves.
Hence, the homological relation 
\bea \label{homrelYuka}
[{\cal C}_{{\bf R}_1}^{a_1}] = [{\cal C}_{{\bf R}_2}^{a_2}] + [{\cal C}_{{\bf R}_3}^{a_3}]    \in H_2(\hat Y_4, \mathbb Z)
\eea
implied by the splitting (\ref{splitting-Yukawa1}) is equivalent to
\bea \label{betasuma}
\beta^{a_1}({\bf R}_1) = \beta^{a_2}({\bf R}_2) + \beta^{a_3}({\bf R}_3) \,.
\eea
This in turn means that in representation theory there exists a triple mapping
\bea \label{tripleR1R2R3}
{\overline {\bf  R}_1} \times {\bf  R}_2 \times  {\bf  R}_3 \rightarrow 1 \,.
\eea
The splitting of the M2-brane state with weight $\beta^{a_1}({\bf R}_1) $ into the M2-brane states with weights $\beta^{a_2}({\bf R}_2)$ and $\beta^{a_3}({\bf R}_3)$ is interpreted \cite{Marsano:2011hv} as a triple interaction 
among the fields associated with the M2-brane excitations in the effective action.   
The homological relation (\ref{homrelYuka}) is equivalent to the existence of a 3-chain $\Gamma$ with boundary
\bea \label{Gammap}
\partial \Gamma = - {\cal C}_{{\bf R}_2}^{a_2}  +  {\cal C}_{{\bf R}_2}^{a_2} + {\cal C}_{{\bf R}_3}^{a_3} \,.
\eea
The fact that the splitting (\ref{splitting-Yukawa1}) occurs in the fiber over a point $p$ means that this 3-chain can be localised at this point. More precisely, the volume of the interpolating 3-chain $\Gamma$ vanishes in the limit of vanishing fiber volume, which is the F-theory limit. The triple interaction localised at $p$ can be viewed as being induced by an M2-instanton wrapping $\gamma$. Since its volume vanishes in the F-theory limit, the coupling is not exponentially  suppressed \cite{Martucci:2015dxa}.

The actual computation of this interaction in the literature proceeds not at the level of M2-brane excitations, but directly in the language of the effective action, using the topologically twisted field theory governing the gauge theory on the 7-branes \cite{Donagi:2008ca,Beasley:2008kw,Hayashi:2008ba,Hayashi:2009ge,Hayashi:2009bt,Hayashi:2010zp}. The triple coupling in question translates into a holomorphic Yukawa coupling in the superpotential of the four-dimensional effective theory. The codimension-three enhancement loci are therefore oftentimes referred to as Yukawa points.

The structure of the codimension-three singularities indicates which of these Yukawa couplings are in fact present in the effective action.
While it is clear from the above that the existence of a splitting (\ref{splitting-Yukawa1}) implies the homological relation (\ref{homrelYuka}) (or, equivalently, (\ref{tripleR1R2R3})), the converse is in general not true:
The fact that a coupling is allowed from the perspective of the group theory in the sense of (\ref{betasuma}) only implies a relation of the type (\ref{homrelYuka}) in the homology group of the full fibration; but this as such does not guarantee that there exists a point $p$ over which a  splitting of the form (\ref{splitting-Yukawa1}) actually occurs. For this to happen, it must be possible to transport the three fiber curves, in a holomorphic way, to the same point such that  a geometric relation of the form (\ref{splitting-Yukawa1}) is realized locally in the fiber over $p$ \cite{Martucci:2015dxa}.
This can indeed fail to happen. In this case, there still exists a 3-chain $\Gamma_{\rm np}$ bounding the three fibral curves as in (\ref{Gammap}), but since there is no point over which the fibral curves merge into each other, the volume of $\Gamma_{\rm np}$ is non-zero even in the F-theory limit. This is because it has '1 leg in the fiber, 2 legs in the base' and its volume consequently stays finite in the limit (\ref{Frescaling4d}). The instantonic coupling induced by an M2-brane wrapping $\Gamma_{\rm np}$ is then non-perturbatively suppressed in the F-theory limit. The coupling mediated by the M2-brane along $\Gamma_{\rm np}$ is not expected to give rise to an F-term, but if it forms a bound state with an M5-brane instanton a holomorphic coupling can be induced \cite{Martucci:2015oaa}, in agreement with expectation from Type IIB theory.

We can illustrate these ideas in our $I_5$ example:
Over a generic base $B_3$, both triple intersection points $p_1$ and $p_2$ exist. A straightforward analysis of the resolved fibers and the splittings of the form (\ref{splitting-Yukawa1}) over these points reveals that at $p_1$ and $p_2$ the two Yukawa couplings ${\bf 10} \, {\bf 10} \, {\bf 5} $ and ${\bf 10} \, {\bf \bar 5} \, {\bf \bar 5}$ are generated, in the above sense. These couplings have been studied extensively in the F-theory literature beginning with \cite{Donagi:2008ca,Beasley:2008kw,Hayashi:2008ba,Hayashi:2009ge,Hayashi:2009bt,Hayashi:2010zp,Marsano:2011hv}.

On the other hand, suppose that on the base $B_3$ the topological intersection number $[w] \cdot [a_1] \cdot [a_{2,1}] = 0$. This means that there are no actual intersection points of the type $p_1$, and hence no coupling of the above form. An example of this type has been constructed in \cite{Martucci:2015dxa,Blumenhagen:2009up}. In the present case this phenomenon occurs for $I_5$ models with a smooth Sen limit. The absence of the coupling  ${\bf 10} \,  {\bf 10} \, {\bf 5} $ in perturbative Type IIB orientifold models is well-known and a consequence of global $U(1)$ symmetries which arise as gauge symmetries whose gauge potential acquires a St\"uckelberg mass.
The natural interpretation is that the absence of a Yukawa point (even though the coupling is compatible with all gauge symmetries - including discrete ones - of the theory) is a geometric manifestation of this effect in F-theory models {\it smoothly} connected to a perturbative Type IIB orientifold.\footnote{More generally, the Sen limit of a generic $I_5$ F-theory model exhibits a non-resolvable conifold singularity at the $E_6$ Yukawa point \cite{Donagi:2009ra}. In \cite{Collinucci:2016hgh} it is shown that a D1-instanton wrapping an associated vanishing cycle localised at this singularity generates the  ${\bf 10} \, {\bf 10} \, {\bf 5} $ coupling in Type IIB, effectively violating the massive $U(1)$ symmetry along the lines of \cite{Blumenhagen:2007zk}. While exponentially suppressed with ${\rm exp}(-{1}/{g_s})$ in Type IIB, this coupling indeed becomes of order 1 in the strong coupling F-theory regime.}  D3-brane instantons (reviewed e.g. in \cite{Blumenhagen:2007zk}) can nonetheless generate such a coupling \cite{Blumenhagen:2006xt,Ibanez:2006da}, albeit with an exponentially suppressed dependence on the K\"ahler moduli. In absence of gauge flux, the D3-brane instanton must in fact carry 'instanton flux' for this effect to occur \cite{Grimm:2011dj}. This translates into the M2-M5 bound state advocated in this context in \cite{Martucci:2015oaa} (see also \cite{Marsano:2011nn,Kerstan:2012cy}).

\subsubsection{Yukawa couplings in the field theoretic picture} \label{sec_YukcouplFT}

From the perspective of the topological field theory governing the dynamics on the 7-branes, the gauge theory along the 7-brane stack with Lie algebra $\mathfrak{g}$ is understood as a deformation of a gauge theory with a higher gauge algebra $\mathfrak{h} \supset \mathfrak{g}$ by suitably rotating out some of the branes. We have already invoked this picture  in codimension two \cite{Katz:1996xe} in section \ref{subsec_gen}, and now apply it to the enhancements in codimension three. 
The rotated branes all intersect at the Yukawa point, which locally remembers the higher gauge algebra $\mathfrak{h}$.
This is the interpretation of the Lie algebra associated with the codimension-three fiber (i.e. the Lie algebra whose extended Dynkin diagram - possibly modulo deleting some of the nodes  - reproduces the fiber topology).
The deformation from the local symmetry group $\mathfrak{h}$ to $\mathfrak{g}$ by rotating away some of the branes intersecting at the Yukawa point is encoded in the vacuum expectation value of a Higgs field $\varphi$ which varies along the 7-brane. The Higgs field is originally taken in the adjoint representation of $\mathfrak{h}$ and its vacuum expectation value over generic points of the 7-brane breaks $\mathfrak{h}$ to $\mathfrak{g}$ \cite{Donagi:2008ca,Beasley:2008kw,Hayashi:2008ba,Hayashi:2009ge,Hayashi:2009bt,Hayashi:2010zp}.
In this picture, the triple Yukawa coupling is a remnant of the triple gauge interaction in a theory associated with the bigger Lie algebra. 
More precisely, the parent gauge theory allows for a triple coupling  
\bea \label{tripleh}
{ \bf adj}(\mathfrak{h}) \times { \bf adj}(\mathfrak{h}) \times { \bf adj}(\mathfrak{h}) \rightarrow 1.
\eea
Decomposing  ${ \bf adj}(\mathfrak{h})$ into representations of ${ \bf adj}(\mathfrak{g})$ and plugging the decomposition into (\ref{tripleh}) contains the coupling (\ref{tripleR1R2R3}).  
Mathematically, the variation of $\varphi$ over the 7-brane defines a Higgs bundle  or spectral cover \cite{Donagi:2009ra}, and the details of this bundle are reflected in the structure of the fiber of  the resolved 4-fold $\hat Y_4$: As explained in \cite{Braun:2013cb}, if the spectral cover is ramified at the Yukawa point $p$, some of the fiber components are missing compared to the full Kodaira typer associated with the Lie algebra $\mathfrak{h}$. 
Ramification of the spectral cover in particular implies that even though the eigenvalues of $\varphi$ vanish at the Yukawa point, some of its components are non-zero. 
However, we stress again that the 'missing' of certain nodes compared to the full affine Dynkin diagram does not affect the appearance of Yukawa couplings according to the arguments above \cite{Marsano:2011hv}.

There have been intensive studies of the structure of the Yukawa couplings in the effective action within this local field theory approach, which we can only very superficially outline here.
Locally, at any single intersection point $p$, the rank of the Yukawa coupling matrix involving the different families of matter is one \cite{Heckman:2008qa,Hayashi:2009ge,Cecotti:2009zf}. This is due to a geometric selection rule acting on the wavefunctions whose overlap at $p$ determines the interaction. In phenomenological applications this means in particular that only one family of massless matter is 'heavy'.  From the perspective of particle physics this is a welcome property given the observed flavour structure in the Standard Model.\footnote{Note, however, that generically the number of Yukawa points is bigger than one and we have to sum over all interaction points \cite{Hayashi:2009bt,Cordova:2009fg,Hayashi:2011aa}.} Subleading corrections can then increase the rank of the coupling matrix \cite{Cecotti:2009zf, Marchesano:2009rz}. 
Detailed evaluations of the Yukawa couplings in benchmark models of $SU(5)$ GUT type have been performed, including \cite{Font:2009gq,Conlon:2009qq,Leontaris:2010zd,Cecotti:2010bp,Chiou:2011js,Aparicio:2011jx,Font:2012wq,Font:2013ida,Marchesano:2015dfa,Carta:2015eoh}.  In particular, the non-perturbative effects of \cite{Marchesano:2009rz} lead to interesting hierarchical structures in the flavour sector, at least at the level of the local theory \cite{Aparicio:2011jx,Font:2012wq,Font:2013ida,Marchesano:2015dfa,Carta:2015eoh}.\footnote{A different approach to account for such hierarchies is via $U(1)$ selection rules and their approximate breaking in Froggat-Nielson-type models, see e.g. \cite{Dudas:2009hu,King:2010mq,Krippendorf:2010hj,Krippendorf:2015kta} and references therein.}

Furthermore we should stress that not only the localised charged modes, but also the bulk modes propagating along the 7-brane can exhibit triple interactions. These likewise follow from  the topologically twisted field theory and are rooted in the cubic gauge interactions of the underlying 8d SYM theory \cite{Donagi:2008ca,Beasley:2008kw}, but are not localised along special points along the 7-branes.

\subsection{Codimension-three and -four singularities on elliptic 5-folds}

Let us briefly address  the situation on Calabi-Yau 5-folds \cite{Schafer-Nameki:2016cfr,Apruzzi:2016iac}.
The codimension-three singularities now occur over complex curves on the base $B_4$ over which two or more surfaces $C_{{\bf R}_i}$ intersect. At the level of the elliptic fiber, the structure of joining and splitting is analogous to the situation on 4-folds and indicates which triple couplings are group theoretically allowed. The actual computation of the couplings is to be performed in the topologically twisted field theory by evaluating the overlap between the wavefunctions in the involved representations ${\bf R}_i$ over the codimension-three curve.
The structure of 2d $(0,2)$ supersymmetry dictates that the holomorphic couplings involve one Fermi multiplet and two chiral multiplets and can be expressed in terms of a so-called $E$- or $J$-type F-term \cite{Witten:1993yc}.
This can be confirmed by analyzing the possible ways to integrate three localised matter zero-mode wavefunctions over a complex curve to form a scalar:
The zero-modes take values in the cohomology groups (\ref{Y5localisedmatter}), and the only possible map between three such wavefunctions   is of the form
\bea
H^1(C_{{\bf \bar R}_1},  \sqrt{K_{C_{{\bf \bar R}_1}}}) \times  H^0(C_{{\bf R}_2},  \sqrt{K_{C_{{\bf R}_2}}})   \times H^0(C_{{\bf R}_3},  \sqrt{K_{C_{{\bf R}_3}}}) \rightarrow \mathbb C \,.
\eea
This assumes of course that at the group theoretic level a coupling of the form (\ref{tripleR1R2R3}) is allowed, as guaranteed geometrically by the codimension-three fiber.
In particular, precisely one Fermi multiplet can participate in the holomorpic triple coupling. At the level of component fields, the coupling itself is between two fermions and one scalar: One of the fermions stems from the Fermi multiplet and another one from one of the two chiral multiplets.
While the general structure has been outlined in \cite{Schafer-Nameki:2016cfr,Apruzzi:2016iac}, explicit quantitative studies comparable to the ones on Calabi-Yau 4-folds haven not yet been performed as of this writing. 

A qualitatively novel phenomenon is that the singularity type can enhance further over codimension-four points on the base $B_4$, where several of the coupling curves intersect. The splitting and joining of fibral curves responsible for this enhancement allows for quartic interactions, which have no analogue in F-theory on 4-folds. This is in agreement with the fact that a scalar field in two dimensions has vanishing mass dimension. Hence a quartic, unsuppressed coupling involving two fermionic and two scalar fields is possible for dimensional reasons. Such holomorphic quartic couplings must again be writable as $E$- or $J$-type F-terms.

\subsection{A comment on terminal singularities in codimension three} \label{sec_N3}

To end this section, note that we have assumed here that the codimension-three singularities allow for a crepant, i.e. Calabi-Yau, resolution.
As in codimension-two - see section \ref{sec_terminalco2} -  this need not be the case in general. The mathematics and physics of potential terminal singularities in codimension three (let alone four) on the base has not been studied fully systematically in the F-theory context yet.
There is, however, at least one important class of non-crepant resolvable singularities on elliptic 4-folds in codimension three on the base which plays an interesting role in F-theory \cite{Garcia-Etxebarria:2015wns}. The singularities in question are isolated, terminal cyclic quotient singularities as classified in \cite{MorrisonStevens} (see also \cite{Anno}).
Both their geometric properties and their physics interpretation differ fundamentally from the Yukawa points reviewed in this section.

Isolated terminal cyclic 4-fold singularities compatible with the structure of an elliptic fibration are locally of the form $\mathbb C^4/ G$ with the orbifold group $G = \mathbb Z_k$ for $k=2,3,4,6$.
 The group $G$ acts simultaneously on the base coordinates and the elliptic fiber, and the restriction to the above small list of possibilities is due to the requirement that $G$ can act consistently on the elliptic fiber.
Locally near the singularity the 4-fold geometry is of the form $\mathbb C^3 \times T^2$ with local coordinates $(z_1, z_2, z_3, z_4)$; $z_4 = x + \tau y$ is the local coordinate on the torus fiber with complex structure $\tau$.
The orbifold action of $G = \mathbb Z_k$ is then defined as \cite{Garcia-Etxebarria:2015wns}
\bea \label{ziaction}
z_i \rightarrow  e^{2 \pi i v^i_k} \, z_ i \,, \qquad v_k = (1,-1,1,-1)/k \,.
\eea
The resulting orbifold singularity for $k=2$ corresponds to the location of a  perturbative O3-plane: The action on the fiber is the F-theory lift of the perturbative orientifold action $\Omega (-1)^{F_L}$, which can be identified with the $SL(2,\mathbb Z)$-action corresponding to the matrix ${\rm diag}(-1,-1)$.  The action on $\mathbb C^3$ is the geometric part of the orientifold involution. Consistently, for $k=2$ the complex structure $\tau$ of the elliptic fiber is not fixed by (\ref{ziaction}), and a weak coupling limit is possible.
 On the other hand, the values $k=3,4,6$  give rise to a non-perturbative generalisation of the concept of an orientifold \cite{Dasgupta:1996ij}.
In particular, the orbifold action on the elliptic fiber fixes the 
axio-dilaton $\tau$ at the non-perturbative values $\tau = e^{i \pi}$ (for $k=3,6$) and $\tau = i$ (for $k=4$). This has already been discussed from the $[p,q]$ 7-brane angle at the end of section \ref{sec_pq }. 

The crucial observation of  \cite{Garcia-Etxebarria:2015wns} is that probing the strongly coupled region near the orbifold singularity with a D3-brane leads to an interesting conformal field theory on the D3-brane worldvolume with (rigid) ${\cal N}=3$ supersymmetry. This F-theoretic construction, in fact, provided the first example of such exotic 
field theories. They evade conventional no-go theorems forbidding the existence of ${\cal N}=3$  field theories by being inherently strongly coupled and non-Lagrangian.
More on the concept of probing a singularity with D3-branes can be found in section \ref{NPQFT}.

\section{The Mordell-Weil group and abelian gauge symmetries } \label{sec_MWGROUP}

Up to now we have exclusively dealt with the structure of non-abelian gauge algebras in F-theory compactifications. The type of Lie algebra, its representations, and holomorphic interactions are determined by local data: this information is encoded in the singularity structure over strata of codimension one, two and three (or four) on the base of the fibration. 
Gauge theories with abelian Lie algebras, on the other hand, are sensitive to data which are intrinsically global. This is true already in perturbative orientifold models and is  even more pronounced in the language of F-theory. 
On an elliptic fibration $\hat Y_{n+1}$, non-Cartan abelian gauge symmetries are associated with rational sections of the fibration. While this general connection was known early on \cite{Morrison:1996pp,Aspinwall:1998xj,Aspinwall:2000kf} and was also at play in \cite{Klemm:1996hh}, a systematic 
investigation both within the globally defined geometry of explicit elliptic fibrations and from a physics perspective has only appeared in the more F-theory literature beginning with \cite{Grimm:2010ez}, motivated originally by particle physics model building.\footnote{In so-called semi-global models, abelian gauge symmetries have been understood in terms of split spectral covers \cite{Donagi:2009ra,Marsano:2009ym,Blumenhagen:2009yv,Marsano:2009gv,Marsano:2009wr,Dolan:2011iu,Marsano:2011nn,Dolan:2011aq}.   }
We begin by first explaining the general connection between the Mordell-Weil group of rational sections and abelian gauge symmetries in F-theory in section \ref{rationalShioda}. A key role is played by the Shioda homomorphism.
As we continue to elaborate in section \ref{app_Gaugecoupl}, its so-called height pairing determines the gauge kinetic matrix of the abelian sector.  
We explain in detail the realisation of  abelian gauge symmetries in F-theory models with a Mordell-Weil group of rank one in section  \ref{subsecU1restr} and  \ref{sec-MPm}.
Section \ref{sec_generalisations} contains a survey of generalizations and in section \ref{sec_combab-nonab} we comment on the combination of abelian with non-abelian gauge algebras. 
The torsional part of the Mordell-Group and its relation to the global structure of the gauge group (as opposed to the gauge algebra) in F-theory is the subject of section \ref{sec_torsion}.

This and the next section have overlap with the lectures by M. Cveti{\v c} at the same TASI school; we include the material here for completeness of the notes and refer to the contribution of M. Cveti{\v c} for a complementary review of this topic.

\subsection{From rational sections to abelian gauge symmetries via Shioda's map} \label{rationalShioda}

The concept of a rational section $s_A$ has already been introduced in (\ref{sigmo0def}) as a meromorphic map from the base $B_n$ of an elliptic fibration $\pi: Y_{n+1} \rightarrow B_n$ to the fiber.
The important point of this definition is that the image of $s_A$ must be defined globally, in the sense that the image does not undergo  monodromies along closed loops on the base. 
A generic Weierstrass model admits only one rational section, which is in fact holomorphic and given by the zero-section (\ref{sigma0Weier}). For non-generic $f$ and $g$, extra rational sections can occur. We will discuss examples of fibrations with extra sections in sections \ref{subsecU1restr} and  \ref{sec-MPm}.

As every elliptic curve, the elliptic fiber $\mathbb E_\tau$ is endowed with an additive group structure which assigns to every pair of points $p_1$, $p_2$ on $\mathbb E_\tau$ a third point via a commutative, associative map
\be
\begin{split} \label{boxplusdef}
\boxplus: \mathbb E_\tau \times \mathbb E_\tau &\rightarrow \mathbb E_\tau \cr
(p_1, p_2) &\mapsto p_1 \boxplus p_2 \,.
\end{split}
\ee
This gives rise to an abelian group law on $\mathbb E_\tau$ with the marked point of the elliptic curve as the zero element.
If we think of $\mathbb E_\tau$ as a lattice $\mathbb C/\Lambda$ as described in section \ref{sec_Frompqtoell}, $\boxplus$ is the obvious addition in $\mathbb C$ modulo the lattice identification and the origin as the zero element. The map (\ref{CtoEtau}) from $\mathbb C/\Lambda$ to the Weierstrass coordinates determines the addition law on the projective coordinates of the Weierstrass model. This addition is equivalent to a more inherently geometric definition given e.g. in standard textbooks on arithmetic geometry such as \cite{Silverman:2008,MR715605}.
According to the Mordell-Weil theorem \cite{Mordell,Weil}, this abelian group is finitely generated.

The group law (\ref{boxplusdef}) induces the notion of an addition on the space of rational sections in the obvious way, by defining 
\be
(s_A \boxplus s_B)(b) := s_A(b) \boxplus s_B(b) \,.
\ee 
Rational sections hence form an abelian group known as the Mordell-Weil group ${\rm MW}(\pi)$. 
The zero element is the zero-section $s_0$ which by definition exists on every elliptic fibration.   According to the Mordell-Weil theorem for function fields, proven in \cite{MR0102520},  ${\rm MW}(\pi)$ is likewise finitely generated abelian, i.e. 
 \bea
 {\rm MW}(\pi) \simeq \IZ^{\oplus r} \oplus {\rm MW}(\pi)_{\rm tor} \,.
 \eea
Here $r = {\rm rk}({\rm MW})$ is called the rank of the Mordell-Weil group and gives the number of independent rational sections (in addition to the zero element). The torsional part ${\rm MW}(\pi)_{\rm tor} = \mathbb Z_{k_1} \oplus \ldots  \oplus \mathbb Z_{k_n} $ is generated by sections such that $k_i \, s_{k_i} \equiv s_{k_i}^{\boxplus_{k_i}}= s_0$ for some finite $k_i \in \mathbb Z$. 

The Mordell-Weil group is the subject of intense studies in algebraic and arithmetic number theory, with fascinating open problems. One of them concerns the existence of an upper limit for its rank.
For example, the current record for the Mordell-Weil rank of an elliptic curve defined over $\mathbb Q$ is $r=28$ \cite{2007arXiv0709.2908E} and it remains open if higher ranks exist. 
Concerning elliptic fibrations, the highest known Mordell-Weil rank on a complex elliptic surface over $\mathbb P^1$ is $r = 68$ as constructed by T. Shioda \cite{2009arXiv0907.0298S}, while the rank of an elliptic K3 is of course at most $r \leq18$ (because $b^2({\rm K3}) = 22$, $h^{2,0}({\rm K3}) = 1$ and $\rm{rank}({\rm NS}({\rm K3})) \leq 20$); in fact the possible Mordell-Weil groups (including their torsion part) of elliptic K3 surfaces are completely known \cite{MR1813537}. Unfortunately, this is not the case for elliptic Calabi-Yau $n$-folds for $n > 2$:
As for elliptic Calabi-Yau 3-folds the current world record for ${\rm rk}({\rm MW})$ is $r=10$ as realized in an elliptic K3-fibration by N. Elkies \cite{Elkies}.

Both the freely generated subgroup $\IZ^{\oplus r}$ and the torsion part  ${\rm MW}(\pi)_{\rm tor}$ have a direct interpretation in F-theory. In the following  sections we focus first on the freely generated Mordell-Weil group, reserving an interpretation of the torsion part to section \ref{sec_torsion}. 
The importance of extra rational sections in the context of abelian gauge symmetries in F-theory lies in the fact that a non-torsional rational section induces an embedding of the base $B_n$ into $Y_{n+1}$ as a non-trivial divisor.
As we will discuss below, the specifications of the Weierstrass model required to engineer extra rational sections typically induce singularities in codimension-two which require again a resolution of the Weierstrass model to a smooth fibration $\hat Y_{n+1}$. On the resolved space $\hat Y_{n+1}$ we denote the divisor associated with a rational section $s_A$ as 
 \bea
 S_A := {\rm div}(s_A)  \in {\rm NS}(\hat Y_{n+1})
 \eea
with $ {\rm NS}(\hat Y_{n+1})$ the N\'eron-Severi (divisor class) group on $\hat Y_{n+1}$ (see Appendix \ref{app_divisors} for a definition). The class in (co)homology will again be denoted by $[S_A]$.
This is in analogy to the notation for $S_0 = {\rm div}(s_0)$ introduced in (\ref{S0divisor}).
It turns out that the divisors associated with the independent rational sections are homologically independent in  $H_n(\hat Y_{n+1})$. By the Shioda-Tate-Wazir theorem\footnote{A more precise formulation will be given in section \ref{sec_torsion}.} \cite{Shioda:1972,Wazir:2001}, $H^{1,1}(\hat Y_{n+1})$ is in fact generated by\footnote{More precisely, the theorem states that the N\'eron-Severi group of a smooth elliptic fibration $X_{n+1}$ (not necessarily Calabi-Yau) is generated by $S_0$, $S_A$, $E_{i_I}$ and $\pi^{-1}(D^{\rm b}_\alpha)$. On a Calabi-Yau $\hat Y_{n+1}$, we can identify ${\rm NS}(\hat Y_{n+1})$ with $H^{1,1}(\hat Y_4)$. See also Appendix \ref{app_divisors} for more information.}
\bea \label{h11generators-a}
H^{1,1}(\hat Y_4) = \langle [S_0], [S_A], [E_{i_I}], [\pi^{-1}(D^{\rm b}_\alpha)] \rangle \,,
\eea
and hence
\bea \label{STW}
h^{1,1}(\hat Y_{n+1}) = h^{1,1}(\hat Y_{n+1}) + {\rm rk}({\mathfrak{g}}) + 1 + {\rm rk}(\rm MW)(\pi) \,.
\eea
This is the generalisation of (\ref{STW-nosection}), which is valid if the trivial Mordell-Weil group is trivial. Note that the  sum $S_A + S_B$ within the group of divisors  is \emph{not} the divisor class associated with the addition of the sections $s_A \boxplus s_B$. A better behaved map from the Mordell-Weil group to  the divisor group will be given momentarily in terms of the so-called Shioda homomorphism.

For later purposes, let us summarize some of the intersection numbers obeyed by the rational section divisors.
To this end recall (in particular from Appendix \ref{app-notation}) that the projection $\pi: \hat Y_{n+1} \rightarrow B_n$ induces a pushforward map in homology
 $\pi_\ast:  H_{k}(\hat Y_{n+1}) \rightarrow H_{k}(B_n)$
and a pullback map in cohomology
$ \pi^\ast:  H^{k}(B_n) \rightarrow  H^{k}(\hat Y_{n+1})$ \,,
as well as an eponymous pushforward and pullback on the underlying cycle classes.
As in the rest of this article we take the liberty of using the same symbol for a cohomology class and its Poincar\'e dual homology class as well as for the intersection product in homology and cohomology.
With this understanding the projection formula can be written (for  $\omega^{\rm b}$  a cycle of real dimension $k$ on $B_n$  with associated homology class $[\omega^{\rm b}] \in H_k(B_n)$ and $\gamma \in H_{2n-k}(\hat Y_{n+1})$ as
\bea \label{projectionformula}
[\pi^{-1}(\omega^{\rm b})] \cdot_{\hat Y_{n+1}} \gamma = [\omega^{\rm b}] \cdot_{B_n} \pi_\ast(\gamma) \,.
\eea
Using this notation the intersection numbers we will need are
\bea
[S_A] \cdot \pi^\ast[D^{\rm b}_{\alpha_1}] \cdot \ldots \cdot \pi^\ast[D^{\rm b}_{\alpha_n}] &=&  [D^{\rm b}_{\alpha_1}] \cdot_{B_n} \ldots \cdot_{B_n} [D^{\rm b}_{\alpha_n}]  \label{SAint1}\\
{} [S_A] \cdot  [S_B]  \cdot     \pi^\ast[D^{\rm b}_{\alpha_1}] \cdot \ldots \cdot \pi^\ast[D^{\rm b}_{\alpha_{n-1}}]   &=&  [\pi_\ast(S_A \cdot S_B)]  \cdot_{B_n}[D^{\rm b}_{\alpha_1}] \cdot_{B_n} \ldots \cdot_{B_n}[D^{\rm b}_{\alpha_{n-1}}]  \label{SAint2} \\
{}  [S_A] \cdot  [S_A] \cdot \pi^\ast[D^{\rm b}_{\alpha_1}] \cdot \ldots \cdot \pi^\ast[D^{\rm b}_{\alpha_{n-1}}]   &=&  -c_1(B_n) \cdot_{B_n} [D^{\rm b}_{\alpha_1}]  \cdot_{B_n}\ldots \cdot_{B_n} [D^{\rm b}_{\alpha_{n-1}}]  \label{SAint3} \\
{} [S_A] \cdot  [E_{i_I}] \cdot  \pi^\ast[D^{\rm b}_{\alpha_1}] \cdot \ldots \cdot \pi^\ast[D^{\rm b}_{\alpha_{n-1}}] &=&    \pi_{A \,  i_I}   [\Sigma_I]   \cdot_{B_n}   [D^{\rm b}_{\alpha_1}] \cdot_{B_n} \ldots \cdot_{B_n} [D^{\rm b}_{\alpha_{n-1}}]  \,.\label{SAEi}
\eea

Relations   (\ref{SAint1}) and   (\ref{SAint3})  are analogous to    (\ref{S0int1}) and  (\ref{SSwint}).
Relation (\ref{SAint3}) is a special case of (\ref{SAint2}), with $\pi_\ast(S_A \cdot S_B)$ a divisor on the base. Intersection number (\ref{SAEi}) is the statement that the section intersects only one of the resolution divisors in one point over each discriminant component in one point. We have denoted this intersection number by  $\pi_{A \,  i_I}$. Note that the 
intersection point with the fiber might lie on the component $\mathbb P^1_{0_I}$ intersected by the zero-section, in which case 
$\pi_{A \,  i_I} = 0$ for $i_{I} = 1, \ldots, {\rm rk}(\mathfrak{g}_I)$.

Instead of introducing the Shioda homomorphism from a purely mathematical perspective, we consider first the relevance of rational sections in physics. As we will see, mathematics and physics will pose the same conditions on the definition of a well-behaved map from the space of sections to the space of divisors.
From the presentation in sections \ref{sec_firstlook} and
\ref{sec_Nonabeliangaugealgebras}  we recall the general relation between a divisor class $[\mathfrak{D}] \in H^{1,1}(\hat Y_{n+1})$ and gauge fields in the M-theory effective action obtained by reducing $C_3$ in terms of the dual harmonic 2-forms as $C_3 = A^{\mathfrak{D}} \wedge [\mathfrak{D}]  + \ldots$. It is then clear that M-theory compactified on an elliptic fibration with ${\rm rk}({\rm MW}) =r$ has $r$ independent abelian gauge fields in addition to the gauge potentials associated, in M-theory,  with the zero-section, the non-abelian resolution divisors  $E_{i_I}$ or the base divisors  $D^{\rm b}_\alpha$. 
In order to guarantee that these lift to abelian gauge fields in F-theory, the 2-form $[\mathfrak{D}]$  must obey the two transversality conditions (\ref{transversal1}) and  (\ref{transversal2}) motivated in section \ref{sec_firstlook}.
Finally, it is useful to normalize the $U(1)$ from the extra sections such that the non-abelian gauge bosons carry no charge under them. The latter coming from M2-branes along the fibral curves in codimension one, one imposes in addition
\bea \label{SAgaugeinvariance}  
[{\mathfrak D}]   \cdot [\mathbb P^1_{i_I}] = 0 \,.
\eea

It is worth stressing that in formulating the conditions (\ref{transversal1}) and (\ref{transversal2}) we have singled out one of the independent sections as the zero-section $S_0$.
This choice is arbitrary and does not affect the physics after taking the F-theory limit. However, intermediate computations, in particular of the loop induced Chern-Simons terms, differ, especially if the chosen zero-section $S_0$ is meromorphic as opposed to holomorphic \cite{Grimm:2013oga,Grimm:2015zea}.

One can make a general ansatz of the form ${\mathfrak D} = S_A + \delta_A$ and determine the correction term $\delta_A$ such that (\ref{transversal1}), (\ref{transversal2})  and (\ref{SAgaugeinvariance}) hold. 
Concerning (\ref{transversal2}) we note that the section divisor $S_A$ intersects the generic fiber exactly once,
\bea
[S_A] \cdot [\mathbb E_\tau] =1 \,,
\eea
and thus (\ref{transversal2}) is satisfied by $\mathfrak{D} = S_A - S_0$. Recalling the projection formula (\ref{projectionformula}), we see that 
the constraints (\ref{transversal1}) and (\ref{transversal2}) are then satisfied by 
${\mathfrak D} = S_A-S_0-\pi^{-1} (\pi_\ast((S_A-S_0)\cdot S_0))$. 

Finally, to implement (\ref{SAgaugeinvariance}) suppose the section divisor $S_A$ intersects the fibral curves in codimension-one with some intersection numbers
\bea \label{defpiAi}
\pi_{A \,  i_I} = [S_A] \cdot [\mathbb P^1_{i_I}] \,.
\eea
These are in fact the same as appearing in (\ref{SAEi}). Introduce furthermore  
\bea \label{lAjik}   
\ell_{A}^{i_I}  = \pi_{A \, j_I} (C^{-1})^{j_I i_I} \in \mathbb Q
\eea
with $(C^{-1})^{j_I i_I}$ the inverse Cartan matrix associated with the Lie algebra $\mathfrak{g}_I$. 
Then (\ref{SAgaugeinvariance}) as well as  (\ref{transversal1}) and (\ref{transversal2}) are fulfilled by $ {\mathfrak{D}} = \sigma(S_A)$ with 
\bea\label{shioda}
\sigma(s_A) = S_A-S_0-\pi^{-1} (\pi_\ast((S_A-S_0)\cdot S_0))+ \sum_I \sum_{i_I}  \ell_{A}^{i_I}   E_{i_I}  \in {\rm NS}(\hat Y_{n+1}) \otimes \mathbb Q \,.
\eea 
This follows from (\ref{defpiAi}) and (\ref{fundrelEalpha2}).

To summarize, expanding the M-theory 3-form as 
\bea \label{C3expU1A}
C_3 = A^A \wedge [\sigma(s_A) ] + \ldots
\eea
gives rise to gauge potentials in M-theory which uplift in F-theory to abelian gauge potentials associated with a (non-Cartan) abelian gauge group $U(1)_A$. 
Matter with  $U(1)_A$ charge $q_A$ originates in M2-branes wrapping a fibral curve ${\cal C}$ with non-zero intersection number with $\sigma(s_A)$ such that
\bea
q_A = [\sigma(s_A)] \cdot [{\cal C}] \,.
\eea 
This includes first the so-called 'charged singlets', by which one means states charged only under the abelian gauge algebra. These appear in fibral curves over codimension-two loci on $B_n$ over which the Weierstrass model $Y_{n+1}$ is singular due to the tuning required to  engineer the extra section $s_A$. 
Resolving these singularities on $\hat Y_{n+1}$ gives rise to a Kodaira fiber of type $I_2$ with charged matter due to M2-branes wrapped on the resolution curve in the fiber.
An interesting exception to the generic existence of such charged singlets are the models studied in \cite{Morrison:2016lix}, where the abelian gauge group is non-Higgsable because no matter fields are charged under it. 
In addition, matter charged under the non-abelian part of the gauge algebra, if present, can also acquire a non-zero $U(1)_A$ charge if the intersection of its associated fibral curve with $\sigma(s_A)$ is non-zero. 
We will present examples of both types in sections \ref{subsecU1restr},  \ref{sec-MPm} and \ref{sec_combab-nonab}.

The map $\sigma(s_A)$ associates to the section $s_A$ a divisor class on $\hat Y_{n+1}$, i.e. an element of ${\rm NS}(\hat Y_{n+1})$, with {\it rational} coefficients.
The latter point is due to the  coefficients $\ell_{A}^{i_I} \in \mathbb Q$, whose rationality is a consequence of the appearance of the inverse Cartan matrix in (\ref{lAjik}).
The map $\sigma(s_A)$ has the important and non-trivial property that 
\bea
\sigma(s_A \boxplus s_B) = \sigma(s_A) + \sigma(s_B)
\eea
with respect to the addition of divisor classes. I.e. $\sigma$ furnishes a group homomorphism
\bea  \label{Shioda-homgen}
\sigma: {\rm MW}(\pi) \to {\rm NS}(\hat Y_{n+1}) \otimes \mathbb Q \,. 
\eea
This is the Shioda homomorphism introduced first in \cite{shioda} on elliptic surfaces and generalized in \cite{Wazir:2001,Park:2011ji} to higher dimensional fibrations.

The relation between the rank of the Mordell-Weil group and the number of abelian gauge group factors in F-theory, more precisely between $\pi_1$ of the gauge group and the Mordell-Weil group, was pointed out first in \cite{Morrison:1996pp}  and elaborated further in \cite{Aspinwall:1998xj,Aspinwall:2000kf}.
The construction underlying (\ref{shioda}) has first been used explicitly in \cite{Grimm:2010ez} (in absence of non-abelian gauge algebra factors) and in~\cite{Braun:2011zm,Krause:2011xj} in presence of non-abelian gauge factors. The relation to the Shioda map has been explained in full detail in \cite{Park:2011ji}.

Finally, let us point out the following important fact: In the presence of certain gauge fluxes in F-theory compactifications on elliptic 4-folds and 5-folds,
the gauge potentials engineered in this section can acquire a mass via the 'flux induced St\"uckelberg mechanism'. In this case, the abelian symmetry is broken at the level of the gauge algebra, but it remains as a global symmetry in the low-energy effective action below the mass scale of vector field, broken only by D3/M5-instanton effects. In this sense, the Mordell-Weil group not only generates extra gauge symmetries, but, in conjunction with suitable fluxes, also (approximative) global abelian symmetries. The flux induced St\"uckelberg mechanism will concern us more in section \ref{sec_fluxes}.

\subsection{Gauge couplings and the height pairing} \label{app_Gaugecoupl}

The difference between the geometry of a non-abelian gauge algebra and an abelian one is quite striking: In the non-abelian case, the gauge degrees of freedom are clearly localized along a component $\Sigma_I$ of the discriminant divisor. 
In the abelian case, by contrast, the discriminant divisor does not split off an extra component after the coefficients of the Weierstrass model are tuned such as to accommodate an extra section. We will see this in the explicit examples of section  \ref{subsecU1restr} and  \ref{sec-MPm}. 
This raises the question if a similar understanding of an abelian gauge algebra is possible in terms of a divisor on the base $B_n$. To some extent, this turns out to be possible. 
The divisor is the so-called height pairing associated with the section \cite{Park:2011ji}. One way to understand its significance is by analyzing the $U(1)_A$ gauge coupling in the F-theory effective action.  

Consider first the effective action obtained by compactifying M-theory on $\hat Y_{n+1}$ with an extra rational section $s_A$. In view of (\ref{C3expU1A}) the $U(1)_A$ gauge kinetic action follows as 
\bea \label{Skin1a}
S_{\rm    kin} = - \frac{2   \pi}{2 } \int_{{\cal M}^{1,10}}  d C_3 \wedge \ast d C_3 = - \frac{2\pi}{2}  f_{A B}  \int_{\mathbb R^{1,8-2n}} dA^A \wedge \ast d A^B 
\eea   
where
\bea
 f_{A B} = \int_{\hat Y_{n+1}}   [\sigma(s_A)] \wedge \ast  [\sigma(s_B)] \,. 
\eea
For simplicity, let us focus for now on M-theory compactified on an elliptic 3-fold $\hat Y_{3}$ with base $B_2$.
If we denote the K\"ahler form of $\hat Y_3$ by $J_{\hat Y_3}$, then the M-theory gauge kinetic matrix is \cite{Candelas:1990pi}
\bea \label{fABM}
f_{AB} =  -  \int_{\hat Y_3}  J_{\hat Y_3} \wedge [\sigma(s_A)] \wedge [\sigma(s_B)]  +   \frac{3}{2}   \left( \int_{\hat Y_3} J_{\hat Y_3}^2 \wedge [\sigma(s_A)] \right)   \left( \int_{\hat Y_3} J_{\hat Y_3}^2 \wedge [\sigma(s_B)]   \right)    \left(  \int_{\hat Y_3}   J_{\hat Y_3}^3  \right)^{-1} \,.
\eea
To extract from this the gauge kinetic matrix in F-theory, we follow the general rules for implementing the F-theory uplift as explained in detail in \cite{Grimm:2010ks,Bonetti:2011mw}.
First, we make a general ansatz 
\bea
J_{\hat Y_3} = \pi^\ast J  +  t^0  [Z] + t^A  [S_A]  + t^{i_I} [E_{i_I}]\,, \qquad  J = t^\alpha \,  [D^{\rm b}_\alpha] \,,
\eea
where $[D^{\rm b}_\alpha]$ is a basis of $H^{1,1}(B_n)$.
Next we rescale the fiber volume to zero while at the same time scaling up the volume of the base divisors such that the volume of the divisor $\pi^{-1}(D^{\rm b}_\alpha)$ stays finite \cite{Witten:1996bn}.
For elliptic 3-folds this requires \cite{Bonetti:2011mw}
\bea
t^\alpha  \rightarrow \epsilon^{-1/2} \,  t^\alpha, \qquad t^0 \rightarrow  \epsilon \,  t^0, \qquad  t^A \rightarrow  \epsilon \,  t^A, \qquad t^{i_I} \rightarrow  \epsilon \,  t^{i_I}
\eea
with the F-theory limit corresponding to $\epsilon \rightarrow 0$. 
Since the Shioda homomorphism satisfies (\ref{transversal2}), the second term vanishes as $\epsilon \rightarrow 0$; the first term reduces, in this limit, to
\bea
f_{AB} \rightarrow \hat f_{AB}    =  - \int_{\hat Y_3} \pi^* J \wedge [\sigma(s_A)] \wedge   [\sigma(s_B)] =  \int_{B_2} J \wedge [ - \pi_\ast (\sigma(s_A) \cdot   \sigma(s_B)) ] \,,
\eea
where we are making use of the projection formula (\ref{projectionformula}).
Similar reasoning can be applied on elliptic fibrations of general dimension $n+1$.
The end result is that  the abelian gauge kinetic matrix  $\hat f_{AB}$ in the F-theory effective action is given by the K\"ahler volume, on the base $B_n$, of the divisor  
\bea\label{heightp}
\Ub_{AB}:=  - \pi_\ast (\sigma(s_A) \cdot  \sigma(s_B)) \in {\rm Cl}(B_n).
\eea
This divisor is indeed effective. It is called the height pairing of the sections $s_A$ and $s_B$.

Before coming to the physics interpretation of this divisor, let us evaluate it further as done in the F-theory literature in \cite{Braun:2011zm,Krause:2011xj,Grimm:2011fx,Morrison:2012ei}.
For brevity we focus on the diagonal terms $\Ub_{AA}$. Consider first the case without a non-abelian gauge algebra and abbreviate the correction term in (\ref{shioda}) as
\bea
D_A = \pi^{-1} \left(\pi_\ast((S_A-Z)\cdot Z) \right) \in H_4(\hat Y_{n+1}) \,.
\eea
Then 
\begin{equation}
\begin{split}
\Ub_{AA}|_{\mathfrak{g}=0} &=-    \pi_\ast \left((S_A-S_0-D_A)   \cdot(S_A-S_0-D_A) \right) \cr \label{Ua}
&= -\pi_\ast (S_A \cdot     S_A)-\pi_\ast (S_0 \cdot S_0) + 2 \pi_\ast (S_A\cdot S_0) - \pi_\ast({D_A}\cdot D_A) + 2 \pi_\ast((S_A-S_0)\cdot D_A) \,.
\end{split}
\end{equation}
The projection formula (\ref{projectionformula}) together with (\ref{SAint1})   and (\ref{S0int1})    shows that $\pi_\ast((S_A-S_0)\cdot D_A)=0$ since $D_A$ is a vertical divisor. Similarly, (\ref{S0int3}) implies that $\pi_\ast({D_A}\cdot D_A) = 0$.
By  (\ref{SAint3}) and  (\ref{SSwint}), 
\bea
\pi_{\ast} (S_A \cdot S_A) = \pi_{\ast} (S_0 \cdot S_0)   =  - \bar K  \,.
\eea
In absence of non-abelian gauge algebra, therefore
\bea \label{hp}
\Ub_{AA}|_{\mathfrak{g}=0}=2\bar K + 2\pi_\ast(S_A \cdot Z) \,.
\eea
It is left as a simple exercise to generalize this, for the full Shioda map (\ref{shioda}), to
\bea \label{hpairing-full}
\Ub_{AA} =2\bar K + 2\pi_\ast(S_A \cdot Z)  - \sum_I    \pi_{A \, k_I}   (C^{-1})^{k_I i_I}   \pi_{A \, i_I} \,  \Sigma_I   \,.
\eea
To arrive at (\ref{hpairing-full}), one uses that the Shioda map satisfies the relations (\ref{SAgaugeinvariance}), (\ref{transversal1}) and   (\ref{transversal2}) together with the intersection numbers (\ref{SAEi}).

The fact that the gauge kinetic function is controlled, via the height pairing, by the volume of the anti-canonical divisor of the base implies that abelian gauge symmetries cannot be consistently decoupled from the gravitational sector, as discussed recently with special emphasis on six-dimensional compactifications in \cite{Lee:2018ihr}. This underlines once more the global nature of abelian gauge symmetries.

The height pairing not only encodes the gauge coupling in the F-theory limit. It also plays a major role for the structure of anomalies and their cancellation via the Green-Schwarz mechanism in F-theory compacitifcations to six \cite{Park:2011ji,Bonetti:2011mw}, to four \cite{Cvetic:2012xn,Bies:2017abs} and to two \cite{Weigand:2017gwb} dimensions.

The fact that a divisor on the base carries information about the dynamics of the abelian gauge groups is not surprising from a Type IIB perspective.
In Type IIB orientifolds, stacks of 7-branes wrap distinct divisors $D_I$ on the Calabi-Yau $X_n$ which is the double cover of the base $B_n$ as sketched in section \ref{sec_PhysicsIIB1}.
While for more background on such constructions we refer to \cite{Blumenhagen:2008zz} and references therein,
let us recall here the following facts: If the divisor is not invariant under the holomorphic involution as a cycle, $D_I \neq \sigma(D_I)$, each such brane stack carries a priori an abelian gauge group factor. More precisely, a stack of $N_I$ 7-branes along $D_I$ carries a gauge group $U(N_I)$. If also the homology class $[D_I]$ is not invariant under the orientifold involution, $[D_I] \neq \sigma^\ast[D_I]$, the diagonal $U(1)_I$ factor acquires a mass via a St\"uckelberg mechanism. Since this St\"uckelberg mechanism operates even in absence of gauge fluxes, it is sometimes referred to as the 'geometric St\"uckelberg mechanism', whose realisation in F-theory invokes non-harmonic forms \cite{Grimm:2011tb}. If the mass matrix for all $U(1)_I$ gauge potentials is not of maximal rank, then  a linear combination $U(1)_A$ of individually massive $U(1)_I$  gauge potentials remains massless. This $U(1)_A$ is then associated with a certain linear combination of divisors $\sum_I  n_A^I D_I$ on $X_n$ such that $[\sum_I n_A^I D_I] = \sigma^\ast [\sum_I n_A^I D_I]$. In this sense the $U(1)_A$ is 'delocalized': It cannot be attributed to one of the 7-brane divisors $D_I$ and hence not to an individual discriminant component $\Sigma_I$ in the F-theory uplift. The linear combination of divisors, in particular the F-theory uplift of their cohomology class, is related to the height pairing divisor $b_{AA}$, which indeed is not a component of the discriminant. The relation between the Type IIB and the F-theory description of abelian gauge fields has been studied in particular in \cite{Krause:2012yh,Braun:2014nva,MayorgaPena:2017eda}.


\subsection{Example 1: The ${\rm Bl}^1 \mathbb P_{231}[6]$-fibration ($U(1)$ restricted Tate model)} \label{subsecU1restr}

Elliptic fibrations with a non-trivial Mordell-Weil group have been studied extensively  in the recent F-theory literature.
As a warmup we start with the simplest example of an elliptic fibration with Mordell-Weil group of rank one, which already contains, despite its simplicity, all relevant features to understand abelian gauge group in F-theory \cite{Grimm:2010ez}. It is also the first example that has been studied explicitly in this context.
The model is sometimes called $U(1)$ restricted Tate model because its
starting point is a generic Weierstrass model in Tate form (\ref{PT}), where, however, we set the polynomial $a_6$ to zero. 
The rational point $[x : y : z] = [0 : 0 : 1]$ now lies on the elliptic fiber.
As a result, the fibration exhibits an extra section
\bea
s_1: b \mapsto s_1(b) = [0 : 0 : 1] \,.
\eea
The price to pay is that $Y_{n+1}$ acquires an $I_2$ singularity in the fiber over the locus 
\bea \label{C34}
C_{34} = \{a_3=0 \} \cap \{a_4=0\} \,,
\eea
where the vanishing orders of $(f,g,\Delta)$ enhance to $(0,0,2)$. The locus (\ref{C34}) occurs at a self-intersection of the discriminant divisor $\Sigma$. Apart from this self-intersection, $\Sigma$ does {\it not} factorise into several irreducible components. This is a general pattern in fibrations with extra sections. The singularity in the fiber over $C_{34}$ is located at $x = y = 0$, where $P_T = dP_T=0$.
This means that the extra section $s_1$ itself is singular: It passes through a singular point in the fiber. As a result, the section is not a holomorphic, but merely a rational section.

In fact, the singularity is a conifold singularity:
The defining hypersurface equation $P_T = 0$ of a generic Tate model can be written in the suggestive form
\bea \label{PT-conifoldtypea}
A B = C D +  a_6 z^6 
\eea 
with 
\bea
A = y, \qquad  B = y + a_1 x z + a_3 z^3, \qquad C = x, \qquad D = (a_2 x z^2 + a_4 z^4) \,.
\eea
When $a_6 \equiv 0$, the hypersurface is of conifold form $A B = C D$, and the singularity at $A = B= C =D =0$ is exactly the singularity we are talking about.
In the language of section \ref{sec_terminalco2}, the conifold singularity is crepant resolvable for $a_6 \equiv 0$ because it is not $\mathbb Q$-factorial: The divisor $\{A = 0\} \cap \{C= 0\}$ on $Y_4$ is not Cartier because it cannot be locally expressed by the vanishing of a single holomorphic function on $Y_4$.

The singularity in the fiber over $C_{34}$ can be resolved by a simple blow-up \cite{Grimm:2010ez,Krause:2011xj}, introducing the blowup coordinate $s$ and replacing
\bea \label{blowups-Tate}
x \rightarrow x \, s\,, \qquad y\rightarrow y \, s \,.
\eea
After this substitution, the defining Tate polynomial is replaced as $P_T \rightarrow s^2 \hat P_T$, with the proper transform $\hat P_T$ 
\bea \label{hatPTU1restr}
\hat P_T =     y^2  s+ a_1 x y z s+ a_3 y z^3 - x^3 s^2 - a_2 x^2 z^2 s - a_4 x z^4 \,.
\eea
The resolved Calabi-Yau is defined as $\hat Y_{n+1}: \hat P_T = 0$.
The equation $\hat P_T$ is an equation in the new fiber coordinates $x$, $y$, $z$ and $s$.
The blowup (\ref{blowups-Tate}) comes together with an extra scaling relation. Including the original scaling of the Weierstrass model, the fiber ambient space coordinates are subject to the identifications
\bea
(x,y,z,s) \simeq ( \lambda^2 x,  \lambda^2 y, \lambda z, s) \simeq (  \mu x, \mu y, z, \mu^{-1} s  ) \,, \qquad \lambda, \mu \in \mathbb C^\ast \,.
\eea
The second scaling follows from the fact that $x s$ and $y s$, which correspond to the coordinates in the Weierstrass model before the blowup, are independent of the new scaling. 
The model can be compactly characterized as a ${\rm Bl}^1 \mathbb P_{231}[6]$ fibration.

The combination of both scaling relations imply that the following combinations of fiber ambient space coordinates are not allowed to vanish simultaneously,
\bea \label{SRU1restr}
{\rm SR} = \{x y, z s \} \,.
\eea
In toric language, they generate the Stanley-Reisner (SR) ideal of the fiber ambient space. 
The fiber over $C_{34}$ splits into two components because of the factorization
\bea
\hat P_T|_{a_3 =a_4=0} =  s  \, (y^2  + a_1 x y z  - x^3 s - a_2 x^2 z^2 ) \,. 
\eea
We hence identify the two fibral curves ${\cal C}_0$ and ${\cal C}_s$, where the latter is along the vanishing of $s$. They manifestly  intersect in two distinct points.
This is indeed the structure of an $I_2$ Kodaira fiber.

On the resolved space, the point $[0 : 0 : 1]$ no longer exists because $x y$ is in the SR ideal (\ref{SRU1restr}).
The role of the section $s_1$ is now played by the section divisor which is in fact given by the resolution divisor
\bea
S: \{s  =0\} \,.
\eea
This divisor on $\hat Y_{n+1}$ takes the form
\bea
\{s  =0\} \cap \{\hat P_T=0 \} = \{s  =0\} \cap \{a_3 y - a_4 x = 0\} \,,
\eea
where we have set $z \equiv 1$ by means of the scaling relations because $s z$ is in the SR ideal (\ref{SRU1restr}). 
Over a generic point of the base, the divisor $S$ describes one point in the fiber, with ambient coordinates $y = \frac{a_4}{a_3} x$ if $a_3 \neq 0$ or $x = \frac{a_3}{a_4} y$
 if $a_4 \neq 0$. The defining equation of this point is a meromorphic function in the coordinate ring, again illustrating that the section is a rational section. 
By contrast, over $C_{34}$ the divisor $S$ wraps the entire component ${\cal C}_s$ of the fiber \cite{Krause:2011xj}: Since along $C_{34}$ $a_3 =a_4=0$, the equation $a_3 y - a_4 x = 0$ is clearly solved for any $x$ and $y$.
Such behaviour is the hallmark of a rational section. 
The zero-section of the Weierstrass model 
\bea
S_0: \{z=0\} 
\eea
continues to play the role of a  holomorphic section.

There are other, equivalent presentations of the same geometry. In \cite{Braun:2011zm}, the conifold singularity is resolved by a small resolution. The resolved space is then presented as a complete intersection rather than a hypersurface in an ambient space. After the resolution, the section intersects the exceptional curve over $C_{34}$ in a point. The physics properties of this fibration are, however, completely identical.

The Shioda map and height pairing are readily evaluated: What makes this model non-generic is that $S \cdot S_0 = 0$ because $s z$ are in the SR ideal.
Hence 
\bea
\sigma(s_1) = S - Z - \bar K, \qquad b_{SS} = 2 \bar K \,.
\eea

It remains to discuss the appearance of $U(1)$ charged matter \cite{Grimm:2010ez,Krause:2011xj}.
To this send let us compute the intersection numbers of ${\cal C}_0$ and ${\cal C}_s$ with $S_0$ and $S$,
\bea
&& [S] \cdot [{\cal C}_s]=-1, \qquad [S] \cdot [{\cal C}_0]=2  \\
&& [S_0] \cdot [{\cal C}_s]=0, \qquad [S_0] \cdot [{\cal C}_0]=1 \,. 
\eea
The second equation follows by noting that $S$ wraps ${\cal C}_s$, which intersects ${\cal C}_0$ in two points as pointed out above. Since $[S] \cdot [\mathbb E]_\tau=1$ and   $[\mathbb E]_\tau =  [{\cal C}_0] +  [{\cal C}_s]$ this explains the first intersection number. Furthermore, the zero-section does not intersect ${\cal C}_s$ because $z s$ are in the SR ideal.

According to our general logic, an M2-brane wrapping ${\cal C}_s$ hence gives rise to the KK zero-modes of charged states in F-theory, with charge
\bea
q  =[\sigma(s_1)] \cdot {\cal C}_s = -1 \,.
\eea
An M2-brane wrapping ${\cal C}_s$ with opposite orientation gives rise to states with charge $+1$. 
The counting of these states proceeds in complete analogy to the counting of charged localised matter in a representation of non-abelian gauge algebras.
For instance, if the base is $B_2$, we find $n = [a_3] \cdot_{B_2} [a_4] = 12 \bar K \cdot_{B_2} \bar K$ hypermultiplets of charge $-1$. On $B_3$ and $B_4$ the numbers of charged multiplets with charge $-1$ is computed as explained in section \ref{subsec-zeromode2d-1} and \ref{subsec-zeromode4d-1} (in absence of gauge flux).

\subsection{Example 2: The ${\rm Bl}^1\mathbb P^1_{112}[4]$-fibration (Morrison-Park model)} \label{sec-MPm}   

The above construction is a special case of a more general type of fibrations with Mordell-Weil group of rank one \cite{Morrison:2012ei}. 
Finding the most general type of fibration with a given number of independent sections is a very hard and in general unsolved problem in arithmetic geometry. But there are a few ways to proceed systematically as follows.

Every elliptic fibration is birationally equivalent to a Weierstrass model and in principle the most general form of such a fibration with Mordell-Weil rank $r$ corresponds to a certain specialization of the Weierstrass polynomials $f$ and $g$. This then gives the fibration in its singular form, prior to resolution. For instance, if the Weierstrass model is to exhibit a special section of the form $[x : y : z] = [a : b : 1]$, it must be possible to write the Weierstrass equation globally as \cite{Morrison:2012ei}
\bea
(y-b)(y+b) = (x-a) (x^2 + a x + c) \,,
\eea
which can be read as a constraint on $f$ and $g$. Note the appearance of a conifold singularity at $\{ y=0\} \cap \{x=a\} \cap \{b=0\} \cap \{c + 2 a^2 =0 \}$ whose resolution is analogous to the process studied in the previous section.

Unfortunately, a section of the form $[x : y : z] = [a : b : 1]$ is still rather special because its coordinate $z =1$ globally. A more general form\footnote{The fact that this does still not give the most general such fibration is due  to the fact that, as stressed in the text, Delgine's argument is a priori true for elliptic curves.} of an elliptic fibration with Mordell-Weil group of rank one can be determined following an argument due to Deligne.
The argument is spelled out in Appendix A of \cite{Morrison:2012ei} and shows that every {\it elliptic curve} with $n$ independent rational points can be written as a  generic hypersurface of the following form:
\bea
&n=1: \qquad &\mathbb E_\tau =  \mathbb P_{231}[6] \\
&n=2: \qquad &\mathbb E_\tau =  {\rm Bl}^1\mathbb P_{112}[4]  \label{n=2Deligne} \\
&n=3: \qquad &\mathbb E_\tau =   {\rm Bl}^2\mathbb P_{111}[3] \,.
\eea
The most generic hypersurface of degree six in $\mathbb P_{231}$ is in fact the generic Tate model, which is equivalent to the generic Weierstrass model (by completing the square in $y$ and the cube in $x$) and gives rise to and elliptic curve with Mordell-Weil rank $r = n-1 =0$.\footnote{More precisely, in order for the hypersurface to describe a genus-one curve, the coefficient of $y^2$ and $x^3$ must be non-vanishing and can then be scaled to one.}
The second model is the $r =1$ model studied in \cite{Morrison:2012ei} and discussed in more detail momentarily, and the third is the $r=2$ model analysed in \cite{Borchmann:2013jwa,Cvetic:2013nia,Cvetic:2013uta,Borchmann:2013hta}.  
For $n=4$, the canonical form for ${\mathbb E}_\tau$ is as a complete intersection in ${\rm Bl}^3\mathbb P^3$ as analysed in \cite{Cvetic:2013qsa}, and beginning with $n =5$ even a description in terms of determinantal or  Grassmannian varieties is required. However, all of these models of elliptic curves are birational to  - very non-generic if $n \geq 2$ - Weierstrass models.

It is important to appreciate that the argument of Deligne is a priori about elliptic curves rather than elliptic fibrations; promoting the elliptic curves implied by Deligne's algorithm to families of elliptic curves leads to elliptic fibrations with a Mordell-Weil group of rank $r= n-1$; however, these are not necessarily the most general ones due to subtle effects in higher codimension of the parameter space. We will come back to this point in section \ref{sec_generalisations} and for now continue studying the elliptic fibration with one extra rational section as motivated by (\ref{n=2Deligne}).


To understand the meaning of  (\ref{n=2Deligne}), consider first the most general hypersurface of degree four in $\mathbb{P}_{112}$  with homogenous coordinates $[u : v :  w]$. The hypersurface is  cut out by the polynomial
\begin{equation}
\begin{aligned} \label{P2withc4}
P =  b_w w^2 + b_0 u^2 w  + b_1 u v w + b_2 v^2 w  + c_0 u^4  + c_1 u^3 v  + c_2 u^2 v^2  + c_3 u v^3  + c_4 v^4 \,.
\end{aligned}
\end{equation}
If we take the coefficients $b_i$ and $c_i$ to be constant, this defines a genus-one curve as long as $b_w$ is non-vanishing; it can hence scaled to one \cite{Morrison:2012ei}.
The most general such genus-one curve does not admit a rational point. However, 
setting $c_4 \equiv 0$ gives rise to the two rational points 
\begin{eqnarray} \label{eq:sec01}
 s_1: [u : v : w] = [0 : 1 : -b_2], \qquad s_2: [u : v : w] = [0 : 1 : 0] \,,
\end{eqnarray}
which indeed generate the Mordell-Weil group of $\mathbb E_\tau$.
With $c_4 \equiv 0$ the above hypersurface is not the most general one of its degree any more, but this can be remedied by introducing an exceptional divisor and hence changing the ambient space to ${\rm Bl}^1\mathbb P_{112}$, the blowup of $\mathbb P_{112}$ in one point. The most general 
hypersurface of degree four  (with respect to the scaling of  $[u : v : w]$) in ${\rm Bl}^1\mathbb P_{112}$ has then two independent rational points. We will describe the blowup in more detail below. This is the result predicted by Deligne's algorithm.

With $c_i$ and $b_i$ sections of appropriate line bundles on $B_n$, (\ref{P2withc4}) defines a Calabi-Yau space which is torus-fibered over $B_n$.
A special case is the one where $b_w \equiv 1$ globally; unlike for the case of a genus-one {\it curve}, as treated by Deligne's argument, this is not strictly necessary to the extent that the fiber of a genus-one {\it fibration} is only required to be a genus-one curve over generic points.
The model of \cite{Morrison:2012ei} makes this restriction. We will stick to this choice for now and comment on generalizations in section \ref{sec_generalisations}.
The classes of $b_i$ and $c_i$ are displayed in table \ref{tab:coeff1} with $\beta$ some class on $B_n$ such that all $[b_i]$ and $[c_i]$ are effective classes. This is required in order for the model to exist with holomorphic polynomials.  
\begin{table} 
\centering
\begin{tabular}{c|c|c|c|c|c|c|c}
 $b_{0}$ & $b_{1}$ & $b_{2}$ & $c_{0}$ & $c_1$ & $c_2$&$c_{3}$ & $c_4$ \\
\hline
 $\beta$& $\bar { K}$ &  $-\beta+2 \bar { K}$ &   $ 2 \beta$ &   $\beta + \bar { K}$ &    $ 2\bar { K}$ &   $-\beta+3\bar {K}$ &   $ -2\beta+4 \bar { K}$\\
\end{tabular}
\caption{Classes of the coefficients entering the hypersurface (\ref{P2withc4}).}\label{tab:coeff1}
\end{table}

In order to obtain an elliptic fibration with two independent sections, we set $c_4 \equiv 0$ globally, which leads to the elliptic fibration
\bea
Y_{n+1}: P_{\rm MP} = 0 
\eea
with
\bea \label{PMP}
P_{\rm MP} =  w^2 + b_0 u^2 w  + b_1 u v w + b_2 v^2 w  + c_0 u^4  + c_1 u^3 v  + c_2 u^2 v^2  + c_3 u v^3  \,.
\eea
The two independent sections are the divisors associated with the two independent rational points (\ref{eq:sec01}). 
The elliptic fibration $Y_{n+1}$ obtained in this way is singular: Its defining equation can be written globally in conifold form
\bea
 w \,  (w + b_0 u^2   + b_1 u v  + b_2 v^2 ) = u \, (c_0 u^3  + c_1 u^2 v  + c_2 u v^2  + c_3  v^3)
\eea
with a conifold singularity at 
\bea
\{u = 0\} \cap \{w = 0 \} \cap \{ b_2 = 0 \} \cap \{ c_3 = 0 \} \,.
\eea
The section $s_2$ passes through the singularity and is hence only a rational section. This is analogous to the behaviour encountered around (\ref{PT-conifoldtypea}).
A blowup is required to remove  this singularity, and after the blowup the section will wrap the exceptional fibral curve over the  base locus
\bea \label{Cb2c3}
C_I :=C_{b_2 c_3} \equiv \{ b_2 = 0 \} \cap \{ c_3 = 0 \} \,.
\eea
Even before coming to the resolution, we need to elaborate on a new phenomenon that we have not seen in the model of section \ref{subsecU1restr}: The unresolved singular fibration (\ref{PMP}) (as well as the smooth genus-one fibration with non-zero $c_4$)
contains {\it smooth} $I_2$ fibers. This is to be contrasted with the situation in a Weierstrass model: The generic Weierstrass model is smooth and has only fibers of Kodaira type $I_0$,  Kodaira type $I_1$ and type $II$. 
A generic $\mathbb P_{112}[4]$-fibration with non-zero $c_4$ is also smooth, but it allows in addition for smooth fibers of $I_2$ consisting of two rational curves intersecting at two points. The appearance of conifold singularities in model (\ref{PMP}) does not affect the existence of such smooth $I_2$ fibers even in the singular model with $c_4 \equiv 0$.
The richer set of smooth fiber types in elliptic fibrations not in Weierstrass form has been stressed, independently of the relation to the Mordell-Weil group, amongst other places in \cite{Aluffi:2009tm}.
It is in perfect agreement with the fact that every elliptic fibration can be cast in Weierstrass form - up to birational equivalence: the degenerate fibers in higher codimension will in general change under the birational map relating both descriptions of the fibration.

To find the smooth $I_2$ fibers we adopt the procedure developed in \cite{Morrison:2012ei}, where we follow the presentation in \cite{Mayrhofer:2014haa}:
At the location of a  smooth $I_2$ fiber, the defining equation (\ref{PMP}) must factorise. 
The hypersurface equation can in fact be written as 
\begin{equation}
\begin{aligned} \label{P1logic}
P_1 = &\left[  w + \frac{1}{2}(b_{0}  u^2 + b_1  u v + b_2 v^2)\right]^2\\ 
&+ (c_{0}-\frac{1}{4}b_{0}^2) u^4 + (c_{1}-\frac{1}{2}b_{0}b_1) u^3v + (c_{2} - \frac{1}{2}b_{0}b_2-\frac{1}{4}b_1^2) u^2 v^2 + (c_3 - \frac{1}{2}b_1 b_2) u v^3 - \frac{1}{4}b_2^2 v^4 \,.
\end{aligned}
\end{equation}
In order for $P_1$ to factorise the expression in the second line must form a perfect square.
Let us therefore make an ansatz of the form $(a u^2 + b u v + C v^2)^2$ for the second line. Comparing coefficient gives five equations. Three of these determine $a$, $b$ and $c$, and the remaining two equations define the codimension-two locus on the base where the factorization occurs. Since in the process of solving for $a$ and $b$ we must divide by $b_2$ and by $2 c_3 - b_1 b_2$, the two equations obtained in this way are only valid away from the locus where the latter vanish. In fact, we know already that over $b_2 = c_3 = 0$ the fibration is singular, and this locus hence does not describe smooth $I_2$ fibers prior to resolution. The factorization locus is given by the vanishing locus $V(f_1, f_2)$ of $f_1$ and $ f_2$ with  
\begin{equation}\label{eq:poly6_singlet_loci}
\begin{aligned}
&f_1 =  - c_1 b_2^4 +b_1b_2^3 c_2+b_0 b_2^3 c_3-b_1^2 b_2^2 c_3-2 b_2^2c_2 c_3+3 b_1 b_2 c_3^2-2c_3^3 \\
& f_2 = -c_3^2 b_0^2 +b_1
   b_2 b_0^2 c_3-b_1 b_2^2 b_0
   c_1+b_2^2 c_1^2+b_1^2 b_2^2
   c_0-4 b_1 b_2 c_0 c_3+4 c_0
   c_3^2 \,,
\end{aligned}
\end{equation}
with the understanding that the locus (\ref{Cb2c3}) must be exempt from this as explained above.
Technically, this is done by performing a primary ideal decomposition, e.g. using the computer algebra programme SAGE, of the ideal generated by $\langle f_1, f_2\rangle$. Each primary ideal describes an irreducible locus on $B_n$. One of them, as it turns out, is the locus  (\ref{Cb2c3}).
The remaining locus is described by a complicated primary ideal which, in this case, has 15 generators. 
We denote this locus as
\bea \label{Vf1f2minusC1}
C_{II}  =V(f_1, f_2)  \setminus C_I \,.
\eea

We can now proceed to resolve the conifold singularity over (\ref{Cb2c3}). 
This is possible again in terms of a blow-up 
\bea
u \rightarrow u \, s \qquad w \rightarrow w \, s \,.
\eea
The resolved space $\hat Y_{n+1}$ is the hypersurface in the ${\rm Bl}^1\mathbb P_{112}[4]$-fibration cut out by
\bea \label{PMP-res}
\hat P_{\rm MP} =  
 s w^2   + b_{0}  s^2 u^2 w  + b_1s u v w  + b_2 v^2 w 
 + c_{0}s^3 u^4   + c_{1} s^2  u^3 v  + c_{2}s u^2 v^2   + c_3 u v^3 \,.
\eea
The scaling relations are\footnote{We are combining here the original scaling of $[u : v : w]$ with the one from the blow-up.}
\begin{eqnarray}
(u,v,w,s) \simeq (\lambda u, \lambda \mu v, \lambda^2  \mu \, w, \mu \, s)
\end{eqnarray}
leading to the SR-ideal 
\bea
{\rm SR} = \{ u w, v s \} \,.
\eea
Over $C_I$ the fiber factorises into the exceptional curve $A_I$ wrapped by the extra section $S_1$ and the remaining fiber component $B_I$ with intersection numbers
\bea \label{CsCrem}
&[S] \cdot [A_I] = -1, \qquad &[S] \cdot [B_I] = 2 \\
&[S_0] \cdot [A_I] = +1, \qquad &[S_0] \cdot [B_I] = 0  \,.
\eea
The fiber over $C_{II}$ continues to factorise as in the singular model prior to resolution into two fibral curves $A_{II}$ and $B_{II}$ with
\bea \label{CACB}
&[S] \cdot [A_{II}] = 0, \qquad &[S] \cdot [B_{II}] = 1 \\
&[S_0] \cdot [A_{II}] = 1, \qquad &[S_0] \cdot [B_{II}] = 0  \,.
\eea

\begin{figure}
\centering
\begin{tikzpicture}
\node at (1.3,2) {fibre over $C_I$};
\node at (-0.2,1) {$A_I$};
\node at (2.8,1) {$B_I$};
\begin{scope}[scale=0.7]
\begin{scope}[xshift=3.6cm]
   \shadedraw[ball color=gray!30!white,xscale=-1] (-0.3,0) .. controls (-0.3,1) and (.3,2) .. (1,2)
               .. controls (1.7,2) and (1.8,1.5) .. (1.8,1.2)
               .. controls (1.8,.5) and (1,.5) .. (1,0)
               .. controls (1,-.5) and (1.8,-.5) .. (1.8,-1.2)
               .. controls (1.8,-1.5) and (1.7,-2) .. (1,-2)
               .. controls (.3,-2) and (-0.3,-1) .. (-0.3,0);
\end{scope}
\shadedraw[ball color=blue!40!white] (-0.3,0) .. controls (-0.3,1) and (.3,2) .. (1,2)
               .. controls (1.7,2) and (1.8,1.5) .. (1.8,1.2)
               .. controls (1.8,.5) and (1,.5) .. (1,0)
               .. controls (1,-.5) and (1.8,-.5) .. (1.8,-1.2)
               .. controls (1.8,-1.5) and (1.7,-2) .. (1,-2)
               .. controls (.3,-2) and (-0.3,-1) .. (-0.3,0);
\draw[green, line width=0.5mm, xshift=.5cm,yshift=1cm] (0,0)--(0.2,0.2);
\draw[green, line width=0.5mm, xshift=.5cm,yshift=1cm] (0,0.2)--(0.2,0);
\end{scope}

\begin{scope}[xshift=4cm]
\node at (1.3,2) {fibre over $C_{II}$};
\node at (-0.3,1) {$A_{II}$};
\node at (2.8,1) {$B_{II}$};
\begin{scope}[scale=0.7]
   \shadedraw[ball color=gray!30!white] (-0.3,0) .. controls (-0.3,1) and (.3,2) .. (1,2)
               .. controls (1.7,2) and (1.8,1.5) .. (1.8,1.2)
               .. controls (1.8,.5) and (1,.5) .. (1,0)
               .. controls (1,-.5) and (1.8,-.5) .. (1.8,-1.2)
               .. controls (1.8,-1.5) and (1.7,-2) .. (1,-2)
               .. controls (.3,-2) and (-0.3,-1) .. (-0.3,0);
\begin{scope}[xshift=3.6cm]
\shadedraw[ball color=gray!80!white,xscale=-1] (-0.3,0) .. controls (-0.3,1) and (.3,2) .. (1,2)
               .. controls (1.7,2) and (1.8,1.5) .. (1.8,1.2)
               .. controls (1.8,.5) and (1,.5) .. (1,0)
               .. controls (1,-.5) and (1.8,-.5) .. (1.8,-1.2)
               .. controls (1.8,-1.5) and (1.7,-2) .. (1,-2)
               .. controls (.3,-2) and (-0.3,-1) .. (-0.3,0);
\end{scope}
\draw[green, line width=0.5mm, xshift=1cm,yshift=1cm] (0,0)--(0.2,0.2);
\draw[green, line width=0.5mm, xshift=1cm,yshift=1cm] (0,0.2)--(0.2,0);
\draw[blue!40!white, line width=0.5mm, xshift=2.5cm,yshift=-1cm] (0,0)--(0.2,0.2);
\draw[blue!40!white, line width=0.5mm, xshift=2.5cm,yshift=-1cm] (0,0.2)--(0.2,0);
\end{scope}
\end{scope}

\end{tikzpicture}
\caption{$I_2$-fibres  over the singlet curves $C_I$ and $C_{II}$ of the ${\rm Bl}^1\mathbb P_{112}[4]$-fibration (\ref{PMP-res}) of \cite{Morrison:2012ei}.  Blue represents  the rational section $S$ and green the zero-section $S_0$. The picture is taken from \cite{Mayrhofer:2014laa}.}\label{fig:fibre}
 \end{figure}
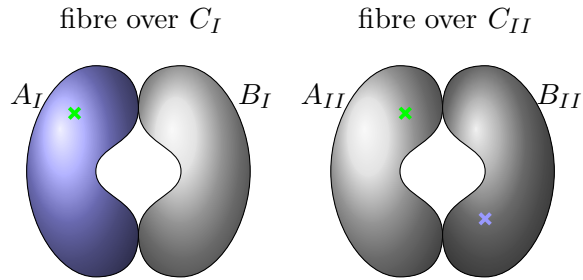

The situation is depicted in Figure \ref{fig:fibre}.
After the blowup, the section divisor associated with $s_1$ is replaced by the holomorphic section $S_0 =\{u=0\}$, and the singular section $s_2$ by $S: \{s=0\}$. It is rational and wraps the exceptional curve $A_I$ in the fiber over $C_I$. Since the section divisors $S_0$ and $S_1$ are given by toric divisors, i.e. by the vanishing locus of toric ambient space coordinates on the hypersurface, the sections are called toric sections. This need not always be the case, and non-toric sections may well exist (possibly in addition to obvious toric ones). In the present model the Mordell-Weil rank is indeed one and hence no independent non-toric sections exist.

As a  novelty compared to the model of section \ref{subsecU1restr}, $S_0$ and $S_1$ intersect over $\{b_2=0\}$ so that 
\bea
\pi_\ast (S \cdot S_0) = b_2 \,,
\eea
where we are denoting by $b_2$ also the divisor associated with $\{b_2=0\}$.
This leads to the Shioda map and height pairing
\bea
\sigma(S) = S - S_0 -( {\bar K + b_2}) \,, \qquad b_{SS} = 2 \bar K + 2 b_2 \,.
\eea

There are now two types of charged matter:
An M2-brane wrapping the exceptional fiber $A_I$ over $C_I$ gives rise to states of charge $|q_1| = 2$, while M2-branes wrapping the fiber $B_{II}$ over $C_{II}$ carry charge $q_2 = 1$. As usual, adding and subtracting the full fiber class reproduces the whole tower of KK states in M-theory.

The discussion of this elliptic fibration has been phrased entirely in the language of the special $\mathbb P_{112}[4]$-fibration (\ref{PMP}) and its resolution (\ref{PMP-res}), because the starting point was Deligne's algorithm to produce elliptic curves with two rational points.
It must of course be possible to bring the fibration into Weierstrass form by a  birational transformation. By definition, after such a change the fibers in codimension one or higher might differ.
The birational map from (\ref{PMP}) to the Weierstrass model has been worked out in \cite{Morrison:2012ei}, leading to
\begin{equation}
\begin{split} \label{WeierMP}
P_{\rm W-U(1)}: y^2 &= x^3 + f x z^4 + g z^6 \cr
f &=  e_1 \, e_3 - \frac{1}{3} e_2^2 - 4 e_0 \, e_4 \,, \quad   g = - e_0 e_3^2 + \frac{1}{3}  e_1 e_2 e_3 - \frac{2}{27} e_2^3 + \frac{8}{3} e_0 e_2 e_4 - e_1^2 e_4 
\end{split}
\end{equation}
with 
\begin{equation}
\begin{split} \label{es}
e_0 &= - c_0 + \frac{1}{4} b_0^2,  \qquad e_1 = - c_1 + \frac{1}{2} b_0 b_1, \nonumber  \cr
e_2 &=  - c_2 + \frac{1}{2} b_0 b_2 + \frac{1}{4} b_1^2,   \qquad e_3 = - c_3 + \frac{1}{2} b_1 b_2, \qquad  e_4 =     \frac{1}{4} b_2^2.   
\end{split}
\end{equation}
Note that \cite{Morrison:2012ei} first set $b_0 \equiv 0$ and $b_1 \equiv 0$ by a change of coordinates in (\ref{PMP}). What is important is that all $e_i$ are generic polynomials except for $e_4$, which is a perfect square. 
In the Weierstrass model the smooth $I_2$ fibers of (\ref{PMP}) over $C_{II}$ are contracted to singular $I_2$ fibers, which, together with the conifold locus at $C_I$, form the singularities of the fibration.

\subsection{Generalisations and systematics} \label{sec_generalisations}

It is natural to wonder if the elliptic fibration (\ref{PMP}) or its associated Weierstrass model (\ref{WeierMP}) represent the most general fibration with a Mordell-Weil group of rank one. In particular, it is of great interest in physics to determine the possible $U(1)$ charges in F-theory. Part of the fascination of these questions from a physics perspective is because they might allow us to distinguish the landscape of consistent (non-perturbative) vacua of string and M-theory from the swampland of seemingly consistent supergravity theories with no quantum gravity completion. Questions concerning abelian gauge symmetries and the allowed charges have been addressed in particular in the context of 6d F-theory compactifications, where the supergravity constraints due to anomalies are particularly stringent, see section \ref{sec_Applications1}. 

While the general answer to this question has not been settled as of this writing, let us briefly summarize the current state of the art: 
The resolved fibration (\ref{PMP-res}) is a hypersurface in a toric ambient space. It has already been pointed out at the end of section \ref{subsec_TateI2-1} that this construction is defined in terms of a two-dimensional reflexive polygon and its dual; they encode, respectively, the toric ambient space coordinates and the monomials appearing in the hypersurface equation. There are, in fact, 16 inequivalent such two-dimensional reflexive polygons (see \cite{Bouchard:2003bu,Grassi:2012qw} for background information and references).
Both (\ref{PMP-res}) and the $U(1)$ restricted model (\ref{hatPTU1restr}) are included in this list.  The toric Mordell-Weil group of these 16 toric hypersurface models has been determined in \cite{Grassi:2012qw,Braun:2013nqa}.
In \cite{Klevers:2014bqa} the genus-one fibrations obtained by fibering all of these 16 toric hypersurfaces over a general base have been analyzed in detail along the lines of the discussion exemplified here for  the models (\ref{hatPTU1restr}) and (\ref{PMP-res}).
This list turns out to include in particular one model with a Mordell-Weil group of rank $r=1$ which contains massless matter of charges $q=1,2,3$. 
In \cite{Klevers:2016jsz} it is observed that the associated Weierstrass model is not the form (\ref{WeierMP}), which can therefore not give the most generic such elliptic fibration. 
Nonentheless, one can associate to the model with $q=3$ matter a non-Calabi-Yau fibration in Weierstrass form of the type  (\ref{WeierMP}) \cite{Morrison:2016xkb}.

The mismatch is due to subtle effects which involve so-called non-UFDs (universal factorization domains).
In \cite{Raghuram:2017qut}, a general and systematic ansatz has been used to derive Weierstrass models with charges $q=3$, which are indeed not of the form (\ref{WeierMP}).
The starting point is a Weierstrass model with a section 
\bea
s_1: [x : y : z] = [\hat x : \hat y : \hat z] \,.
\eea
Note that  the coordinates of the section, $\hat x$, $\hat y$ and $\hat z$, are functions in the coordinate ring $R$ of the base $B_n$.
Since $s_1$ must lie on the elliptic fiber, the functions $f$ and $g$ defining the Weierstrass model are constrained as
\bea \label{hatWeier}
\hat y^2 - \hat x^3 = \hat z^4 ( f \hat x + g \hat z^2) \,.
\eea
The idea is then to make a general ansatz for $\hat x$ and $\hat y$ as a power series in $\hat z$ such as to satisfy (\ref{hatWeier}) and to read off the corresponding form of $f$ and $g$.
The crucial point is now whether the divisor $\hat z = 0$ defines a smooth or a singular variety on the base $B_n$ \cite{Raghuram:2017qut}.\footnote{At the end of section \ref{subsec_gen} we have noted that it makes an important difference if a discriminant component is smooth when it comes to classifying the possible non-abelian representations with respect to the associated non-abelian gauge algebra. Here smoothness of $\hat z=0$ poses restrictions on the abelian models for very similar reasons.} If it is smooth, the quotient $R/\langle z \rangle$ is a so-called UFD (unique factorization domain).
The general ansatz for $\hat x$ and $\hat y$ in terms of $\hat z$ then reproduces precisely the form (\ref{WeierMP}). If $\hat z=0$ is a singular variety, $R/\langle z \rangle$ is a non-UFD, and more general solutions to the ansatz, making use of techniques in \cite{Klevers:2017aku}, are possible.
The model of \cite{Klevers:2014bqa} with a $q=3$ state has exactly the property that $\hat z=0$ is singular, in agreement with the findings of \cite{Klevers:2016jsz}. 
More general models with $q=3$ are likewise found in this way \cite{Raghuram:2017qut}, but the most general such model is still not determined.
Another remaining challenge is that these highly non-generic Weierstrass models are not yet resolved.

 Ref. \cite{Raghuram:2017qut} also finds the first example of an elliptic fibration with Mordell-Weil rank $r=1$ and a $q=4$ states (albeit with a different technique).  As of this writing it remains an open question how to generalize these results to higher charges in practice and what the upper limit of possible charges in such F-theory construction might be.

Another line of generalisation is to increase the rank of the Mordell-Weil group.
Promoting the elliptic curve (\ref{n=2Deligne}) obtained via Deligne's argument to an elliptic fibration produces a Mordell-Weil  group of rank $r=2$, corresponding to an F-theory gauge group $U(1) \times U(1)$.
These geometries have been investigated in \cite{Borchmann:2013jwa,Cvetic:2013nia,Cvetic:2013uta,Borchmann:2013hta}. In particular, there are six types of charged singlets. 
Again the manipulations, in particular the fibral coordinate transformations, leading to the specific form of these fibrations are the most general ones for elliptic {\it curves} with three independent rational points, but not necessarily for the associated {\it fibrations}.
This is explained in \cite{Cvetic:2015ioa}, which provides a generalisation of these fibrations: In the more general models, there are even nine types of charged singlets and correspondingly a richer Higgs branch.

As we increase the rank of the Mordell-Weil group, we are bound to leave the regime where the elliptic fiber is merely a hypersurface in a toric ambient space - at least if we are trying to find the most generic, i.e. 'normal forms' for the fibration as opposed to highly constrained and hence somewhat unwieldy Weierstrass models.
For instance, extrapolating again Deligne's result, such a {\it normal form} for a Mordell-Weil group of rank $r=3$ involves a complete intersection within a toric hypersurface \cite{Cvetic:2013qsa}.
Similar in spirit to the classification of the genus-one fibrations obtained as hypersurfaces in toric ambient space, \cite{Braun:2014qka} analyses all possible 3134 complete intersection representations of genus-one fibers and discusses the toric Mordell-Weil group. The maximal rank found is $r=4$. By contrast, the maximal rank which can be obtained by hypersurfaces in toric ambient spaces is $r=3$ \cite{Grassi:2012qw,Braun:2013nqa,Klevers:2014bqa}.

Up to this point we have tried to explicitly construct an elliptic fibration with extra sections or certain properties. 
The inverse problem starts with a given Calabi-Yau and asks whether this geometry admits for an elliptic fibration structure, and if so whether the fibration has a non-trivial Mordell-Weil rank.
As explained at this TASI school in \cite{Anderson:2018pui}, a criterion when a fibration is at least a genus one fibration (not necessarily with a section) was given by
Oguiso \cite{Oguiso} and Wilson \cite{Wilson} for Cablabi-Yau 3-folds; it is conjectured by Koll\'ar \cite{Kollar} to hold also for higher Calabi-Yau $n$-folds.
Interestingly, almost all complete intersection Calabi-Yau threefolds \cite{Anderson:2016cdu,Anderson:2017aux} and fourfolds \cite{Gray:2014fla} in projective spaces admit for (oftentimes even multiple) genus-one fibrations, and most toric hypersurface Calabi-Yau varieties seem to share this property (see e.g. \cite{Johnson:2014xpa,Johnson:2016qar}). For details and more references we recommend sections 3.5 and 3.6 of \cite{Anderson:2018pui}.
To specify if the genus-one fibration is elliptic and whether it has extra independent sections, an algorithm is needed to determine if one (or several) of the divisors of the Calabi-Yau can play the role of a meromorphic section. In particular, a putative section must satisfy  the intersection numbers (\ref{SAint1}) - (\ref{SAEi}). 
Tools to systematically study this problem have been developed in \cite{Anderson:2016ler}.

\subsection{Combining abelian and non-abelian gauge algebras} \label{sec_combab-nonab}

Rational sections can of course be combined with fiber degenerations in codimension-one leading to non-abelian gauge algebras. 
Such constructions are, in fact, of great importance in model building given the structure of the Standard Model gauge algebra or the relevance of extra abelian gauge symmetries (once rendered massive by a St\"uckelberg mechanism) as extra selection rules. This in fact partly motivated the interest in the F-theory community in elliptic fibrations with sections.

Localised charged matter in representation ${\bf R}$ under the non-abelian part of the gauge algebra now carries in addition charges with respect to the abelian gauge algebra generated by the section\footnote{The non-localised matter along the bulk of the 7-branes remains massless due to the normalization of the $U(1)_A$ such that $[\sigma(s_A)] \cdot \mathbb [\mathbb P^1_{i_I}]=0$}.
In particular, different abelian charges can occur for the same non-abelian representation ${\bf R}$.
Compared to generic elliptic fibrations with the same non-abelian enhancement, this leads to a refinement of the codimension-two enhancement loci corresponding to the abelian charges.
If the base has dimension $n=3$ or $n=4$, engineering a section then splits (some of) the codimension-two matter curves or surfaces, respectively, into various irreducible components. The mathematical reason for this is the specialization of the Weierstrass polynomials responsible for the appearance of a section.

If the elliptic fibration is given in Weierstrass or Tate form, the same algorithms to engineer non-abelian gauge algebras can be applied.
The first systematic analysis has appeared in the context of the $U(1)$ restricted Tate model of section \ref{subsecU1restr} combined with an enhancement to gauge algebra $\mathfrak{su}(5)$ over a divisor \cite{Krause:2011xj} (generalized to $I_n$ with $n \leq 5$ in \cite{Krause:2012yh}).
At the level of the Tate model the enhancement from the generic $I_5$-model (\ref{TateSU5ai}) to its $U(1)$ restricted version proceeds again by setting $a_6 \equiv 0$.
As is apparent from (\ref{defPSU5}), this leads to a factorization of the codimension-two locus carrying the ${\bf 5}$-representation of $\mathfrak{su}(5)$. The two loci
\bea
C_{{\bf 5}_{-3/5}} = \{w=0\} \cap \{a_{3,2} = 0\}        \,, \qquad C_{{\bf 5}_{2/5}} = \{w=0\} \cap \{a_{2,1} a_{3,2} - a_1 a_{4,3} = 0\}
\eea
are now distinguished by their $U(1)$ charges. 
To compute these, an explicit resolution must be analyzed. For the $I_5$ $U(1)$-restricted model, there are six birationally equivalent resolutions \cite{Krause:2011xj}.
The Shioda map takes a slightly different form in each of these resolutions due to the different intersection pattern of the section with the resolution divisors. 
Nonetheless, the final result for the matter charges is the same 
\bea \label{chargeassU1restr}
{\bf 10}_{-1/5}, \qquad {\bf 5}_{-3/5}, \qquad {\bf 5}_{2/5},  \qquad {\bf 1}_{-1} \,.
\eea
Note that the curve $C_{\bf 10}$ is unchanged by the engineering of the extra section. Furthermore the singlets ${\bf 1}_{-1}$ are localised along $C_{{\bf 1}_{-1} } = \{a_{3,2} =0\} \cap \{a_{4,3} =0\}$ and hence not contained in the discriminant divisor $\Sigma_1$ wrapped the $\mathfrak{su}(5)$  7-brane stack. Its intersection with  $\Sigma_1$ forms a triple intersection with $C_{{\bf 5}_{-3/5}} $ and $C_{{\bf 5}_{2/5}}$: Here triple Yukawa couplings of the form ${\bf 1}_1 \, {\bf 5}_{-3/5}  {\bf \bar 5}_{-2/5} + c.c$ appear, which are, of course, absent in the generic $I_5$-model.

To get a better handle on the systematics, observe first that 
the $I_5$ $U(1)$-restricted Tate model is an example of a toric model: The fiber ambient space ${\rm Bl}^1 \mathbb P_{231}$ is a toric space and the specification of the non-abelian gauge algebra is due to certain vanishing orders of the polynomials defining the hypersurface along a divisor in the base.
The situation is therefore largely parallel to that of the Tate model without the $U(1)$ restriction  $a_6\equiv 0$, as discussed in section \ref{subsec_TateI2-1}.
The vanishing orders leading to a certain gauge algebra of the subset of toric models are the same as the ones in table 2 of \cite{Bershadsky:1996nh} for the model $a_6 \neq 0$, modulo one caveat: Since $a_6$ vanishes to all orders, if two models differ only by the vanishing order of $a_6$, then the model with the lower vanishing order of $a_6$ is automatically enhanced to the higher one. 
In the present case, the vanishing orders are obtained by imposing the form $a_i = a_{i,j} w^j$ as in (\ref{TateSU5ai}) (with $a_6 \equiv 0$). This can be enforced torically by setting all monomials $a_{i,k} w^k$ with $k <j$ to zero. In fact the polygon of the toric ambient space of the $U(1)$ restricted model is polygon 11 in the classification of \cite{Bouchard:2003bu}. We recall from section  \ref{subsec_TateI2-1} that the toric non-abelian enhancements are given in terms of so-called toric tops \cite{Candelas:1996su}. The toric tops of all 16 hypersurface polygons have been classified in  \cite{Bouchard:2003bu}.
For instance, for $I_n$ models within the polygon 11 there exists only one such top in the list of \cite{Bouchard:2003bu}, and hence only one type of toric model, in agreement with application of Tate's algorithm.

For some of the other 16 polygons there exist several inequivalent toric tops, or equivalently toric models, realising a given non-abelian gauge algebra. These typically differ in the $U(1)$ charges of the representations.
All toric tops with gauge algebra $\mathfrak{su}(5)$ within polygon 6, corresponding to the ${\rm Bl}^1 \mathbb P_{112}[4]$ fibration reviewed in section \ref{sec-MPm}, have been worked out in \cite{Borchmann:2013jwa,Borchmann:2013hta}, leading to more general charge assignments compared to (\ref{chargeassU1restr}).
The toric tops now describe all possible ways to engineer a non-abelian gauge algebra by constraining the polynomials in (\ref{PMP}) to take the form
$b_i = b_{i,j} w^j$ and $c_{i,k} w^k$, with $b_{i,j}$ and  $c_{i,k}$ otherwise generic. 
\cite{Borchmann:2013hta} also provides an analysis of the $\mathfrak{su}(5)$-tops of various other polygons with an extra abelian gauge factor including the resulting Shioda-map and charge assignments. The toric $SU(5)$ tops for all 16 hypersurfaces are independently analyzed in \cite{Braun:2013nqa}. 

In the context of the ${\rm Bl}^2 \mathbb P_{111}[3]$ elliptic fibrations with Mordell-Weil rank $r=2$, all toric enhancements to gauge algebra $\mathfrak{su}(5)$ have been studied in full detail in \cite{Borchmann:2013jwa,Borchmann:2013hta} (see also \cite{Cvetic:2013uta}). This corresponds to polygon 5 in the list of \cite{Bouchard:2003bu}. 
The same fibration has been combined with gauge algebra  $\mathfrak{su}(3) \oplus \mathfrak{su}(2)$ in \cite{Lin:2014qga}, where all toric tops of type $I_3$ and $I_2$ are worked out.

Some of the toric hypersurfaces of \cite{Klevers:2014bqa} generically exhibit extra non-abelian enhancements without the need of restricting the defining polynomial. This includes one with total gauge algebra $\mathfrak{su}(3) \oplus \mathfrak{su}(2) \oplus \mathfrak{u}(1)$, as analysed further in \cite{Cvetic:2015txa}. The same phenomenon also occurs in  some of the complete intersection models analyzed in \cite{Braun:2014qka}.

The toric models form a subset of the possible gauge enhancements. More generally, there can be non-trivial relations between the non-vanishing monomials. 
The first example where this has been put to use in the presence of abelian gauge symmetry is \cite{Mayrhofer:2012zy}, which considers ${\rm Bl}^1 \mathbb P_{112}[4]$ with an $\mathfrak{su}(5)$ gauge algebra realized in a non-generic way by exploiting exactly such relations among the monomials. 
The explicit resolution of this model requires representing the fiber ambient space as a complete intersection \cite{Mayrhofer:2012zy}.
Unlike the toric hypersurfaces models, this induces for instance a splitting of the ${\bf 10}$ curve as envisaged in the phenomenologically appealing semi-global split spectral covers of \cite{Marsano:2009wr,Dolan:2011aq}. Other examples with this property are provided, amongst other things, in \cite{Braun:2013yti}, and more complete intersections with this property are discussed in \cite{Braun:2014qka}.

 A systematic investigation of Tate's algorithm in the context of the ${\rm Bl}^1 \mathbb P_{112}[4]$ fibration reviewed in section \ref{sec-MPm} has been provided in \cite{Kuntzler:2014ila}. This reference first classifies all possible  {\it canonical Tate forms} in the presence of an extra abelian gauge group factor realized via the ${\rm Bl}^1 \mathbb P_{112}[4]$ construction. 
The canonical models coincide with the toric models described above in that the singularity is engineered entirely on the basis of the vanishing orders of the polynomials. 
All possible canonical models within the ${\rm Bl}^1 \mathbb P_{112}[4]$ fibration are classified in \cite{Kuntzler:2014ila}. 
In addition, in {\it non-canonical models}, the $b_{i,j}$ and $c_{i,k}$ may satisfy relations such that non-trivial cancellations are responsible for the appearance of the gauge algebra.
 Examples of this phenomenon are also analysed in \cite{Kuntzler:2014ila}. These models cannot be realized torically.
An analysis of both canonical (or toric) and non-canonical Tate forms and the resulting charges for the ${\rm Bl}^2 \mathbb P_{111}[3]$-fibration of  \cite{Borchmann:2013jwa,Cvetic:2013nia,Cvetic:2013uta,Borchmann:2013hta} with two abelian gauge factors has been provided in  \cite{Lawrie:2014uya}.
A detailed recent comparison between toric methods and Tate tunings, including situations with abelian gauge groups, is carried out in \cite{Huang:2018gpl}.

Independently of concrete constructions, it is an important question which abelian charges can occur even in principle for a given non-abelian gauge algebra, and coexist at the same time. This question is answered in \cite{Lawrie:2015hia} for non-abelian algebra $\mathfrak{su}(5)$ under the hypothesis that both the section generating the abelian part and also  $\mathfrak{su}(5)$ discriminant divisor are smooth (the latter is required to constrain the non-abelian representations occurring). The analysis is very general in that it proceeds entirely in terms of the possible intersection structure of the extra section with the fibral curves in codimension-two. It reproduces all $U(1)$ charge patterns in previously constructed $\mathfrak{su}(5)$ fibrations, along with more general charge assignments without a concrete realisation as of this writing.

\subsection{Torsional sections and the global structure of the gauge group} \label{sec_torsion}

The torsional part of the Mordell-Weil group has an interpretation rather different from the free subgroup: It is related to the global structure of the gauge group \cite{Aspinwall:1998xj,Mayrhofer:2014opa}.
The argument presented in \cite{Mayrhofer:2014opa} is as follows: Consider first the situation where the gauge algebra consists only of one single non-abelian summand $\mathfrak{g}_{I}$, and the Mordell-Weil group of the fibration $\hat Y_{n+1}$ is generated by a $k$-torsional section $s_k$. By definition $s_k^{\boxplus_k} = s_k \boxplus \ldots \boxplus s_k = s_0$.
Consider the associated divisor $S_k = {\rm div}(s_k)$ and define the object
\bea \label{sigmast}
\sigma(s_k) =  S_k-S_0-\pi^{-1} (D_k)+  \sum_{i_I}  \ell_{k}^{i_I}   E_{i_I}  \in {\rm NS}(\hat Y_{n+1}) \otimes \mathbb Q    \,.
\eea
This is the analogue of the Shioda map  (\ref{shioda}) for non-torsional sections. The coefficients $\ell_{k}^{i_I}$ are computed as in  (\ref{lAjik}), and we assume for now at least one of them to be non-zero. We have furthermore introduced the notation $D_k = \pi_\ast (S_k-S_0)\cdot S_0$. 
Since the Shioda map is a group homomorphism from ${\rm MW}(\pi)$ to ${\rm NS}(\hat Y_{n+1}) \otimes \mathbb Q$, it is clear that 
\bea
0 = \sigma(s_0)  =  \sigma(s_k^{\boxplus_k}) =  k \,  \sigma(s_k) \,.
\eea
Since ${\rm NS}(\hat Y_{n+1}) \otimes \mathbb Q$ is torsion-free, this implies that 
\bea
\sigma(s_k) = 0 \in {\rm NS}(\hat Y_{n+1}) \otimes \mathbb Q \,.
\eea
First of all, this shows that torsional sections do not give rise to abelian gauge factors because the Shioda map leads to a divisor which is trivial in ${\rm NS}(\hat Y_{n+1}) \otimes \mathbb Q$.
Instead, the existence of the $k$-torsional section affects the gauge theory as follows \cite{Mayrhofer:2014opa}:
The coefficients $\ell_{k}^{i_I} $ in the linear combination (\ref{sigmast}) are in general rational due to the appearance of the inverse Cartan matrix in their definition (\ref{lAjik}).
As we will justify below,  $k \, \ell_{k}^{i_I} \in \mathbb Z$, which allows us to write
\bea
 \ell_{k}^{i_I} = \frac{1}{k} \hat \ell_{k}^{i_I} \,, \qquad \hat \ell_{t}^{i_I} \in \mathbb Z \,.
\eea
It follows that 
\bea \label{SkS0a}
k  \, ( S_k-S_0-\pi^{-1} (D_k)) = -  \sum_{i_I}  \hat \ell_{k}^{i_I}   E_{i_I} \in \mathbb {\rm NS}(\hat Y_{n+1}) \,,
 \eea
 or
 \bea \label{SkS0b}
  S_k-S_0-\pi^{-1} (D_k) = - \frac{1}{k} \sum_{i_I}  \hat \ell_{k}^{i_I}   E_{i_I}  \,.
 \eea
The right-hand side of (\ref{SkS0b}) is therefore an integer cycle even though it is a linear combination of divisors with $\mathbb Q$-coefficients. 
One way to read  (\ref{SkS0a}) is to state that  $S_k-S_0-\pi^{-1} (D_k)$ defines a $k$-torsional cycle in ${\rm NS}(\hat Y_{n+1})/\langle E_{i_I}\rangle_{\mathbb Z}$.
A more useful conclusion is the following:
The intersection number of the RHS of (\ref{SkS0a})  with any fibral curve must lie in $k \,  \mathbb Z$. Given the general relation (\ref{ECbeta}) between such intersection numbers and the weights of the representations, this acts as a constraint on the weight lattice of the fibration: Every representation ${\bf R}$ which does not satisfy the constraint 
\bea
\frac{1}{k}  \sum_{i_I}  \hat \ell_{k}^{i_I}    \beta^a_{i_I}({\bf R}) \in \mathbb Z
\eea
 is absent from the spectrum. This has implications for the global structure of the gauge group.

Let us denote by $G^{(0)}_I = {\rm exp}({\mathfrak{g}_I})$ the universal cover group of the Lie algebra $\mathfrak{g}_I$. By definition, this group is simply connected. The resolution divisors $E_{i_I}$ span to coweight lattice $\Lambda^\vee$ of the gauge group. This is just the familiar statement that their intersection numbers with the fibral curves produce the weights. Since the fractional linear combination $\frac{1}{k} (\sum_{i_I}  \hat \ell_{k}^{i_I}   E_{i_I}) $ is integral, the coweight lattice is finer by order $k$ compared to the coweight lattice of $G_I^{(0)}$. 
In group theory this means that the gauge group is not $G^{(0)}_I$, but 
$G_I = G^{(0)}_I/\mathbb Z_k$. This group is not simply-connect and  has $\pi_1(G_I) = \mathbb Z_k$. 

The converse statement is also true: Whenever the gauge group is of the form $G_I = G^{(0)}_I/\mathbb Z_k$ (with $G^{(0)}_I$ the covering group of a non-abelian Lie algebra $\mathfrak{g}_I$), the fibration exhibits a $k$-torsional section whose Shioda map involves the resolution divisors associated
with $\mathfrak{g}_I$. 
To see this, observe that $G_I = G^{(0)}_I/\mathbb Z_k$ means that there exist integers $a_{i_I}$ (not all zero)  such that for all weight vectors $\frac{1}{k} \sum_{i_I} a_{i_I} \beta^a_{i_I}({\bf R}) \in \mathbb Z$.
Given the relation between the weights and the intersection numbers of the $E_{i_I}$ with the fibral curves in codimension-two, this implies that $\frac{1}{k} \sum_{i_I} a_{i_I} E_{i_I}$ is an integer  
cycle. Let us write 
\bea \label{kXi}
k \,  \Xi_k =  \sum_{i_I} a_{i_I} E_{i_I} \,.
\eea
To proceed, we need to recall a more precise formulation of the Shioda-Tate-Wazir theorem \cite{Shioda:1972,Wazir:2001} encountered before:
First, there exists a surjective homomorphism
\bea
\psi: {\rm NS}(\hat Y_{n+1}) &\rightarrow& {\rm MW}(\mathbb E_\tau) \\
D &\mapsto& D|_{{\mathbb E}_\tau} - (D \cdot {\mathbb E}_\tau) \, O  \,,
\eea
where  ${\rm MW}(\mathbb E_\tau)$ is the Mordell-Weil group of the generic fiber, defined as the additive group of rational points on the fiber with zero-element  $O$. 
According to the Shioda-Tate-Wazir theorem the kernel ${\cal T}$ of this map within ${\rm NS}(\hat Y_{n+1})$ is the subgroup of  ${\rm NS}(\hat Y_{n+1})$ spanned by the zero-section $s_0$, all vertical divisors pulled back from the base, and the resolution divisors $E_{i_I}$.
Now, according to (\ref{kXi}), the divisor $k \, \Xi_k$ lies in the kernel ${\cal T}$ because the objects on the right are divisors of the right type with integer coefficients.\footnote{Note that $k \,  \Xi_k$ cannot be a vertical divisor because any vertical  linear combination of the resolution divisors must include the fibration of the affine node $E_{0_I}$ as in (\ref{piSigmaI}).} 
At the same time, $\Xi_k$ only lies in the span of these divisors with $\mathbb Q$ coefficients. Hence, possibly after subtracting a suitable vertical divisor, $\Xi_k  = \tilde \Xi_k + \pi^{-1}(\delta)$, $\tilde \Xi_k$ must be a torsional section, i.e. $\psi( k \, \tilde\Xi) = 0 
\in  {\rm MW}(\mathbb E_\tau)$ but $\psi(\tilde \Xi)$ as such is torsion. This reasoning also explains why the fraction in (\ref{SkS0b}) must agree with the order of the torsional section $k$. If the equation were satisfied with $k$ replaced by $q$, then $q(S_k-S_0)$ would be in the kernel of $\psi$ within ${\rm NS}(\hat Y_{n+1})$ and $S_k$ would hence be a $q$-torsional section.

We can generalize the construction by starting with a non-abelian gauge algebra
\bea
\mathfrak{g} = \mathfrak{g}_1 \oplus \mathfrak{g}_2 \,,
\eea
where both  $\mathfrak{g}_1$ and $ \mathfrak{g}_2$ can be taken to be semi-simple. Suppose again that ${\rm MW}_{\rm tors}(\pi) = \mathbb Z_k$
and that the Shioda map associated with the $k$-torsional generator involves only the resolution divisors associated with $ \mathfrak{g}_2$.
Then by the above reasoning the total gauge group is
\bea \label{Gtotala}
G = G_1^{(0)} \times (G_2^{(0)}/\mathbb Z_k) \,.
\eea
In particular,
\bea \label{pi1G}
\pi_1(G)_{\rm tors} = {\rm MW}_{\rm tors}(\pi) \,.
\eea
The conclusion (\ref{pi1G}) continues  for general ${\rm MW}_{\rm tors}(\pi)  =\mathbb Z_{k_1} \oplus \ldots \oplus \mathbb Z_{k_n}$.
This interpretation of the torsional Mordell-Weil group has been illustrated, in \cite{Mayrhofer:2014opa}, in many examples taken from \cite{Aspinwall:1998xj} (some of which were also studied form a slightly different perspective in \cite{Berglund:1998va}) by explicitly constructing the torsional section in the resolution and constructing the weight lattice. The Mordell-Weil torsion encountered in the realizations of the elliptic fiber as a hypersurface in toric ambient space has been classified in \cite{Grassi:2012qw,Braun:2013nqa,Klevers:2014bqa}, and the torsion associated with the complete intersection fibers in \cite{Braun:2014qka}.
The maximal torsion group  which can occur in hypersurfaces and complete intersections is $\mathbb Z_3$ and, respectively, $\mathbb Z_4$.
Another interesting phenomenon is that Higgsing phenomena can interpolate between non-torsional and torsional sections \cite{Baume:2017hxm}.

We have been careful to stress that we have been working, up to now, in a context where the gauge algebra is purely non-abelian.
The reason is that in the presence of abelian gauge factors, the nature of the Shioda map implies that the global structure of the gauge group is more subtle, as discussed independently and with complementary methods in \cite{Grimm:2015wda} and \cite{Cvetic:2017epq}.
Consider a Lie algebra $\mathfrak{g}_I \oplus \mathfrak{u}(1)$ with $\mathfrak{g}_I$ semi-simple. Then even in absence of Mordell-Weil torsion, the global structure of the gauge group is \cite{Cvetic:2017epq}
\bea
(G_I^{(0)} \times U(1)_A)/ \mathbb Z_{n_A}
\eea
where the integer $n_A$ is the smallest integer such that the coefficients $n_A  \, \ell_A^{i_I}$ in the Shioda map are integer. This is in fact a statement about the quantization of the $U(1)_A$ charge carried by a non-abelian representation ${\bf R}$ (rather than about the absence of certain representations from the non-abelian weight lattice as before).
This effect must be taken into account in the presence of both torsional and non-torsional sections as illustrated in \cite{Cvetic:2017epq}.

\section{Genus-one fibrations and discrete gauge symmetries} \label{sec_genusone}

Given that we have spent so much time on the origin of continuous gauge symmetries in F-theory, what about discrete symmetries? 
Such symmetries are very well-motivated from the perspective of particle physics applications, where discrete symmetries are oftentimes invoked to control dangerous couplings such as dimension-four and -five proton decay operators or act as flavour symmetries. The most famous example is probably matter parity in the MSSM (see e.g. \cite{Dreiner:2005rd} and references therein).
Discrete symmetries are also fascinating to study from a fundamental point of view:
According to a general conjecture, {\it continuous} global symmetries cannot exist in any consistent quantum gravity of dimension at least four. This is concisely summarized in \cite{Banks:2010zn}, which also contains a review  of much of the original literature.
Even though black hole arguments are less compelling in this case, it is believed that the same is true for {\it discrete} global symmetries in quantum gravity \cite{Banks:2010zn}.

Irrespective of these quantum gravity considerations, it is quite generally the case in string theory that what appears as a global symmetry in the low-energy effective action originates in a gauge symmetry in the ultra-violet.
F-theory is no exception here.
In particular, as we will review in this section, abelian discrete symmetries are really to be thought of as 'massive' gauge symmetries in F-theory, in perfect agreement with quantum gravity reasoning.

The systematic investigation of $\mathbb Z_k$ gauge symmetries in F-theory began with an analysis of genus-one fibrations with a $k$-section \cite{Braun:2014oya}, whose role in the context of $\mathbb Z_k$ symmetries in F-theory was subsequently better understood in \cite{Morrison:2014era}\cite{Anderson:2014yva}\cite{Klevers:2014bqa} \cite{Garcia-Etxebarria:2014qua,Mayrhofer:2014haa,Mayrhofer:2014laa,Cvetic:2015moa}. 
For pedagogical reasons we 
present the material not following this historical path, but start by explaining the origin of discrete symmetries in terms of torsional cohomology \cite{Mayrhofer:2014laa}, which, in our opinion, is the simplest and most direct picture from a physics points of view. 
The subtle relation to genus-one fibrations and the Tate-Shafarevich group will then be reviewed in section \ref{sec_genusone-b}.

\subsection{Discrete gauge groups and torsional cohomology}

According to the general lore applied many times in these lectures, a massless gauge potential in M-theory and its dual F-theory is obtained by reducing the M-theory 3-form $C_3$ as $C_3 = A_X \wedge {\rm w}_X + \ldots$ with ${\rm w}_X$ a harmonic 2-form on $\hat Y_{n+1}$ representing a non-trivial cohomology class in  $H^{2}(\hat Y_4,\mathbb R)$. On the other hand, if the 2-form is non-harmonic, the 1-form $A$ will be massive and hence its gauge invariance is broken.
More precisely, suppose there exist a non-harmonic 2-form ${\rm w}_2$ and a non-harmonic 3-form $\alpha_3$ on $\hat Y_{n+1}$, related by
\bea \label{w2a3ansatz}
d  {\rm w}_2 = k \, \alpha_3 \,.
\eea
Consider the Kaluza-Klein ansatz
\bea \label{w2a3ansatz-b}
C_3 = A \wedge {\rm w}_2 + c \,  \alpha_3
\eea
with $A$ a 1-form and $c$ a scalar field. The derivative $d C_3 = dA \wedge  {\rm w}_2 +   (k A + d c) \wedge \alpha_3 $, once inserted into the 11d kinetic term of $C_3$, yields a kinetic term in the effective action 
\bea
S \simeq  \int (d c + k A) \wedge \ast (d c + k A) + \ldots
\eea
The 1-form potential hence acquires a mass term in the M-theory effective action by absorbing the real axionic scalar $c$. 
The kinetic term is invariant under a simultaneous gauge transformation
\bea
A \rightarrow A + d \chi \qquad \quad c \rightarrow c - k \, \chi \,,
\eea
which allows us to go to a gauge with only a massive vector field. This is nothing but the M/F-theoretic version \cite{Grimm:2011tb} of the celebrated St\"uckelberg mechanism. It is a consequence of a gauging of the axionic shift symmetry $c \rightarrow c + {\rm const}$.
The charge of the axion $c$ is  $k$.

As is known on general grounds \cite{Banks:2010zn}, a St\"uckelberg mechanism with a charge $k$ axion breaks a $U(1)$ gauge symmetry to a $\mathbb Z_k$ gauge symmetry.
The realisation of a discrete $\mathbb Z_k$ symmetry via a reduction ansatz of the form (\ref{w2a3ansatz}), (\ref{w2a3ansatz-b}) was first described in the Type II RR sector and in M-theory in \cite{Camara:2011jg} \footnote{A systematic investigation of discrete gauge symmetries in open string Type II models can be found e.g. in \cite{BerasaluceGonzalez:2011wy,Ibanez:2012wg,Honecker:2013sww}. }. This is also the  description of $\mathbb Z_k$ gauge symmetries in F-theory which is perhaps the most immediate from a physical perspective \cite{Mayrhofer:2014laa}. 

The $3$-form  $\alpha_3$ has the property that $k \, \alpha_3$ is exact and hence defines a trivial cohomology class. By definition $\alpha_3$ therefore takes its value in the torsion cohomology group  ${\rm Tor}(H^3(\hat Y_{n+1},\mathbb Z))$. In particular, a $\mathbb Z_k$ discrete gauge symmetry appears both in F-theory {\it and simultaneously its M-theory dual} if \cite{Mayrhofer:2014laa}
\bea
{\rm Tor}(H^3(\hat Y_{n+1}, \mathbb Z)) = \mathbb Z_k \,.
\eea
Many useful properties of torsional cohomology groups are reviewed in \cite{Camara:2011jg}, including the relations 
\bea
{\rm Tor}(H^k(X_{d}, \mathbb Z)) \simeq {\rm Tor}(H^{2d-k}(X_d,\mathbb Z)), \qquad  {\rm Tor}(H_k(X_{d},, \mathbb Z)) \simeq {\rm Tor}(H^{k+1}(X_{d}, \mathbb Z))  \,,
\eea
where $X_d$ is a complex variety of complex dimension $d$. Here  ${\rm Tor}(H_k(X_{d},\mathbb Z)) $ denotes the group of torsional $k$-cycles, i.e. $k$-chains $\gamma_k$ such that $k \, \gamma_k = \partial \Sigma_{k+1}$.
The first relation is Poincar\'e duality, and the second follows from the so-called universal coefficient theorem. 

A special case of this construction occurs for $k=1$. In this situation, which was studied in  \cite{Grimm:2011tb,Braun:2014nva}, the $U(1)$ gauge symmetry is broken completely, up to the effects analyzed in \cite{Martucci:2015dxa,Martucci:2015oaa}.

Of particular interest are of course massless states which are charged under the $\mathbb Z_k$ symmetry. Given the origin of the discrete gauge symmetry in modes of $C_3$, these must be due to M2-branes wrapping suitable 2-chains.
We are most interested in the discrete symmetries in M-theory which uplift to  discrete 1-form symmetries in F-theory; in this case, the relevant M2-branes wrap chains in the fiber over codimension-two loci on the base. This leads to an immediate complication: As we recall from previous sections, when the states charged under a  $U(1)$ gauge symmetry  become massless, the fibral curve wrapped by the M2-brane shrinks to zero volume and creates a singularity; resolving this singularity, in turn, corresponds to moving along the Coulomb branch of the $U(1)$ gauge symmetry.
In the case of a discrete gauge symmetry, we should think of the gauge potential as having a mass, as motivated above. This clearly lifts the Coulomb branch. If we start from the origin of the Coulomb branch, where the charged particle is massless, a quadratic potential prevents us from continuously deforming the theory and hence resolving the singularity in a supersymmetric way. Put differently, moving out on the Coulomb branch by resolving the singularity costs energy and thus breaks supersymmetry. As a result, the codimension-two singularity in the fiber cannot be crepant resolvable \cite{Braun:2014nva,Braun:2014oya,Arras:2016evy}. In the language of section \ref{sec_terminalco2}, it must be a $\mathbb Q$-factorial
terminal singularity. 

A simple procedure to engineer a Weierstrass model $Y_{n+1}$ with a $\mathbb Z_k$ discrete gauge symmetry is to start with a model with a massless $U(1)$ gauge symmetry and a charge $k$ massless state and to perform a Higgsing $U(1) \rightarrow \mathbb Z_k$ \cite{Morrison:2014era}. 
As always we are working not directly in F-theory, but in M-theory.
If the elliptic fibration with Mordell-Weil group of rank $1$ is given as a smooth manifold $\hat Y_{n+1}$, we must first take the limit of vanishing fibral curve associated with the charge $k$-state. More precisely, we are moving to the origin of the Coulomb branch, where all $U(1)$ charged singlets (with vanishing M-theory KK charge) become massless.
Let us denote the resulting singular fibration again as $Y_{n+1}$. As the $k$-charged matter is massless, we can give it a vacuum expectation value, i.e. we move from the M-theory Coulomb branch onto the Higgs branch. 
Geometrically, the Higgsing corresponds to a complex structure deformation smoothing out the singularity in codimension-two associated with the former charge $k$ states. This is, in fact, a generalisation of a conifold transition \cite{Braun:2011zm,Krause:2012yh,Intriligator:2012ue}.  
After the deformation the singular elliptic fibration $Y_{n+1}$ must have trivial Mordell-Weil group, but $\mathbb Q$-factorial terminal singularities of Kodaira fiber type $I_2$ over codimension-two loci. 
The  states with former $U(1)$ charge $q$ now have $\mathbb Z_k$ charge $q\, \,  {\rm mod} \, k$ and are trapped in the singular fibers. 

Such a transition has first been described in  \cite{Morrison:2014era} by Higgsing the ${\rm Bl}^1\mathbb P_{112}[4]$ fibration of section \ref{sec-MPm} with the state with charge $q=2$. 
To realize the Higgsing outlined above we first pass from the smooth ${\rm Bl}^1\mathbb P_{112}[4]$-fibration $\hat P_{\rm MP}$, ({\ref{PMP-res}), to the birational Weierstrass model $P_{\rm W-U(1)}$, (\ref{WeierMP}), describing the origin of the $U(1)$ Coulomb branch in M-theory. Recall that the Weierstrass model $P_{\rm W-U(1)}$ has two types of $I_2$ singularities  - one over the locus $C_{II}$ in (\ref {Vf1f2minusC1}) with matter of charge $q=1$ and another over $C_I$, given in (\ref{Cb2c3}), with matter of charge $q=2$.
More precisely, moving to the Weierstrass model contracts both the curves $B_I$ and $B_{II}$  in the fibers over $C_I$ and $C_{II}$ which are not intersected by the zero-section.\footnote{Recall that in a Weierstrass model, singularities in the fiber are always away from the holomophic zero-section $z=0$.} Their intersection numbers are listed in (\ref{CACB}) and  (\ref{CsCrem}).  
Importantly, the massless states after the contraction hence carry vanishing Kaluza-Klein charge in M-theory  \cite{Mayrhofer:2014haa,Mayrhofer:2014laa}.

To Higgs $U(1) \rightarrow \mathbb Z_2$ we must deform the model such as to smoothen out the singularity over the charge-two locus $C_I$. The resulting Weierstrass model is called $P_{{\rm W}-{\mathbb Z}_2}$.  It turns out \cite{Morrison:2014era} that the correct transformation is by growing back a non-zero coefficient $c_4$, which had been set to zero in ({\ref{PMP}). To see the effect of this, one has to repeat the analysis of \cite{Morrison:2012ei} and map the full model  with $c_4 \neq 0$ into Weierstrass form. The result is to render the polynomial $e_4$ appearing in (\ref{WeierMP}) and (\ref{es}) generic,
\bea \label{PWZ2}
P_{{\rm W}-{\mathbb Z}_2} = P_{\rm W-U(1)}     \qquad     {\rm with} \qquad  e_4 \rightarrow -c_4 + \frac{1}{4} b_2^2  
\eea     

The singular $I_2$ fiber over $C_{II}$ is essentially unaffected by this deformation, but it now corresponds, as discussed above, to 
 a $\mathbb Q$-factorial terminal singularity, which is in fact of (non crepant resolvable) conifold type. 
 
What remains to show is that ${\rm Tor}(H^3(P_{\rm W-U(1)},\mathbb Z)) = \mathbb Z_2$.
 This is argued to be the case in \cite{Mayrhofer:2014laa} (possibly up to subtle effects associated with the terminal singularities), focusing for simplicity on 3-folds. The strategy is to follow various divisors and curves through the generalized conifold transition to observe the emergence of torsional cycles and their dual torsional cohomology groups. In fact, \cite{1992alg.geom.10009D} shows in a model with $\mathbb Z_3$ symmetry that the Weierstrass fibration has, after a blowup of the base to remove the terminal singularities and a resolution of the fiber, $\mathbb Z_3$ torsional 3-cohomology. This fact is also stressed in \cite{Braun:2014oya}. It would be very desirable to develop techniques to directly compute this torsional cohomology on the singular space
described by $P_{\rm W-\mathbb Z_2}$, and we conjecture that the physical effect of torsion cohomology should be detectable also in the singular geometry.

Barring this subtle point, as a result of the torsional cohomology group the effective action in M-theory and F-theory both contain a discrete $\mathbb Z_2$ gauge group \cite{Mayrhofer:2014laa}.
We are summarizing the transition in (\ref{transition1}), where the two last lines refer to the gauge group in M-theory and F-theory. By $U(1)_{S-S_0}$ we refer to the abelian group generated by the Shioda map $\sigma(S) = S - S_0 + \ldots$ of the rational section $S$, and $U(1)_{S_0}$ is the Kaluza-Klein $U(1)$ in the $S^1$-reduction of F-theory to M-theory.

\be \label{transition1}
\begin{array}{lllll}
{}\underline{{\hat P_{\rm MP}}  }     &  \multirow{6}{*}{  $\longrightarrow$  }       & \underline{P_{\rm W- U(1)} }   &  \multirow{6}{*}{ $\longrightarrow$  }  & \underline{ P_{{\rm W}-{\mathbb Z}_2} } \vspace{1mm} \\ 
{}{\rm rk(MW)} = 1  &                                                                                   &   {\rm rk(MW)} = 1              &                                                    			         &    {\rm rk(MW)} = 0    \vspace{1mm}   \\
{} {\rm Tor}(H^3) = 0  &                                                                                   &   {\rm Tor}(H^3) =0            &                                                    			         &    {\rm Tor}(H^3) = \mathbb Z_2     \vspace{1mm} \\
{\rm M}: U(1)_{S_0} \times U(1)_{S-S_0}                            &                                                                                   &   {\rm M}: U(1)_{\rm S_0} \times U(1)_{S-S_0}            &                                                    			         &   {\rm M}: U(1)_{S_0} \times \mathbb Z_2  \vspace{1mm} \\
{\rm F}:    U(1)_{S-S_0}                            &                                                                                   &   {\rm F}: U(1)_{S-S_0}      &                                                    			         &   {\rm F}:  \mathbb Z_2
\end{array}
\ee

 \subsection{Genus-one fibrations without sections} \label{sec_genusone-b}

 To recap, the  transition (\ref{transition1}) produces  a singular Weierstrass model which exhibits a $\mathbb Z_k$ gauge symmetry both in the F-theory and the M-theory effective action, and the $\mathbb Z_k$ gauge theory is to be interpreted as a $U(1)$ gauge theory with a mass term for the gauge potential. 
 In the M-theory effective action, the mass of the gauge potential lifts the M-theory Coulomb branch and obstructs small continuous deformations away from its origin. However, by general field theory reasoning such a theory still allows for another $k-1$ degenerate vacua on the Coloumb branch which are macroscopically far away from its origin. These correspond to a vacuum expectation value
 \bea
 \xi = n \qquad {\rm mod}\, \,  k \,
 \eea 
 for the scalar field $\xi$ in the (massive) M-theory vector multiplet. In other words, there exist $k$ supersymmetric isolated vacua along the M-theory Coulomb branch which are identified modulo $k$ \cite{Morrison:2014era}. 
 From an F-theory perspective these $k$ inequivalent vacua correspond to $k$ different values for the Wilson lines of the $\mathbb Z_k$ F-theory gauge field $\mathbb A$  along the F-theory circle, according to the usual identification (\ref{Axigen}),
 \be
 \oint_{\tilde S^1_B} \mathbb A = \xi \,.
 \ee
 All of these $k$ different M-theory vacua map to the same F-theory vacuum upon decompactification of the F-theory circle $\tilde S^1_B$.
 This prompts the question which geometries describe the remaining $k-1$ inequivalent M-theory vacua associated with an F-theory model with discrete gauge group $\mathbb Z_k$.
 The other vacua in M-theory have the property that the $\mathbb Z_k$ charged states acquire a mass due to the discrete vacuum expectation value of the Wilson line scalar even though this vacuum expectation value cannot be reached continuously starting from the origin of the Coloumb branch. 
This means in particular that the associated geometry is smooth as the terminal $I_2$ fibers are replaced by smooth $I_2$ fibers. 

To stay in the example of the $\mathbb Z_2$-model, the natural candidate for the geometry describing the second possible phase of the Coloumb branch
is the geometry described by 
\bea \label{PMPZ2}
P_{{\rm MP}-\mathbb Z_2} = w^2 + b_0 u^2 w  + b_1 u v w + b_2 v^2 w  + c_0 u^4  + c_1 u^3 v  + c_2 u^2 v^2  + c_3 u v^3 + c_4 v^4 \,.
\eea
This is the singular ${\rm Bl}^1 \mathbb P_{112}[4]$-fibration with $c_4$ grown back such as to  render the fibration maximally generic.
The resulting elliptic fibration has vanishing Mordell-Weil rank and no degenerate fibers over $C_{I}$, and it is smooth. By similar logic as spelled out around (\ref{P1logic}),
the fiber over the analogue of the locus $C_{II}$ is a {\it smooth} $I_2$ fiber.
However, the fibration has not only no extra rational sections, it has no rational sections at all! It is an example of a genus-one fibration with no section \cite{Braun:2014oya}.

Before addressing the absence of a zero-section in more detail, let us first verify via the Higgsing perspective that (\ref{PMPZ2}) describes the other Coulomb branch phase in M-theory.  
Starting from the $U(1)$ model (\ref{PMP-res}) the first step is to blow down the resolution divisor $S$. This process contracts the fibral curve $A_I$ over $C_I$ wrapped by $S$ (recall figure \ref{fig:fibre}).
This is notably different from the contraction summarized in (\ref{transition1}), where the fibral curve $B_I$ over $C_I$ is blown-down, along with $B_{II}$ over $C_{II}$.
After the contraction, the deformation by growing $c_4$ breaks $U(1)_{S-S_0}$. This time the state which obtains a VEV is the M2-brane wrapping the contracted fiber $A_I$ over $C_I$ (indeed this is the massless state). As can be seen from (\ref{CsCrem}), the associated  state $\varphi_{A_I}$ is charged under the M-theory KK $U(1)_{S_0}$, but uncharged under the linear combination $U(1)_{S-S_0} + 2 \, U(1)_{S_0}$ \cite{Mayrhofer:2014haa,Mayrhofer:2014laa}.
 In terms of the original gauge symmetries a Higgs-VEV $\langle \varphi_{A_I}\rangle \neq 0$ hence induces  the breaking 
\bea \label{U1tildeU}
U(1)_{S_0} \times U(1)_{S-S_0}   \longrightarrow U(1)_{\tilde U} = U(1)_{S-S_0} + 2 \, U(1)_{S_0} \,,
\eea 
with no {\it no additional} $\mathbb Z_2$ gauge group in M-theory. 
Indeed, the geometry (\ref{PMPZ2}) has no torsional cohomology and hence cannot describe an extra $\mathbb Z_2$ gauge symmetry in M-theory in addition to  $U(1)_{\tilde U}$. Fortunately this is not necessary because the $\mathbb Z_2$ symmetry which appears in F-theory is a subgroup of $U(1)_{\tilde U}$ \cite{Mayrhofer:2014laa}. This is important because from the perspective of the Wilson line, the other vacuum with $\xi = 1$ should not break the discrete symmetry. 

As noted already, (\ref{PMPZ2}) has no globally defined rational section \cite{Braun:2014oya}. Rather, if we consider the analogue of the zero-section $u=0$ on the model (\ref{PMP}) with $c_4=0$, we observe that 
\bea
P_{{\rm MP}-\mathbb Z_2}  |_{u=0} = w^2 + b_2 v^2 w + c_4 v^4  = w \pm \frac{1}{2}\sqrt{b_2 v^2 - 4 c_4 v^4} \,.
\eea
Locally, the equation describes two points in the fiber, which are  exchanged by a global monodromy as a result of the square-root. 
This is what defines a bisection, and we denote the associated divisor by $\tilde U$.
As we tune $c_4\rightarrow 0$ the two points in the fiber asymptote to the two rational points $s_1$ and $s_2$ in (\ref{eq:sec01}). Conversely, the two independent rational sections of the elliptic fibration (\ref{PMP}) combine into a bisection in (\ref{PWZ2}). The correspondence between the bisection $\tilde U$ and $S + S_0$ prior to Higgsing is of course  in exact agreement with (\ref{U1tildeU}).
As we go from M to F-theory, $U(1)_{\tilde U}$ becomes part of the higher-dimensional Poincar\'e symmetry (which is the usual behaviour for a KK $U(1)$), but a $\mathbb Z_2$ subgroup is realized in addition as the F-theory discrete symmetry. In this sense one can interpret the $U(1)_{\tilde U}$ in M-theory compactification with a bisection $\tilde U$ as a
 mixture of the KK $U(1)$ and the $\mathbb Z_2$ gauge field in F-theory \cite{Mayrhofer:2014laa}. 

There is another view on the Higgsing process: As discussed, the Higgs field prior to Higgsing carries KK charge and hence has a varying field profile along the circle $\tilde S^1_B$. Its derivative can be identified with a flux along the circle. This makes contact with the analysis of \cite{Anderson:2014yva}, which interprets the genus-one fibration in terms of a St\"uckelberg mechanism. The connection between both pictures is  to view the St\"uckelberg axion $c$ as the argument of the complex Higgs field $\varphi_{A_I} = |\varphi_{A_I} | e^{i c}$ which triggers the transition  \cite{Mayrhofer:2014laa}.  


The second transition and its F and M-theory interpretation can be summarized as follows:

\be \label{transition2}
\begin{array}{lllll}
{}\underline{{\hat P_{\rm MP}}}       &  \multirow{6}{*}{  $\longrightarrow$  }       & \underline{ P_{\rm MP}    }&  \multirow{6}{*}{ $\longrightarrow$  }  & \underline { P_{{\rm MP}-\mathbb Z_2} } \vspace{1mm} \\ 
{}{\rm rk(MW)} = 1  &                                                                                   &   {\rm rk(MW)} = 1              &                                                    			         &    {\rm rk(MW)} = 0    \vspace{1mm}   \\
{} {\rm Tor}(H^3) = 0  &                                                                                   &   {\rm Tor}(H^3) =0            &                                                    			         &    {\rm Tor}(H^3) = 0     \vspace{1mm} \\
{\rm M}: U(1)_{S_0} \times U(1)_{S-S_0}                            &                                                                                   &   {\rm M}: U(1)_{S_0} \times U(1)_{S-S_0}               &                                                    			         &   {\rm M}: U(1)_{\tilde U}  \vspace{1mm} \\
{\rm F}:    U(1)_{S-S_0}                            &                                                                                   &   {\rm F}: U(1)_{S-S_0}      &                                                    			         &   {\rm F}:  \mathbb Z_2
\end{array}
\ee

This pattern is expected to generalize to general $\mathbb Z_k$ theories. In \cite{Cvetic:2015moa}, the case of a $\mathbb Z_3$ symmetry has been investigated. The starting point is
 the toric hypersurface fibration with ${\rm rk}({\rm MW}) = 1$ and charged matter with $q=1,2,3$ found in \cite{ Klevers:2014bqa}. Upon Higgsing with the charged $3$ states 
 in three possible ways as dictated  by the above reasoning, perfect match with the geometry is found.

Mathematically, the $k$ different geometries associated with a $\mathbb Z_k$ symmetry in F-theory generate the Tate-Shafarevich group associated with a Jacobian fibration, which in this case is $\mathbb Z_k$. 
A summary of its main properties in the present context can be found in \cite{Braun:2014oya}. 
The Tate-Shafarevich group can be thought of as all genus-one fibrations with the same Jacobian fibration \cite{1992alg.geom.10009D}.
The zero-element is the Jacobian itself, i.e. the Weierstrass model with $k$-torsional homology (again possibly up to subtle effects from the singularities) and $\mathbb Q$-factorial terminal singularities \cite{1992alg.geom.10009D}. It is the only geometry which gives rise to a separate $\mathbb Z_k$ gauge symmetry (not contained in an abelian group) both in F-theory and M-theory. The remaining $k-1$ geometries are smooth, non-elliptic, $k$-section fibrations. They describe a $U(1)$ gauge theory in M-theory (containing the $\mathbb Z_k$ as a subgroup), whose subtle F-theory uplift gives the same effective action in F-theory as the singular Weierstrass model. 

Various other examples of $k$-section geometries have been studied in the F-theory literature. The highest $\mathbb Z_k$ symmetry obtained in this way in toric hypersurface models over generic bases is $\mathbb Z_3$ \cite{Klevers:2014bqa} and in complete intersection fibrations over generic bases $\mathbb Z_4$ \cite{Braun:2014qka}. Other examples of multi-section fibrations include \cite{Kimura:2015qpz} \cite{Kimura:2016crs}.
The physical effect of a $k$-section (leading to $\mathbb Z_k$ discrete symmetry) is very different from that of a $k$-torsional section (leading to $\pi_1(G) = \mathbb Z_k$ as described in section \ref{sec_torsion}).
Even more strikingly, both types of geometries are related by mirror symmetry in the fiber, as explained and verified  both for hypersurface and complete intersection fibrations in \cite{Klevers:2014bqa,Oehlmann:2016wsb}. The relation between genus-one fibrations and the appearance of multiple fibers has been discussed, along with interesting physics applications, in \cite{Anderson:2018heq}.

We have up to here studied $k$-section fibrations without non-abelian gauge symmetries. In presence of such gauge symmetries, a number of interesting new effects occur, some of which have already been pointed out in \cite{Braun:2014oya}. From a model building perspective, the relevance of a $\mathbb Z_k$ discrete symmetry becomes particularly evident as a selection rules governing the structure of Yukawa couplings \cite{Garcia-Etxebarria:2014qua,Mayrhofer:2014haa,Mayrhofer:2014laa,Lin:2015qsa}.

\section{Gauge backgrounds and zero-mode counting} \label{sec_Gaugebackgrounds}

We have, up to this point, focused on the geometry of the elliptic or torus fibration underlying an F-theory compactification.
The definition of the vacuum, however, depends on additional data. The missing ingredient is a choice of background value for the M-theory 3-form potential $C_3$ and its field strength $G_4$, which enter the M-theory effective action as in (\ref{Mtheoryaction}).
In Type IIB language, these data encode both the background values of the 2-form potentials $B_2$ and $C_2$ and the gauge background along the 7-branes. This reflects the general picture that what appears, to leading order, as a separate closed and open string sector from the perturbative Type IIB perspective is in fact unified into a common moduli space in F-theory. 
After some introductory remarks in \ref{sec_FluxversusDeligne} regarding the very nature of the 3-form background we first focus on its discrete part in section \ref{sec_fluxes} and \ref{Fluxes-Examples}. A finer analysis of the gauge backgrounds is possible in terms of the Deligne cohomology group, which in turn can be parametrized by elements of the Chow group, as we describe in section \ref{sec_Chowgroups}. This geometric formalism allows us to compute the massless spectrum of charged matter in F-theory beyond the chiral index (cf. section \ref{sec_cohos}).

\subsection{Flux versus Deligne cohomology} \label{sec_FluxversusDeligne}

An M-theory 3-form background involves two types of data: The background value of the field strength $G_4$ is called 4-form flux and is discrete in nature.
More precisely, $G_4$ takes values in the cohomology group $H^4(\hat Y_{n+1}, \mathbb R)$ and is in general half-integer quantized in such a way that \cite{ Witten:1996md}
\bea \label{quantisation}
G_4 + \frac{1}{2} c_2(\hat Y_{n+1}) \in H^4(\hat Y_4, \mathbb Z) \,.
\eea
While every element of this type represents a flux background in M-theory, in order for it to lift to a background flux in a Lorentz invariant F-theory vacuum additional conditions must be imposed, as will be discussed in the next section.
A $G_4$ background is incompatible with supersymmetry in F-theory compactifications to six dimensions, i.e. on $\hat Y_3$, but plays a crucial part in 4d/2d F-theory compactifications on 4-/5-folds.
In both situations, superymmetry requires that \cite{Becker:1996gj,Gukov:1999ya,Haack:2001jz,Haupt:2008nu}
\bea \label{G4alignment}
G_4 \in H^{2,2}(\hat Y_{n+1}, \mathbb R) \cap H^4(\hat Y_{n+1}, \mathbb Z/2) =: H^{2,2}_{\mathbb Z/2}(\hat Y_{n+1})
\eea
together with a primitivity condition $J \wedge G_4 = 0$. For definiteness we will focus on the four-dimensional case in the sequel. 

There can, however, exist flat, but topologically non-trivial 3-form backgrounds even in the absence of $G_4$-fluxes. Such backgrounds correspond to a non-zero vacuum expectation value of $C_3$ with $\langle d C_3\rangle = 0$.
By gauge invariance we identify $C_3 \simeq C_3 + d \chi$ for a 2-form $\chi$ so that the flat background values of $C_3$ take values in $H^3(\hat Y_4,\mathbb R)$. Large gauge transformations identify backgrounds differing by elements in $H^{3}(\hat Y_4, \mathbb Z)$. Taking into account that $H^{i,0}(\hat Y_4)$ for $i=1,2,3$ due to the Calabi-Yau condition, this leads us to identifying the 'Wilson line background' with elements in the intermediate Jacobian 
\bea
J^2(\hat Y_4) = H^3(\hat Y_4, \mathbb C) / (H^{2,1}(\hat Y_4, \mathbb C) + H^3(\hat Y_4,\mathbb C)) \,.
\eea
Such continuous flat gauge backgrounds are therefore possible if $\hat Y_4$ has a non-vanishing cohomology group $H^{2,1}(\hat Y_4,\mathbb C)$. This cohomology group has been studied in detail in \cite{Greiner:2015mdm,Greiner:2017ery}.  

The information about both the discrete 4-form flux and the flat Wilson line background is conveniently encoded in the so-called Deligne cohomology group $H^4(\hat Y_4, \mathbb Z(2))$.\footnote{We are ignoring here, for simplicity, the potentially half-integer shift in the quantization condition (\ref{quantisation}).} It has the property that it fits into the short exact sequence
\bea \label{Deligneexactseq}
\begin{tikzcd}
0 \arrow{r} & \underbrace{J^2(\hat Y_4)}_{ \oint C_3 \,  {\rm 'Wilson \,  lines'}} \arrow{r} & \underbrace{H^{4}_D(\hat Y_4,\mathbb{Z}(2))}_{\rm  Deligne \, cohomology}
&  \hspace{-2mm} \xrightarrow[{\rm  onto}]{\hat{c}_2 }  \hspace{-3mm}  & \underbrace{ H^{2,2}_{\mathbb{Z}}(\hat Y_4)}_{{\rm field \, strength} \, G_4} \arrow{r} & 0 \,.
\end{tikzcd}
\eea
This just means that there exists a \emph{surjective} map $\hat c_2$ which maps each element ${\cal A} \in H^4(\hat Y_4, \mathbb Z(2))$ to a flux configuration $G_4 = \hat c_2({\cal A}) \in H^{2,2}_{\mathbb{Z}}(\hat Y_4)$.
This map is in general not injective, and its kernel is given exactly by the intermediate Jacobian $J^2(\hat Y_4)$. Indeed the elements of $J^2(\hat Y_4)$ correspond precisely to the 3-form backgrounds whose associated flux $G_4$ vanishes.

For many purposes it suffices to consider exclusively the information encoded in the field strength $G_4$. These include the computation of the flux induced F-term and D-term potential as well as the computation of the chiral index of the charged massless spectrum. 
The non-chiral part of the charged massless spectrum, by contrast, requires finer information and makes connection with the more sophisticated description of the gauge background as in (\ref{Deligneexactseq}). 
Even if $h^{2,1}(\hat Y_4)=0$ so that there are not flat Wilson lines on $\hat Y_4$, such a refined description is necessary because non-trivial Wilson lines can occur on the matter loci.
The fact that the Deligne cohomology is the correct object characterizing the gauge background was first pointed out, via duality with the heterotic string, in \cite{Curio:1998bva}. 
More information on its definition can be found in \cite{Bies:2014sra} and the original literature referenced therein.

In the sequel we will mostly be working, as our notation suggests, on a smooth resolution $\hat Y_4$ of the elliptic fibration. The reason is that this avoids the intricacies of which cohomology theory to use on singular spaces.
The price to pay is the following limitation: Since the process of resolving the singularities corresponds to moving to the Coulomb branch of the gauge theory (in M-theory), the Deligne cohomology (or ordinary cohomology, if we restrict ourselves to the flux part) on the resolved space can only detect the gauge backgrounds in the Cartan of the F-theory gauge algebra. For many purposes, this is sufficient, in particular to engineer a chiral spectrum. 
Nonetheless, it is important to keep in mind that this approach misses truly non-abelian data such as non-abelian vector bundles on brane stacks. A related piece of information that cannot be described in this way is 
the so-called T-brane data \cite{Donagi:2003hh,Hayashi:2009bt,Cecotti:2010bp,Donagi:2011jy,Donagi:2011dv,Collinucci:2014qfa,Collinucci:2016hpz,Bena:2016oqr,Marchesano:2016cqg,Mekareeya:2016yal,Marchesano:2017kke}. In principle the Deligne cohomology can also be defined on singular spaces, first directions having been taken in the context of describing T-branes in \cite{Anderson:2013rka,Anderson:2017rpr}. In \cite{Collinucci:2014taa}, gauge backgrounds are addressed directly on the singular space.

\subsection{Discrete flux data: Constraints and chirality} \label{sec_fluxes}

The structure of the middle cohomology group $H^4(\hat Y_4,\mathbb C)$ of a Calabi-Yau 4-fold is rather complicated.
Its so-called \emph{horizontal} piece
\bea
H^4(\hat Y_4,\mathbb C) = H^{4,0}(\hat Y_4,\mathbb C)  \oplus H^{3,1}(\hat Y_4,\mathbb C) \oplus H^{2,2}_{\rm hor}(\hat Y_4,\mathbb C) \oplus  H^{1,3}(\hat Y_4,\mathbb C) \oplus H^{0,4}(\hat Y_4,\mathbb C) 
\eea
contains the cohomology group spanned by the unique harmonic $(4,0)$ form $\Omega_4$. Under variation of complex structure, $\Omega_4$ picks up, to first order, components along $H^{3,1}(\hat Y_4, \mathbb C)$, to second order components along  $H^{2,2}_{\rm hor}(\hat Y_4,\mathbb C)$ etc. 
For given complex structure, $H^{2,2}(\hat Y_4)$ itself enjoys a decomposition into orthogonal subspaces \cite{Greene:1993vm,Braun:2014xka}
\bea \label{22decomp}
H^{2,2}(\hat Y_4,\mathbb C) = H^{2,2}_{\rm hor}(\hat Y_4,\mathbb C) \oplus H^{2,2}_{\rm vert}(\hat Y_4,\mathbb C) \oplus H^{2,2}_{\rm rem}(\hat Y_4,\mathbb C) \,.
\eea
The primary vertical subspace is spanned by the product of $(1,1)$ forms,
\bea \label{primvert}
H^{2,2}_{\rm vert}(\hat Y_4,\mathbb C) = \langle  H^{1,1}(\hat Y_4, \mathbb C) \wedge  H^{1,1}(\hat Y_4, \mathbb C)  \rangle \,.
\eea
In addition there exists a remainder piece $H^{2,2}_{\rm rem}(\hat Y_4,\mathbb C)$ which neither descends from $H^{4,0}(\hat Y_4, \mathbb C)$ by variation of Hodge structure nor does it lie in (\ref{primvert}) \cite{Braun:2014xka}.
Note that both  $H^{2,2}_{\rm vert}(\hat Y_4,\mathbb C)$ and  $H^{2,2}_{\rm rem}(\hat Y_4,\mathbb C)$ are of $(2,2)$ Hodge type for every value of the complex structure moduli while primitivity is a non-trivial constraint only for fluxes in $H^{2,2}_{\rm vert}(\hat Y_4,\mathbb C)$.

As noted already, in order for $G_4$ not only to represent a valid 4-form flux within the context of M-theory compactified on $\hat Y_4$, but also in the dual F-theory vacuum, additional constraints must be imposed.
There are various equivalent ways of understanding the origin of these constraints. 
According to \cite{Dasgupta:1999ss}, in order for an M-theory flux not to spoil Poincar\'e invariance in the dual F-theory vacuum, the associated harmonic 4-form should have '1 leg in the fiber', i.e. it should neither be the pullback of a 4-form entirely defined on the base of the elliptic fibration nor should it give a non-zero value upon integration along the full elliptic fiber. 
These conditions are derived in \cite{Dasgupta:1999ss} by applying the simple rules reviewed in section (\ref{sec_Mtheorytoell}) for the definition of the F/M-theory duality. They can be imposed by requiring the transversality conditions
\bea \label{transversality-a}
[G_4] \cdot [S_0] \cdot \pi^\ast[D_\alpha^{\rm b}] = 0, \qquad [G_4] \cdot \pi^\ast[D_\alpha^{\rm b}] \cdot \pi^\ast[D_\beta^{\rm b}]  = 0
\eea
for every $D_\alpha^{\rm b} \in H^{1,1}(B_3)$. 
These two conditions have also been recovered for elliptic fibrations  in \cite{Grimm:2011fx,Cvetic:2012xn} by matching the flux-induced Chern-Simons terms in the effective action of M-theory on $\hat Y_4$ with the Chern-Simons terms induced at 1-loop level in the circle  reduction of the dual F-theory effective action. 
We will understand these constraints, in section \ref{sec_cohos}, as the statement that the tower of KK modes and the M2-brane states wrapped on curves in the base are not affected by the flux background.  
On genus-one fibrations without section, the class of the multi-section replaces $[S_0]$ in the first constraint \cite{Lin:2015qsa}. 

Apart from these two 'kinematical' conditions, the flux induces a dynamical potential on the moduli. 
If we consider M-theory on Calabi-Yau 4-folds,
the presence of flux in general generates a Gukov-Vafa-Witten superpotential  \cite{Gukov:1999ya,Haack:2001jz}
\bea \label{WGVW}
W = W_1 + W_2 \, \qquad  
W_1 = \int_{\hat Y_4} \Omega_4 \wedge G_4 \,, \qquad W_2 = \int_{\hat Y_4} J \wedge J \wedge G_4 \,.
\eea
The scalar potential resulting from $W_1$ involves the complex structure moduli and enforces the condition (\ref{G4alignment}). 
This can be viewed as a result of the stabilization of (part of) the complex structure moduli such that the flux aligns along the (2,2) component of the middle cohomology.
According to our discussion above, this results in a non-trivial constraint only for fluxes in $H^{2,2}_{\rm hor}(\hat Y_4, \mathbb C)$.
Note again that the M-theory complex structure moduli describe both the closed string Type IIB complex structure moduli and the open string D-brane moduli (see, e.g.,\cite{Alim:2009bx,Grimm:2009ef,Jockers:2009ti} in the present context).
The scalar potential from $W_2$ induces a D-term in the dual F-theory proportional to \cite{Lerche:1997zb,Grimm:2010ks,Grimm:2011sk,Grimm:2011tb}
\bea
V_D  \simeq \int_{\hat Y_4} \pi^\ast J \wedge w_\Lambda \wedge G_4 \,,
\eea
where $w_\Lambda$ refers both to the resolution divisors $E_{i_I}$ and the Shioda map $\sigma(s_A)$ in the presence of extra $U(1)_A$ gauge factors and $J_B$ is the K\"ahler form on the base.
For vanishing charged matter field VEVs this D-term potential must be zero, thereby constraining the K\"ahler moduli. This is the effective action realisation of the primitivity condition on the fluxes.
Note that $V_D$ vanishes identically for fluxes in $H^{2,2}_{\rm hor}(\hat Y_4,\mathbb C)$ and $H^{2,2}_{\rm rem}(\hat Y_4,\mathbb C)$ due to the orthogonality, with respect to the intersection product on $\hat Y_4$, of the decomposition (\ref{22decomp}). If non-zero, $V_D$ plays the role of what is sometimes called 'field dependent Fayet-Iliopoulos term' for the Cartan $U(1)_{i_I}$ or the non-Cartan $U(1)_A$. At the same time, the gauging of the axionic partners of the K\"ahler moduli induces a St\"uckelberg mechanism for the respective abelian gauge boson. This is precisely the 'flux-induced'  St\"uckelberg mechanism we have alluded to at the end of section \ref{rationalShioda} and in various other places.

Finally, the Chern-Simons coupling in the effective action (\ref{Mtheoryaction}) shows that non-vanishing 4-form flux contributes to a net M2-brane tadpole on $\hat Y_4$. Tadpole cancellation is equivalent to the integrability condition for the Bianchi identity for $G_4$ and implies  \cite{Sethi:1996es,Dasgupta:1996yh,Dasgupta:1999ss}
\bea
 - \frac{1}{2} \int_{\hat Y_4} G_4 \wedge G_4 + \frac{1}{24} \chi(\hat Y_4) = N_{M2} \stackrel{!}{\geq} 0 \,.
\eea
The second term involving the Euler characteristic $\chi(\hat Y_4)  = \int_{\hat Y_4} c_4(\hat Y_4)$ of the elliptic fibration represents the M2-brane charge induced by the curvature dependent part of the M-theory Chern-Simons action. The number $N_{M_2}$ of spacetime-filling M2-branes in M-theory equals the number of spacetime-filling D3-branes in the dual F-theory vacuum and must be non-negative if the vacuum is to preserve supersymmetry.
The fact that the right-hand side is integer is a non-trivial consequence of the quantization condition (\ref{quantisation}).

Apart from inducing a non-trivial F- or D-term potential for the complex structure or K\"ahler moduli, respectively, an important effect of flux is to generate a non-trivial chiral index for massless charged matter in F-theory. The chiral index admits an intuitive expression given by integrating $G_4$ over the 'matter surface' associated with the multiplet.
Consider first the case of localised matter in representation ${\bf R}$ along an irreducible matter curve $C_{\bf R}$ on $B_3$, in the notation introduced in section \ref{subsec-zeromode4d-1}.
Recall that to each element of the weight vector $\beta^a({\bf R})$, $a=1, \ldots, {\rm dim}({\bf R})$ one associates a rational curve in the fiber over $C_{\bf R}$ such that an M2-brane wrapping this combination of curves gives rise to matter with Cartan charges $\beta^a({\bf R})$.
The fibration of this curve of the matter allows us to define a surface $S^a({\bf R})$ called 'matter surface'.
 The final result is that the chiral index of massless matter associated with weight $\beta^a({\bf R})$ can be computed as \cite{Donagi:2008ca} \cite{Braun:2011zm} \cite{Marsano:2011hv} \cite{Krause:2011xj} \cite{Grimm:2011fx}
\bea \label{chibeta-formula}
\chi(\beta^a({\bf R})) = \int_{S^a({\bf R})} G_4 \,.
\eea
From this we conclude that a necessary condition for the flux $G_4$ to leave the non-abelian gauge group along the 7-branes unbroken in the F-theory limit is that
\bea \label{gaugeinvariance-a}
G_4 \cdot [E_{i_I}] \cdot \pi^\ast[D^{\rm b}_{\alpha}] = 0 \,.
\eea
This guarantees that the chiral index for all weights of a given representation are the same, and we can hence write
\bea \label{G4indexa}
\chi({\bf R}) = \int_{S^a({\bf R})} G_4
\eea
for any choice of $S^a({\bf R})$. We will present a one-line derivation of  the formula (\ref{chibeta-formula}) in section (\ref{sec_cohos}), equ. (\ref{derivchi}), and also see that the condition (\ref{gaugeinvariance-a}) is indeed only necessary, not sufficient. 
The correct condition is (\ref{gaugeinvariance-finer}).

The same formula counts chiral massless matter associated with some of the roots, i.e. the weights of the adjoint representation of gauge algebra $\mathfrak{g}_I$. The associated matter is localized along the entire 7-brane divisor. Nonetheless, one formally define a matter surface whose overlap with the gauge flux, (\ref{gaugeinvariance-a}), counts the chiral index of such states.
 The matter surface is a suitable linear combination of 4-cycles obtained by restricting the resolution divisors $E_{i_I}$ to the anti-canonical divisor $\bar K_{\Sigma_I}$, see
equ. (\ref{chi-bulk}) in section \ref{sec_cohos} \cite{Bies:2017fam}.
The expression for the chiral index of such states is only non-zero if (\ref{gaugeinvariance-a}) is not obeyed.

\subsection{Examples of fluxes} \label{Fluxes-Examples}

Let us now provide a few examples  of background fluxes with values in each of the three orthogonal subspaces of the decomposition (\ref{22decomp}).

\subsubsection{Horizontal fluxes} 

By definition, the horizontal subspace  $H^{2,2}_{\rm hor}(\hat Y_4)$ is generated by cohomology classes which are of (2,2) Hodge type only on a special subspace of the complex structure moduli space of $\hat Y_4$. The dual homology class is hence the class of a complex 2-cycle which is algebraic only for certain complex structure moduli. 
To arrive at a simple example of such a situation, we follow \cite{Braun:2011zm} and start from a generic Tate model (\ref{PT}) over a base $B_3$. The elliptic 4-fold $\hat Y_4$ is embedded into an ambient 5-fold $\hat X_5$ obtained by fibering the fiber ambient space $\mathbb P_{231}$
over $B_3$. 
We have already observed before that the hypersurface can be written as  $A B = C D +  a_6 z^6$ as discussed around (\ref{PT-conifoldtypea}).
 Suppose now that the polynomial $a_6$ factorises as  $a_6= \rho \,  \tau$, where $\rho$ and $\tau$ are two holomorphic polynomials on $B_3$ whose classes add up to $6 \bar K_{B_3}$. 
The complete intersection within the ambient 5-fold $\hat X_5$ given by
\bea \label{Gamma1}
\Gamma = V(A,C,\rho) = \{A = 0 \} \cap  \{C = 0\} \cap \{\rho = 0\} \subset \hat X_5
\eea
lies on $\hat Y_4$ and defines an algebraic complex 2-cycle thereon - at least as long as $a_6= \rho \,  \tau$. Away from this special locus in complex structure moduli space, $\Gamma$ is no longer algebraic as a complex 2-cycle on $\hat Y_4$. Its Poincar\'e dual 4-form is therefore exactly of the type we are after  \cite{Braun:2011zm}. Note that even for $a_6= \rho \,  \tau$, the surface  $\Gamma$ is represented by a complete intersection only on the ambient space $\hat X_5$, but not on  $\hat Y_4$ because it cannot be written as the vanishing locus of the defining hypersurface equation (\ref{PT-conifoldtypea}) with two further polynomials on $\hat X_5$.

Suppose now that we switch on a background flux $G_4 = [\Gamma]$ on $\hat Y_4$. This flux is of $(2,2)$ Hodge type only for $a_6 = \rho \,  \tau$. Since fluxes which are not of $(2,2)$ Hodge type break supersymmetry by an F-term, such flux dynamically drives $\hat Y_4$ to the critical locus in complex moduli space along which it is $(2,2)$. This is precisely a consequence of the Gukov-Vafa-Witten superpotential $W_1$ in (\ref{WGVW}).    
The flux dual to (\ref{Gamma1}) and its generalizations  in fact play an important role in conifold transitions in 4-folds including genus one fibrations with and without section \cite{Braun:2011zm,Krause:2012yh,Intriligator:2012ue,Mayrhofer:2014haa,Lin:2015qsa,Jockers:2016bwi}. 

A systematic analysis of the complex structure moduli stabilizing effect of horizontal gauge fluxes requires the computation of the 4-form periods which enter the Gukov-Vafa-Witten superpotential. The state of the art as of this writing can be found in \cite{Braun:2014ola,Bizet:2014uua,Cota:2017aal} and references therein.

\subsubsection{Vertical fluxes}

Fluxes in $H^{2,2}_{\rm vert}(\hat Y_4)$ have been constructed and analysed in the F-theory context, for instance, in \cite{Grimm:2010ez,Braun:2011zm,Marsano:2011hv,Krause:2011xj,Krause:2012yh,Borchmann:2013hta,Bizet:2014uua,Lin:2015qsa,Cvetic:2013uta,Cvetic:2015txa,Lin:2016vus,Lin:2016zha,Bies:2017fam}. Many detailed explanations and more references can be found in \cite{Lin:2016zha}.
Conceptually, the primary vertical fluxes are perhaps the most immediate to construct, and at the same time they are the only ones generating a chiral index.\footnote{This follows from orthogonality of the decomposition (\ref{22decomp}) and is true provided the matter surfaces $[S^a({\bf R})]$ do not receive contributions from $H^{2,2}_{\rm rem}(\hat Y_4)$.}
Recall that according to the Tate-Shioda-Wazir theorem $H^{1,1}(\hat Y_4)$ is given by the span
\bea \label{h11generators}
H^{1,1}(\hat Y_4) = \langle [S_0], [S_A], [E_{i_I}], \pi^\ast[D^{\rm b}_\alpha] \rangle \,,
\eea
and $H^{2,2}_{\rm vert}(\hat Y_4)$ is generated by all linear combinations of products of two such elements. 
In general, the resulting products obey a number of cohomological relations within $H^{2,2}(\hat Y_4)$ so that the dimension of $H^{2,2}_{\rm vert}(\hat Y_4)$ is considerably smaller than the naive value $\frac{1}{2} h^{1,1}(\hat Y_4) (h^{1,1}(\hat Y_4)+1)$. 
On a concrete elliptic fibration $\hat Y_4$ the intersection numbers $H^{2,2}(\hat Y_4) \times H^{2,2}(\hat Y_4) \rightarrow \mathbb Z$ can be evaluated explicitly and expressed in terms of intersection numbers of divisors on the base.
By dividing out the so-obtained numerical relations one arrives at a generating set of $H^{2,2}_{\rm vert}(\hat Y_4)$. A detailed explanation how to systematically reduce these intersection numbers to intersections on $B_3$ in a base-independent manner can be found in \cite{Lin:2016vus,Lin:2016zha}.

By construction, since the generators in (\ref{h11generators}) are elements of $H^2(\hat Y_4, \mathbb Z)$ the resulting generating set of $H^{2,2}_{\rm vert}(\hat Y_4)$ is guaranteed to be integer, but the real challenge lies in finding the \emph{minimal} integral basis of $H^{2,2}_{\rm vert}(\hat Y_4)$. This is important  when it comes to implementing the quantization condition (\ref{quantisation}) on the fluxes \cite{Bizet:2014uua}. 
Oftentimes in the literature, necessary conditions for  (\ref{quantisation})  to hold are checked by verifying intersection numbers with explicitly known algebraic cycles. 
To ensure that the resulting 4-forms uplift to suitable fluxes in F-theory, it  remains to implement the transversality conditions (\ref{transversality-a}). This is achieved by making an ansatz for a linear combination in terms of the generating set of $H^{2,2}_{\rm vert}(\hat Y_4)$ and imposing the intersections (\ref{transversality-a}).

While in principle one can simply run this algorithm to systematically arrive at a basis of valid vertical F-theory fluxes, it is useful to have an intuition for what types of vertical fluxes can occur.
These are, in fact, of two types:

First, in the presence of a non-trivial Mordell-Weil group, from each independent rational section $S_A$ and its image under the Shioda map $U_A$ one can construct a so-called $U(1)_A$ flux \cite{Grimm:2010ez}
\bea \label{U1Aflux}
 \pi^\ast F \wedge [\sigma(s_A)] \qquad {\rm with} \quad  F \in H^{1,1}(B_3) \,. 
\eea
Note that  by virtue of the Shioda map, this flux automatically satisfies the transversality condition (\ref{transversality-a}) and the gauge invariance condition (\ref{gaugeinvariance-a}).  
Similarly,
\bea \label{Cartanflux}
\pi^\ast F \wedge [E_{i_I}] \qquad {\rm with} \quad  F \in H^{1,1}(B_3)
\eea
represents the gauge flux associated with the $U(1)_{i_I}$ Cartan subalgebras. It is transversal, but by construction  breaks the Lie algebra $\mathfrak{g}_I \rightarrow \mathfrak{h}_{i_I} \oplus \mathfrak{u}(1)_{i_I}$ with $\mathfrak{h}_{i_I} $ the commutant of $\mathfrak{u}(1)_{i_I}$ within $\mathfrak{g}_I$.
This type of flux is localized in a manifest manner in the sense that $F|_{\Sigma_{I}}$ represents the corresponding Cartan gauge flux along the 7-brane stack $\Sigma_{I}$ in Type IIB language. Such an interpretation is less immediate for the non-Cartan $U(1)_A$ flux (\ref{U1Aflux}). In this case, what is relevant to understand the massless spectrum is the restriction of $F$ to the matter curves on the base, as will be discussed below.

Even in absence of non-Cartan $U(1)_A$ gauge groups, extra types of vertical gauge fluxes are possible. These are related to the Poincar\'e dual cohomology classes of the matter surfaces $S^a_{\bf R}$ introduced before (\ref{gaugeinvariance-a}), as first exemplified in \cite{Marsano:2011hv } and systematized in \cite{Borchmann:2013hta,Bies:2017fam}.
The cohomology class $[S^a_{\bf R}]$ satisfies the transversality conditions (\ref{transversality-a}) by construction and is hence a candidate for a 4-form flux. 
With one exception \cite{Braun:2014pva}, in all cases studied in the literature to date $[S^a_{\bf R}]$ lies in the vertical part of the middle cohomology (as opposed to the remainder).
In order to implement also the gauge invariance condition (\ref{gaugeinvariance-a}), one has to add a vertical correction term. The result is the so-called matter surface flux
\bea \label{mattersurfacefluxdef}
[A^a({\bf R})] = [S^a_{\bf R}] + [\Delta^a({\bf R})], \qquad  \Delta^a({\bf R}) = \beta^a({\bf R})^T_{i_I} (C^{-1})^{i_I j_J} \, E_{j_J} |_{C_{\bf R}} \,.
\eea
Note that this object is independent of the choice of weight $a$: Since two weights $\beta^a({\bf R})$ and $\beta^b({\bf R})$ differ by a root,   the difference $[S^a_{\bf R}]  - [S^b_{\bf R}] $ is given simply by the restriction 
of a linear combination of resolution divisors $E_{i_I} |_{C_{\bf R}}$. This difference is then offset by the correct terms and overall $[A^a({\bf R})]  = [A^b({\bf R})]$. 

In general, the fluxes associated with all matter surfaces are not cohomologically independent. In fact, some of the cohomological relations are a consequence of anomaly cancellation in the 4-dimensional effective action, as described in \cite{Bies:2017abs}. The relations following from anomaly cancellation take the form
\begin{empheq}
{align}
\sum_{\mathbf{R}} \sum_{a} n^a_{i_I j_J k_K} \left( \mathbf{R} \right) \, \left[ A^a(\mathbf{R}) \right]_\mathrm{vert} \, & = 0 \in H^{2,2} ( \hat Y_4 ) \label{naijkA-1} \\
\sum_{\mathbf{R}} \sum_{a} n^a_{A \Sigma \Gamma} \left( \mathbf{R} \right) \, \left[ A^a \left( \mathbf{R} \right) \right]_{\mathrm{vert}} \, - 3\, \left[ U_{(A} \right] \cdot \left[ {\pi}_* \left( F_{\Sigma} \cdot F_{\Gamma)} \right) \right] &= 0 \in H^{2,2} ( \hat Y_4 ) \label{naijkA-2} \\
\sum_{\mathbf{R}} \sum_{a} q_A \, \left[ A^a \left( \mathbf{R} \right) \right]_{\mathrm{vert}} \, + 6 \,\left[ U_A \right] \cdot \left[ \overline{K}_{B_3} \right] &= 0 \in H^{2,2} ( \hat Y_4 ) \,. \label{naijkA-3}
\end{empheq}
Here $F_\Sigma \in \{\sigma(s_A), E_{i_I}\}$ refers to any of the Cartan or non-Cartan generators and  
\be
n^a_{\Sigma \Lambda \Gamma} = \beta^a_{\Sigma}({\bf R}) \beta^a_{\Lambda}({\bf R})   \beta^a_{\Gamma} ({\bf R})
\ee
 with the understanding that $\beta^a_A({\bf R}) = q_A$ is the $U(1)_A$ charged of  representation ${\bf R}$. 
 The second term in (\ref{naijkA-2}) and (\ref{naijkA-3}) represent special types of $U(1)_A$ and Cartan fluxes. Relation (\ref{naijkA-1}) can be derived from the requirement that all cubic non-abelian anomalies must cancel, whereas (\ref{naijkA-2}) and  (\ref{naijkA-3}) follow from the cancellation of the mixed abelian-non-abelian and mix abelian-gravitational anomalies via the Green-Schwarz mechanism. In particular the second term in (\ref{naijkA-2}) and (\ref{naijkA-3}) follows from the Green-Schwarz counter-terms derived in \cite{Cvetic:2012xn}. 

As an example for the construction of vertical fluxes, consider the $SU(2)$ Tate model of section \ref{SU2Tatematter}. In absence of non-Cartan abelian gauge symmetries, a candidate for a vertical gauge flux which does not break the $SU(2)$ gauge symmetry in the F-theory limit is the matter surface flux associated with the representation ${\bf R} = {\bf 2}$. As it turns out, this flux is in fact trivial:
Indeed, fibering each of the two curves $C_{\rm sp}^{(1)}$ and $C_{\rm sp}^{(2)}$ over the matter curve $C_{\bf 2}$ in the base gives rise to a surface which we call $\hat C_{\rm sp}^{(1)}$ and $\hat C_{\rm sp}^{(2)}$.
As a result of (\ref{Csp1homol2}), these are homologous. Furthermore 
\bea \label{C1C2rel}
[\hat C_{\rm sp}^{(1)}] + [\hat C_{\rm sp}^{(2)}] = 2 [\hat C_{\rm sp}^{(1)}]  = 2 [\hat C_{\rm sp}^{(2)}] = [E_1 |_{C_{\bf 2}}] \,.
\eea
According to (\ref{SU2weights1}) we can define the classes of the matter surface associated with the weights $\beta^1({\bf 2})$ and $\beta^2({\bf 2})$  as 
\bea
[S^1({\bf R})] = - [\hat C_{\rm sp}^{(2)}], \qquad [S^2({\bf R})] =  [\hat C_{\rm sp}^{(1)}]  \,.
\eea
Evaluating (\ref{mattersurfacefluxdef}), with Cartan matrix $C_{11} = -2$, we deduce from  (\ref{C1C2rel}) that the flux is indeed trivial. 
Hence in this model, the only possible vertical gauge flux corresponds to the Cartan flux $E_1 \wedge \pi^\ast F$ for some $F \in H^{1,1}(\hat Y_4)$, which breaks the non-abelian gauge symmetry in the F-theory limit to the Cartan subgroup. This is in agreement with the intuition that vertical gauge fluxes induce chirality in the massless charged spectrum. But the fundamental representation ${\bf 2}$ of $SU(2)$ is pseudo-real and hence there exists no notion of chirality compatible with an unbroken $SU(2)$ gauge group. 
In Type IIB language, all chirality inducing candidate fluxes are ruled out by the D5 tadpole cancellation condition \cite{Krause:2012yh}. 

In more complicated fibrations the matter surface fluxes can well be non-trivial. For the $I_n$ series realized as a Tate model, the first model where this is the case is the $I_5$ Tate model with gauge algebra $\mathfrak{su}(5)$. For $n=3$ and $n=4$, the triviality of the matter surface fluxes is in fact a consequence of the relations (\ref{naijkA-1}) \cite{Bies:2017abs}.
For the generic $I_5$ Tate model, on the other hand, there exist a priori two different matter surface fluxes $[A({\bf 10})]$ and $[A({\bf 5})]$ associated with the two representations of localized charged matter. These satisfy the cohomological relation
\bea
[A({\bf 10})] + [A({\bf 5})] = 0 \,,
\eea
in agreement with (\ref{naijkA-1}).

\subsubsection{Fluxes in $H^{2,2}_{\rm rem}(\hat Y_4)$}

As a starting point to exemplify a flux in the remainder piece $H^{2,2}_{\rm rem}(\hat Y_4)$ \cite{Braun:2014xka}, consider a component $\Sigma$ of the discriminant divisor associated with the some Lie algebra $\mathfrak{g}$. The embedding $\iota: \Sigma \rightarrow B_3$ embeds all curves on $\Sigma$ into $B_3$ and hence the full fibration $\hat Y_4$. In general, two curves on $\Sigma$ which are independent in the homology of $\Sigma$ need not be homologically independent on $B_3$. In particular, we can consider a curve $C$ on $\Sigma$ with $[C] \neq 0 \in H_2(\Sigma,\mathbb Z)$, but $\iota_\ast [C] = 0 \in H_2(B_3,\mathbb Z)$. Such a curve on $\Sigma$ cannot arise by intersecting $\Sigma$ with a divisor on $B_3$. 

Consider now the resolution divisors $E_i$, which by construction are fibered over $\Sigma$. We can hence restrict each $E_i$ to $C$ on $\Sigma$ and obtain a non-trivial surface $E_i |_C$ \cite{Mayrhofer:2013ara,Braun:2014pva}.  Its cohomology group $[E_i |_C]$ is generally non-zero within $H^{2,2}(\hat Y_4)$,\footnote{For example, its self-intersection is generically non-vanishing as long as $[C]$ has a non-zero self-intersection number within $\Sigma$ because $\int_{\hat Y_4} [E_i |_C] \wedge [E_j |_C]  = -\mathfrak{C}_{ij} \int_\Sigma [C] \wedge [C]$ by means of (\ref{fundrelEalpha1}).} but it does not lie within $H^{2,2}_{\rm vert}(\hat Y_4)$ because it cannot be written as the linear combination of an intersection of divisors on $\hat Y_4$. On the other hand, the surface $E_i |_C$ is algebraic for every choice of complex structure of $\hat Y_4$. Its class is hence always of $(2,2)$ Hodge type and therefore $[E_i |_C]$ is not in $H^{2,2}_{\rm hor}(\hat Y_4)$ either. We are forced to conclude that $[E_i |_C] \in H^{2,2}_{\rm rem}(\hat Y_4)$. Explicit examples of this type have been provided  in \cite{Braun:2014pva}; in these examples, the 4-cycle classes can be written as a complete intersection of three divisors in a  ambient complex 5-fold into which $\hat Y_4$ is embedded as a hypersurface, even though they are no complete intersections on $\hat Y_4$ itself.

From the perspective of the gauge theory along $\Sigma$, this flux corresponds to a line bundle $L$ on $\Sigma$ with $c_1(L) = [C] \in H^2(\Sigma)$ and structure group the Cartan factor $U(1)_i$. Such a line bundle breaks the gauge group along $\Sigma$ to the commutant of $U(1)_i$ in the F-theory limit. 
At the same time, the flux obeys the necessary condition (\ref{gaugeinvariance-a}) for gauge invariance because of the orthogonality of $H^{2,2}_{\rm vert}(\hat Y_4)$ and $H^{2,2}_{\rm rem}(\hat Y_4)$. This condition must therefore be modified as will be discussed in section \ref{sec_cohos}. 
In fact, in the context of $SU(5)$ GUT models fluxes of this type \cite{Mayrhofer:2013ara,Braun:2014pva} can be invoked to break the GUT group to the Standard Model  gauge group by choosing the Cartan subgroup to be the hypercharge group $U(1)_Y \subset SU(5)$. The condition $\iota_\ast [C] = 0 \in H_2(B_3,\mathbb Z)$ ensures that no St\"uckelberg mechanism renders the $U(1)_Y$ gauge potential massive \cite{Buican:2006sn}. In this sense, fluxes in $H^{2,2}_{\rm rem}(\hat Y_4)$ lie at the heart of the $SU(5)$ F-theory GUT paradigm of \cite{Donagi:2008ca,Beasley:2008dc,Beasley:2008kw,Donagi:2008kj}.

\subsection{Chow groups and gauge backgrounds}\label{sec_Chowgroups}

If we are interested in determining not only the chiral index of charged matter zero modes, but the exact number of massless chiral and anti-chiral multiplets, we must specify the gauge background beyond the field strength $G_4$. To this end we will first describe a practical parametrization  of the Deligne cohomology group encoding the full gauge background and then extract a formula counting the massless matter states \cite{Bies:2014sra,Bies:2017fam}.

At the level of the gauge flux, $G_4$ is specified by an element in the middle cohomology group. By Poincar\'e duality this defines a 4-cycle class in $H_4(\hat Y_4)$. The Hodge conjecture states that every element in $H_4(\hat Y_4, \mathbb Q)$ is in fact dual to the homology class of an algebraic complex 2-cycle. Assuming this for now, a natural way to think about the flux background is therefore in terms of complex  2-cycle classes modulo homological equivalence \cite{Braun:2011zm}. 
The chiral index (\ref{G4indexa}) can in particular be understood as the topological intersection number between a matter surface  class and the flux cycle class, up to homological equivalence. 
Indeed, changing e.g. the class of the matter surface by a homology transformation leaves this intersection product invariant and hence does not change the chiral index.

As it turns out, identifying complex 2-cycles up to homological relations loses in general too much information to capture not only the field strength $G_4$, but the full information encoded in $H^4_{\cal D}(\hat Y_4, \mathbb Z(2))$. 
A more refined notion of equivalence suitable for our purposes is given by rational equivalence. More explanations of the following summary and references to the mathematics literature can be found in \cite{Bies:2014sra}.

Two complex $p$-cycles $Z_1$ and $Z_2$ are called rationally equivalent if they are two members of  a rationally parametrized family of $p$-cycles, i.e. if there exists a family of $p$-cycles $Z(t)$ with $t \in \mathbb P^1$ such that $Z_1 = Z(t_1)$ and $Z_2 = Z(t_2)$. The equivalence class of algebraic cycles of complex dimension $p$ (or of complex codimension $p$) modulo rational equivalence is called the Chow group ${\rm CH}_p(\hat Y_4)$ (or, respectively, ${\rm CH}^p(\hat Y_4)$). In particular, for algebraic cycles of complex codimension $p=1$, i.e. for Weil divisors, rational equivalence coincides with the perhaps more familiar notion of linear equivalence, and ${\rm CH}^1(\hat Y_4)$ is the group of Weil divisors modulo linear equivalence. These facts have been collected for the reader's convenience in Appendix \ref{app_divisors}. 

To every such Weil divisor class $D$ one can associate a sheaf ${\cal O}(D)$. If the space is smooth (or more generally has only 'factorial' singularities), every Weil divisor is in fact Cartier (meaning that it can be locally expressed as the zeroes or poles of a single meromorphic function), and the associated sheaf is a line bundle. In any event, the sheaf or line bundle can be interpreted as a gauge bundle encoding the gauge background data of a 1-form gauge theory. 

In our context, we are not dealing with a 1-form gauge theory, but with a 3-form gauge theory (whose gauge potential is $C_3$). Nonetheless, it is still true that a suitable Chow group parametrizes the gauge background data of this theory, in the following sense:
 There exists a so-called refined cycle map $\hat \gamma_2$ which maps equivalence classes of complex 2-cycles on $\hat Y_4$ to elements of $H^4_{\cal D}(\hat Y_4, \mathbb Z(2))$, i.e.
\bea
\hat \gamma_2: {\rm CH}^2(\hat Y_4) \rightarrow H^4_{\cal D}(\hat Y_4, \mathbb Z(2)) \,.
\eea
Most importantly, the map is well-defined on  ${\rm CH}^2(\hat Y_4)$, i.e. if we encode a gauge background given by an element in the image of $\hat \gamma_2$ in $H^4_{\cal D}(\hat Y_4, \mathbb Z(2))$ by a Chow-class with representative $Z$, then changing $Z$ modulo rational equivalence (hence leaving its class in   ${\rm CH}^2(\hat Y_4)$ unchanged) does not alter the gauge background. 
The refined cycle map $\hat \gamma_2$ is surjective if the Hodge conjecture holds, meaning that in this case \emph{every} gauge background can be encoded by a Chow class. It is in general not injective, i.e. there might in general be some redundancy in our geometric description of gauge backgrounds via Chow groups. 
The relation between the cycle classes and the gauge backgrounds is summarized in more detail in the following diagram:
\bea
\begin{tikzcd}
  0 \arrow{r} & {\rm CH}^2_{\text{hom}}(\hat Y_4) \arrow{r} \arrow{d}{AJ} & \overbrace{  {  {\rm CH}^2(\hat Y_4)}}^{\rm  geometry}
  \arrow{r}{\gamma_2} \arrow{d}{\bf  \hat{\gamma}_2} & H^{2,2}_{alg}(\hat Y_4)
  \arrow{r} \arrow[hookrightarrow]{d} & 0 \\
  0 \arrow{r} & \underbrace{J^2(\hat Y_4)}_{\oint C_3 \,  \rm 'Wilson \, lines' } \arrow{r} & \underbrace{{  H^{4}_D(\hat Y_4,\mathbb{Z}(2))}}_{\rm  full  \, gauge \, data}
  \arrow{r}{\hat{c}_2}  & \underbrace{H^{2,2}_{\mathbb{Z}}(\hat Y_4)}_{{\rm field \, strength} \, G_4} \arrow{r} & 0   
\end{tikzcd}
\eea
Here $\gamma_2$ is the cycle map which assigns to a Chow class its associated cohomology class. Its kernel  ${\rm CH}^2_{\text{hom}}(\hat Y_4)$ maps to the flat gauge backgrounds, i.e. the elements in the Jacobian
$J^2(\hat Y_4)$, via the Abel-Jacobi map. 
Note that given a Chow class $A \in {\rm CH}^2(\hat Y_4)$ the composition 
\bea \label{composition}
\hat c_2 \circ  \hat \gamma_2 (A) = [A] \in H^{2,2}_{\mathbb Z}(\hat Y_4)
\eea
is the gauge flux associated with $A$.

The advantage of this parametrization of the gauge background is that we can proceed by explicitly constructing complex 2-cycles and considering operations modulo rational equivalence. 
In fact, to each of the fluxes constructed in section \ref{Fluxes-Examples} we can associate its underlying Chow class in the sense of (\ref{composition}). 
For instance, the cycle $\Gamma$ defined in (\ref{Gamma1}) can be viewed as a representative of a certain Chow class, and we will denote this element of ${\rm CH}^2(\hat Y_4)$ by the same symbol.
Its image under the composition of the refined cycle map $\hat \gamma_2$ and $\hat c_2$ is the horizontal flux $[\Gamma]$. 
The same logic can be applied to all other types of fluxes of  \ref{Fluxes-Examples}, which are under good  computational control.

\subsection{Cohomology formulae counting zero-modes} \label{sec_cohos}

We are finally in a position to approach the zero-mode counting in global F-theory compactifications to 4d in more detail, following the formalism of \cite{Bies:2014sra,Bies:2017fam}.


\subsubsection{Localised charged matter} \label{sec_localchargedmatterb}

It has already been  described in section \ref{subsec_Countingloc-1} that the massless matter in representation ${\bf R}$ localised along an irreducible (self-)intersection curve $C_{\bf R}$ of the discriminant is counted by certain cohomology groups. In absence of any gauge background along the 7-branes, these groups are given in (\ref{H0H1a}), and they have been derived in the framework of  the topologically twisted local field theory describing the dynamics of the modes along a 7-brane \cite{Beasley:2008dc,Donagi:2008ca}.  
More generally, a gauge background along the 7-branes induces a corresponding gauge background also along the matter curves $C_{\bf R}$, and the charged zero-modes will couple to it in a manner dictated by the representation ${\bf R}$. 

As before, our notation is that to each weight $\beta^a({\bf R})$ we associate the matter surface $S^a_{\bf R}$ obtained by fibering a rational curve over $C_{\bf R}$.
Let us furthermore denote by $L^{(a)}$ the gauge background to which the charged matter associated with weight $\beta^a({\bf R})$ couples.
In the simplest situation this can be a line bundle on $C_{\bf R}$, but more generally we can consider coherent sheaves $L^{(a)}$.  
The question is now how to extract the object $L^{(a)}$ on $C_{\bf R}$ from the gauge background on a globally defined fibration. The massless states are the fluctuations of M2-branes wrapped the fiber of the matter surface, which couple to the 3-form background according to the standard coupling (setting $\ell_{11} \equiv 1$)
\bea
S_{\rm M2} \supset  2 \pi \int_{\rm M2} C_3 \,.
\eea
The gauge potential on $C_{\bf R}$ along which the states propagate is hence obtained by integrating the 3-form background over the fiber of $S^a({\bf R})$ along which the M2-brane is wrapped.

Integration along the fiber translates into the following operation on the complex 2-cycle class $A \in {\rm CH}^2(\hat Y_4)$ representing the gauge background:
First we consider the pullback of $A$ onto $S^a({\bf R})$. If we denote by $\iota_a: S^a({\bf R}) \rightarrow \hat Y_4$ the inclusion of the matter surface into the total space, then the pullback is given by the intersection
$A \cdot_{\iota_a} S^a_{\bf R}$, where we are now interpreting $S^a({\bf R})$ as the representative of the eponymous Chow class, i.e. as an elelemt of ${\rm CH}^2(\hat Y_4)$.
Indeed this intersection product is well-defined within the Chow ring.\footnote{Sometimes we will omit the subscript in $\cdot_{\iota_a}$ if the context is clear. In this sense our notation does not distinguish between the intersection product in (co)homology and in the Chow ring. When we refer to cohomological objects and their intersection, we will denote this by a square bracket of the form e.g. $[S^a_{\bf R}] \in H^4(\hat Y_4)$.} This means in particular that we are allowed to use manipulations modulo rational equivalence without changing the result within the Chow ring. This is particularly important when we are to perform non-transverse  intersections: These can be rewritten as a sum of transverse intersections by exploiting linear relations within the Chow ring.

We interpret $A \cdot_{\iota_a} S^a({\bf R})$ as an element within ${\rm CH}_0(S^a_{\bf R})$, the class of points on $S^a_{\bf R}$ modulo rational equivalence. 
The actual integration along the fiber then consists in projecting this point class onto the base, i.e. onto the curve $C_{\bf R}$.
This operation amounts to considering the pushforward with respect to the projection $\pi: \hat Y_4 \rightarrow B_3$, restricted to the fibration over $C_{\bf R}$. Let us denote this map as
\bea
\pi_a:  S^a_{\bf R} \rightarrow C_{\bf R} \,.
\eea
In all, we obtain the object
\bea \label{paclass}
p_a :=\pi_{a \ast} (A \cdot_{\iota_a} S^a_{\bf R}) \in {\rm CH_0}(C_{\bf R}) \,.
\eea
Note that  ${\rm CH_0}(C_{\bf R}) \simeq {\rm CH^1}(C_{\bf R})$ because the complex curve $C_{\bf R}$ is of complex dimension one. As pointed out before, this is the group of divisors on $C_{\bf R}$ modulo rational equivalence. It is well-known that to each element in the divisor group we associate a sheaf on $C_{\bf R}$. This sheaf is the gauge background we are after,
\bea \label{Opa}
L^{(a)} = {\cal O}_{C_{\bf R}}(p_a) \,,
\eea
and  we find the following cohomology groups counting massless matter:
\begin{equation}
\begin{split} \label{HiLa}  
{\rm chiral \, \,  multiplets}&: \qquad   H^0(C_{\bf R}, L^{(a)} \otimes \sqrt{K_{C_{\bf R}}}) \cr
{\rm anti-chiral \,\,  multiplets}&: \qquad   H^1(C_{\bf R}, L^{(a)} \otimes \sqrt{K_{C_{\bf R}}}) \,.
\end{split}
\end{equation}

Recall furthermore from above that  in general  ${\rm CH^1}(C_{\bf R})$ is the group of Weil divisors and that  if $C_{\bf R}$ is smooth (or more generally has only singularities which leave it factorial as a variety), then this equals the group ${\rm Pic}(C_{\bf R})$ of Cartier divisors. In this case,  $L^{(a)}$ is a line bundle on  $C_{\bf R}$ as opposed to merely a coherent sheaf on $C_{\bf R}$. 
 
In any given application, the representative of the Chow class  $p_a$ (which we again denote by the same symbol) is constructed very explicitly in terms of the vanishing locus of certain functions along $C_{\bf R}$.
Suppose first that $p_a$ is effective, i.e. it consists of points with positive multiplicity, and that these points are the vanishing locus of the functions $f_1, \ldots, f_n$ on $C_{\bf R}$. These need not be complete intersections, and in general generate an ideal within the ring of functions in the coordinates on $C_{\bf R}$.
There is a standard procedure in algebraic geometry to associate to the ideal  $\langle f_1, \ldots, f_n \rangle$ a sheaf ${\cal I}$, the so-called ideal sheaf.\footnote{The sheaf is defined such that its restriction to any open neighborhood ${\cal U}$ has as its stalk the restriction of the ideal to ${\cal U}$, i.e. ${\cal I}({\cal U}) = \langle {f_1}|_{\cal U}, \ldots,  {f_n}|_{\cal U}  \rangle$. } This sheaf is precisely ${\cal I} = {\cal O}_{C_{\bf R}}( - p_a)$, and the sought-after sheaf (\ref{Opa})  is obtained by a dualisation procedure as ${\cal O}_{C_{\bf R}}(p_a) = {Hom}({\cal O}_{C_{\bf R}}, {\cal I})$. 
If $p_a$ is anti-effective, then there exists an ideal $\langle g_1, \ldots, g_n \rangle$ describing the effective object $-p_a$ and ${\cal O}_{C_{\bf R}}(p_a)$ is the  ideal sheaf associated with  $\langle g_1, \ldots, g_n \rangle$.
More generally, if $p_a$ contains both effective and anti-effective cycles, i.e. $p_a = r_a - s_a$ with $r_a$ and $s_a$ effective, one obtains   (\ref{Opa}) as the tensor product of the ideal sheaves associated with $r_a$ and $s_a$ according to the above logic.
More details on this standard procedure can be found e.g. in section 6  of \cite{Bies:2017fam}.

Having determined the cohomology groups (\ref{HiLa}) counting chiral and anti-chiral multiplets, 
we can immediately compute the associated chiral index. According to the Riemann-Roch-Hirzebruch index theorem,
\bea \label{derivchi}
\chi(\beta^a({\bf R})) = {\rm deg}(L^{(a)}) = {\rm deg}(\pi_{a \ast} (A \cdot_{\iota_a} S^a_{\bf R}) ) =  {\rm deg} (A \cdot_{\iota_a} S^a_{\bf R})   = [A] \cdot [S^a_{\bf R}] = \int_{S^a({\bf R})} [A] \,.
\eea
 This exactly reproduces, and in fact derives, the expression (\ref{chibeta-formula}).

It is high time to illustrate this general procedure.
As our first example let us consider a non-Cartan $U(1)$ gauge background. If we denote by $F \in {\rm CH}^1(B_3)$ and $\sigma(s_A) \in {\rm CH}^1(\hat Y_4)$ the divisor classes whose associated homology classes define the $U(1)$ gauge flux (\ref{U1Aflux}), then at the level of Chow groups $A = \pi^\ast F \cdot \sigma(s_A)$. Here $\cdot$ refers to the intersection product within the Chow ring of $\hat Y_4$. The operation $A \cdot_{\iota_a} S^a_{\bf R}$ splits into a base and a fiber part, as explained in more detail in \cite{Bies:2014sra}, and after projection onto the base one arrives at 
\bea \label{U1AChowclass}
p_a= \pi_{a \ast} (A \cdot_{\iota_a} S^a_{\bf R}) = q_A({\bf R}) (F \cdot_{\iota_{C_{\bf R}}} C_{\bf R}) \,.
\eea
Here $q_A({\bf R})$ is the $U(1)_A$ charge of representation ${\bf R}$ which is reproduced by the intersection in the fiber,  and $\iota_{C_{\bf R}}: C_{\bf R} \rightarrow B_3$ is the inclusion for the matter curve $C_{\bf R}$. 
The intersection $F \cdot_{\iota_{C_{\bf R}}} C_{\bf R}$ defines a class of points on $C_{\bf R}$, obtained by intersecting the curve $C_{\bf R}$ with the divisor $F$ on $B_3$ modulo rational equivalence.
The degree of this point class is the topological intersection number $\int_{C_{\bf R}} [F]$, but it contains more information beyond this cohomological intersection number.
The point class (\ref{U1AChowclass}) then defines a sheaf  ${\cal O}_{C_{\bf R}}(p_a)$ on $C_{\bf R}$. If $C_{\bf R}$ is smooth, this sheaf is a line bundle on $C_{\bf R}$, given by $L^{\otimes q_A({\bf R})}$ with $L = {\cal O}(F)|_{C_{\bf R}}$ the line bundle obtained by pulling back the line bundle on $B_3$ associated with the divisor class $F$. This is a special situation valid for abelian gauge backgrounds, though.

For matter surface fluxes, the intersection $A \cdot_{\iota_a} S^a_{\bf R}$ can again be performed separately in the fiber and the base. This is explained in detail in \cite{Bies:2017fam}, to which we refer the interested reader.
Since the complex 2-cycle underlying such gauge background is itself a matter surface, the structure of intersections is in fact governed by the Yukawa points. In general, the resulting sheaf on $C_{\bf R}$ is {\it not} the pullback of a line bundle to the matter curve.  

Finally, what remains to be understood is how to evaluate the cohomology groups (\ref{HiLa}) explicitly.  Our input is a point class $p_a$ on $C_{\bf R}$ which defines the sheaf ${\cal O}_{C_{\bf R}}(p_a)$. If this sheaf is the pullback of a line bundle from $B_3$, as is the case e.g. for the $U(1)_A$ gauge background above, then the cohomology groups can be computed by restricting cohomology groups of the line bundle on $B_3$ to $C_{\bf R}$ via the Koszul sequence. If $B_3$ is itself a toric space, or embedded into a toric ambient space as a hypersurface or complete intersection, the cohomology groups on $B_3$ in turn are obtained via the Koszul sequence from the cohomology groups on the ambient space. The CohomCalg algorithm developed in \cite{Blumenhagen:2010pv} and implemented in \cite{Blumenhagen:2010ed,Blumenhagen:2011xn} performs precisely this task. 
Unfortuntately, in most situations the sheaf on the matter curve is not a pullback line bundle, but  the object $L^{(a)}$ on $ C_{\bf R}$ (or rather its pushorward onto $B_3$) really defines a coherent sheaf on $B_3$. The computation of sheaf cohomology groups on toric spaces has been implemented in computational algebraic geometry in \cite{2010arXiv1003.1943B,2012arXiv1202.3337B,2014arXiv1409.6100B}. In \cite{Bies:2017fam,Bies:2018uzw}  it is explained in detail and illustrated how this machinery can be applied to compute the sheaf cohomology groups (\ref{HiLa}). In particular, it is applicable also in situations where $C_{\bf R}$ as a curve is singular.

Instead of repeating this discussion in detail here, let us stress an interesting property of the resulting matter spectrum: The dimensions of the sheaf cohomology groups (\ref{HiLa}) explicitly depend on the choice of complex structure moduli defining the matter curves $C_{\bf R}$ \cite{Bies:2017fam,Bies:2018uzw}. These enter via the explicit functions whose poles or zeroes cut out the point class (\ref{paclass}) on $C_{\bf R}$. This is a notable difference from the chiral index, which is a topological invariant. Changing the complex structure moduli in general leads to jumps in the number of massless vectorlike pairs. The smallest number of such massless vectorlike pairs is found at the most generic point in moduli space, and along loci in higher codimension in moduli space extra massless zero modes can appear.

 \subsubsection{ Charged bulk matter}

Let us now address the bulk modes propagating along the surface $\Sigma_I$ wrapped by a stack of 7-branes. 
In the expression (\ref{4dbulknoflux}) for the cohomology groups counting the massless bulk modes in absence of gauge flux, the trivial bundle ${\cal O}_I$ must be replaced accordingly.

To this end consider a state in the adjoint representation associated with the weight
\bea \label{weightbeta-bulk}
\beta =  - \sum_{i_I=1}^{{\rm rk} (\mathfrak{g}_I)  }  b_{i_I}  \,   \alpha_{i_I} \,.
\eea
Here $\alpha_{i_I}$ denotes the positive simple roots of the algebra $\mathfrak{g}_{i_I}$. We focus for simplicity on  a simply-laced algebra.
From the discussion in section \ref{sec_singcodim1} we know that the zero-modes carrying this weight vector are the massless fluctuations  of M2-branes wrapping the linear combination
$\sum_{i_I} b_{i_I} \mathbb P^1_{i_I}$ in the fiber over $\Sigma_I$. 
Each $\mathbb P^1_{i_I}$ is the fiber of a resolution divisor $E_{i_I}$.


For each of the resolution divisors $E_{i_I}$ fibered over $\Sigma_I$ with inclusion map $i_I: E_{i_I} \rightarrow \hat Y_4$, we can form the intersection product $A \cdot_{i_I} E_{i_I} \in {\rm CH}^2(E_{i_I})$. On the complex 3-dimensional divisor $E_{i_I}$,  ${\rm CH}^2(E_{i_I}) \simeq {\rm CH}_1(E_{i_I}) $, and projecting the above intersection to the base of $E_{i_I}$ gives an element
\bea \label{LiIdefinition}
\pi_{i_I \ast} (A \cdot_{i_I} E_{i_I}) \in {\rm CH}_1(\Sigma_I) \,.
\eea
Since ${\rm CH}_1(\Sigma_I) \simeq {\rm CH}^1(\Sigma_I)$ this again produces an object in the divisor class of $\Sigma_I$, which defines a sheaf (or line bundle in the smooth context)  $L_{i_I}$ on $\Sigma_I$. 
The sheaf to which the state associated carrying the weight (\ref{weightbeta-bulk}) couples is then
\bea
L^{(\beta)} = \otimes_{i_I} L^{b_{i_I}}_{i_I} \,.
\eea
Generalising (\ref{4dbulknoflux}), the massless matter carrying weight $\beta$ organizes into ${\cal N}=1$ chiral and anti-chiral  multiplets counted by the following cohomology groups:
\begin{equation}
\begin{split} \label{4dbulkWITHflux}
        {\rm chiral}: \quad &  H^1(\Sigma_I, L^{(\beta)})     \oplus   H^0(\Sigma_I, L^{(\beta)} \otimes K_{\Sigma_I}) \cr
 {\rm anti-chiral}:\quad &  H^2(\Sigma_I, L^{(\beta)})     \oplus   H^1(\Sigma_I, L^{(\beta)} \otimes K_{\Sigma_I})  \,.
\end{split}
\end{equation}

The chiral index associated to this matter can be computed by noting that if supersymmetry is unbroken, $H^0(\Sigma_I, L^{(\beta)}) = 0$ and likewise  
$ H^2(\Sigma_I, L^{(\beta)} \otimes K_{\Sigma_I}) = 0$.\footnote{The precise assumption is that the line bundle $L^{(\beta)}$ allows for a solution to the D-term equation inside the K\"ahler cone without the need to turn on charged matter field VEVs. This is equivalent to  $H^0(\Sigma_I, L^{(\beta)}) = 0$ and $ H^2(\Sigma_I, L^{(\beta)} \otimes K_{\Sigma_I}) = 0$ \cite{Blumenhagen:2008zz}. A non-zero matter VEV would break the gauge algebra and hence modify also the zero-mode counting.} The chiral index then follows as
\bea \label{index-bulk}
\chi(\beta) &=& \sum_{i=0}^2 h^i(\Sigma_I, L^{(\beta)}) - \sum_{i=0}^2 h^i(\Sigma_I, L^{(\beta)}\otimes K_{\Sigma_I})  \\
&=& -  \int_{\Sigma_I} \, c_1(\Sigma_I) \,  c_1(L^{(\beta)}) \,,
\eea
where we have used to the Atiyah-Singer index theorem (see e.g. \cite{Beasley:2008dc,Donagi:2008ca,Blumenhagen:2008zz} for details).
The index can in fact be written directly in terms of the gauge flux $G_4 = [A]$ in a manner which makes contact with the expression (\ref{chibeta-formula}) for localised matter:
The role of the matter surface for bulk matter is now taken by 
\bea \label{Sbeta}
S(\beta) = \sum b_{i_I} E_{i_I} |_{\bar K_{\Sigma_I}} \,.
\eea
Indeed, (\ref{index-bulk}) is identical to
\bea \label{chi-bulk}
\chi(\beta)  =  \int_{S(\beta)} G_4 \,.
\eea

Whenever (\ref{LiIdefinition}) is non-trivial, the gauge background in fact breaks the gauge algebra $\mathfrak{g}_I$ in the F-theory limit. 
The condition for gauge invariance is therefore
\bea \label{gaugeinvariance-finer}
\pi_{i_I \ast} (A \cdot_{i_I} E_{i_I}) = 0  \qquad   \forall \quad i_I \,.
\eea
This condition certainly implies (\ref{gaugeinvariance-a}), but is stronger. For instance, fluxes in the remainder piece $H^{2,2}_{\rm rem}(\hat Y_4)$
always satisfy (\ref{gaugeinvariance-a}), but they may well break the non-abelian gauge algebra, the prime example being the hypercharge flux in F-theory GUTs.
The correct condition (\ref{gaugeinvariance-finer}) is sensitive to this effect.

\section{Applications} \label{sec_applications}

One of virtues of F-theory, and the guideline of these lectures so far, is the fruitful combination of physical and geometric reasoning.
We have seen how this establishes a clear physics interpretation of many advanced concepts in algebraic geometry.
We would like to conclude these lectures by giving an admittedly rather brief outlook on some of the applications of this dictionary between geometry and physics in F-theory. We will focus on three different aspects, which oftentimes go hand in hand and inspire each other: Applications to string model building and questions of the string landscape, applications to formal questions in Quantum Field Theory, and applications to mathematics.

\subsection{F-theory model building and landscape reasoning} \label{sec_Applications1}

Beginning with \cite{Donagi:2008ca,Beasley:2008dc, Beasley:2008kw,Donagi:2008kj}, F-theory has been established as  a fruitful framework for particle physics oriented model building, in particular in the context of Grand Unified Model Building (GUTs).
Many of the early developments in this very active field have already been surveyed in the reviews \cite{Weigand:2010wm,Heckman:2010bq,Maharana:2012tu,Leontaris:2012mh}, to which we refer for a more detailed account of the key ideas and a more exhaustive list of references. Here we would like to stress some of the more recent results in the context of the technical framework laid out in these lectures.

What makes F-theory so attractive with respect to model building is the combination, mentioned already in the introduction, of localisation of gauge degrees of freedom, matter, and Yukawa interactions with the appearance of symmetry groups of exceptional type, which are otherwise realized  only in the heterotic string (at a perturbative level). 
The localisation of gauge degrees of freedom implies that many physics questions decouple - at least to leading order - from global considerations. This has been the main viewpoint taken in GUT model building. A key idea is that the UV completeness of a 4d GUT theory reflects in the fact that gravity can be decoupled by placing the GUT 7-brane on a divisor which can shrink within the base of the fibration. 
Such reasoning has inspired the detailed development of local or semi-local techniques for the analysis of the gauge theory on the 7-brane in the context of a spectral cover or Higgs bundles \cite{Donagi:2008ca,Beasley:2008kw,Hayashi:2008ba,Hayashi:2009ge,Donagi:2009ra,Hayashi:2009bt,Hayashi:2010zp,Marsano:2009ym,Blumenhagen:2009yv,Marsano:2009gv,Marsano:2009wr,Dolan:2011iu,Marsano:2011nn,Dolan:2011aq}.
A local approach is certainly justified for those aspects which only involve the non-abelian degrees of the freedom and their interactions. Among them are the Yukawa couplings between matter fields in different non-abelian representations, which, as described in section \ref{Codim3Yuk}, are localised at isolated points along the 7-branes.
For a rather incomplete list of further phenomenological studies using a local technique see e.g. \cite{Leontaris:2010zd,King:2010mq,Callaghan:2012rv,Callaghan:2013kaa} and references therein.

This is not to say, though, that global effects are irrelevant. On the contrary, we have seen that essentially all physics associated with abelian or discrete gauge symmetries is global in nature.  Clearly  abelian gauge symmetries lie at the heart of model building applications  when it comes to realising the gauge group of the Standard Model. Indeed, in $SU(5)$ GUTs the quest for the hypercharge abelian $U(1)_Y$ factor is a global question: As one of the hallmarks of F-theory GUTs, the very attractive scenario of breaking the gauge group $SU(5)$ via a nontrivial $U(1)_Y$ gauge background \cite{Beasley:2008kw,Donagi:2008kj} is sensitive to information about the embedding of the GUT 7-brane cycle into the base $B_3$ \cite{Buican:2006sn}. This is realized for instance in the global models \cite{Blumenhagen:2008zz,Marsano:2009ym,Blumenhagen:2009yv}. 
Another key idea in the context of F-theory GUTs is the elegant solution to the doublet-triplet splitting problem and suppression of proton decay operators via suitable Peccei-Quinn type symmetries \cite{Beasley:2008kw,Marsano:2009wr,Dolan:2011aq}. Implementing this and other effects into a globally defined compactification has been one of the motivations for and driving forces behind the systematic exploration of abelian gauge symmetries in F-theory. 
 As of this writing it is indeed possible to obtain a charge assignment of the form envisaged in \cite{Beasley:2008kw,Marsano:2009wr,Dolan:2011aq} in global $SU(5)$ models, including e.g. a split of the ${\bf 10}$ representation \cite{Mayrhofer:2012zy,Braun:2013yti,Braun:2014qka,Kuntzler:2014ila}. However, the mechanism of hypercharge flux breaking requires that both Higgs curves (on which ${\bf 5}_{H_u}$ and  ${\bf 5}_{H_d}$ localize)  are not by themselves realized as the pullback of a divisor to the GUT surface. In the language of section \ref{Fluxes-Examples},  the classes of the associated matter surfaces must have contributions in $H^{2,2}_{\rm rem}(\hat Y_4)$. 
 While models of this type are guaranteed to exist because they can be constructed in principle at the perturbative level \cite{Mayrhofer:2013ara},
 this property has not yet been combined with the requirement that the matter curves in question carry different abelian charges  (see, however, \cite{Braun:2014pva} for the split of a matter surface into two matter surfaces with components in $H^{2,2}_{\rm rem}(\hat Y_4)$, where the two associated matter fields carry the same $U(1)$ charge).

More generally, it would be highly desirable for model building applications to {\it classify} the possible abelian or discrete charges which non-abelian matter can attain in a consistent F-theory compactification. This would allow us to single out which phenomenologically attractive bottom-up scenarios for selection rules can be realized after taking quantum gravity (or stringy) constraints into account. 
Such a classification of possible charges in $SU(5)$ GUT models has been given in \cite{Lawrie:2015hia}, subject to the assumption that the rational section underlying the abelian gauge group factor and the GUT divisor are both smooth. It would be extremely important for model building to push this classification further without this technical assumption.

Overall, two different approaches to flavour have been taken in the F-theory literature, again each with certain global implications.
As already described at the end of section \ref{Codim3Yuk}, at a single Yukawa point the coupling matrix between the different families of matter localised on the intersecting matter curves has rank one \cite{Heckman:2008qa,Hayashi:2009ge,Cecotti:2009zf}.
Higher order effects \cite{Marchesano:2009rz} can produce subleading corrections with excellent phenomenological properties, see in particular the detailed computations \cite{Aparicio:2011jx,Font:2012wq,Font:2013ida,Marchesano:2015dfa,Carta:2015eoh} within a local framework. Implementing these into global models requires knowledge of global data such as the spectrum of instantons available in the model. 
Alternatively, Froggat-Nielson type models invoke global symmetries to act as flavour symmetries distinguishing between different families, which necessarily localize on different matter curves \cite{Dudas:2009hu,King:2010mq,Krippendorf:2010hj,Baume:2015wia,Krippendorf:2015kta}. This line of reasoning is again directly tied to the formal developments in the context of abelian or discrete symmetries discussed in sections \ref{sec_MWGROUP} and \ref{sec_genusone}. In order for an abelian symmetry to act as a global symmetry a flux-induced St\"uckelberg mechanism must render the gauge boson massive.

The idea of Grand Unification hinges, to considerable extent, on low-scale supersymmetry. It is therefore important to assess the status of model building if supersymmetry were not to be found at low energies.
First, gauge threshold corrections  might come to rescue even if supersymmetry is broken at an intermediate scale.
The hypercharge flux breaking mechanism comes with its own type of such threshold corrections \cite{Donagi:2008kj,Blumenhagen:2008aw}.
Possible scenarios for F-theory GUTs with intermediate scale supersymmetry have been discussed 
 from various viewpoints in \cite{Ibanez:2012zg,Hebecker:2014uaa}.

A different approach to engineering the Standard Model is to bypass any intermediate level grand unified gauge group \cite{Lin:2014qga}. This is certainly the philosophy underlying perturbative Type II model building (see e.g. \cite{Blumenhagen:2006ci,Ibanez:2012zz} and references therein); the motivation to take this route in the context of F-theory 
is its generality; F-theory models with gauge group $SU(3) \times SU(2) \times U(1)_Y$ might therefore include possibilities that cannot be obtained perturbatively. Clearly, compared to e.g. toroidal orientifolds, this comes at the expensive of a lack of an explicit worldsheet theory which would allow for the evaluation of stringy effects.
Non-perturbative genericity, on the other hand, pays off in particular in the context of strongly coupled models, where the $SU(2)$ and $SU(3)$ factors are engineered as Kodaira type $III$ and $IV$ enhancements \cite{Grassi:2014sda,Grassi:2014zxa}. 
Global Standard-like models with three chiral generations have been constructed in \cite{Cvetic:2015txa,Lin:2016vus}.


\subsubsection*{Landscape versus swampland}

A question of fundamental importance for theoretical high energy physics is how to constrain the vast set of possible low energy theories which appear to comply with all known consistency conditions.  
Given its genericity, F-theory is an ideal framework  for a systematic study of the low energy effective theories 
at least in a large subclass of string compactifications, including non-perturbative effects in the string coupling. 
In eight dimensions, F-theory  exhausts the list of consistent gauge theories \cite{Garcia-Etxebarria:2017crf} (modulo the potential caveat in footnote \ref{FootnoteG2}). The analogue of this question in six dimensions has received considerable attention \cite{Taylor:2011wt}: In chiral $N=(1,0)$ supergravities constraints from gauge and gravitational anomaly cancellation are particularly strong and already allow one to considerably constrain the possible supergravity theories \cite{Kumar:2009ac,Kumar:2010am,Seiberg:2011dr,Park:2011wv,Monnier:2017oqd}. These are then to be compared with known string theory realizations \cite{Kumar:2009ac,Grimm:2012yq}. As a recent surprise an infinite set of abelian charges seems to be compatible with 6d anomaly cancellation in supergravity models \cite{Taylor:2018khc}, and it remains, as of today, an interesting question if such supergravities violate as yet unknown low-energy consistency conditions of if they can, on the contrary, be realized as consistent string or F-theory compactifications. Before conclusive statements can be made it is important to develop an (even) better understanding of the classifications of charges and representations which can occur in F-theory. The developments outlined in sections \ref{sec_generalisations} and \ref{sec_combab-nonab} are particularly relevant in this context.

\subsection{Non-perturbative Quantum Field Theories from F-theory} \label{NPQFT}

F-theory epitomizes the idea of geometrisation of Quantum Field Theory: Fundamental concepts of gauge theories are translated into geometric properties of an elliptic fibration, and we can thus use geometry to {\it define} a quantum field theory via F-theory. This becomes even more attractive in strongly coupled situations and more generally in the context of field theories which do not admit a Lagrangian description. In many cases, a geometric definition via string or F-theory is the best indication for the very existence of such theories in the first place.

We have already stressed in this context the recent classification of 6d $N=(1,0)$ superconformal field theories in F-theory \cite{Heckman:2013pva,Heckman:2015bfa}, and since this vast topic has been reviewed in detail in \cite{Heckman:2018jxk}, we can afford being rather brief. 
The key insight is to identify the tensor branch of a 6d $N=(1,0)$  supersymmetric gauge theory with the K\"ahler moduli space controlling the volumes of holomorphic curves in the two-dimensional base $B_2$ of an elliptic fibration (while leaving the base at finite volume).
Given a curve $\Sigma_I$, its K\"ahler volume ${\rm Vol}(\Sigma_I) = \int_{\Sigma_I} J$ is identified with a real scalar field. It sits in the same $N=(1,0)$ multiplet as the anti-self-dual 2-form originating from reduction of the Type IIB 4-form $C_4$ along the same curve $\Sigma_I$. The volume of $\Sigma_I$ controls two important physical quantities: First, the gauge coupling of a 7-brane wrapping $\Sigma_I$ is given by $1/g^2_I = {\rm Vol}(\Sigma_I)$. Second, wrapping a D3-brane along $\Sigma_I$ leads to a string in 6d which couples to the anti-self-dual tensor in the same multiplet as the volume of $\Sigma_I$. The tension of this string is likewise set by ${\rm Vol}(\Sigma_I)$.
At the origin of the tensor branch, i.e. in the limit where  ${\rm Vol}(\Sigma_I) \rightarrow 0$, the 7-brane theory along $\Sigma_I$ becomes strongly coupled and the string from the D3-brane along $\Sigma_I$ acquires zero tension. Due to the amount of supersymmetry in 6d, there are no quantum corrections to both statements.
The zero-tension limit results in a strongly coupled theory with infinitely many massless degrees of freedom. Such a theory is believed to represent a non-trivial superconformal field theory (SCFT). 
The classification of the SCFTs which can be obtained in this way via F-theory amounts to the classification of the possible configurations of shrinkable curves on an F-theory base $B_2$, along with all possible enhancements of the gauge algebra beyond the minimal type.
 In general, holomorphic curves on a K\"ahler surface which can shrink to zero volume (while keeping the volume of the embedding surface fixed) must have negative self-intersection if they are irreducible; configurations of several such curves must have a negative semi-definite intersection matrix $A_{IJ} = \Sigma_I \cdot \Sigma_J$ to be simultaneously shrinkable. Extra constraints arise on a base $B_2$ suitable for F-theory from the requirement that the functions $f$ and $g$ defining a Weierstrass model over $B_2$ do not vanish beyond order   $(4,6)$ in codimension one. This implies that a shrinkable curve in the above sense must in addition be rational and its self-intersection is constrained as \cite{Morrison:2012np}
 \be
 \Sigma_I \cdot \Sigma_I = - n \, \qquad n = 1, \ldots, 12 \,.
 \ee
For $n \geq 3$, $\Sigma_I$ must necessarily be a component of the discriminant divisor, i.e. it must be wrapped by a 7-brane whose minimal gauge algebra cannot be higgsed further. For $n=1$ and $n=2$, the gauge algebra along $\Sigma_I$ can be trivial, but nonetheless strings wrapping the curve become tensionless in the limit of vanishing volume and furnish 6d SCFTs (of E-string type for $n=1$ and with enhanced $(2,0)$ symmetry for $n=2$).
The key result is that shrinking the curves to zero volume leads to a canonical singularity (as defined in section \ref{sec_terminalco2}) on $B_2$ of the local form $\mathbb C^2/G$ with $G \subset U(2)$ \cite{Heckman:2013pva,Heckman:2015bfa}.

A conclusion to be drawn from the above is that not only 7-branes, but also D3-branes are key players in the engineering of interesting non-perturbative field theories in F-theory.
A D3-brane which is pointlike on the base $B_n$ probes the singularities of the F-theory elliptic fibration and can give rise, under suitable conditions, to four-dimensional strongly coupled field theories.
Wrapping the D3-brane on a curve or surface on $B_n$, on the other hand, engineers a supersymmetric gauge theory with varying gauge coupling.
The key to both types of constructions is to identify the axio-dilaton $\tau$ of Type IIB string theory in $\mathbb R^{1,9}$ with the complexified gauge coupling of the 4d ${\cal N}=4$ supersymmetric gauge theory in the worldvolume of a D3-brane,
\be \label{couplingD3}
\left(\frac{\theta}{2\pi} + \frac{4 \pi i}{g^2} \right)_{\rm D3} = \tau \,.
\ee
Suppose first that the D3-brane is pointlike on the F-theory base $B_n$.
In the vicinity of a 7-brane, $(p,q)$ strings between the 3-brane and the 7-brane give rise to light matter charged under the D3-brane gauge group.
The gauge group on the 7-brane hence appears as the flavor symmetry group of the 3-brane theory. 
These identifications open up a number of beautiful connections to supersymmetric field theories.
For instance, according to a celebrated result
D3-branes probing an $I_0^\ast$ singularity in 
 F-theory compactified on K3 engineer 4d ${\cal N}=2$ Seiberg-Witten theory with gauge group $SU(2)$ and $N_f=4$ fundamental hypermultiplets \cite{Sen:1996vd,Banks:1996nj}.
The gauge group along the 3-brane can be derived by analyzing the 3-3 strings including possibly monodromies along paths encircling the singularities. If the D3-brane is on top of the $I_0^\ast$ singularity, 3-3 strings encircling the latter become massless and enhance the gauge group $U(1)$ of a single D3 to $SU(2)$. The position of the D3-brane in the one complex direction $w$ normal to the singularity on the base of the K3 therefore translates into the Coulomb branch parameter of the D3-brane gauge theory.
The fact that the axio-dilaton in the vicinity of an $I_0^\ast$ singularity on K3 is constant (see the discussion around (\ref{So8Ek})) reflects the conformality of the ${\cal N}=2$ $SU(2)$ theory on the probe D3-brane with $N_f=4$.
Deforming the  $I_0^\ast$ singularity breaks the $SO(8)$ flavour symmetry; the resulting holomorphic variation of $\tau(w)$ in the directions normal to the D3-brane quantitatively matches the behaviour of the complexified gauge coupling on the Coulomb branch of Seiberg-Witten theory, and the Seiberg-Witten geometry {\it is} the elliptic fibration. 
The gauge instantons correcting the classical gauge coupling on the D3-brane are identified by $D(-1)$ instantons in Type IIB theory, whose effect is automatically included in the profile of $\tau(w)$ as determined by the elliptic fibration in F-theory on K3 \cite{Billo:2010mg}.

 There are various generalizations of this construction, including the possibility of engineering 4d ${\cal N}=1$ SCFTs along D3-branes probing codimension-three enhancement loci in F-theory on elliptic fourfolds \cite{Heckman:2010qv,Heckman:2011hu}. An important part of the technical analysis is to determine the spectrum of 3-7 strings from string junctions. The analysis differs from that of 7-7 strings in that the D3-brane is an $SL(2,\mathbb Z)$ singlet because it couples to the invariant 4-form $C_4$. Recent advances, and a list of earlier references on this topic, can be found in \cite{Grassi:2016bhs,Grassi:2018wfy}.

A novel and unexpected result is that D3-branes probing, in a similar way, $\mathbb Q$-factorial terminal singularities in codimension three on the base of an F-theory Calabi-Yau 4-fold lead to strongly coupled field theories with ${\cal N}=3$
supersymmetry \cite{Garcia-Etxebarria:2015wns}. See section \ref{sec_N3} for more information.

Next, consider a 
 D3-brane along a curve $C$ or even a surface $S$ on the base $B_n$ of an elliptic fibration.
The restriction of the fibration to $C$ or $S$ is itself an elliptic fibration, which is non-trivial if and only if $C$ or $S$ intersect the discriminant locus. 
The identification (\ref{couplingD3}) implies that the gauge coupling varies along the worldvolume of the D3-brane. Examples of such configurations are the $(1,0)$ SCFT strings from D3-branes wrapping shrinkable curves, as described at the beginning of this section, but more generally $C$ need not be of negative self-intersection.
Such non-perturbative theories can be described by combining the usual topological twist (which is comparable with the one along the 7-branes in F-theory, see section \ref{subsec_Zeromodeson7brane}), with an additional twist which was called topological duality twist in \cite{Martucci:2014ema}. The duality twist is described in \cite{Martucci:2014ema} for D3-branes on a surface on a base $B_3$ and in \cite{Lawrie:2016axq} for D3-branes wrapping a curve $C$ on $B_n$ for $n=1,2,3,4$ (see also \cite{Haghighat:2015ega}). The duality twist is best understood for abelian gauge theories, but extensions to non-abelian settings are possible by duality with M-theory, mapping the D3-brane stack to a stack of M5-branes \cite{Assel:2016wcr}.
In the case of a surface, the resulting topologically duality twisted field theory is the effective action along a D3-brane instanton. For a curve $C$, we get again effective string theories along $\mathbb R^{1,1}$. The amount of supersymmetry depends on the dimension of the embedding base. Holographic duals with varying axio-dilaton have been constructed in \cite{Couzens:2017way,Couzens:2017nnr}.

D3-branes wrapping curves are also an important ingredient in the recently studied class of compactifications of F-theory to two dimensions with $N=(0,2)$ supersymmetry \cite{Schafer-Nameki:2016cfr,Apruzzi:2016iac,Apruzzi:2016nfr,Lawrie:2016axq,Lawrie:2016rqe,Weigand:2017gwb}.

\subsection{From physics back to mathematics}

The gauge theory - geometry correspondence in F-theory is of obvious practical use for physics. It enables us to employ mathematical insights into the structure of elliptic fibrations to deduce properties of the associated effective action.
Reading the dictionary backwards opens up the possibility of obtaining  new and perhaps unexpected insights into the geometry of elliptic fibrations by relying on intuition on the physics side of the correspondence.
This is, of course, a common theme in string theory more generally. Some of the most spectacular examples are the prediction of mirror symmetry from the behaviour of conformal field theories describing string propagation on Calabi-Yau spaces, or highly non-trivial results in enumerative geometry by counting BPS invariants in topological string theory.
F-theory adds many more chapters to this success story.

The very idea of assigning a gauge algebra to the codimension-one strata of the discriminant and a weight lattice of representations to the fibers in codimension two is highly non-trivial from the perspective of geometry alone.
F-theory provides a welcome source of intuition and, in fact, deeper explanation for this beautiful result.
Another source of physics intuition comes from the fact, stressed in section \ref{sec_Coulomb}, that resolving the fibral singularities corresponds to moving along the Coulomb branch of the gauge theory in M-theory. This identifies different, birationally equivalent resolutions of the same singular Weierstrass model as the various Coulomb phases of one and the same gauge theory. In particular, it is clear that in all these birational geometries we should assign the same representation to the degenerate fibers in codimension two, as can be checked in all concrete examples. This fact is a priori non-trivial from a purely mathematical point of view, and indeed not even proven in full generality.

Similar physical reasoning can be invoked in many more cases to either explain ex post or to  predict  the behaviour of the geometry. 
An example of the second type appears in the context of quantum anomalies: 
The absence of net local anomalies in 10-dimensional string theory implies that all consistent compactifications of F-theory must automatically lead to an anomaly free theory. 
The field theoretic anomalies have in general two contributions: A 1-loop induced anomaly from chiral states running in the loop, and a Green-Schwarz counterterm \cite{Green:1984sg,Green:1984bx,Sagnotti:1992qw}, first computed in F-theory in \cite{Sadov:1996zm}, which contributes a classical variance of the action. Both types of sources for the anomalies can be computed in purely geometric terms on the elliptic fibration, and must exactly cancel each other.
This leads to a number of topological identities for any elliptically fibered Calabi-Yau.
Anomaly cancellation was first used in \cite{Grassi:2000we,Grassi:2011hq} to establish a non-trivial relation between the Euler characteristic of a smooth elliptic Calabi-Yau 3-fold and the data associated with the codimension-one and two degenerate fibers, and further extended in compactifications to six dimensions in \cite{Park:2011ji,Grassi:2018rva}.
A number of identities can be deduced from the cancellation of anomalies which must hold as identities in the cohomology ring $H^{2,2}(\hat Y_{n+1})$ (or possibly even at the level of the Chow ring) and which exhibit a universal structure across dimensions \cite{Cvetic:2012xn,Bies:2017abs,Weigand:2017gwb}. A subset these have been collected in the previous section, eqns (\ref{naijkA-1}) - (\ref{naijkA-3}).   
A first principle derivation of the cancellation of anomalies in F-theory via M-theory \cite{Corvilain:2017luj} turns, via this connection, into a physics proof of a number of highly non-trivial such topological identities whose general proof based solely on geometric reasoning is yet to be achieved.

\subsection*{Acknowledgements}

I would like to thank the organizers of TASI 2017 "Physics at the Fundamental Frontier", Mirjam Cveti{\v c} and Igor Klebanov as well as Senarath de Alwis, Thomas DeGrand and Oliver DeWolfe, for inviting me to lecture in such an inspiring atmosphere, and the participants of the TASI school for their 
excellent questions, remarks, comments and their impressively lively interaction even in the final week  of an intense programme.
I am deeply indebted to my collaborators on F-theory and related matters of the past years,
P. Arras, M. Bies, R. Blumenhagen, J. Borchmann, M. Cveti{\v c}, A. Grassi, T. Grimm, A. Hebecker, B. Jurke, M. Kerstan, S. Krause, C. Lawrie, S.-J. Lee, W. Lerche, L. Lin, D. L{\"u}st,  L. Martucci, C. Mayrhofer, D. Morrison, E. Palti, C. Pehle, D. Regalado, C. Reichelt, O. Till, S. Sch\"afer-Nameki and F. Xu. This work was supported in part by DFG TR33 'The Dark Universe' and by DFG GK 'Particle Physics Beyond the Standard Model'.



\appendix

\section{Divisors, cycles, and equivalence relations} \label{app_divisors}

For the reader's convenience, this appendix reviews a few standard definitions and facts concerning various equivalence relations between complex cycles on a complex projective variety $X$ of complex dimension $n$. 

\begin{definition}
A Weil divisor is a formal linear combination of irreducible hypersurfaces, i.e. of complex codimension-one cycles, of $X$. The group of Weil divisors is called ${\rm Div}(X)$. 
\end{definition}

\begin{definition}
A principal divisor can be written as the zeroes and poles of a globally defined meromorphic function on $X$.
\end{definition}

\begin{definition}
 Weil divisors $Z_1$ and $Z_2$ are linearly equivalent, $Z_1 \sim Z_2$, if they differ by a principal divisor. The group of Weil divisors modulo linear equivalence is the divisor class group ${\rm Cl}(X)= {\rm Div}(X)/{\sim{}}$.
\end{definition}

Linear equivalence for divisors is the same as rational equivalence for complex codimension-one cycles. Two complex $p$-cycles are rationally equivalent if they belong to a family of cycles parametrized by a rational curve $\mathbb P^1$:
\begin{definition}
Two complex $p$-cycles $Z_1$ and $Z_2$ are rationally equivalent if there exists a cycle $V$ on $X \times \mathbb P^1$ such that $V \cap (X \times \{t_1\}) - V \cap (X \times \{t_2\}) = Z_1 - Z_2$ for $t_1, t_2 \in \mathbb P^1$.
The group of complex $p$ cycles modulo rational equivalence is the Chow group ${\rm CH}_p(X)$. In particular, ${\rm Cl}(X) = {\rm CH}_{n-1}(X)$.
The group of complex codimension $p$ cycles modulo rational equivalence is denoted by ${\rm CH}^p(X)$.
\end{definition}

A second notion of divisor is that of a Cartier divisor:
\begin{definition}
A Weil divisor which can be {\it locally} expressed as the zeroes or poles of a single meromorphic function on $X$ is called a Cartier divisor. 
The group of Cartier divisors modulo linear equivalence is the Picard group ${\rm Pic}(X)$. 
\end{definition}

The first Chern class map $c_1$ associates to each Cartier divisor class a cohomology class in $H^2(X,\mathbb Z)$,
\bea
c_1: {\rm Pic}(X) \rightarrow H^2(X,\mathbb Z) \,,
\eea
and its kernel is the  component ${\rm Pic}_0(X)$ connected to the zero element in ${\rm Pic}(X)$:
\bea
{\rm Pic}_0(X) = {\rm ker}(c_1) \,.
\eea

If $X$ is smooth every Weil divisor is also Cartier, and in this case ${\rm Cl}(X) = {\rm Pic}(X)$.
More generally, this remains true if $X$ is a complex projective variety with only factorial singularities.

A stronger equivalence relation than rational equivalence is given by algebraic equivalence. The intuition is again that two cycles are algebraically equivalent if they are members of the same family parametrized by an algebraic curve: 
\begin{definition}
Two complex $p$-cycles $Z_1$ and $Z_2$ are algebraically equivalent 
if there exists a cycle $V$ on $X \times C$ with $C$ an algebraic curve such that $V \cap (X \times \{t_1\}) - V \cap (X \times \{t_2\}) = Z_1 - Z_2$ for $t_1, t_2 \in C$.
\end{definition}

\begin{definition}
The group of Weil divisors modulo algebraic equivalence is called N\'eron-Severi group ${\rm NS}(X)$.
\end{definition}

Note that for complex codimension-one cycles, i.e. for divisors, on a complex projective variety $X$, the notion of homological equivalence and of algebraic equivalence are the same. This fails to be correct for higher codimension cycles.

If $X$ is smooth, the N\'eron-Severi group is related to the Picard group as follows:
Since according to the above homological and algebraic equivalence are the same for divisors, and furthermore ${\rm Pic}(X) = {\rm Cl}(X)$ if $X$ is smooth, the N\'eron-Severi group is
\bea
{\rm NS}(X) = {\rm Pic}(X) /{\rm Pic}_0(X) \,.
\eea
This equals
\bea
{\rm NS}(X) =  {\rm Pic}(X) /{\rm Pic}_0(X) =  {\rm Pic}(X) /{\rm ker}(c_1) = {\rm im} (c_1) \,.
\eea
Note furthermore that if $H^1(X,{\cal O}) = 0$, as is the case for all simply connected varieties, ${\rm Pic}_0(X) = 0$. In this case ${\rm NS}(X) = {\rm Pic}(X)$. In particular, if $X$ is smooth and $H^1(X,{\cal O}) = 0$, as is the case for a smooth Calabi-Yau variety, then ${\rm NS}(X) = {\rm Pic}(X) = {\rm Cl}(X) = {\rm CH}^1(X)$.

\section{Notation and Conventions}  \label{app-notation}

Given a complex variety, in our case mostly the resolved elliptic fibration $\hat Y_{n+1}$, we denote the vanishing locus of a set of meromorphic functions $f_1, \ldots, f_n$ as 
\bea
V(f_1, \ldots, f_n) := \{f_1=0\} \cap \{f_2=0\}  \cap \ldots \cap \{ f_n=0\}     \,.
\eea
Mathematically, we are dealing with the variety associated with the ideal $\langle  f_1, \ldots, f_n \rangle$ generated by the functions $f_1, \ldots, f_n$.

We typically use capital letters to denote this vanishing locus. The same letter is used to refer to the complex $p$-cycle class (modulo rational equivalence) on $\hat Y_{n+1}$. Depending on the context it will be clear if we are having the cycle class or a specific representative (i.e. the specific vanishing locus) in mind.
For example, given a (local) holomorphic coordinate $z$, we denote by $Z$ as in 
\bea
Z := V(z)= \{z=0\}
\eea
both the vanishing locus in $z$ and the associated divisor class in ${\rm Cl}(\hat Y_{n+1})$. 

The homology class of a complex $p$-cycle $C$ will usually be denoted by $[C] \in H_{2p}(\hat Y_{n+1})$. We use the same notation for its Poincar\'e dual cohomology class in $H^{2n-2p}(\hat Y_{n+1})$.
In particular, we therefore denote by $[Z]$
 both the class in $H_{2n}(\hat Y_{n+1})$ and its dual in $H^{2}(\hat Y_{n+1})$. For divisors on a smooth Calabi-Yau the distinction between the element in  ${\rm Cl}(\hat Y_{n+1})$ and its cohomology class in $H^{2}(\hat Y_{n+1})$ would strictly speaking not be necessary according to the remark at the end of Appendix \ref{app_divisors}. The reader may forgive us for sticking to this redundant notation also in the case of divisors.

The projection $\pi: \hat Y_{n+1} \rightarrow B_n$ induces a pushforward map on the space of (complex) $p$-cycles modulo rational equivalence, 
\bea
\pi_\ast: {\rm CH}_p(\hat Y_{n+1}) \rightarrow {\rm CH}_p(B_n) 
\eea
and a pullback map on the space of complex codimension $p$-cycles modulo rational equivalence
\bea
\pi^\ast: {\rm CH}^p(B_n)  \rightarrow  {\rm CH}^p(\hat Y_{n+1}) \,.
\eea
By abuse of notation we used the same symbol $\pi_\ast$ to denote the induced push-forward map in homology,
\bea
\pi_\ast: {H}_{2p}(\hat Y_{n+1}) \rightarrow {H}_{2p}(B_n) 
\eea
and $\pi^\ast$ for the pullback map in cohomology 
\bea
\pi^\ast: {H}^{2p}(B_n)  \rightarrow  {H}^{2p}(\hat Y_{n+1}) \,.
\eea

The intersection product between (co)homology classes is denoted by the same symbol $"\cdot"$. If necessary, we indicate with a subscript on which space the intersection product is to be evaluated.

The intersection product yielding a top-form is interpreted in the sense of integrating the latter over the full space to give a number.
This is sometimes also written in form of an integral, and by abuse of notation with $"\cdot"$ replaced by $"\wedge"$. E.g. we sometimes write
\bea
 [S_0] \cdot  [S_0] \cdot  \pi^\ast (w^{\rm b}_{2n-2})   \equiv \int_{\hat Y_{n+1}} [S_0] \wedge  [S_0] \wedge  \pi^\ast (w^{\rm b}_{2n-2})  =  - \int_{B_n}  c_1(B_n) \wedge  w^{\rm b}_{2n-2} \equiv - c_1(B_n) \cdot_{B_n} w^{\rm b}_{2n-2}  \nonumber \,.
\eea



\newpage
\bibliography{papers}

\providecommand{\href}[2]{#2}\begingroup\raggedright\begin{thebibliography}{100}

\bibitem{Vafa:1996xn}
C.~Vafa, {\it {Evidence for F theory}},  {\em Nucl. Phys.} {\bf B469} (1996)
  403--418, [\href{http://xxx.lanl.gov/abs/hep-th/9602022}{{\tt
  hep-th/9602022}}].

\bibitem{Morrison:1996na}
D.~R. Morrison and C.~Vafa, {\it {Compactifications of F theory on Calabi-Yau
  threefolds. 1}},  {\em Nucl. Phys.} {\bf B473} (1996) 74--92,
  [\href{http://xxx.lanl.gov/abs/hep-th/9602114}{{\tt hep-th/9602114}}].

\bibitem{Morrison:1996pp}
D.~R. Morrison and C.~Vafa, {\it {Compactifications of F theory on Calabi-Yau
  threefolds. 2.}},  {\em Nucl. Phys.} {\bf B476} (1996) 437--469,
  [\href{http://xxx.lanl.gov/abs/hep-th/9603161}{{\tt hep-th/9603161}}].

\bibitem{Schwarz:1995dk}
J.~H. Schwarz, {\it {An SL(2,Z) multiplet of type IIB superstrings}},  {\em
  Phys. Lett.} {\bf B360} (1995) 13--18,
  [\href{http://xxx.lanl.gov/abs/hep-th/9508143}{{\tt hep-th/9508143}}].
  [Erratum: Phys. Lett.B364,252(1995)].

\bibitem{Witten:1996bn}
E.~Witten, {\it {Nonperturbative superpotentials in string theory}},  {\em
  Nucl. Phys.} {\bf B474} (1996) 343--360,
  [\href{http://xxx.lanl.gov/abs/hep-th/9604030}{{\tt hep-th/9604030}}].

\bibitem{Donagi:2008ca}
R.~Donagi and M.~Wijnholt, {\it {Model Building with F-Theory}},  {\em Adv.
  Theor. Math. Phys.} {\bf 15} (2011), no.~5 1237--1317,
  [\href{http://xxx.lanl.gov/abs/0802.2969}{{\tt 0802.2969}}].

\bibitem{Beasley:2008dc}
C.~Beasley, J.~J. Heckman, and C.~Vafa, {\it {GUTs and Exceptional Branes in
  F-theory - I}},  {\em JHEP} {\bf 01} (2009) 058,
  [\href{http://xxx.lanl.gov/abs/0802.3391}{{\tt 0802.3391}}].

\bibitem{Beasley:2008kw}
C.~Beasley, J.~J. Heckman, and C.~Vafa, {\it {GUTs and Exceptional Branes in
  F-theory - II: Experimental Predictions}},  {\em JHEP} {\bf 01} (2009) 059,
  [\href{http://xxx.lanl.gov/abs/0806.0102}{{\tt 0806.0102}}].

\bibitem{Donagi:2008kj}
R.~Donagi and M.~Wijnholt, {\it {Breaking GUT Groups in F-Theory}},  {\em Adv.
  Theor. Math. Phys.} {\bf 15} (2011), no.~6 1523--1603,
  [\href{http://xxx.lanl.gov/abs/0808.2223}{{\tt 0808.2223}}].

\bibitem{Malmendier:2014uka}
A.~Malmendier and D.~R. Morrison, {\it {K3 surfaces, modular forms, and
  non-geometric heterotic compactifications}},  {\em Lett. Math. Phys.} {\bf
  105} (2015), no.~8 1085--1118, [\href{http://xxx.lanl.gov/abs/1406.4873}{{\tt
  1406.4873}}].

\bibitem{Font:2016odl}
A.~Font, I.~Garcia-Etxebarria, D.~Lust, S.~Massai, and C.~Mayrhofer, {\it
  {Heterotic T-fects, 6D SCFTs, and F-Theory}},  {\em JHEP} {\bf 08} (2016)
  175, [\href{http://xxx.lanl.gov/abs/1603.09361}{{\tt 1603.09361}}].

\bibitem{Garcia-Etxebarria:2016ibz}
I.~Garcia-Etxebarria, D.~Lust, S.~Massai, and C.~Mayrhofer, {\it {Ubiquity of
  non-geometry in heterotic compactifications}},  {\em JHEP} {\bf 03} (2017)
  046, [\href{http://xxx.lanl.gov/abs/1611.10291}{{\tt 1611.10291}}].

\bibitem{Font:2017cya}
A.~Font and C.~Mayrhofer, {\it {Non-geometric vacua of the
  $\mathbf{\text{Spin}(32)/\mathbb Z_2}$ heterotic string and little string
  theories}},  {\em JHEP} {\bf 11} (2017) 064,
  [\href{http://xxx.lanl.gov/abs/1708.05428}{{\tt 1708.05428}}].

\bibitem{Morrison:2012np}
D.~R. Morrison and W.~Taylor, {\it {Classifying bases for 6D F-theory models}},
   {\em Central Eur. J. Phys.} {\bf 10} (2012) 1072--1088,
  [\href{http://xxx.lanl.gov/abs/1201.1943}{{\tt 1201.1943}}].

\bibitem{Heckman:2013pva}
J.~J. Heckman, D.~R. Morrison, and C.~Vafa, {\it {On the Classification of 6D
  SCFTs and Generalized ADE Orbifolds}},  {\em JHEP} {\bf 05} (2014) 028,
  [\href{http://xxx.lanl.gov/abs/1312.5746}{{\tt 1312.5746}}]. [Erratum:
  JHEP06,017(2015)].

\bibitem{Heckman:2015bfa}
J.~J. Heckman, D.~R. Morrison, T.~Rudelius, and C.~Vafa, {\it {Atomic
  Classification of 6D SCFTs}},  {\em Fortsch. Phys.} {\bf 63} (2015) 468--530,
  [\href{http://xxx.lanl.gov/abs/1502.05405}{{\tt 1502.05405}}].

\bibitem{Heckman:2018jxk}
J.~J. Heckman and T.~Rudelius, {\it {Top Down Approach to 6D SCFTs}},
  \href{http://xxx.lanl.gov/abs/1805.06467}{{\tt 1805.06467}}.

\bibitem{Taylor:2011wt}
W.~Taylor, {\it {TASI Lectures on Supergravity and String Vacua in Various
  Dimensions}},  \href{http://xxx.lanl.gov/abs/1104.2051}{{\tt 1104.2051}}.

\bibitem{Brennan:2017rbf}
T.~D. Brennan, F.~Carta, and C.~Vafa, {\it {The String Landscape, the
  Swampland, and the Missing Corner}},
  \href{http://xxx.lanl.gov/abs/1711.00864}{{\tt 1711.00864}}.

\bibitem{Grassi:2000we}
A.~Grassi and D.~R. Morrison, {\it {Group representations and the Euler
  characteristic of elliptically fibered Calabi-Yau threefolds}},
  \href{http://xxx.lanl.gov/abs/math/0005196}{{\tt math/0005196}}.

\bibitem{Grassi:2011hq}
A.~Grassi and D.~R. Morrison, {\it {Anomalies and the Euler characteristic of
  elliptic Calabi-Yau threefolds}},  {\em Commun. Num. Theor. Phys.} {\bf 6}
  (2012) 51--127, [\href{http://xxx.lanl.gov/abs/1109.0042}{{\tt 1109.0042}}].

\bibitem{Bies:2017abs}
M.~Bies, C.~Mayrhofer, and T.~Weigand, {\it {Algebraic Cycles and Local
  Anomalies in F-Theory}},  {\em JHEP} {\bf 11} (2017) 100,
  [\href{http://xxx.lanl.gov/abs/1706.08528}{{\tt 1706.08528}}].

\bibitem{Grassi:2018rva}
A.~Grassi and T.~Weigand, {\it {On topological invariants of algebraic
  threefolds with ($\mathbb Q$-factorial) singularities}},
  \href{http://xxx.lanl.gov/abs/1804.02424}{{\tt 1804.02424}}.

\bibitem{Lerche:1999de}
W.~Lerche, {\it {On the heterotic / F theory duality in eight-dimensions}},
\newblock \href{http://xxx.lanl.gov/abs/hep-th/9910207}{{\tt hep-th/9910207}}.

\bibitem{Johnson:2003gi}
C.~V. Johnson, {\em {D-branes}}.
\newblock Cambridge University Press, 2005.

\bibitem{Denef:2008wq}
F.~Denef, {\it {Les Houches Lectures on Constructing String Vacua}},  {\em Les
  Houches} {\bf 87} (2008) 483--610,
  [\href{http://xxx.lanl.gov/abs/0803.1194}{{\tt 0803.1194}}].

\bibitem{Weigand:2010wm}
T.~Weigand, {\it {Lectures on F-theory compactifications and model building}},
  {\em Class. Quant. Grav.} {\bf 27} (2010) 214004,
  [\href{http://xxx.lanl.gov/abs/1009.3497}{{\tt 1009.3497}}].

\bibitem{Blumenhagen:2013fgp}
R.~Blumenhagen, D.~L{\"u}st, and S.~Theisen, {\em {Basic concepts of string
  theory}}.
\newblock Springer, Heidelberg, Germany, 2013.

\bibitem{Greene:1989ya}
B.~R. Greene, A.~D. Shapere, C.~Vafa, and S.-T. Yau, {\it {Stringy Cosmic
  Strings and Noncompact Calabi-Yau Manifolds}},  {\em Nucl. Phys.} {\bf B337}
  (1990) 1--36.

\bibitem{Witten:1995im}
E.~Witten, {\it {Bound states of strings and p-branes}},  {\em Nucl. Phys.}
  {\bf B460} (1996) 335--350,
  [\href{http://xxx.lanl.gov/abs/hep-th/9510135}{{\tt hep-th/9510135}}].

\bibitem{Douglas:1996du}
M.~R. Douglas and M.~Li, {\it {D-brane realization of N=2 superYang-Mills
  theory in four-dimensions}},
  \href{http://xxx.lanl.gov/abs/hep-th/9604041}{{\tt hep-th/9604041}}.

\bibitem{Sen:1996vd}
A.~Sen, {\it {F theory and orientifolds}},  {\em Nucl. Phys.} {\bf B475} (1996)
  562--578, [\href{http://xxx.lanl.gov/abs/hep-th/9605150}{{\tt
  hep-th/9605150}}].

\bibitem{Gaberdiel:1997ud}
M.~R. Gaberdiel and B.~Zwiebach, {\it {Exceptional groups from open strings}},
  {\em Nucl. Phys.} {\bf B518} (1998) 151--172,
  [\href{http://xxx.lanl.gov/abs/hep-th/9709013}{{\tt hep-th/9709013}}].

\bibitem{Bhardwaj:2018jgp}
L.~Bhardwaj, D.~R. Morrison, Y.~Tachikawa, and A.~Tomasiello, {\it {The frozen
  phase of F-theory}},  \href{http://xxx.lanl.gov/abs/1805.09070}{{\tt
  1805.09070}}.

\bibitem{Johansen:1996am}
A.~Johansen, {\it {A Comment on BPS states in F theory in eight-dimensions}},
  {\em Phys. Lett.} {\bf B395} (1997) 36--41,
  [\href{http://xxx.lanl.gov/abs/hep-th/9608186}{{\tt hep-th/9608186}}].

\bibitem{Dasgupta:1996ij}
K.~Dasgupta and S.~Mukhi, {\it {F theory at constant coupling}},  {\em Phys.
  Lett.} {\bf B385} (1996) 125--131,
  [\href{http://xxx.lanl.gov/abs/hep-th/9606044}{{\tt hep-th/9606044}}].

\bibitem{2009arXiv0907.0298S}
M.~{Schuett} and T.~{Shioda}, {\it {Elliptic Surfaces}},  {\em ArXiv e-prints}
  (July, 2009) [\href{http://xxx.lanl.gov/abs/0907.0298}{{\tt 0907.0298}}].

\bibitem{Bianchi:2011qh}
M.~Bianchi, A.~Collinucci, and L.~Martucci, {\it {Magnetized E3-brane
  instantons in F-theory}},  {\em JHEP} {\bf 12} (2011) 045,
  [\href{http://xxx.lanl.gov/abs/1107.3732}{{\tt 1107.3732}}].

\bibitem{Douglas:2014ywa}
M.~R. Douglas, D.~S. Park, and C.~Schnell, {\it {The Cremmer-Scherk Mechanism
  in F-theory Compactifications on K3 Manifolds}},  {\em JHEP} {\bf 05} (2014)
  135, [\href{http://xxx.lanl.gov/abs/1403.1595}{{\tt 1403.1595}}].

\bibitem{Green:1987mn}
M.~B. Green, J.~H. Schwarz, and E.~Witten, {\em {Superstring Theory. Vol. 2}}.
\newblock 1988.

\bibitem{Billo:2010mg}
M.~Billo, L.~Gallot, A.~Lerda, and I.~Pesando, {\it {F-theoretic versus
  microscopic description of a conformal N=2 SYM theory}},  {\em JHEP} {\bf 11}
  (2010) 041, [\href{http://xxx.lanl.gov/abs/1008.5240}{{\tt 1008.5240}}].

\bibitem{Duff:1995wd}
M.~J. Duff, J.~T. Liu, and R.~Minasian, {\it {Eleven-dimensional origin of
  string-string duality: A One loop test}},  {\em Nucl. Phys.} {\bf B452}
  (1995) 261--282, [\href{http://xxx.lanl.gov/abs/hep-th/9506126}{{\tt
  hep-th/9506126}}]. [,142(1995)].

\bibitem{Klemm:1996hh}
A.~Klemm, P.~Mayr, and C.~Vafa, {\it {BPS states of exceptional noncritical
  strings}},  {\em Nucl. Phys. Proc. Suppl.} {\bf 58} (1997) 177,
  [\href{http://xxx.lanl.gov/abs/hep-th/9607139}{{\tt hep-th/9607139}}].
  [177(1996)].

\bibitem{Berglund:1998va}
P.~Berglund, A.~Klemm, P.~Mayr, and S.~Theisen, {\it {On type IIB vacua with
  varying coupling constant}},  {\em Nucl. Phys.} {\bf B558} (1999) 178--204,
  [\href{http://xxx.lanl.gov/abs/hep-th/9805189}{{\tt hep-th/9805189}}].

\bibitem{Aluffi:2009tm}
P.~Aluffi and M.~Esole, {\it {New Orientifold Weak Coupling Limits in
  F-theory}},  {\em JHEP} {\bf 02} (2010) 020,
  [\href{http://xxx.lanl.gov/abs/0908.1572}{{\tt 0908.1572}}].

\bibitem{Sen:1997gv}
A.~Sen, {\it {Orientifold limit of F theory vacua}},  {\em Phys. Rev.} {\bf
  D55} (1997) R7345--R7349, [\href{http://xxx.lanl.gov/abs/hep-th/9702165}{{\tt
  hep-th/9702165}}].

\bibitem{Collinucci:2008pf}
A.~Collinucci, F.~Denef, and M.~Esole, {\it {D-brane Deconstructions in IIB
  Orientifolds}},  {\em JHEP} {\bf 02} (2009) 005,
  [\href{http://xxx.lanl.gov/abs/0805.1573}{{\tt 0805.1573}}].

\bibitem{Clingher:2012rg}
A.~Clingher, R.~Donagi, and M.~Wijnholt, {\it {The Sen Limit}},  {\em Adv.
  Theor. Math. Phys.} {\bf 18} (2014), no.~3 613--658,
  [\href{http://xxx.lanl.gov/abs/1212.4505}{{\tt 1212.4505}}].

\bibitem{Collinucci:2008zs}
A.~Collinucci, {\it {New F-theory lifts}},  {\em JHEP} {\bf 08} (2009) 076,
  [\href{http://xxx.lanl.gov/abs/0812.0175}{{\tt 0812.0175}}].

\bibitem{Collinucci:2009uh}
A.~Collinucci, {\it {New F-theory lifts. II. Permutation orientifolds and
  enhanced singularities}},  {\em JHEP} {\bf 04} (2010) 076,
  [\href{http://xxx.lanl.gov/abs/0906.0003}{{\tt 0906.0003}}].

\bibitem{Blumenhagen:2009up}
R.~Blumenhagen, T.~W. Grimm, B.~Jurke, and T.~Weigand, {\it {F-theory uplifts
  and GUTs}},  {\em JHEP} {\bf 09} (2009) 053,
  [\href{http://xxx.lanl.gov/abs/0906.0013}{{\tt 0906.0013}}].

\bibitem{Esole:2011cn}
M.~Esole, J.~Fullwood, and S.-T. Yau, {\it {$D_5$ elliptic fibrations:
  non-Kodaira fibers and new orientifold limits of F-theory}},  {\em Commun.
  Num. Theor. Phys.} {\bf 09} (2015), no.~3 583--642,
  [\href{http://xxx.lanl.gov/abs/1110.6177}{{\tt 1110.6177}}].

\bibitem{Esole:2012tf}
M.~Esole and R.~Savelli, {\it {Tate Form and Weak Coupling Limits in
  F-theory}},  {\em JHEP} {\bf 06} (2013) 027,
  [\href{http://xxx.lanl.gov/abs/1209.1633}{{\tt 1209.1633}}].

\bibitem{Krause:2012yh}
S.~Krause, C.~Mayrhofer, and T.~Weigand, {\it {Gauge Fluxes in F-theory and
  Type IIB Orientifolds}},  {\em JHEP} {\bf 08} (2012) 119,
  [\href{http://xxx.lanl.gov/abs/1202.3138}{{\tt 1202.3138}}].

\bibitem{MayorgaPena:2017eda}
D.~K. Mayorga~Pena and R.~Valandro, {\it {Weak coupling limit of F-theory
  models with MSSM spectrum and massless U(1)'s}},
  \href{http://xxx.lanl.gov/abs/1708.09452}{{\tt 1708.09452}}.

\bibitem{Park:2011ji}
D.~S. Park, {\it {Anomaly Equations and Intersection Theory}},  {\em JHEP} {\bf
  01} (2012) 093, [\href{http://xxx.lanl.gov/abs/1111.2351}{{\tt 1111.2351}}].

\bibitem{Bonetti:2011mw}
F.~Bonetti and T.~W. Grimm, {\it {Six-dimensional (1,0) effective action of
  F-theory via M-theory on Calabi-Yau threefolds}},  {\em JHEP} {\bf 05} (2012)
  019, [\href{http://xxx.lanl.gov/abs/1112.1082}{{\tt 1112.1082}}].

\bibitem{Grimm:2010ks}
T.~W. Grimm, {\it {The N=1 effective action of F-theory compactifications}},
  {\em Nucl. Phys.} {\bf B845} (2011) 48--92,
  [\href{http://xxx.lanl.gov/abs/1008.4133}{{\tt 1008.4133}}].

\bibitem{Grimm:2011sk}
T.~W. Grimm and R.~Savelli, {\it {Gravitational Instantons and Fluxes from
  M/F-theory on Calabi-Yau fourfolds}},  {\em Phys. Rev.} {\bf D85} (2012)
  026003, [\href{http://xxx.lanl.gov/abs/1109.3191}{{\tt 1109.3191}}].

\bibitem{Witten:1996qb}
E.~Witten, {\it {Phase transitions in M theory and F theory}},  {\em Nucl.
  Phys.} {\bf B471} (1996) 195--216,
  [\href{http://xxx.lanl.gov/abs/hep-th/9603150}{{\tt hep-th/9603150}}].

\bibitem{Intriligator:1997pq}
K.~A. Intriligator, D.~R. Morrison, and N.~Seiberg, {\it {Five-dimensional
  supersymmetric gauge theories and degenerations of Calabi-Yau spaces}},  {\em
  Nucl.Phys.} {\bf B497} (1997) 56--100,
  [\href{http://xxx.lanl.gov/abs/hep-th/9702198}{{\tt hep-th/9702198}}].

\bibitem{Aharony:1997bx}
O.~Aharony, A.~Hanany, K.~A. Intriligator, N.~Seiberg, and M.~Strassler, {\it
  {Aspects of N=2 supersymmetric gauge theories in three-dimensions}},  {\em
  Nucl.Phys.} {\bf B499} (1997) 67--99,
  [\href{http://xxx.lanl.gov/abs/hep-th/9703110}{{\tt hep-th/9703110}}].

\bibitem{Grimm:2011fx}
T.~W. Grimm and H.~Hayashi, {\it {F-theory fluxes, Chirality and Chern-Simons
  theories}},  {\em JHEP} {\bf 1203} (2012) 027,
  [\href{http://xxx.lanl.gov/abs/1111.1232}{{\tt 1111.1232}}]. 53 pages, 5
  figures/ v2: typos corrected, minor improvements.

\bibitem{Bonetti:2013ela}
F.~Bonetti, T.~W. Grimm, and S.~Hohenegger, {\it {One-loop Chern-Simons terms
  in five dimensions}},  {\em JHEP} {\bf 07} (2013) 043,
  [\href{http://xxx.lanl.gov/abs/1302.2918}{{\tt 1302.2918}}].

\bibitem{Bonetti:2013cza}
F.~Bonetti, T.~W. Grimm, and S.~Hohenegger, {\it {Exploring 6D origins of 5D
  supergravities with Chern-Simons terms}},  {\em JHEP} {\bf 05} (2013) 124,
  [\href{http://xxx.lanl.gov/abs/1303.2661}{{\tt 1303.2661}}].

\bibitem{Kodaira2}
K.~Kodaira, {\it {On compact analytic surfaces, II}},  {\em Annals of Math.}
  {\bf 77} (1964) 563--626.

\bibitem{Neron}
A.~N{\'e}ron, {\it {Mod\`{e}les minimaux des vari{\'e}t{\'e}s ab{\'e}liennes
  sur les corps locaux et globaux}},  {\em Annals of Math.} {\bf 82} (1965)
  249--331.

\bibitem{Katz:2011qp}
S.~Katz, D.~R. Morrison, S.~Schafer-Nameki, and J.~Sully, {\it {Tate's
  algorithm and F-theory}},  {\em JHEP} {\bf 08} (2011) 094,
  [\href{http://xxx.lanl.gov/abs/1106.3854}{{\tt 1106.3854}}].

\bibitem{Tate}
J.~Tate, {\it {Algorithm for Determining the Type of a Singular Fiber in an
  Elliptic Pencil}},  {\em Lecture Notes in Math., vol. 476, Springer-Verlag}
  (1975).

\bibitem{Bershadsky:1996nh}
M.~Bershadsky, K.~A. Intriligator, S.~Kachru, D.~R. Morrison, V.~Sadov, and
  C.~Vafa, {\it {Geometric singularities and enhanced gauge symmetries}},  {\em
  Nucl. Phys.} {\bf B481} (1996) 215--252,
  [\href{http://xxx.lanl.gov/abs/hep-th/9605200}{{\tt hep-th/9605200}}].

\bibitem{Esole:2017rgz}
M.~Esole, P.~Jefferson, and M.~J. Kang, {\it {The Geometry of F$_4$-Models}},
  \href{http://xxx.lanl.gov/abs/1704.08251}{{\tt 1704.08251}}.

\bibitem{Esole:2017qeh}
M.~Esole, R.~Jagadeesan, and M.~J. Kang, {\it {The Geometry of G$_2$, Spin(7),
  and Spin(8)-models}},  \href{http://xxx.lanl.gov/abs/1709.04913}{{\tt
  1709.04913}}.

\bibitem{Esole:2018mqb}
M.~Esole and M.~J. Kang, {\it {The Geometry of the SU(2)$\times$ G$_2$-model}},
   \href{http://xxx.lanl.gov/abs/1805.03214}{{\tt 1805.03214}}.

\bibitem{Grassi:2018wfy}
A.~Grassi, J.~Halverson, C.~Long, J.~L. Shaneson, and J.~Tian, {\it
  {Non-simply-laced Symmetry Algebras in F-theory on Singular Spaces}},
  \href{http://xxx.lanl.gov/abs/1805.06949}{{\tt 1805.06949}}.

\bibitem{Morrison:2014lca}
D.~R. Morrison and W.~Taylor, {\it {Non-Higgsable clusters for 4D F-theory
  models}},  {\em JHEP} {\bf 05} (2015) 080,
  [\href{http://xxx.lanl.gov/abs/1412.6112}{{\tt 1412.6112}}].

\bibitem{Halverson:2015jua}
J.~Halverson and W.~Taylor, {\it {$ {\mathrm{\mathbb{P}}}^1 $-bundle bases and
  the prevalence of non-Higgsable structure in 4D F-theory models}},  {\em
  JHEP} {\bf 09} (2015) 086, [\href{http://xxx.lanl.gov/abs/1506.03204}{{\tt
  1506.03204}}].

\bibitem{Taylor:2015ppa}
W.~Taylor and Y.-N. Wang, {\it {A Monte Carlo exploration of threefold base
  geometries for 4d F-theory vacua}},  {\em JHEP} {\bf 01} (2016) 137,
  [\href{http://xxx.lanl.gov/abs/1510.04978}{{\tt 1510.04978}}].

\bibitem{Halverson:2017ffz}
J.~Halverson, C.~Long, and B.~Sung, {\it {Algorithmic universality in F-theory
  compactifications}},  {\em Phys. Rev.} {\bf D96} (2017), no.~12 126006,
  [\href{http://xxx.lanl.gov/abs/1706.02299}{{\tt 1706.02299}}].

\bibitem{Taylor:2017yqr}
W.~Taylor and Y.-N. Wang, {\it {Scanning the skeleton of the 4D F-theory
  landscape}},  {\em JHEP} {\bf 01} (2018) 111,
  [\href{http://xxx.lanl.gov/abs/1710.11235}{{\tt 1710.11235}}].

\bibitem{Wang:2018rkk}
Y.-N. Wang and Z.~Zhang, {\it {Learning non-Higgsable gauge groups in 4D
  F-theory}},  \href{http://xxx.lanl.gov/abs/1804.07296}{{\tt 1804.07296}}.

\bibitem{Halverson:2018xge}
J.~Halverson and P.~Langacker, {\it {TASI Lectures on Remnants from the String
  Landscape}},
\newblock \href{http://xxx.lanl.gov/abs/1801.03503}{{\tt 1801.03503}}.

\bibitem{Aspinwall:2000kf}
P.~S. Aspinwall, S.~H. Katz, and D.~R. Morrison, {\it {Lie groups, Calabi-Yau
  threefolds, and F theory}},  {\em Adv. Theor. Math. Phys.} {\bf 4} (2000)
  95--126, [\href{http://xxx.lanl.gov/abs/hep-th/0002012}{{\tt
  hep-th/0002012}}].

\bibitem{Katz:1996ht}
S.~H. Katz, D.~R. Morrison, and M.~R. Plesser, {\it {Enhanced gauge symmetry in
  type II string theory}},  {\em Nucl. Phys.} {\bf B477} (1996) 105--140,
  [\href{http://xxx.lanl.gov/abs/hep-th/9601108}{{\tt hep-th/9601108}}].

\bibitem{Garcia-Etxebarria:2017crf}
I.~Garcia-Etxebarria, H.~Hayashi, K.~Ohmori, Y.~Tachikawa, and K.~Yonekura,
  {\it {8d gauge anomalies and the topological Green-Schwarz mechanism}},  {\em
  JHEP} {\bf 11} (2017) 177, [\href{http://xxx.lanl.gov/abs/1710.04218}{{\tt
  1710.04218}}].

\bibitem{Cvetic:2012xn}
M.~Cveti{\v c}, T.~W. Grimm, and D.~Klevers, {\it {Anomaly Cancellation And
  Abelian Gauge Symmetries In F-theory}},  {\em JHEP} {\bf 02} (2013) 101,
  [\href{http://xxx.lanl.gov/abs/1210.6034}{{\tt 1210.6034}}].

\bibitem{Corvilain:2017luj}
P.~Corvilain, T.~W. Grimm, and D.~Regalado, {\it {Chiral anomalies on a circle
  and their cancellation in F-theory}},
  \href{http://xxx.lanl.gov/abs/1710.07626}{{\tt 1710.07626}}.

\bibitem{Schafer-Nameki:2016cfr}
S.~Sch{\"a}fer-Nameki and T.~Weigand, {\it {F-theory and 2d $(0, 2)$
  theories}},  {\em JHEP} {\bf 05} (2016) 059,
  [\href{http://xxx.lanl.gov/abs/1601.02015}{{\tt 1601.02015}}].

\bibitem{Lawrie:2016rqe}
C.~Lawrie, S.~Schafer-Nameki, and T.~Weigand, {\it {The gravitational sector of
  2d (0, 2) F-theory vacua}},  {\em JHEP} {\bf 05} (2017) 103,
  [\href{http://xxx.lanl.gov/abs/1612.06393}{{\tt 1612.06393}}].

\bibitem{Witten:1993yc}
E.~Witten, {\it {Phases of N=2 theories in two-dimensions}},  {\em Nucl. Phys.}
  {\bf B403} (1993) 159--222,
  [\href{http://xxx.lanl.gov/abs/hep-th/9301042}{{\tt hep-th/9301042}}].
  [AMS/IP Stud. Adv. Math.1,143(1996)].

\bibitem{Candelas:1996su}
P.~Candelas and A.~Font, {\it {Duality between the webs of heterotic and type
  II vacua}},  {\em Nucl. Phys.} {\bf B511} (1998) 295--325,
  [\href{http://xxx.lanl.gov/abs/hep-th/9603170}{{\tt hep-th/9603170}}].

\bibitem{Candelas:1997eh}
P.~Candelas, E.~Perevalov, and G.~Rajesh, {\it {Toric geometry and enhanced
  gauge symmetry of F theory / heterotic vacua}},  {\em Nucl. Phys.} {\bf B507}
  (1997) 445--474, [\href{http://xxx.lanl.gov/abs/hep-th/9704097}{{\tt
  hep-th/9704097}}].

\bibitem{Bouchard:2003bu}
V.~Bouchard and H.~Skarke, {\it {Affine Kac-Moody algebras, CHL strings and the
  classification of tops}},  {\em Adv. Theor. Math. Phys.} {\bf 7} (2003),
  no.~2 205--232, [\href{http://xxx.lanl.gov/abs/hep-th/0303218}{{\tt
  hep-th/0303218}}].

\bibitem{Blumenhagen:2009yv}
R.~Blumenhagen, T.~W. Grimm, B.~Jurke, and T.~Weigand, {\it {Global F-theory
  GUTs}},  {\em Nucl. Phys.} {\bf B829} (2010) 325--369,
  [\href{http://xxx.lanl.gov/abs/0908.1784}{{\tt 0908.1784}}].

\bibitem{Esole:2011sm}
M.~Esole and S.-T. Yau, {\it {Small resolutions of SU(5)-models in F-theory}},
  {\em Adv. Theor. Math. Phys.} {\bf 17} (2013), no.~6 1195--1253,
  [\href{http://xxx.lanl.gov/abs/1107.0733}{{\tt 1107.0733}}].

\bibitem{Marsano:2011hv}
J.~Marsano and S.~Schafer-Nameki, {\it {Yukawas, G-flux, and Spectral Covers
  from Resolved Calabi-Yau's}},  {\em JHEP} {\bf 11} (2011) 098,
  [\href{http://xxx.lanl.gov/abs/1108.1794}{{\tt 1108.1794}}].

\bibitem{Krause:2011xj}
S.~Krause, C.~Mayrhofer, and T.~Weigand, {\it {$G_4$ flux, chiral matter and
  singularity resolution in F-theory compactifications}},  {\em Nucl.Phys.}
  {\bf B858} (2012) 1--47, [\href{http://xxx.lanl.gov/abs/1109.3454}{{\tt
  1109.3454}}].

\bibitem{Lawrie:2012gg}
C.~Lawrie and S.~SchŠfer-Nameki, {\it {The Tate Form on Steroids: Resolution
  and Higher Codimension Fibers}},  {\em JHEP} {\bf 04} (2013) 061,
  [\href{http://xxx.lanl.gov/abs/1212.2949}{{\tt 1212.2949}}].

\bibitem{Hayashi:2013lra}
H.~Hayashi, C.~Lawrie, and S.~Schafer-Nameki, {\it {Phases, Flops and F-theory:
  SU(5) Gauge Theories}},  {\em JHEP} {\bf 10} (2013) 046,
  [\href{http://xxx.lanl.gov/abs/1304.1678}{{\tt 1304.1678}}].

\bibitem{Braun:2014kla}
A.~P. Braun and S.~Schafer-Nameki, {\it {Box Graphs and Resolutions I}},  {\em
  Nucl. Phys.} {\bf B905} (2016) 447--479,
  [\href{http://xxx.lanl.gov/abs/1407.3520}{{\tt 1407.3520}}].

\bibitem{Braun:2015hkv}
A.~P. Braun and S.~Schafer-Nameki, {\it {Box Graphs and Resolutions II: From
  Coulomb Phases to Fiber Faces}},  {\em Nucl. Phys.} {\bf B905} (2016)
  480--530, [\href{http://xxx.lanl.gov/abs/1511.01801}{{\tt 1511.01801}}].

\bibitem{Huang:2018gpl}
Y.-C. Huang and W.~Taylor, {\it {Comparing elliptic and toric hypersurface
  Calabi-Yau threefolds at large Hodge numbers}},
  \href{http://xxx.lanl.gov/abs/1805.05907}{{\tt 1805.05907}}.

\bibitem{Apruzzi:2016iac}
F.~Apruzzi, F.~Hassler, J.~J. Heckman, and I.~V. Melnikov, {\it {UV Completions
  for Non-Critical Strings}},  {\em JHEP} {\bf 07} (2016) 045,
  [\href{http://xxx.lanl.gov/abs/1602.04221}{{\tt 1602.04221}}].

\bibitem{Weigand:2017gwb}
T.~Weigand and F.~Xu, {\it {The Green-Schwarz Mechanism and Geometric Anomaly
  Relations in 2d (0,2) F-theory Vacua}},  {\em JHEP} {\bf 04} (2018) 107,
  [\href{http://xxx.lanl.gov/abs/1712.04456}{{\tt 1712.04456}}].

\bibitem{Morrison:2011mb}
D.~R. Morrison and W.~Taylor, {\it {Matter and singularities}},  {\em JHEP}
  {\bf 01} (2012) 022, [\href{http://xxx.lanl.gov/abs/1106.3563}{{\tt
  1106.3563}}].

\bibitem{Klevers:2017aku}
D.~Klevers, D.~R. Morrison, N.~Raghuram, and W.~Taylor, {\it {Exotic matter on
  singular divisors in F-theory}},  {\em JHEP} {\bf 11} (2017) 124,
  [\href{http://xxx.lanl.gov/abs/1706.08194}{{\tt 1706.08194}}].

\bibitem{Katz:1996xe}
S.~H. Katz and C.~Vafa, {\it {Matter from geometry}},  {\em Nucl. Phys.} {\bf
  B497} (1997) 146--154, [\href{http://xxx.lanl.gov/abs/hep-th/9606086}{{\tt
  hep-th/9606086}}].

\bibitem{Anderson:2015cqy}
L.~B. Anderson, J.~Gray, N.~Raghuram, and W.~Taylor, {\it {Matter in
  transition}},  {\em JHEP} {\bf 04} (2016) 080,
  [\href{http://xxx.lanl.gov/abs/1512.05791}{{\tt 1512.05791}}].

\bibitem{Grassi:2013kha}
A.~Grassi, J.~Halverson, and J.~L. Shaneson, {\it {Matter From Geometry Without
  Resolution}},  {\em JHEP} {\bf 10} (2013) 205,
  [\href{http://xxx.lanl.gov/abs/1306.1832}{{\tt 1306.1832}}].

\bibitem{Grassi:2014sda}
A.~Grassi, J.~Halverson, and J.~L. Shaneson, {\it {Non-Abelian Gauge Symmetry
  and the Higgs Mechanism in F-theory}},  {\em Commun. Math. Phys.} {\bf 336}
  (2015), no.~3 1231--1257, [\href{http://xxx.lanl.gov/abs/1402.5962}{{\tt
  1402.5962}}].

\bibitem{Grassi:2014ffa}
A.~Grassi, J.~Halverson, and J.~L. Shaneson, {\it {Geometry and Topology of
  String Junctions}},  \href{http://xxx.lanl.gov/abs/1410.6817}{{\tt
  1410.6817}}.

\bibitem{Grassi:2016bhs}
A.~Grassi, J.~Halverson, F.~Ruehle, and J.~L. Shaneson, {\it {Dualities of
  deformed $ \mathcal{N}=2 $ SCFTs from link monodromy on D3-brane states}},
  {\em JHEP} {\bf 09} (2017) 135,
  [\href{http://xxx.lanl.gov/abs/1611.01154}{{\tt 1611.01154}}].

\bibitem{Sadov:1996zm}
V.~Sadov, {\it {Generalized Green-Schwarz mechanism in F theory}},  {\em Phys.
  Lett.} {\bf B388} (1996) 45--50,
  [\href{http://xxx.lanl.gov/abs/hep-th/9606008}{{\tt hep-th/9606008}}].

\bibitem{Kumar:2010am}
V.~Kumar, D.~S. Park, and W.~Taylor, {\it {6D supergravity without tensor
  multiplets}},  {\em JHEP} {\bf 04} (2011) 080,
  [\href{http://xxx.lanl.gov/abs/1011.0726}{{\tt 1011.0726}}].

\bibitem{Cvetic:2015ioa}
M.~Cveti{\v c}, D.~Klevers, H.~Piragua, and W.~Taylor, {\it {General $U(1)
  xU(1)$ F-theory compactifications and beyond: geometry of unHiggsings and
  novel matter structure}},  {\em JHEP} {\bf 11} (2015) 204,
  [\href{http://xxx.lanl.gov/abs/1507.05954}{{\tt 1507.05954}}].

\bibitem{Klevers:2016jsz}
D.~Klevers and W.~Taylor, {\it {Three-Index Symmetric Matter Representations of
  SU(2) in F-Theory from Non-Tate Form Weierstrass Models}},  {\em JHEP} {\bf
  06} (2016) 171, [\href{http://xxx.lanl.gov/abs/1604.01030}{{\tt
  1604.01030}}].

\bibitem{Hayashi:2014kca}
H.~Hayashi, C.~Lawrie, D.~R. Morrison, and S.~Schafer-Nameki, {\it {Box Graphs
  and Singular Fibers}},  {\em JHEP} {\bf 05} (2014) 048,
  [\href{http://xxx.lanl.gov/abs/1402.2653}{{\tt 1402.2653}}].

\bibitem{Esole:2014bka}
M.~Esole, S.-H. Shao, and S.-T. Yau, {\it {Singularities and Gauge Theory
  Phases}},  {\em Adv. Theor. Math. Phys.} {\bf 19} (2015) 1183--1247,
  [\href{http://xxx.lanl.gov/abs/1402.6331}{{\tt 1402.6331}}].

\bibitem{Esole:2014hya}
M.~Esole, S.-H. Shao, and S.-T. Yau, {\it {Singularities and Gauge Theory
  Phases II}},  {\em Adv. Theor. Math. Phys.} {\bf 20} (2016) 683--749,
  [\href{http://xxx.lanl.gov/abs/1407.1867}{{\tt 1407.1867}}].

\bibitem{Cattaneo:2013vda}
A.~Cattaneo, {\it {Crepant resolutions of Weierstrass threefolds and
  non-Kodaira fibres}},  \href{http://xxx.lanl.gov/abs/1307.7997}{{\tt
  1307.7997}}.

\bibitem{Miranda83}
R.~Miranda, {\it {Smooth models for elliptic threefolds. In: The birational
  geometry of degeneration (Cambridge, Mass., 1981)}},  {\em Progr. Math.} {\bf
  29} (1983) 85--133.

\bibitem{Morrison:1996xf}
D.~R. Morrison and N.~Seiberg, {\it {Extremal transitions and five-dimensional
  supersymmetric field theories}},  {\em Nucl. Phys.} {\bf B483} (1997)
  229--247, [\href{http://xxx.lanl.gov/abs/hep-th/9609070}{{\tt
  hep-th/9609070}}].

\bibitem{deBoer:1997kr}
J.~de~Boer, K.~Hori, and Y.~Oz, {\it {Dynamics of N=2 supersymmetric gauge
  theories in three-dimensions}},  {\em Nucl. Phys.} {\bf B500} (1997)
  163--191, [\href{http://xxx.lanl.gov/abs/hep-th/9703100}{{\tt
  hep-th/9703100}}].

\bibitem{Diaconescu:1998ua}
D.-E. Diaconescu and S.~Gukov, {\it {Three-dimensional N=2 gauge theories and
  degenerations of Calabi-Yau four folds}},  {\em Nucl. Phys.} {\bf B535}
  (1998) 171--196, [\href{http://xxx.lanl.gov/abs/hep-th/9804059}{{\tt
  hep-th/9804059}}].

\bibitem{Katz:2002gh}
S.~H. Katz and E.~Sharpe, {\it {D-branes, open string vertex operators, and Ext
  groups}},  {\em Adv. Theor. Math. Phys.} {\bf 6} (2003) 979--1030,
  [\href{http://xxx.lanl.gov/abs/hep-th/0208104}{{\tt hep-th/0208104}}].

\bibitem{Blumenhagen:2008zz}
R.~Blumenhagen, V.~Braun, T.~W. Grimm, and T.~Weigand, {\it {GUTs in Type IIB
  Orientifold Compactifications}},  {\em Nucl. Phys.} {\bf B815} (2009) 1--94,
  [\href{http://xxx.lanl.gov/abs/0811.2936}{{\tt 0811.2936}}].

\bibitem{Intriligator:2012ue}
K.~Intriligator, H.~Jockers, P.~Mayr, D.~R. Morrison, and M.~R. Plesser, {\it
  {Conifold Transitions in M-theory on Calabi-Yau Fourfolds with Background
  Fluxes}},  {\em Adv. Theor. Math. Phys.} {\bf 17} (2013), no.~3 601--699,
  [\href{http://xxx.lanl.gov/abs/1203.6662}{{\tt 1203.6662}}].

\bibitem{Bies:2014sra}
M.~Bies, C.~Mayrhofer, C.~Pehle, and T.~Weigand, {\it {Chow groups, Deligne
  cohomology and massless matter in F-theory}},
  \href{http://xxx.lanl.gov/abs/1402.5144}{{\tt 1402.5144}}.

\bibitem{Blumenhagen:2010pv}
R.~Blumenhagen, B.~Jurke, T.~Rahn, and H.~Roschy, {\it {Cohomology of Line
  Bundles: A Computational Algorithm}},  {\em J. Math. Phys.} {\bf 51} (2010)
  103525, [\href{http://xxx.lanl.gov/abs/1003.5217}{{\tt 1003.5217}}].

\bibitem{Grassi1991}
A.~Grassi, {\it {On minimal models of elliptic threefolds}},  {\em Math. Ann.}
  {\bf 290} (1991), no.~2 287--301.

\bibitem{GrassiEqui}
A.~Grassi, {\it {Log contractions and equidimensional models of elliptic
  threefolds}},  {\em J. Algebraic Geom.} {\bf 4} (1995), no.~2 255--276.

\bibitem{DelZotto:2014hpa}
M.~Del~Zotto, J.~J. Heckman, A.~Tomasiello, and C.~Vafa, {\it {6d Conformal
  Matter}},  {\em JHEP} {\bf 02} (2015) 054,
  [\href{http://xxx.lanl.gov/abs/1407.6359}{{\tt 1407.6359}}].

\bibitem{Arras:2016evy}
P.~Arras, A.~Grassi, and T.~Weigand, {\it {Terminal Singularities, Milnor
  Numbers, and Matter in F-theory}},  {\em J. Geom. Phys.} {\bf 123} (2018)
  71--97, [\href{http://xxx.lanl.gov/abs/1612.05646}{{\tt 1612.05646}}].

\bibitem{Ishi}
S.~Ishii, {\it {Introduction to singularities}},  {\em Springer, Tokyo} {\bf 9}
  (2014) viii+223.

\bibitem{Braun:2014oya}
V.~Braun and D.~R. Morrison, {\it {F-theory on Genus-One Fibrations}},  {\em
  JHEP} {\bf 08} (2014) 132, [\href{http://xxx.lanl.gov/abs/1401.7844}{{\tt
  1401.7844}}].

\bibitem{Morrison:2016lix}
D.~R. Morrison, D.~S. Park, and W.~Taylor, {\it {Non-Higgsable abelian gauge
  symmetry and F-theory on fiber products of rational elliptic surfaces}},
  \href{http://xxx.lanl.gov/abs/1610.06929}{{\tt 1610.06929}}.

\bibitem{Braun:2014nva}
A.~P. Braun, A.~Collinucci, and R.~Valandro, {\it {The fate of U(1)'s at strong
  coupling in F-theory}},  {\em JHEP} {\bf 07} (2014) 028,
  [\href{http://xxx.lanl.gov/abs/1402.4054}{{\tt 1402.4054}}].

\bibitem{Martucci:2015dxa}
L.~Martucci and T.~Weigand, {\it {Non-perturbative selection rules in
  F-theory}},  {\em JHEP} {\bf 09} (2015) 198,
  [\href{http://xxx.lanl.gov/abs/1506.06764}{{\tt 1506.06764}}].

\bibitem{Hayashi:2008ba}
H.~Hayashi, R.~Tatar, Y.~Toda, T.~Watari, and M.~Yamazaki, {\it {New Aspects of
  Heterotic--F Theory Duality}},  {\em Nucl. Phys.} {\bf B806} (2009) 224--299,
  [\href{http://xxx.lanl.gov/abs/0805.1057}{{\tt 0805.1057}}].

\bibitem{Hayashi:2009ge}
H.~Hayashi, T.~Kawano, R.~Tatar, and T.~Watari, {\it {Codimension-3
  Singularities and Yukawa Couplings in F-theory}},  {\em Nucl. Phys.} {\bf
  B823} (2009) 47--115, [\href{http://xxx.lanl.gov/abs/0901.4941}{{\tt
  0901.4941}}].

\bibitem{Hayashi:2009bt}
H.~Hayashi, T.~Kawano, Y.~Tsuchiya, and T.~Watari, {\it {Flavor Structure in
  F-theory Compactifications}},  {\em JHEP} {\bf 08} (2010) 036,
  [\href{http://xxx.lanl.gov/abs/0910.2762}{{\tt 0910.2762}}].

\bibitem{Hayashi:2010zp}
H.~Hayashi, T.~Kawano, Y.~Tsuchiya, and T.~Watari, {\it {More on Dimension-4
  Proton Decay Problem in F-theory -- Spectral Surface, Discriminant Locus and
  Monodromy}},  {\em Nucl. Phys.} {\bf B840} (2010) 304--348,
  [\href{http://xxx.lanl.gov/abs/1004.3870}{{\tt 1004.3870}}].

\bibitem{Martucci:2015oaa}
L.~Martucci and T.~Weigand, {\it {Hidden Selection Rules, M5-instantons and
  Fluxes in F-theory}},  {\em JHEP} {\bf 10} (2015) 131,
  [\href{http://xxx.lanl.gov/abs/1507.06999}{{\tt 1507.06999}}].

\bibitem{Donagi:2009ra}
R.~Donagi and M.~Wijnholt, {\it {Higgs Bundles and UV Completion in F-Theory}},
   {\em Commun. Math. Phys.} {\bf 326} (2014) 287--327,
  [\href{http://xxx.lanl.gov/abs/0904.1218}{{\tt 0904.1218}}].

\bibitem{Collinucci:2016hgh}
A.~Collinucci and I.~Garcia-Etxebarria, {\it {E$_{6}$ Yukawa couplings in
  F-theory as D-brane instanton effects}},  {\em JHEP} {\bf 03} (2017) 155,
  [\href{http://xxx.lanl.gov/abs/1612.06874}{{\tt 1612.06874}}].

\bibitem{Blumenhagen:2007zk}
R.~Blumenhagen, M.~Cvetic, D.~Lust, R.~Richter, and T.~Weigand, {\it
  {Non-perturbative Yukawa Couplings from String Instantons}},  {\em Phys. Rev.
  Lett.} {\bf 100} (2008) 061602,
  [\href{http://xxx.lanl.gov/abs/0707.1871}{{\tt 0707.1871}}].

\bibitem{Blumenhagen:2006xt}
R.~Blumenhagen, M.~Cvetic, and T.~Weigand, {\it {Spacetime instanton
  corrections in 4D string vacua: The Seesaw mechanism for D-Brane models}},
  {\em Nucl. Phys.} {\bf B771} (2007) 113--142,
  [\href{http://xxx.lanl.gov/abs/hep-th/0609191}{{\tt hep-th/0609191}}].

\bibitem{Ibanez:2006da}
L.~E. Ibanez and A.~M. Uranga, {\it {Neutrino Majorana Masses from String
  Theory Instanton Effects}},  {\em JHEP} {\bf 03} (2007) 052,
  [\href{http://xxx.lanl.gov/abs/hep-th/0609213}{{\tt hep-th/0609213}}].

\bibitem{Grimm:2011dj}
T.~W. Grimm, M.~Kerstan, E.~Palti, and T.~Weigand, {\it {On Fluxed Instantons
  and Moduli Stabilisation in IIB Orientifolds and F-theory}},  {\em Phys.
  Rev.} {\bf D84} (2011) 066001, [\href{http://xxx.lanl.gov/abs/1105.3193}{{\tt
  1105.3193}}].

\bibitem{Marsano:2011nn}
J.~Marsano, N.~Saulina, and S.~SchŠfer-Nameki, {\it {G-flux, M5 instantons, and
  U(1) symmetries in F-theory}},  {\em Phys. Rev.} {\bf D87} (2013) 066007,
  [\href{http://xxx.lanl.gov/abs/1107.1718}{{\tt 1107.1718}}].

\bibitem{Kerstan:2012cy}
M.~Kerstan and T.~Weigand, {\it {Fluxed M5-instantons in F-theory}},  {\em
  Nucl. Phys.} {\bf B864} (2012) 597--639,
  [\href{http://xxx.lanl.gov/abs/1205.4720}{{\tt 1205.4720}}].

\bibitem{Braun:2013cb}
A.~P. Braun and T.~Watari, {\it {On Singular Fibres in F-Theory}},  {\em JHEP}
  {\bf 07} (2013) 031, [\href{http://xxx.lanl.gov/abs/1301.5814}{{\tt
  1301.5814}}].

\bibitem{Heckman:2008qa}
J.~J. Heckman and C.~Vafa, {\it {Flavor Hierarchy From F-theory}},  {\em Nucl.
  Phys.} {\bf B837} (2010) 137--151,
  [\href{http://xxx.lanl.gov/abs/0811.2417}{{\tt 0811.2417}}].

\bibitem{Cecotti:2009zf}
S.~Cecotti, M.~C.~N. Cheng, J.~J. Heckman, and C.~Vafa, {\it {Yukawa Couplings
  in F-theory and Non-Commutative Geometry}},
  \href{http://xxx.lanl.gov/abs/0910.0477}{{\tt 0910.0477}}.

\bibitem{Cordova:2009fg}
C.~Cordova, {\it {Decoupling Gravity in F-Theory}},  {\em Adv. Theor. Math.
  Phys.} {\bf 15} (2011), no.~3 689--740,
  [\href{http://xxx.lanl.gov/abs/0910.2955}{{\tt 0910.2955}}].

\bibitem{Hayashi:2011aa}
H.~Hayashi, T.~Kawano, and T.~Watari, {\it {Constraints on GUT 7-brane Topology
  in F-theory}},  {\em Phys. Lett.} {\bf B708} (2012) 191--194,
  [\href{http://xxx.lanl.gov/abs/1112.2032}{{\tt 1112.2032}}].

\bibitem{Marchesano:2009rz}
F.~Marchesano and L.~Martucci, {\it {Non-perturbative effects on seven-brane
  Yukawa couplings}},  {\em Phys. Rev. Lett.} {\bf 104} (2010) 231601,
  [\href{http://xxx.lanl.gov/abs/0910.5496}{{\tt 0910.5496}}].

\bibitem{Font:2009gq}
A.~Font and L.~E. Ibanez, {\it {Matter wave functions and Yukawa couplings in
  F-theory Grand Unification}},  {\em JHEP} {\bf 09} (2009) 036,
  [\href{http://xxx.lanl.gov/abs/0907.4895}{{\tt 0907.4895}}].

\bibitem{Conlon:2009qq}
J.~P. Conlon and E.~Palti, {\it {Aspects of Flavour and Supersymmetry in
  F-theory GUTs}},  {\em JHEP} {\bf 01} (2010) 029,
  [\href{http://xxx.lanl.gov/abs/0910.2413}{{\tt 0910.2413}}].

\bibitem{Leontaris:2010zd}
G.~K. Leontaris and G.~G. Ross, {\it {Yukawa couplings and fermion mass
  structure in F-theory GUTs}},  {\em JHEP} {\bf 02} (2011) 108,
  [\href{http://xxx.lanl.gov/abs/1009.6000}{{\tt 1009.6000}}].

\bibitem{Cecotti:2010bp}
S.~Cecotti, C.~Cordova, J.~J. Heckman, and C.~Vafa, {\it {T-Branes and
  Monodromy}},  {\em JHEP} {\bf 07} (2011) 030,
  [\href{http://xxx.lanl.gov/abs/1010.5780}{{\tt 1010.5780}}].

\bibitem{Chiou:2011js}
C.-C. Chiou, A.~E. Faraggi, R.~Tatar, and W.~Walters, {\it {T-branes and Yukawa
  Couplings}},  {\em JHEP} {\bf 05} (2011) 023,
  [\href{http://xxx.lanl.gov/abs/1101.2455}{{\tt 1101.2455}}].

\bibitem{Aparicio:2011jx}
L.~Aparicio, A.~Font, L.~E. Ibanez, and F.~Marchesano, {\it {Flux and Instanton
  Effects in Local F-theory Models and Hierarchical Fermion Masses}},  {\em
  JHEP} {\bf 08} (2011) 152, [\href{http://xxx.lanl.gov/abs/1104.2609}{{\tt
  1104.2609}}].

\bibitem{Font:2012wq}
A.~Font, L.~E. Ibanez, F.~Marchesano, and D.~Regalado, {\it {Non-perturbative
  effects and Yukawa hierarchies in F-theory SU(5) Unification}},  {\em JHEP}
  {\bf 03} (2013) 140, [\href{http://xxx.lanl.gov/abs/1211.6529}{{\tt
  1211.6529}}]. [Erratum: JHEP07,036(2013)].

\bibitem{Font:2013ida}
A.~Font, F.~Marchesano, D.~Regalado, and G.~Zoccarato, {\it {Up-type quark
  masses in SU(5) F-theory models}},  {\em JHEP} {\bf 11} (2013) 125,
  [\href{http://xxx.lanl.gov/abs/1307.8089}{{\tt 1307.8089}}].

\bibitem{Marchesano:2015dfa}
F.~Marchesano, D.~Regalado, and G.~Zoccarato, {\it {Yukawa hierarchies at the
  point of E$_{8}$ in F-theory}},  {\em JHEP} {\bf 04} (2015) 179,
  [\href{http://xxx.lanl.gov/abs/1503.02683}{{\tt 1503.02683}}].

\bibitem{Carta:2015eoh}
F.~Carta, F.~Marchesano, and G.~Zoccarato, {\it {Fitting fermion masses and
  mixings in F-theory GUTs}},  {\em JHEP} {\bf 03} (2016) 126,
  [\href{http://xxx.lanl.gov/abs/1512.04846}{{\tt 1512.04846}}].

\bibitem{Dudas:2009hu}
E.~Dudas and E.~Palti, {\it {Froggatt-Nielsen models from E(8) in F-theory
  GUTs}},  {\em JHEP} {\bf 01} (2010) 127,
  [\href{http://xxx.lanl.gov/abs/0912.0853}{{\tt 0912.0853}}].

\bibitem{King:2010mq}
S.~F. King, G.~K. Leontaris, and G.~G. Ross, {\it {Family symmetries in
  F-theory GUTs}},  {\em Nucl. Phys.} {\bf B838} (2010) 119--135,
  [\href{http://xxx.lanl.gov/abs/1005.1025}{{\tt 1005.1025}}].

\bibitem{Krippendorf:2010hj}
S.~Krippendorf, M.~J. Dolan, A.~Maharana, and F.~Quevedo, {\it {D-branes at
  Toric Singularities: Model Building, Yukawa Couplings and Flavour Physics}},
  {\em JHEP} {\bf 06} (2010) 092,
  [\href{http://xxx.lanl.gov/abs/1002.1790}{{\tt 1002.1790}}].

\bibitem{Krippendorf:2015kta}
S.~Krippendorf, S.~Schafer-Nameki, and J.-M. Wong, {\it {Froggatt-Nielsen meets
  Mordell-Weil: A Phenomenological Survey of Global F-theory GUTs with U(1)s}},
   {\em JHEP} {\bf 11} (2015) 008,
  [\href{http://xxx.lanl.gov/abs/1507.05961}{{\tt 1507.05961}}].

\bibitem{Garcia-Etxebarria:2015wns}
I.~Garcia-Etxebarria and D.~Regalado, {\it {$ \mathcal{N}=3 $ four dimensional
  field theories}},  {\em JHEP} {\bf 03} (2016) 083,
  [\href{http://xxx.lanl.gov/abs/1512.06434}{{\tt 1512.06434}}].

\bibitem{MorrisonStevens}
D.~Morrison and G.~Stevens, {\it {Terminal quotient singularities in dimensions
  three and four}},  {\em Proc. Amer. Math. Soc.} {\bf 90, no. 1} (1984) 15Ð20.

\bibitem{Anno}
R.~E. Anno, {\it {Four-Dimensional Terminal Gorenstein Quotient
  Singularities}},  {\em Mathematical Notes} {\bf 73} (2003) 769.

\bibitem{Aspinwall:1998xj}
P.~S. Aspinwall and D.~R. Morrison, {\it {Nonsimply connected gauge groups and
  rational points on elliptic curves}},  {\em JHEP} {\bf 07} (1998) 012,
  [\href{http://xxx.lanl.gov/abs/hep-th/9805206}{{\tt hep-th/9805206}}].

\bibitem{Grimm:2010ez}
T.~W. Grimm and T.~Weigand, {\it {On Abelian Gauge Symmetries and Proton Decay
  in Global F-theory GUTs}},  {\em Phys. Rev.} {\bf D82} (2010) 086009,
  [\href{http://xxx.lanl.gov/abs/1006.0226}{{\tt 1006.0226}}].

\bibitem{Marsano:2009ym}
J.~Marsano, N.~Saulina, and S.~Schafer-Nameki, {\it {F-theory Compactifications
  for Supersymmetric GUTs}},  {\em JHEP} {\bf 08} (2009) 030,
  [\href{http://xxx.lanl.gov/abs/0904.3932}{{\tt 0904.3932}}].

\bibitem{Marsano:2009gv}
J.~Marsano, N.~Saulina, and S.~Schafer-Nameki, {\it {Monodromies, Fluxes, and
  Compact Three-Generation F-theory GUTs}},  {\em JHEP} {\bf 08} (2009) 046,
  [\href{http://xxx.lanl.gov/abs/0906.4672}{{\tt 0906.4672}}].

\bibitem{Marsano:2009wr}
J.~Marsano, N.~Saulina, and S.~Schafer-Nameki, {\it {Compact F-theory GUTs with
  $U(1)_{PQ}$}},  {\em JHEP} {\bf 04} (2010) 095,
  [\href{http://xxx.lanl.gov/abs/0912.0272}{{\tt 0912.0272}}].

\bibitem{Dolan:2011iu}
M.~J. Dolan, J.~Marsano, N.~Saulina, and S.~Schafer-Nameki, {\it {F-theory GUTs
  with U(1) Symmetries: Generalities and Survey}},  {\em Phys. Rev.}

\bibitem{Dolan:2011aq}
M.~J. Dolan, J.~Marsano, and S.~Schafer-Nameki, {\it {Unification and
  Phenomenology of F-Theory GUTs with $U(1)_{PQ}$}},  {\em JHEP} {\bf 12}
  (2011) 032, [\href{http://xxx.lanl.gov/abs/1109.4958}{{\tt 1109.4958}}].

\bibitem{Silverman:2008}
J.~H. Silverman, {\em {The Arithmetic of Elliptic Curves}}.
\newblock Springer, 2nd~ed., 2008.

\bibitem{MR715605}
S.~Lang, {\em Fundamentals of {D}iophantine geometry}.
\newblock Springer-Verlag, New York, 1983.

\bibitem{Mordell}
L.~Mordell, {\it {On the rational solutions of the indeterminate equation of
  the 3rd and 4th degrees}},  {\em Proc. Cambridge Philosophical Society} {\bf
  21} (1922) 179--192.

\bibitem{Weil}
A.~Weil, {\it {LÕarithm\'etique sur les courbes alg\'ebriques}},  {\em Acta
  Math} {\bf 52} (1929) 281Ð315.

\bibitem{MR0102520}
S.~Lang and A.~N{\'e}ron, {\it Rational points of abelian varieties over
  function fields},  {\em Amer. J. Math.} {\bf 81} (1959) 95--118.

\bibitem{2007arXiv0709.2908E}
N.~D. {Elkies}, {\it {Three lectures on elliptic surfaces and curves of high
  rank}},  {\em ArXiv e-prints} (Sept., 2007)
  [\href{http://xxx.lanl.gov/abs/0709.2908}{{\tt 0709.2908}}].

\bibitem{MR1813537}
I.~Shimada, {\it On elliptic {$K3$} surfaces},  {\em Michigan Math. J.} {\bf
  47} (2000) 423--446, [\href{http://xxx.lanl.gov/abs/math.AG/0505140}{{\tt
  math.AG/0505140}}].

\bibitem{Elkies}
N.~D. {Elkies}, {\it {K3 surfaces and elliptic fibrations in number theory}},
  {\em Talk at 2018 Banff Workshop "Geometry and Physics of F-theory",
  https://www.birs.ca/workshops/2018/}.

\bibitem{Shioda:1972}
T.~Shioda, {\it {On elliptic modular surfaces}},  {\em J. Math. Soc. Japan}
  {\bf 24} (1972).

\bibitem{Wazir:2001}
R.~Wazir, {\it Arithmetic on elliptic threefolds},  {\em Compos.Math.} {\bf
  140} (2001) 567--580, [\href{http://xxx.lanl.gov/abs/math.NT/0112259}{{\tt
  math.NT/0112259}}].

\bibitem{Grimm:2013oga}
T.~W. Grimm, A.~Kapfer, and J.~Keitel, {\it {Effective action of 6D F-Theory
  with U(1) factors: Rational sections make Chern-Simons terms jump}},  {\em
  JHEP} {\bf 07} (2013) 115, [\href{http://xxx.lanl.gov/abs/1305.1929}{{\tt
  1305.1929}}].

\bibitem{Grimm:2015zea}
T.~W. Grimm and A.~Kapfer, {\it {Anomaly Cancelation in Field Theory and
  F-theory on a Circle}},  {\em JHEP} {\bf 05} (2016) 102,
  [\href{http://xxx.lanl.gov/abs/1502.05398}{{\tt 1502.05398}}].

\bibitem{shioda}
T.~Shioda, {\it {Mordell-Weil Lattices and Galois Representation. I}},  {\em
  Proc. Japan Acad.} {\bf A65} (1989) 268--271.

\bibitem{Braun:2011zm}
A.~P. Braun, A.~Collinucci, and R.~Valandro, {\it {G-flux in F-theory and
  algebraic cycles}},  {\em Nucl. Phys.} {\bf B856} (2012) 129--179,
  [\href{http://xxx.lanl.gov/abs/1107.5337}{{\tt 1107.5337}}].

\bibitem{Candelas:1990pi}
P.~Candelas and X.~de~la Ossa, {\it {Moduli Space of {Calabi-Yau} Manifolds}},
  {\em Nucl. Phys.} {\bf B355} (1991) 455--481.

\bibitem{Morrison:2012ei}
D.~R. Morrison and D.~S. Park, {\it {F-Theory and the Mordell-Weil Group of
  Elliptically-Fibered Calabi-Yau Threefolds}},  {\em JHEP} {\bf 1210} (2012)
  128, [\href{http://xxx.lanl.gov/abs/1208.2695}{{\tt 1208.2695}}].

\bibitem{Lee:2018ihr}
S.-J. Lee, D.~Regalado, and T.~Weigand, {\it {6d SCFTs and U(1) Flavour
  Symmetries}},  \href{http://xxx.lanl.gov/abs/1803.07998}{{\tt 1803.07998}}.

\bibitem{Grimm:2011tb}
T.~W. Grimm, M.~Kerstan, E.~Palti, and T.~Weigand, {\it {Massive Abelian Gauge
  Symmetries and Fluxes in F-theory}},  {\em JHEP} {\bf 12} (2011) 004,
  [\href{http://xxx.lanl.gov/abs/1107.3842}{{\tt 1107.3842}}].

\bibitem{Borchmann:2013jwa}
J.~Borchmann, C.~Mayrhofer, E.~Palti, and T.~Weigand, {\it {Elliptic fibrations
  for $SU(5)\times U(1)\times U(1)$ F-theory vacua}},  {\em Phys. Rev.} {\bf
  D88} (2013), no.~4 046005, [\href{http://xxx.lanl.gov/abs/1303.5054}{{\tt
  1303.5054}}].

\bibitem{Cvetic:2013nia}
M.~Cveti{\v c}, D.~Klevers, and H.~Piragua, {\it {F-Theory Compactifications
  with Multiple U(1)-Factors: Constructing Elliptic Fibrations with Rational
  Sections}},  {\em JHEP} {\bf 06} (2013) 067,
  [\href{http://xxx.lanl.gov/abs/1303.6970}{{\tt 1303.6970}}].

\bibitem{Cvetic:2013uta}
M.~Cveti{\v c}, A.~Grassi, D.~Klevers, and H.~Piragua, {\it {Chiral
  Four-Dimensional F-Theory Compactifications With SU(5) and Multiple
  U(1)-Factors}},  {\em JHEP} {\bf 04} (2014) 010,
  [\href{http://xxx.lanl.gov/abs/1306.3987}{{\tt 1306.3987}}].

\bibitem{Borchmann:2013hta}
J.~Borchmann, C.~Mayrhofer, E.~Palti, and T.~Weigand, {\it {SU(5) Tops with
  Multiple U(1)s in F-theory}},  {\em Nucl. Phys.} {\bf B882} (2014) 1--69,
  [\href{http://xxx.lanl.gov/abs/1307.2902}{{\tt 1307.2902}}].

\bibitem{Cvetic:2013qsa}
M.~Cvetic, D.~Klevers, H.~Piragua, and P.~Song, {\it {Elliptic fibrations with
  rank three Mordell-Weil group: F-theory with U(1) x U(1) x U(1) gauge
  symmetry}},  {\em JHEP} {\bf 03} (2014) 021,
  [\href{http://xxx.lanl.gov/abs/1310.0463}{{\tt 1310.0463}}].

\bibitem{Mayrhofer:2014haa}
C.~Mayrhofer, E.~Palti, O.~Till, and T.~Weigand, {\it {Discrete Gauge
  Symmetries by Higgsing in four-dimensional F-Theory Compactifications}},
  {\em JHEP} {\bf 12} (2014) 068,
  [\href{http://xxx.lanl.gov/abs/1408.6831}{{\tt 1408.6831}}].

\bibitem{Mayrhofer:2014laa}
C.~Mayrhofer, E.~Palti, O.~Till, and T.~Weigand, {\it {On Discrete Symmetries
  and Torsion Homology in F-Theory}},  {\em JHEP} {\bf 06} (2015) 029,
  [\href{http://xxx.lanl.gov/abs/1410.7814}{{\tt 1410.7814}}].

\bibitem{Grassi:2012qw}
A.~Grassi and V.~Perduca, {\it {Weierstrass models of elliptic toric $K3$
  hypersurfaces and symplectic cuts}},  {\em Adv. Theor. Math. Phys.} {\bf 17}
  (2013), no.~4 741--770, [\href{http://xxx.lanl.gov/abs/1201.0930}{{\tt
  1201.0930}}].

\bibitem{Braun:2013nqa}
V.~Braun, T.~W. Grimm, and J.~Keitel, {\it {Geometric Engineering in Toric
  F-Theory and GUTs with U(1) Gauge Factors}},  {\em JHEP} {\bf 12} (2013) 069,
  [\href{http://xxx.lanl.gov/abs/1306.0577}{{\tt 1306.0577}}].

\bibitem{Klevers:2014bqa}
D.~Klevers, D.~K. Mayorga~Pena, P.-K. Oehlmann, H.~Piragua, and J.~Reuter, {\it
  {F-Theory on all Toric Hypersurface Fibrations and its Higgs Branches}},
  {\em JHEP} {\bf 01} (2015) 142,
  [\href{http://xxx.lanl.gov/abs/1408.4808}{{\tt 1408.4808}}].

\bibitem{Morrison:2016xkb}
D.~R. Morrison and D.~S. Park, {\it {Tall sections from non-minimal
  transformations}},  {\em JHEP} {\bf 10} (2016) 033,
  [\href{http://xxx.lanl.gov/abs/1606.07444}{{\tt 1606.07444}}].

\bibitem{Raghuram:2017qut}
N.~Raghuram, {\it {Abelian F-theory Models with Charge-3 and Charge-4 Matter}},
   \href{http://xxx.lanl.gov/abs/1711.03210}{{\tt 1711.03210}}.

\bibitem{Braun:2014qka}
V.~Braun, T.~W. Grimm, and J.~Keitel, {\it {Complete Intersection Fibers in
  F-Theory}},  {\em JHEP} {\bf 03} (2015) 125,
  [\href{http://xxx.lanl.gov/abs/1411.2615}{{\tt 1411.2615}}].

\bibitem{Anderson:2018pui}
L.~B. Anderson and M.~Karkheiran, {\it {TASI Lectures on Geometric Tools for
  String Compactifications}},
\newblock \href{http://xxx.lanl.gov/abs/1804.08792}{{\tt 1804.08792}}.

\bibitem{Oguiso}
K.~Oguiso, {\it {On algebraic fiber space structures on a Calabi-Yau 3-fold}},
  {\em Internat. J. Math. 4} {\bf 3} (1993) 439--465.

\bibitem{Wilson}
P.~Wilson, {\it {The existence of elliptic fibre space structures on Calabi-Yau
  threefolds}},  {\em Math. Ann.} {\bf 300} (1994) 693.

\bibitem{Kollar}
J.~Kollar, {\it {Deformations of elliptic Calabi-Yau manifolds}},
  \href{http://xxx.lanl.gov/abs/1206.5721 [math.AG]}{{\tt 1206.5721
  [math.AG]}}.

\bibitem{Anderson:2016cdu}
L.~B. Anderson, X.~Gao, J.~Gray, and S.-J. Lee, {\it {Multiple Fibrations in
  Calabi-Yau Geometry and String Dualities}},  {\em JHEP} {\bf 10} (2016) 105,
  [\href{http://xxx.lanl.gov/abs/1608.07555}{{\tt 1608.07555}}].

\bibitem{Anderson:2017aux}
L.~B. Anderson, X.~Gao, J.~Gray, and S.-J. Lee, {\it {Fibrations in CICY
  Threefolds}},  {\em JHEP} {\bf 10} (2017) 077,
  [\href{http://xxx.lanl.gov/abs/1708.07907}{{\tt 1708.07907}}].

\bibitem{Gray:2014fla}
J.~Gray, A.~S. Haupt, and A.~Lukas, {\it {Topological Invariants and Fibration
  Structure of Complete Intersection Calabi-Yau Four-Folds}},  {\em JHEP} {\bf
  09} (2014) 093, [\href{http://xxx.lanl.gov/abs/1405.2073}{{\tt 1405.2073}}].

\bibitem{Johnson:2014xpa}
S.~B. Johnson and W.~Taylor, {\it {Calabi-Yau threefolds with large
  $h^{2,1}$}},  {\em JHEP} {\bf 10} (2014) 23,
  [\href{http://xxx.lanl.gov/abs/1406.0514}{{\tt 1406.0514}}].

\bibitem{Johnson:2016qar}
S.~B. Johnson and W.~Taylor, {\it {Enhanced gauge symmetry in 6D F-theory
  models and tuned elliptic Calabi-Yau threefolds}},  {\em Fortsch. Phys.} {\bf
  64} (2016) 581--644, [\href{http://xxx.lanl.gov/abs/1605.08052}{{\tt
  1605.08052}}].

\bibitem{Anderson:2016ler}
L.~B. Anderson, X.~Gao, J.~Gray, and S.-J. Lee, {\it {Tools for CICYs in
  F-theory}},  {\em JHEP} {\bf 11} (2016) 004,
  [\href{http://xxx.lanl.gov/abs/1608.07554}{{\tt 1608.07554}}].

\bibitem{Lin:2014qga}
L.~Lin and T.~Weigand, {\it {Towards the Standard Model in F-theory}},  {\em
  Fortsch. Phys.} {\bf 63} (2015), no.~2 55--104,
  [\href{http://xxx.lanl.gov/abs/1406.6071}{{\tt 1406.6071}}].

\bibitem{Cvetic:2015txa}
M.~Cvetic, D.~Klevers, D.~K.~M. Pe–a, P.-K. Oehlmann, and J.~Reuter, {\it
  {Three-Family Particle Physics Models from Global F-theory
  Compactifications}},  {\em JHEP} {\bf 08} (2015) 087,
  [\href{http://xxx.lanl.gov/abs/1503.02068}{{\tt 1503.02068}}].

\bibitem{Mayrhofer:2012zy}
C.~Mayrhofer, E.~Palti, and T.~Weigand, {\it {U(1) symmetries in F-theory GUTs
  with multiple sections}},  {\em JHEP} {\bf 03} (2013) 098,
  [\href{http://xxx.lanl.gov/abs/1211.6742}{{\tt 1211.6742}}].

\bibitem{Braun:2013yti}
V.~Braun, T.~W. Grimm, and J.~Keitel, {\it {New Global F-theory GUTs with U(1)
  symmetries}},  {\em JHEP} {\bf 09} (2013) 154,
  [\href{http://xxx.lanl.gov/abs/1302.1854}{{\tt 1302.1854}}].

\bibitem{Kuntzler:2014ila}
M.~Kuntzler and S.~Schafer-Nameki, {\it {Tate Trees for Elliptic Fibrations
  with Rank one Mordell-Weil group}},
  \href{http://xxx.lanl.gov/abs/1406.5174}{{\tt 1406.5174}}.

\bibitem{Lawrie:2014uya}
C.~Lawrie and D.~Sacco, {\it {TateÕs algorithm for F-theory GUTs with two
  U(1)s}},  {\em JHEP} {\bf 03} (2015) 055,
  [\href{http://xxx.lanl.gov/abs/1412.4125}{{\tt 1412.4125}}].

\bibitem{Lawrie:2015hia}
C.~Lawrie, S.~Schafer-Nameki, and J.-M. Wong, {\it {F-theory and All Things
  Rational: Surveying U(1) Symmetries with Rational Sections}},  {\em JHEP}
  {\bf 09} (2015) 144, [\href{http://xxx.lanl.gov/abs/1504.05593}{{\tt
  1504.05593}}].

\bibitem{Mayrhofer:2014opa}
C.~Mayrhofer, D.~R. Morrison, O.~Till, and T.~Weigand, {\it {Mordell-Weil
  Torsion and the Global Structure of Gauge Groups in F-theory}},  {\em JHEP}
  {\bf 10} (2014) 16, [\href{http://xxx.lanl.gov/abs/1405.3656}{{\tt
  1405.3656}}].

\bibitem{Baume:2017hxm}
F.~Baume, M.~Cvetic, C.~Lawrie, and L.~Lin, {\it {When rational sections become
  cyclic Ñ Gauge enhancement in F-theory via Mordell-Weil torsion}},  {\em
  JHEP} {\bf 03} (2018) 069, [\href{http://xxx.lanl.gov/abs/1709.07453}{{\tt
  1709.07453}}].

\bibitem{Grimm:2015wda}
T.~W. Grimm, A.~Kapfer, and D.~Klevers, {\it {The Arithmetic of Elliptic
  Fibrations in Gauge Theories on a Circle}},  {\em JHEP} {\bf 06} (2016) 112,
  [\href{http://xxx.lanl.gov/abs/1510.04281}{{\tt 1510.04281}}].

\bibitem{Cvetic:2017epq}
M.~Cvetic and L.~Lin, {\it {The Global Gauge Group Structure of F-theory
  Compactification with U(1)s}},  {\em JHEP} {\bf 01} (2018) 157,
  [\href{http://xxx.lanl.gov/abs/1706.08521}{{\tt 1706.08521}}].

\bibitem{Dreiner:2005rd}
H.~K. Dreiner, C.~Luhn, and M.~Thormeier, {\it {What is the discrete gauge
  symmetry of the MSSM?}},  {\em Phys. Rev.} {\bf D73} (2006) 075007,
  [\href{http://xxx.lanl.gov/abs/hep-ph/0512163}{{\tt hep-ph/0512163}}].

\bibitem{Banks:2010zn}
T.~Banks and N.~Seiberg, {\it {Symmetries and Strings in Field Theory and
  Gravity}},  {\em Phys. Rev.} {\bf D83} (2011) 084019,
  [\href{http://xxx.lanl.gov/abs/1011.5120}{{\tt 1011.5120}}].

\bibitem{Morrison:2014era}
D.~R. Morrison and W.~Taylor, {\it {Sections, multisections, and U(1) fields in
  F-theory}},  \href{http://xxx.lanl.gov/abs/1404.1527}{{\tt 1404.1527}}.

\bibitem{Anderson:2014yva}
L.~B. Anderson, I.~Garcia-Etxebarria, T.~W. Grimm, and J.~Keitel, {\it {Physics
  of F-theory compactifications without section}},  {\em JHEP} {\bf 12} (2014)
  156, [\href{http://xxx.lanl.gov/abs/1406.5180}{{\tt 1406.5180}}].

\bibitem{Garcia-Etxebarria:2014qua}
I.~Garcia-Etxebarria, T.~W. Grimm, and J.~Keitel, {\it {Yukawas and discrete
  symmetries in F-theory compactifications without section}},  {\em JHEP} {\bf
  11} (2014) 125, [\href{http://xxx.lanl.gov/abs/1408.6448}{{\tt 1408.6448}}].

\bibitem{Cvetic:2015moa}
M.~Cveti{\v c}, R.~Donagi, D.~Klevers, H.~Piragua, and M.~Poretschkin, {\it
  {F-theory vacua with $\mathbb Z_3$ gauge symmetry}},  {\em Nucl. Phys.} {\bf
  B898} (2015) 736--750, [\href{http://xxx.lanl.gov/abs/1502.06953}{{\tt
  1502.06953}}].

\bibitem{Camara:2011jg}
P.~G. Camara, L.~E. Ibanez, and F.~Marchesano, {\it {RR photons}},  {\em JHEP}
  {\bf 09} (2011) 110, [\href{http://xxx.lanl.gov/abs/1106.0060}{{\tt
  1106.0060}}].

\bibitem{BerasaluceGonzalez:2011wy}
M.~Berasaluce-Gonzalez, L.~E. Ibanez, P.~Soler, and A.~M. Uranga, {\it
  {Discrete gauge symmetries in D-brane models}},  {\em JHEP} {\bf 12} (2011)
  113, [\href{http://xxx.lanl.gov/abs/1106.4169}{{\tt 1106.4169}}].

\bibitem{Ibanez:2012wg}
L.~E. Ibanez, A.~N. Schellekens, and A.~M. Uranga, {\it {Discrete Gauge
  Symmetries in Discrete MSSM-like Orientifolds}},  {\em Nucl. Phys.} {\bf
  B865} (2012) 509--540, [\href{http://xxx.lanl.gov/abs/1205.5364}{{\tt
  1205.5364}}].

\bibitem{Honecker:2013sww}
G.~Honecker and W.~Staessens, {\it {To Tilt or Not To Tilt: Discrete Gauge
  Symmetries in Global Intersecting D-Brane Models}},  {\em JHEP} {\bf 10}
  (2013) 146, [\href{http://xxx.lanl.gov/abs/1303.4415}{{\tt 1303.4415}}].

\bibitem{1992alg.geom.10009D}
I.~{Dolgachev} and M.~{Gross}, {\it {Elliptic Three-folds I: Ogg-Shafarevich
  Theory}},  in {\em eprint arXiv:alg-geom/9210009}, Oct., 1992.

\bibitem{Kimura:2015qpz}
Y.~Kimura, {\it {Gauge Groups and Matter Fields on Some Models of F-theory
  without Section}},  {\em JHEP} {\bf 03} (2016) 042,
  [\href{http://xxx.lanl.gov/abs/1511.06912}{{\tt 1511.06912}}].

\bibitem{Kimura:2016crs}
Y.~Kimura, {\it {Discrete Gauge Groups in F-theory Models on Genus-One Fibered
  Calabi-Yau 4-folds without Section}},  {\em JHEP} {\bf 04} (2017) 168,
  [\href{http://xxx.lanl.gov/abs/1608.07219}{{\tt 1608.07219}}].

\bibitem{Oehlmann:2016wsb}
P.-K. Oehlmann, J.~Reuter, and T.~Schimannek, {\it {Mordell-Weil Torsion in the
  Mirror of Multi-Sections}},  {\em JHEP} {\bf 12} (2016) 031,
  [\href{http://xxx.lanl.gov/abs/1604.00011}{{\tt 1604.00011}}].

\bibitem{Anderson:2018heq}
L.~B. Anderson, A.~Grassi, J.~Gray, and P.-K. Oehlmann, {\it {F-theory on
  Quotient Threefolds with (2,0) Discrete Superconformal Matter}},
  \href{http://xxx.lanl.gov/abs/1801.08658}{{\tt 1801.08658}}.

\bibitem{Lin:2015qsa}
L.~Lin, C.~Mayrhofer, O.~Till, and T.~Weigand, {\it {Fluxes in F-theory
  Compactifications on Genus-One Fibrations}},  {\em JHEP} {\bf 01} (2016) 098,
  [\href{http://xxx.lanl.gov/abs/1508.00162}{{\tt 1508.00162}}].

\bibitem{Witten:1996md}
E.~Witten, {\it {On flux quantization in M theory and the effective action}},
  {\em J. Geom. Phys.} {\bf 22} (1997) 1--13,
  [\href{http://xxx.lanl.gov/abs/hep-th/9609122}{{\tt hep-th/9609122}}].

\bibitem{Becker:1996gj}
K.~Becker and M.~Becker, {\it {M theory on eight manifolds}},  {\em Nucl.
  Phys.} {\bf B477} (1996) 155--167,
  [\href{http://xxx.lanl.gov/abs/hep-th/9605053}{{\tt hep-th/9605053}}].

\bibitem{Gukov:1999ya}
S.~Gukov, C.~Vafa, and E.~Witten, {\it {CFT's from Calabi-Yau four folds}},
  {\em Nucl. Phys.} {\bf B584} (2000) 69--108,
  [\href{http://xxx.lanl.gov/abs/hep-th/9906070}{{\tt hep-th/9906070}}].
  [Erratum: Nucl. Phys.B608,477(2001)].

\bibitem{Haack:2001jz}
M.~Haack and J.~Louis, {\it {M theory compactified on Calabi-Yau fourfolds with
  background flux}},  {\em Phys. Lett.} {\bf B507} (2001) 296--304,
  [\href{http://xxx.lanl.gov/abs/hep-th/0103068}{{\tt hep-th/0103068}}].

\bibitem{Haupt:2008nu}
A.~S. Haupt, A.~Lukas, and K.~S. Stelle, {\it {M-theory on Calabi-Yau
  Five-Folds}},  {\em JHEP} {\bf 05} (2009) 069,
  [\href{http://xxx.lanl.gov/abs/0810.2685}{{\tt 0810.2685}}].

\bibitem{Greiner:2015mdm}
S.~Greiner and T.~W. Grimm, {\it {On Mirror Symmetry for Calabi-Yau Fourfolds
  with Three-Form Cohomology}},  {\em JHEP} {\bf 09} (2016) 073,
  [\href{http://xxx.lanl.gov/abs/1512.04859}{{\tt 1512.04859}}].

\bibitem{Greiner:2017ery}
S.~Greiner and T.~W. Grimm, {\it {Three-form periods on Calabi-Yau fourfolds:
  Toric hypersurfaces and F-theory applications}},  {\em JHEP} {\bf 05} (2017)
  151, [\href{http://xxx.lanl.gov/abs/1702.03217}{{\tt 1702.03217}}].

\bibitem{Curio:1998bva}
G.~Curio and R.~Y. Donagi, {\it {Moduli in N=1 heterotic / F theory duality}},
  {\em Nucl. Phys.} {\bf B518} (1998) 603--631,
  [\href{http://xxx.lanl.gov/abs/hep-th/9801057}{{\tt hep-th/9801057}}].

\bibitem{Donagi:2003hh}
R.~Donagi, S.~Katz, and E.~Sharpe, {\it {Spectra of D-branes with higgs vevs}},
   {\em Adv. Theor. Math. Phys.} {\bf 8} (2004), no.~5 813--859,
  [\href{http://xxx.lanl.gov/abs/hep-th/0309270}{{\tt hep-th/0309270}}].

\bibitem{Donagi:2011jy}
R.~Donagi and M.~Wijnholt, {\it {Gluing Branes, I}},  {\em JHEP} {\bf 05}
  (2013) 068, [\href{http://xxx.lanl.gov/abs/1104.2610}{{\tt 1104.2610}}].

\bibitem{Donagi:2011dv}
R.~Donagi and M.~Wijnholt, {\it {Gluing Branes II: Flavour Physics and String
  Duality}},  {\em JHEP} {\bf 05} (2013) 092,
  [\href{http://xxx.lanl.gov/abs/1112.4854}{{\tt 1112.4854}}].

\bibitem{Collinucci:2014qfa}
A.~Collinucci and R.~Savelli, {\it {T-branes as branes within branes}},  {\em
  JHEP} {\bf 09} (2015) 161, [\href{http://xxx.lanl.gov/abs/1410.4178}{{\tt
  1410.4178}}].

\bibitem{Collinucci:2016hpz}
A.~Collinucci, S.~Giacomelli, R.~Savelli, and R.~Valandro, {\it {T-branes
  through 3d mirror symmetry}},  {\em JHEP} {\bf 07} (2016) 093,
  [\href{http://xxx.lanl.gov/abs/1603.00062}{{\tt 1603.00062}}].

\bibitem{Bena:2016oqr}
I.~Bena, J.~BlŒbŠck, R.~Minasian, and R.~Savelli, {\it {There and back again: A
  T-brane's tale}},  {\em JHEP} {\bf 11} (2016) 179,
  [\href{http://xxx.lanl.gov/abs/1608.01221}{{\tt 1608.01221}}].

\bibitem{Marchesano:2016cqg}
F.~Marchesano and S.~Schwieger, {\it {T-branes and $\alpha'$-corrections}},
  {\em JHEP} {\bf 11} (2016) 123,
  [\href{http://xxx.lanl.gov/abs/1609.02799}{{\tt 1609.02799}}].

\bibitem{Mekareeya:2016yal}
N.~Mekareeya, T.~Rudelius, and A.~Tomasiello, {\it {T-branes, Anomalies and
  Moduli Spaces in 6D SCFTs}},  {\em JHEP} {\bf 10} (2017) 158,
  [\href{http://xxx.lanl.gov/abs/1612.06399}{{\tt 1612.06399}}].

\bibitem{Marchesano:2017kke}
F.~Marchesano, R.~Savelli, and S.~Schwieger, {\it {Compact T-branes}},  {\em
  JHEP} {\bf 09} (2017) 132, [\href{http://xxx.lanl.gov/abs/1707.03797}{{\tt
  1707.03797}}].

\bibitem{Anderson:2013rka}
L.~B. Anderson, J.~J. Heckman, and S.~Katz, {\it {T-Branes and Geometry}},
  {\em JHEP} {\bf 05} (2014) 080,
  [\href{http://xxx.lanl.gov/abs/1310.1931}{{\tt 1310.1931}}].

\bibitem{Anderson:2017rpr}
L.~B. Anderson, J.~J. Heckman, S.~Katz, and L.~Schaposnik, {\it {T-Branes at
  the Limits of Geometry}},  {\em JHEP} {\bf 10} (2017) 058,
  [\href{http://xxx.lanl.gov/abs/1702.06137}{{\tt 1702.06137}}].

\bibitem{Collinucci:2014taa}
A.~Collinucci and R.~Savelli, {\it {F-theory on singular spaces}},  {\em JHEP}
  {\bf 09} (2015) 100, [\href{http://xxx.lanl.gov/abs/1410.4867}{{\tt
  1410.4867}}].

\bibitem{Greene:1993vm}
B.~R. Greene, D.~R. Morrison, and M.~R. Plesser, {\it {Mirror manifolds in
  higher dimension}},  {\em Commun. Math. Phys.} {\bf 173} (1995) 559--598,
  [\href{http://xxx.lanl.gov/abs/hep-th/9402119}{{\tt hep-th/9402119}}].
  [AMS/IP Stud. Adv. Math.1,745(1996)].

\bibitem{Braun:2014xka}
A.~P. Braun and T.~Watari, {\it {The Vertical, the Horizontal and the Rest:
  anatomy of the middle cohomology of Calabi-Yau fourfolds and F-theory
  applications}},  {\em JHEP} {\bf 01} (2015) 047,
  [\href{http://xxx.lanl.gov/abs/1408.6167}{{\tt 1408.6167}}].

\bibitem{Dasgupta:1999ss}
K.~Dasgupta, G.~Rajesh, and S.~Sethi, {\it {M theory, orientifolds and G -
  flux}},  {\em JHEP} {\bf 08} (1999) 023,
  [\href{http://xxx.lanl.gov/abs/hep-th/9908088}{{\tt hep-th/9908088}}].

\bibitem{Alim:2009bx}
M.~Alim, M.~Hecht, H.~Jockers, P.~Mayr, A.~Mertens, and M.~Soroush, {\it {Hints
  for Off-Shell Mirror Symmetry in type II/F-theory Compactifications}},  {\em
  Nucl. Phys.} {\bf B841} (2010) 303--338,
  [\href{http://xxx.lanl.gov/abs/0909.1842}{{\tt 0909.1842}}].

\bibitem{Grimm:2009ef}
T.~W. Grimm, T.-W. Ha, A.~Klemm, and D.~Klevers, {\it {Computing Brane and Flux
  Superpotentials in F-theory Compactifications}},  {\em JHEP} {\bf 04} (2010)
  015, [\href{http://xxx.lanl.gov/abs/0909.2025}{{\tt 0909.2025}}].

\bibitem{Jockers:2009ti}
H.~Jockers, P.~Mayr, and J.~Walcher, {\it {On N=1 4d Effective Couplings for
  F-theory and Heterotic Vacua}},  {\em Adv. Theor. Math. Phys.} {\bf 14}
  (2010), no.~5 1433--1514, [\href{http://xxx.lanl.gov/abs/0912.3265}{{\tt
  0912.3265}}].

\bibitem{Lerche:1997zb}
W.~Lerche, {\it {Fayet-Iliopoulos potentials from four folds}},  {\em JHEP}
  {\bf 11} (1997) 004, [\href{http://xxx.lanl.gov/abs/hep-th/9709146}{{\tt
  hep-th/9709146}}].

\bibitem{Sethi:1996es}
S.~Sethi, C.~Vafa, and E.~Witten, {\it {Constraints on low dimensional string
  compactifications}},  {\em Nucl. Phys.} {\bf B480} (1996) 213--224,
  [\href{http://xxx.lanl.gov/abs/hep-th/9606122}{{\tt hep-th/9606122}}].

\bibitem{Dasgupta:1996yh}
K.~Dasgupta and S.~Mukhi, {\it {A Note on low dimensional string
  compactifications}},  {\em Phys. Lett.} {\bf B398} (1997) 285--290,
  [\href{http://xxx.lanl.gov/abs/hep-th/9612188}{{\tt hep-th/9612188}}].

\bibitem{Bies:2017fam}
M.~Bies, C.~Mayrhofer, and T.~Weigand, {\it {Gauge Backgrounds and Zero-Mode
  Counting in F-Theory}},  {\em JHEP} {\bf 11} (2017) 081,
  [\href{http://xxx.lanl.gov/abs/1706.04616}{{\tt 1706.04616}}].

\bibitem{Jockers:2016bwi}
H.~Jockers, S.~Katz, D.~R. Morrison, and M.~R. Plesser, {\it {SU(N) Transitions
  in M-Theory on CalabiÐYau Fourfolds and Background Fluxes}},  {\em Commun.
  Math. Phys.} {\bf 351} (2017), no.~2 837--871,
  [\href{http://xxx.lanl.gov/abs/1602.07693}{{\tt 1602.07693}}].

\bibitem{Braun:2014ola}
A.~P. Braun, Y.~Kimura, and T.~Watari, {\it {The Noether-Lefschetz problem and
  gauge-group-resolved landscapes: F-theory on K3 $\times$ K3 as a test case}},
   {\em JHEP} {\bf 04} (2014) 050,
  [\href{http://xxx.lanl.gov/abs/1401.5908}{{\tt 1401.5908}}].

\bibitem{Bizet:2014uua}
N.~Cabo~Bizet, A.~Klemm, and D.~Vieira~Lopes, {\it {Landscaping with fluxes and
  the E8 Yukawa Point in F-theory}},
  \href{http://xxx.lanl.gov/abs/1404.7645}{{\tt 1404.7645}}.

\bibitem{Cota:2017aal}
C.~F. Cota, A.~Klemm, and T.~Schimannek, {\it {Modular Amplitudes and
  Flux-Superpotentials on elliptic Calabi-Yau fourfolds}},  {\em JHEP} {\bf 01}
  (2018) 086, [\href{http://xxx.lanl.gov/abs/1709.02820}{{\tt 1709.02820}}].

\bibitem{Lin:2016vus}
L.~Lin and T.~Weigand, {\it {G 4 -flux and standard model vacua in F-theory}},
  {\em Nucl. Phys.} {\bf B913} (2016) 209--247,
  [\href{http://xxx.lanl.gov/abs/1604.04292}{{\tt 1604.04292}}].

\bibitem{Lin:2016zha}
L.~Lin, {\em {Gauge fluxes in F-theory compactifications}}.
\newblock PhD thesis, Inst. Appl. Math., Heidelberg, 2016.

\bibitem{Braun:2014pva}
A.~P. Braun, A.~Collinucci, and R.~Valandro, {\it {Hypercharge flux in F-theory
  and the stable Sen limit}},  {\em JHEP} {\bf 07} (2014) 121,
  [\href{http://xxx.lanl.gov/abs/1402.4096}{{\tt 1402.4096}}].

\bibitem{Mayrhofer:2013ara}
C.~Mayrhofer, E.~Palti, and T.~Weigand, {\it {Hypercharge Flux in IIB and
  F-theory: Anomalies and Gauge Coupling Unification}},  {\em JHEP} {\bf 09}
  (2013) 082, [\href{http://xxx.lanl.gov/abs/1303.3589}{{\tt 1303.3589}}].

\bibitem{Buican:2006sn}
M.~Buican, D.~Malyshev, D.~R. Morrison, H.~Verlinde, and M.~Wijnholt, {\it
  {D-branes at Singularities, Compactification, and Hypercharge}},  {\em JHEP}
  {\bf 01} (2007) 107, [\href{http://xxx.lanl.gov/abs/hep-th/0610007}{{\tt
  hep-th/0610007}}].

\bibitem{Blumenhagen:2010ed}
R.~Blumenhagen, B.~Jurke, T.~Rahn, and H.~Roschy, {\it {Cohomology of Line
  Bundles: Applications}},  {\em J. Math. Phys.} {\bf 53} (2012) 012302,
  [\href{http://xxx.lanl.gov/abs/1010.3717}{{\tt 1010.3717}}].

\bibitem{Blumenhagen:2011xn}
R.~Blumenhagen, B.~Jurke, and T.~Rahn, {\it {Computational Tools for Cohomology
  of Toric Varieties}},  {\em Adv. High Energy Phys.} {\bf 2011} (2011) 152749,
  [\href{http://xxx.lanl.gov/abs/1104.1187}{{\tt 1104.1187}}].

\bibitem{2010arXiv1003.1943B}
M.~Barakat and M.~Lange-Hegermann, {\it An axiomatic setup for algorithmic
  homological algebra and an alternative approach to localization},  {\em
  J.~Algebra Appl.} {\bf 10} (2011), no.~2 269--293.
  (\href{http://arxiv.org/abs/1003.1943}{\texttt{arXiv:1003.1943}}).

\bibitem{2012arXiv1202.3337B}
M.~Barakat and M.~Lange-Hegermann, {\it On monads of exact reflective
  localizations of {A}belian categories},  {\em Homology Homotopy Appl.} {\bf
  15} (2013), no.~2 145--151.
  (\href{http://arxiv.org/abs/1202.3337}{\texttt{arXiv:1202.3337}}).

\bibitem{2014arXiv1409.6100B}
M.~Barakat and M.~Lange-Hegermann, {\it A constructive approach to the module
  of twisted global sections on relative projective spaces},  in {\em
  Algorithmic and Experimental Methods in Algebra, Geometry, and Number
  Theory}, Springer.
\newblock 2017.
\newblock To appear,
  \href{http://arxiv.org/abs/1409.6100}{\texttt{arXiv:1409.6100}}.

\bibitem{Bies:2018uzw}
M.~Bies, {\em {Cohomologies of coherent sheaves and massless spectra in
  F-theory}}.
\newblock PhD thesis, Heidelberg U., 2018-02.
\newblock \href{http://xxx.lanl.gov/abs/1802.08860}{{\tt 1802.08860}}.

\bibitem{Heckman:2010bq}
J.~J. Heckman, {\it {Particle Physics Implications of F-theory}},  {\em Ann.
  Rev. Nucl. Part. Sci.} {\bf 60} (2010) 237--265,
  [\href{http://xxx.lanl.gov/abs/1001.0577}{{\tt 1001.0577}}].

\bibitem{Maharana:2012tu}
A.~Maharana and E.~Palti, {\it {Models of Particle Physics from Type IIB String
  Theory and F-theory: A Review}},  {\em Int. J. Mod. Phys.} {\bf A28} (2013)
  1330005, [\href{http://xxx.lanl.gov/abs/1212.0555}{{\tt 1212.0555}}].

\bibitem{Leontaris:2012mh}
G.~K. Leontaris, {\it {Aspects of F-Theory GUTs}},  {\em PoS} {\bf CORFU2011}
  (2011) 095, [\href{http://xxx.lanl.gov/abs/1203.6277}{{\tt 1203.6277}}].

\bibitem{Callaghan:2012rv}
J.~C. Callaghan and S.~F. King, {\it {E6 Models from F-theory}},  {\em JHEP}
  {\bf 04} (2013) 034, [\href{http://xxx.lanl.gov/abs/1210.6913}{{\tt
  1210.6913}}].

\bibitem{Callaghan:2013kaa}
J.~C. Callaghan, S.~F. King, and G.~K. Leontaris, {\it {Gauge coupling
  unification in $E_6$ F-theory GUTs with matter and bulk exotics from flux
  breaking}},  {\em JHEP} {\bf 12} (2013) 037,
  [\href{http://xxx.lanl.gov/abs/1307.4593}{{\tt 1307.4593}}].

\bibitem{Baume:2015wia}
F.~Baume, E.~Palti, and S.~Schwieger, {\it {On $E_8$ and F-Theory GUTs}},  {\em
  JHEP} {\bf 06} (2015) 039, [\href{http://xxx.lanl.gov/abs/1502.03878}{{\tt
  1502.03878}}].

\bibitem{Blumenhagen:2008aw}
R.~Blumenhagen, {\it {Gauge Coupling Unification in F-Theory Grand Unified
  Theories}},  {\em Phys. Rev. Lett.} {\bf 102} (2009) 071601,
  [\href{http://xxx.lanl.gov/abs/0812.0248}{{\tt 0812.0248}}].

\bibitem{Ibanez:2012zg}
L.~E. Ibanez, F.~Marchesano, D.~Regalado, and I.~Valenzuela, {\it {The
  Intermediate Scale MSSM, the Higgs Mass and F-theory Unification}},  {\em
  JHEP} {\bf 07} (2012) 195, [\href{http://xxx.lanl.gov/abs/1206.2655}{{\tt
  1206.2655}}].

\bibitem{Hebecker:2014uaa}
A.~Hebecker and J.~Unwin, {\it {Precision Unification and Proton Decay in
  F-Theory GUTs with High Scale Supersymmetry}},  {\em JHEP} {\bf 09} (2014)
  125, [\href{http://xxx.lanl.gov/abs/1405.2930}{{\tt 1405.2930}}].

\bibitem{Blumenhagen:2006ci}
R.~Blumenhagen, B.~Kors, D.~Lust, and S.~Stieberger, {\it {Four-dimensional
  String Compactifications with D-Branes, Orientifolds and Fluxes}},  {\em
  Phys. Rept.} {\bf 445} (2007) 1--193,
  [\href{http://xxx.lanl.gov/abs/hep-th/0610327}{{\tt hep-th/0610327}}].

\bibitem{Ibanez:2012zz}
L.~E. Ibanez and A.~M. Uranga, {\em {String theory and particle physics: An
  introduction to string phenomenology}}.
\newblock Cambridge University Press, 2012.

\bibitem{Grassi:2014zxa}
A.~Grassi, J.~Halverson, J.~Shaneson, and W.~Taylor, {\it {Non-Higgsable QCD
  and the Standard Model Spectrum in F-theory}},  {\em JHEP} {\bf 01} (2015)
  086, [\href{http://xxx.lanl.gov/abs/1409.8295}{{\tt 1409.8295}}].

\bibitem{Kumar:2009ac}
V.~Kumar, D.~R. Morrison, and W.~Taylor, {\it {Mapping 6D N = 1 supergravities
  to F-theory}},  {\em JHEP} {\bf 02} (2010) 099,
  [\href{http://xxx.lanl.gov/abs/0911.3393}{{\tt 0911.3393}}].

\bibitem{Seiberg:2011dr}
N.~Seiberg and W.~Taylor, {\it {Charge Lattices and Consistency of 6D
  Supergravity}},  {\em JHEP} {\bf 06} (2011) 001,
  [\href{http://xxx.lanl.gov/abs/1103.0019}{{\tt 1103.0019}}].

\bibitem{Park:2011wv}
D.~S. Park and W.~Taylor, {\it {Constraints on 6D Supergravity Theories with
  Abelian Gauge Symmetry}},  {\em JHEP} {\bf 01} (2012) 141,
  [\href{http://xxx.lanl.gov/abs/1110.5916}{{\tt 1110.5916}}].

\bibitem{Monnier:2017oqd}
S.~Monnier, G.~W. Moore, and D.~S. Park, {\it {Quantization of anomaly
  coefficients in 6D $\mathcal{N}=(1,0)$ supergravity}},
  \href{http://xxx.lanl.gov/abs/1711.04777}{{\tt 1711.04777}}.

\bibitem{Grimm:2012yq}
T.~W. Grimm and W.~Taylor, {\it {Structure in 6D and 4D N=1 supergravity
  theories from F-theory}},  {\em JHEP} {\bf 10} (2012) 105,
  [\href{http://xxx.lanl.gov/abs/1204.3092}{{\tt 1204.3092}}].

\bibitem{Taylor:2018khc}
W.~Taylor and A.~P. Turner, {\it {An infinite swampland of U(1) charge spectra
  in 6D supergravity theories}},
  \href{http://xxx.lanl.gov/abs/1803.04447}{{\tt 1803.04447}}.

\bibitem{Banks:1996nj}
T.~Banks, M.~R. Douglas, and N.~Seiberg, {\it {Probing F theory with branes}},
  {\em Phys. Lett.} {\bf B387} (1996) 278--281,
  [\href{http://xxx.lanl.gov/abs/hep-th/9605199}{{\tt hep-th/9605199}}].

\bibitem{Heckman:2010qv}
J.~J. Heckman, Y.~Tachikawa, C.~Vafa, and B.~Wecht, {\it {N = 1 SCFTs from
  Brane Monodromy}},  {\em JHEP} {\bf 11} (2010) 132,
  [\href{http://xxx.lanl.gov/abs/1009.0017}{{\tt 1009.0017}}].

\bibitem{Heckman:2011hu}
J.~J. Heckman, C.~Vafa, and B.~Wecht, {\it {The Conformal Sector of F-theory
  GUTs}},  {\em JHEP} {\bf 07} (2011) 075,
  [\href{http://xxx.lanl.gov/abs/1103.3287}{{\tt 1103.3287}}].

\bibitem{Martucci:2014ema}
L.~Martucci, {\it {Topological duality twist and brane instantons in
  F-theory}},  {\em JHEP} {\bf 06} (2014) 180,
  [\href{http://xxx.lanl.gov/abs/1403.2530}{{\tt 1403.2530}}].

\bibitem{Lawrie:2016axq}
C.~Lawrie, S.~Schafer-Nameki, and T.~Weigand, {\it {Chiral 2d theories from N =
  4 SYM with varying coupling}},  {\em JHEP} {\bf 04} (2017) 111,
  [\href{http://xxx.lanl.gov/abs/1612.05640}{{\tt 1612.05640}}].

\bibitem{Haghighat:2015ega}
B.~Haghighat, S.~Murthy, C.~Vafa, and S.~Vandoren, {\it {F-Theory, Spinning
  Black Holes and Multi-string Branches}},  {\em JHEP} {\bf 01} (2016) 009,
  [\href{http://xxx.lanl.gov/abs/1509.00455}{{\tt 1509.00455}}].

\bibitem{Assel:2016wcr}
B.~Assel and S.~SchŠfer-Nameki, {\it {Six-dimensional origin of $ \mathcal{N} =
  4$ SYM with duality defects}},  {\em JHEP} {\bf 12} (2016) 058,
  [\href{http://xxx.lanl.gov/abs/1610.03663}{{\tt 1610.03663}}].

\bibitem{Couzens:2017way}
C.~Couzens, C.~Lawrie, D.~Martelli, S.~Schafer-Nameki, and J.-M. Wong, {\it
  {F-theory and AdS$_{3}$/CFT$_{2}$}},  {\em JHEP} {\bf 08} (2017) 043,
  [\href{http://xxx.lanl.gov/abs/1705.04679}{{\tt 1705.04679}}].

\bibitem{Couzens:2017nnr}
C.~Couzens, D.~Martelli, and S.~Schafer-Nameki, {\it {F-theory and
  AdS$_3$/CFT$_2$ (2,0)}},  \href{http://xxx.lanl.gov/abs/1712.07631}{{\tt
  1712.07631}}.

\bibitem{Apruzzi:2016nfr}
F.~Apruzzi, F.~Hassler, J.~J. Heckman, and I.~V. Melnikov, {\it {From 6D SCFTs
  to Dynamic GLSMs}},  {\em Phys. Rev.} {\bf D96} (2017), no.~6 066015,
  [\href{http://xxx.lanl.gov/abs/1610.00718}{{\tt 1610.00718}}].

\bibitem{Green:1984sg}
M.~B. Green and J.~H. Schwarz, {\it {Anomaly Cancellation in Supersymmetric
  D=10 Gauge Theory and Superstring Theory}},  {\em Phys. Lett.} {\bf B149}
  (1984) 117--122.

\bibitem{Green:1984bx}
M.~B. Green, J.~H. Schwarz, and P.~C. West, {\it {Anomaly Free Chiral Theories
  in Six-Dimensions}},  {\em Nucl. Phys.} {\bf B254} (1985) 327--348.

\bibitem{Sagnotti:1992qw}
A.~Sagnotti, {\it {A Note on the Green-Schwarz mechanism in open string
  theories}},  {\em Phys. Lett.} {\bf B294} (1992) 196--203,
  [\href{http://xxx.lanl.gov/abs/hep-th/9210127}{{\tt hep-th/9210127}}].

\end{thebibliography}\endgroup
\bibliographystyle{JHEP}

\end{document}